# Detecting Multiple Change Points in Distributional Sequences Derived from Structural Health Monitoring Data: An Application to Bridge Damage Detection

Xinyi Lei[a,b,c,1] and Zhicheng Chen[a,b,c,1,*]

[a]*Key Lab of Smart Prevention and Mitigation of Civil Engineering Disasters of the Ministry of Industry and Information Technology, Harbin Institute of Technology, Harbin, 150090, China*

[b]*Key Lab of Structures Dynamic Behavior and Control of the Ministry of Education, Harbin Institute of Technology, Harbin, 150090, China*

[c]*School of Civil Engineering, Harbin Institute of Technology, Harbin, 150090, China*

*CONTACT Zhicheng Chen; e-mail address: zhichengchen@hit.edu.cn; Key Lab of Smart Prevention and Mitigation of Civil Engineering Disasters of the Ministry of Industry and Information Technology, Harbin Institute of Technology, Harbin, 150090, China.

[1]The two authors contributed equally to this work.

# Detecting Multiple Change Points in Distributional Sequences Derived from Structural Health Monitoring Data: An Application to Bridge Damage Detection

Detecting damage in critical structures using monitored data is a fundamental task of structural health monitoring, which is extremely important for maintaining structures' safety and life-cycle management. Based on statistical pattern recognition paradigm, damage detection can be conducted by assessing changes in the distribution of properly extracted damage-sensitive features (DSFs). This can be naturally formulated as a distributional change-point detection problem. A good change-point detector for damage detection should be scalable to large DSF datasets, applicable to different types of changes, and capable of controlling for false-positive indications. This study proposes a new distributional change-point detection method for damage detection to address these challenges. We embed the elements of a DSF distributional sequence into the Wasserstein space and construct a moving sum (MOSUM) multiple change-point detector based on Fréchet statistics and establish theoretical properties. Extensive simulation studies demonstrate the superiority of our proposed approach against other competitors to address the aforementioned practical requirements. We apply our method to the cable-tension measurements monitored from a long-span cable-stayed bridge for cable damage detection. We conduct a comprehensive change-point analysis for the extracted DSF data, and reveal interesting patterns from the detected changes, which provides valuable insights into cable system damage.

## 1. Introduction

Bridges are a critical component of transport infrastructure; therefore, conditions and safety are of considerable concern. During its service life, a bridge will suffer from deterioration caused by aging, environmental stressors, and natural/human hazards (Frangopol et al. 2017), leading to damage accumulation and even collapse. Bridge damage refers to any condition change that occurs to the structural system that affects its capacity (Farrar and Worden 2013). To ensure structural safety and serviceability, accurately identifying and monitoring changes in bridge conditions is highly in demand, making civil structural health monitoring (SHM) a crucial tool for this task. Civil SHM is a branch of civil engineering that is concerned with inferring structural condition changes from monitored information collected via a sensing system installed on the structure, and damage detection is one of its basic and challenging objectives.

We seek to develop a statistical tool for detecting changes in the distribution of damage-sensitive features derived from SHM data for damage detection. In real circumstances, valid change information for damage detection is hidden in raw measurements and masked by the confounding effects of operational/environmental variations. To enhance damage detectability, features that are sensitive to structural condition change but insensitive to operational/environmental variability, known as damage-sensitive features (DSFs), are required to be extracted from raw measurements; then, detecting damage can be shifted to detecting changes in the DSF data using appropriate statistical tools. Such a data-driven damage detection philosophy is known as the statistical pattern recognition paradigm (e.g., Farrar and Worden 2013). Accordingly, the damage detection process can be divided into: (1) data cleaning, (2) DSF data extraction, and (3) feature change detection for damage decision. The principle behind such a strategy is that the structural changes can manifest themselves as changes to the distribution of DSF data (Fugate et al. 2001). This fact has been revealed by many researchers through simulation, experiment, and *in situ* damage studies (e.g., see Lakshmi and Rao 2014; Li et al. 2018; Balafas et al. 2018; Ferguson et al. 2022, among others). Previous studies on damage detection have been heavily concentrated on feature extraction/selection (Sun et al. 2020); however, studies on distributional change detection for DSF data are limited. In a majority of the existing literature, DSF change detection has been cast as outlier detection (e.g., see Fugate et al.

2001; Sun et al. 2020) or two-sample testing (e.g., see Limongelli 2010; Balafas et al. 2018). Notably, such strategies can only infer the presence of damage, but are limited in locating the timing of its occurrence; thus, failing to provide critical information for analyzing the structures' deterioration process.

From a statistical perspective, it is natural to treat the above feature change detection problem as a change-point detection problem (Wan and Ni 2019; Lei et al. 2023a, 2023b), but this approach has received limited attention from the SHM community. The appeal of change-point detection is its ability to simultaneously test for and locate the changes. Identifying changes in the distribution of DSF data is crucial for damage detection; therefore, we seek to develop a multiple change-point detection tool for such an application. Practically, change points in the distribution of DSF data may signify damage events (Lei et al. 2023b); thus, they are desired to be identified for further investigating potential causes. If no evidence emerges indicating that a detected change is caused by operational/environmental disturbance, then it can be deemed to be a damage event (Fugate et al. 2001). Additionally, the timing information of damage events is essential for calibrating/updating shock-based deterioration models (SDMs) of a structure. A shock is an event that introduces sudden damage to a structure that is usually caused by extreme events such as impacts, blasts, and earthquakes (Sánchez-Silva et al. 2016). Degradation models play a pivotal role in time-variant reliability analyses, remaining life predictions, optimal maintenance scheduling for structures' life-cycle management; for example, see Kumar et al. (2015), Sánchez-Silva et al. (2016), Biondini and Frangopol (2016) and Yang and Frangopol (2019) for details. Previous studies conducting structural life-cycle analyses have primarily been conducted from theoretical perspectives, with SDMs modeled as a compound point process with known parameters (Sánchez-Silva et al. 2011; Kumar et al. 2015; Yang and Frangopol 2019). In applications, such SDMs should be calibrated/updated using shock samples (i.e., sudden damage) identified from SHM data; however, it is seldom conducted in practice partially due to a lack of shock samples as identifying the timing of damage from massive SHM data is challenging. Therefore, reliable change-point detectors for such tasks are urgently needed.

Effectively and reliably detecting change points from DSF data based on long-term monitoring data involves various challenges. One of the main challenges stems from the massive amount of data. For instance, a load cell with a sampling frequency of 10Hz (the case in our study) can produce 864,000 data points per day. SHM systems for large-scale structures deploy multiple sensors in continuous operation for decades or even longer. Consequently, a derived DSF dataset can be extremely large, posing tremendous challenges to standard change-point detection tools. Moreover, DSF data often contain multiple change points, and changes are not limited to a single type. Developing multiple change-point detection methods that can accommodate different change types is also technically challenging. Finally, a good damage detector should control the risk of false-positive indications (Farrar and Worden 2013), introducing additional challenges to related change point analyses.

The most straightforward strategy for DSF data change-point detection is directly applying standard multiple change-point detectors to raw DSF data. Available detection tools include binary segmentation (BS) (Vostrikova 1981), WBS (Fryzlewicz 2014), PELT (Killick et al. 2012) and ECP (Matteson and James 2014), to name a few. The idea of BS has been applied to damage detection by Wan and Ni (2019). However, the direct detection strategy would (a) be computationally prohibitive for massive DSF data and (b) seriously overestimate the number of change points as DSF data inevitably contain uncertain influences caused by external disturbance. An alternative strategy has employed the idea of functional change-point analysis (Lei et al. 2023a, 2023b), splitting DSF data into segments and estimating their corresponding probability density functions (PDFs) to produce a functional sequence composed of PDFs. In this setting, the distributional change detection problem can be treated conceptually as a functional change-point detection (FCPD) problem. This strategy can significantly (a) reduce the number of data objects and (b) suppress uncertain influences (Fang et al. 2023).

A growing number of statistical studies on FCPD have emerged in the past decade, with detecting single change point as the dominant topic (e.g., Berkes et al. 2009; Aston and Kirch 2012; Gromenko et al. 2017; Aue et al. 2018). However, methods targeting multiple functional change points remain scarce and have been primarily focused on mean change detection (e.g., Chiou et al. 2019; Rice and Zhang 2022; Chen et al. 2023). Moreover, commonly used methods have predominantly dealt with ordinary functional data in linear space. However, PDFs are a special type of functional data with nonlinear constraints, causing the PDF space not to form a linear space. Therefore, directly applying the ordinary FCPD methods to PDF-valued data may yield suboptimal or even erroneous results. Although statistical methods for distributional data (data in the form of PDFs or distributions) have recently attracted growing attention (Petersen et al, 2022), change-point analyses of such data have rarely emerged in the literature (Padilla et al 2019; Horváth et al. 2021; Lei et al. 2023a, 2023b). To our knowledge, Lei et al. (2023b) conducted the only study targeting a multiple change-point setting; however, it suffers from several limitations, including, but not limited to, (a) being computationally prohibitive for a large functional dataset, (b) incurring severe information loss in dimension reduction, and (c) the difficulty of controlling the type I error.

We propose a flexible and highly efficient multiple change-point detector with a linear computational cost for detecting changes in the distributional sequence of DSF data for damage detection. Distributional data do not reside in a linear space, but lie in a metric space known as Wasserstein space (e.g., Petersen et al. 2022). This motivates us to use the embedding of DSF distributions into Wasserstein space to develop a multiple change-point detector for distributional change detection. For this purpose, we extend the moving sum (MOSUM) detection procedure (Eichinger and Kirch 2018) for scalar data to Wasserstein-space-valued data. This extension is challenging, primarily because the metric space lacks the linear structure on which the standard MOSUM procedure is based. We also consider scenarios of mean and variance changes, whereas the standard MOSUM detector only considers mean change. To achieve this, we define the test statistic in our MOSUM procedure based on the Fréchet statistics in Dubey and Müller (2020) and develop our change-point estimators accordingly.

Our contributions are multifold. In terms of SHM applications, we develop a novel distributional change-point detection method for massive DSF data, providing an extremely useful tool for damage detection. Our method has appealing properties in (a) computational efficiency, (b) accounting for different types of changes, and (c) controlling the risk of false-positive damage indications. We demonstrate the superiority of our method for addressing the aforementioned challenges in SHM applications using extensive simulation studies. Employing this new tool, we perform a comprehensive distributional change-point analysis on the DSF datasets from a real bridge for stay cable damage detection. To our knowledge, this is the first attempt to systematically test DSF distributional changes for the large-scale cable system of a long-span cable-stayed bridge using long-term SHM data. The results reveal notable co-occurrence and spatial clustering in the distributional changes of the DSF data, providing valuable insights into the dependence of damage in the cable system. Methodologically, this study is among the first to develop a multiple change-point detection method for complex data in a nonlinear metric space. Although the focus herein is distributional data, our method is suitable for direct generalization to other types of data residing in a proper metric space. We also demonstrate the theoretical consistency of our change-point estimators.

## 2. Data and Problem Description

The data for our analysis are cable-tension measurements collected by a series of sensors instrumented on the stay cables of a long-span bridge crossing the Yangtze River in China. The bridge is an arch-shaped steel pylon cable-stayed bridge with a central span of 648m and two lateral spans of 257 + 63m (Li et al. 2018) (see Supplementary Figure S.1 for a front view of the bridge). A total of 168 stay cables are symmetrically anchored at the two edges of the bridge deck, forming a stay cable system for dispersing the loads from the deck to pylons. The SHM system was installed on the bridge and has been in operation since 2006. Each stay cable was instrumented with a load cell to collect data on cable tension at a sampling frequency of 10 Hz.

Stay cables are crucial load-bearing components of a cable-stayed bridge. Damage of stay cables would adversely affect the bridge's safety and reliability; thus, it is crucial to identify such damage at an early stage. However, it is difficult to detect the presence of damage in stay cables as stay cables' condition is hidden from on-site inspection by a protective sheath (Mehrabi 2006). Using the data-driven damage detection strategy described in the introduction, damage can be identified by detecting the distributional changes in properly selected DSF data. An effective DSF that has been empirically proven for cable damage detection is the ratio of cable tension between a pair of cables anchored at the two edges of the same transverse cross-section of the bridge (Li et al. 2018), which we adopt in this study. Our dataset is obtained from the cable-tension measurements acquired from 2006 to 2012. For cable-tension ratio (CTR) computation, we need use the data simultaneously measured by the two load cells installed on the same pair of cables. After excluding some severely corrupted data, we choose the data with complete records from both of the load cells for CTR data extraction. Due to sensor malfunctions, the load cells on some cables produced meaningless records and some produced no records over long periods of time. Therefore, a total of 50 cable pairs with good measurements over long time periods, relabeled as RCP1–RCP50 (Supplementary Table S.1), are selected for our investigation. We then extract the CTR data for each cable pair using a procedure similar to Lei et al. (2023a) that includes (a) data preprocessing to remove environmental/operational effects, (b) CTR computation, and (c) outlier exclusion. The details of this procedure are omitted here for brevity, and we refer readers to Lei et al. (2023a).

Given a cable pair, the extracted CTR dataset is a scalar sequence denoted as $\boldsymbol{X} = \{X_1, X_2, \cdots, X_m\}$, with $m$ being the sample size. Following Lei et al. (2023a), we separate $\boldsymbol{X}$ into daily sequences denoted as $\boldsymbol{X}_1^d = \{X_1, \cdots, X_{m_1}\}$, $\boldsymbol{X}_2^d = \{X_{m_1+1}, \cdots, X_{m_2}\}$, $\cdots$, $\boldsymbol{X}_n^d = \{X_{m_{n-1}+1}, \cdots, X_m\}$, where $\boldsymbol{X}_j^d (j = 1, \cdots, n)$ represents the sub-sequence of $\boldsymbol{X}$ on the $j$th day and $n$ is the number of days. We then estimate daily CTR distributions using a kernel density estimator, producing a distributional sequence represented as $\{\hat{f}_1, \cdots, \hat{f}_j, \cdots, \hat{f}_n\}$, with $\hat{f}_j$ representing the PDF estimated from the data in $\boldsymbol{X}_j^d$. CTR distributions associated with other cable pairs are obtained in a similar manner. A sudden change that occurs in the CTR distribution sequence of a cable pair can signify the presence of cable damage as demonstrated by Li et al. (2018).

## 3. Methodology

This section presents our proposed multiple change-point detection method for the CTR distributional sequence. We embed the distributional data into a metric space called Wasserstein space, and develop our change-point detector based on the Wasserstein metric. For ease of notation, $\{\hat{f}_1, \cdots, \hat{f}_j, \cdots, \hat{f}_n\}$ is represented as $\{f_1, \cdots, f_j, \cdots, f_n\}$ throughout the rest of this article. The random variables associated with the PDFs $f_1, \cdots, f_n$ are assumed to have finite second moments (FSMs). This assumption is mild and can be automatically satisfied in our analysis as the CTR data have finite variance.

*3.1. Wasserstein Space Embedding for the CTR Distributional Data*

Given a PDF $f_j(x), x \in \mathcal{D}$, there exists a unique probability measure ($v_j$) on $\mathcal{D}$ that possesses a density function of $f_j(x)$. Using one-to-one correspondence between probability measures and PDFs, the CTR distributional sequence $\{f_1, \cdots, f_n\}$ can be equivalently represented as a sequence of probability measures denoted by $\{v_1, \cdots, v_n\}$. Based on the FSM assumption, CTR distributions can be embedded into a metric space called the Wasserstein space. Formally, let $\mathcal{W}_2(\mathcal{D})$ be the set of probability measures defined on a compact interval ($\mathcal{D} \subset \mathbb{R}$) with FSMs, and the Wasserstein space is $\mathcal{W}_2(\mathcal{D})$ equipped with the Wasserstein metric $d_{\mathcal{W}}(v_1, v_2) = (\inf_{X \sim v_1, Y \sim v_2} E(X - Y)^2)^{1/2}$, $\forall v_1, v_2 \in \mathcal{W}_2(\mathcal{D})$ (e.g., Petersen et al. 2022). For univariate distributions (e.g., CTR distributions), $d_{\mathcal{W}}$ takes the following form:

$$d_{\mathcal{W}}^2(v_1, v_2) = \int_0^1 \left(F_1^{-1}(t) - F_2^{-1}(t)\right)^2 dt \text{ with } F_j^{-1}(t) = \inf\{y \in \mathcal{D}: F_j(y) \geq t\} \quad (1)$$

where $F_j^{-1}$ and $F_j$ are the quantile and distribution functions associated with $v_j$ ($j = 1, 2$), respectively. Obviously, CTR distributions can be treated as elements of $\mathcal{W}_2(\mathcal{D})$, or equivalently $\mathcal{W}_2(\mathcal{D})$-valued data. Hereafter, the term distributional data is used interchangeably with $\mathcal{W}_2(\mathcal{D})$-valued data. Consequently, the subsequent change-point analysis can be conducted in a natural space of distributional data. Compared with an alternative strategy that transforms the distributional data into another space for change-point analysis (Lei et al. 2023b), our proposal can provide ease of interpretation. We impose a boundedness assumption for $\mathcal{W}_2(\mathcal{D})$; namely $\sup_{v_1, v_2 \in \mathcal{W}_2(D)} d_{\mathcal{W}_2}(v_1, v_2) < \infty$.

*3.2. Problem Setting for Multiple Change-Point Analysis*

Consider a CTR distributional sequence denoted by $\Gamma = \{v_1, \cdots, v_n\}$. As noted above, the distributional data can be treated as $\mathcal{W}_2(\mathcal{D})$-valued random objects. A change in location (or scale) of these data can be characterized by the change in Fréchet mean (or Fréchet variance), which is defined as follows (Dubey and Müller 2020):

$$\mu = \underset{\omega \in \mathcal{W}_2(\mathcal{D})}{\mathrm{argmin}} E\left(d_{\mathcal{W}}^2(\nu, \omega)\right), V = \min_{\omega \in \mathcal{W}_2(\mathcal{D})} E\left(d_{\mathcal{W}}^2(\nu, \omega)\right), \nu \in \mathcal{W}_2(\mathcal{D}) \quad (2)$$

where $\mu$ and $V$ are the Fréchet mean and Fréchet variance of $\nu \in \mathcal{W}_2(\mathcal{D})$, respectively.

This study considers scenarios of location and scale changes in a CTR distributional sequence, wherein at certain unknown positions called change points the distributional sequence may undergo abrupt changes in the Fréchet mean, variance, or both. If the distributional sequence $\Gamma = \{\nu_1, \cdots, \nu_n\}$ contains $q(>0)$ change points at $k_1^*, k_2^*, \cdots, k_q^*$ such that $0 < k_1^* < k_2^* < \cdots < k_q^* < n$, then we can divide $\Gamma$ into the following $(q+1)$ distinct segments denoted by $\Gamma_1, \Gamma_2, \cdots, \Gamma_{q+1}$:

$$\underbrace{\nu_1, \cdots, \nu_{k_1^*}}_{\Gamma_1}, \underbrace{\nu_{k_1^*+1}, \cdots, \nu_{k_2^*}}_{\Gamma_2}, \nu_{k_2^*+1} \cdots, \nu_{k_q^*}, \underbrace{\nu_{k_q^*+1}, \cdots, \nu_n}_{\Gamma_{q+1}} \quad (3)$$

The data within the same segment are assumed to share common Fréchet mean and Fréchet variance, whereas either or both of the Fréchet mean and Fréchet variance between two adjacent segments are assumed to differ. Moreover, the other distributional features of the $\mathcal{W}_2(\mathcal{D})$-valued data are assumed to be unchanged. Consequently, the data between two adjacent change points are identically distributed.

Our goal is to estimate the number and locations of change points. In our testing problem, the null hypothesis of no change is formulated as follows:

$$H_0: k_1^* = \cdots = k_q^* = n$$

and the alternative is the following:

$H_A: k_1^* \leq \cdots \leq k_q^*$ with at least one $j \in \{2, \cdots, q\}$ such that $k_{j-1}^* < k_j^*$ holds.

If $H_A$ holds, this indicates that at least one change point exists in the data sequence.

We neglect the dependence between the $\mathcal{W}_2(\mathcal{D})$-valued data. These data are non-Euclidean with inherent nonlinear constraints, and it is rather challenging to develop a multiple change-point detection method for them even in the independent case. Further consideration of dependence would be much more challenging from methodological and theoretical perspectives.

*3.3. Multiple Change-Point Detection Method*

To address the distributional change-point detection problem, we develop a novel method for the $\mathcal{W}_2(\mathcal{D})$-valued data based on the moving sum (MOSUM) technique (Eichinger and Kirch 2018). The MOSUM procedure uses local data isolated by a sliding window for change-point detection. The major advantages of this approach include (a) low computational cost (the computational complexity is of order $O(n)$) and (b) ease of controlling the false-positive error rate (Cho and Kirch 2021), making this strategy appealing for addressing the challenges faced by our SHM application. Notably, the standard MOSUM procedure (Eichinger and Kirch 2018) and its variants (e.g., Messer 2022; Chen et al. 2022) are designed for Euclidean data, the limit results for constructing the change-point estimators are established on the basis of the linear operations of the Euclidean space. In contrast, the $\mathcal{W}_2(\mathcal{D})$-valued data are non-Euclidean and reside in a nonlinear space, we cannot perform linear operations on such data. Moreover, the standard MOSUM procedure only considers changes in mean; however, we also consider variance changes. Therefore, developing the MOSUM-type change-point detector in our setting is fundamentally different from previous literature.

Generally, the MOSUM procedure can be summarized as follows (Eichinger and Kirch 2018): (a) scan the data sequence using a sliding window to compute the values of a properly defined scan statistic (SS), yielding a SS sequence; (b) pick the blocks of over-threshold values from the SS sequence based on a threshold determined by the null limit distribution of the maximized SS; and (c) estimate the number and locations of change points based on the picked over-threshold blocks. The SS, combined with the limit distribution of its maximized form, plays a pivotal role in the MOSUM procedure. We next construct our SS for the $\mathcal{W}_2(\mathcal{D})$-valued data based on the Fréchet statistics defined by Dubey and Müller (2020), then derive the related limit distribution and present our multiple change-point estimators. For convenience, our method is referred to as the Fréchet-MOSUM procedure.

Before presenting our results, we introduce some additional notations. Let $\Gamma_k^G = \{v_{k-G+1}, \cdots, v_k, \cdots, v_{k+G}\}$ be the sub-sequence of the considered CTR distributional sequence $\Gamma$ that is isolated by a symmetric window with fixed bandwidth $G$ at $k \in [G, n-G]$. Let $\Gamma_{l,k}^G = \{v_{k-G+1}, \cdots, v_k\}$ and $\Gamma_{r,k}^G = \{v_{k+1}, \cdots, v_{k+G}\}$ be the two segments of $\Gamma_k^G$ delimited by $k$. Denote with $\hat{\mu}_{[k-G+1,k]} = \underset{\omega \in \mathcal{W}_2(D)}{\text{argmin}} \frac{1}{G} \sum_{i=k-G+1}^{k} d_{\mathcal{W}}^2(v_i, \omega)$ and $\hat{\mu}_{[k+1,k+G]} = \underset{\omega \in \mathcal{W}_2(D)}{\text{argmin}} \frac{1}{G} \sum_{i=k+1}^{k+G} d_{\mathcal{W}}^2(v_i, \omega)$ the sample Fréchet means of $\Gamma_{l,k}^G$ and $\Gamma_{r,k}^G$, respectively. Denote with $\hat{V}_{[k-G+1,k]} = \frac{1}{G} \sum_{i=k-G+1}^{k} d_{\mathcal{W}}^2(v_i, \hat{\mu}_{[k-G+1,k]})$ and $\hat{V}_{[k+1,k+G]} = \frac{1}{G} \sum_{i=k+1}^{k+G} d_{\mathcal{W}}^2(v_i, \hat{\mu}_{[k+1,k+G]})$ the corresponding sample Fréchet variances. Denote with $\sigma^2 = \text{var}\{d_{\mathcal{W}}^2(\mu, v)\}$ the asymptotic variance of the Fréchet variance (AVFV) (Dubey and Müller 2019), where $\mu = \underset{\omega \in \mathcal{W}_2(D)}{\text{argmin}} E\left(d_{\mathcal{W}}^2(v, \omega)\right)$ is the Fréchet mean of $v \in \mathcal{W}_2(D)$. Let $\hat{\sigma}_{k,n}^2 = h(v_{k-G+1}, \cdots, v_{k+G})$ be a local estimator of $\sigma^2$ using the data in $\Gamma_k^G$ (the specific form of $\hat{\sigma}_{k,n}^2$ will be presented later in Section 4).

We construct the following test statistic for our testing problem:

$$T_n(G) = \max_{G \leq k \leq n-G} T_n^G(k), \text{ with} \tag{4}$$

$$T_n^G(k) = \left(2\hat{\sigma}_{k,n}^2/G\right)^{-1/2} \left|\hat{V}_{[k+1,k+G]} - \hat{V}_{[k-G+1,k]}\right|$$
$$+ \left(2\hat{\sigma}_{k,n}^2/G\right)^{-1/2} \left|\hat{V}_{[k+1,k+G]}^C - \hat{V}_{[k+1,k+G]} + \hat{V}_{[k-G+1,k]}^C - \hat{V}_{[k-G+1,k]}\right|, \ k \in [G, n-G] \tag{5}$$

where $T_n^G(k)$ is the SS, and $\hat{V}_{[k+1,k+G]}^C$ and $\hat{V}_{[k-G+1,k]}^C$ are the "contaminated" versions of $\hat{V}_{[k+1,k+G]}$ and $\hat{V}_{[k-G+1,k]}$, respectively. $\hat{V}_{[k+1,k+G]}^C = \frac{1}{G} \sum_{i=k+1}^{k+G} d_{\mathcal{W}}^2(v_i, \hat{\mu}_{[k-G+1,k]})$ is computed by replacing $\hat{\mu}_{[k+1,k+G]}$ in $\hat{V}_{[k+1,k+G]}$ with $\hat{\mu}_{[k-G+1,k]}$. $\hat{V}_{[k-G+1,k]}^C$ is computed in a similar manner. We construct $T_n^G(k)$ based on the Wasserstein metric $d_{\mathcal{W}}$, avoiding the application of linear operations on $\mathcal{W}_2(D)$. $T_n^G(k)$ can be considered as a plug-in MOSUM-type statistic (see Supplementary Section S.2.1 for more details).

Similar to the test statistic in Dubey and Müller (2020) for single change-point detection, the SS $T_n^G$ given in equation (5) also tends to be small under $H_0$, as the sample Fréchet means (or variances) tend to exhibit small differences between $\Gamma_{l,k}^G$ and $\Gamma_{r,k}^G$ under $H_0$. Under $H_A$, if a true change point ($k^*$) is located within the interval $[k-G+1, k+G]$, then $T_n^G$ will tend to be large around $k = k^*$ and peak at $k^*$. The first term of $T_n^G(k)$, namely $(2\hat{\sigma}_{k,n}^2/G)^{-1/2}|\hat{V}_{[k+1,k+G]} - \hat{V}_{[k-G+1,k]}|$, is primarily for capturing changes in Fréchet variance, while the remaining term of $T_n^G(k)$ is primarily for capturing changes in Fréchet mean.

To establish the null limit distribution of the test statistic $T_n(G)$, we impose the following assumptions on $G = G(n)$ (bandwidth) and $\hat{\sigma}_{k,n}^2$ (estimator of $\sigma^2$):

(A1) For some $\Delta > 0$, $G = G(n)$ fulfills $n^{\frac{2}{2+\Delta}} \log n / G \to 0$ and $n/G \to \infty$, as $n \to \infty$.

(A2) $\hat{\sigma}_{k,n}^2 (G \leq k \leq n-G)$ fulfill $\max_{G \leq k \leq n-G} |\hat{\sigma}_{k,n}^2 - \sigma^2| = o_p(1/\sqrt{\log(n/G)})$ under $H_0$.

(A3) The quantile functions $F_i^{-1}$ associated with $v_i \in \Gamma$ fulfill condition (SC1), which is provided in Supplementary Section S.2.2.

Condition (A1) has been commonly used in previous MOSUM literature (e.g., Eichinger and Kirch 2018; Kirch and Reckruehm 2022), meaning that $G$ diverges to infinity at a rate faster than $n^{\frac{2}{2+\Delta}} \log n$ but slower than $n$. Condition (A2) imposes a constraint on the uniform convergence rate of the local estimators $\hat{\sigma}_{k,n}^2 (G \leq k \leq n-G)$. Later in Section 4, we will present the specific form of $\hat{\sigma}_{k,n}^2$ and verify condition (A2).

We next determine the null limit distribution of $T_n(G)$, and the result is given as the following theorem, with the proof provided in Supplementary Section S.3.1:

**Theorem 1** Let $\Gamma = \{v_1, \cdots, v_n\}$ be an independent $\mathcal{W}_2(D)$-valued sequence, and suppose conditions (A1)–(A3) hold; then, under $H_0$ the test statistic $T_n(G)$ obeys:

$$\lim_{n\to\infty} P(\gamma_1(n/G)T_n(G) - \gamma_2(n/G) \leq x) = \exp(-2\exp(-x)) \quad (6)$$

with $\gamma_1(x) = \sqrt{2\log x}$ and $\gamma_2(x) = 2\log x + 0.5\log\log x + \log(1.5) - 0.5\log\pi$.

With the above limit results, we can implement an asymptotic level $\alpha$ test to determine whether to reject $H_0$, and the asymptotic rejection region is

$$R_{n,\alpha} = \left\{T_n(G) > D_n(G;\alpha) = \left(-\log\log(1/\sqrt{1-\alpha}) + \gamma_2(n/G)\right)/\gamma_1(n/G)\right\} \quad (7)$$

where $D_n(G;\alpha)$ is the critical value for the level $\alpha$ test.

With the calculated SS sequence $T_n^G = \{T_n^G(k): G \leq k \leq n - G\}$ and the critical value $D_n(G;\alpha)$, change points can be estimated based on an over-threshold block picking strategy, similar to that in Eichinger and Kirch (2018). For an intuitive interpretation of this strategy, we illustrate a distributional sequence with three change points in Supplementary Figure S.2 along with the calculated SS sequence $T_n^G$. The result demonstrates that $T_n^G$ exceeds the threshold (i.e., $D_n(G;\alpha)$) within an interval around each of the change points and peaks near each change point, verifying that change points can be manifested as local maxima of the SS sequence. To pick intervals containing potential change points, we build indexing blocks on which $T_n^G$ takes over-threshold values. Specifically, the $j$th indexing block, which is denoted by $IB_j = \{k \in \mathbb{N}: s_j \leq k \leq e_j\}$, is constructed as follows:

$$\begin{cases} T_n^G(k) \geq D_n(G;\alpha) \text{ for } s_j \leq k \leq e_j \\ T_n^G(k) < D_n(G;\alpha) \text{ for } k = s_j - 1, e_j + 1 \end{cases} \text{ with } s_j, e_j \in \mathbb{N} \text{ and } e_j - s_j \geq \varepsilon G \quad (8)$$

where $\varepsilon \in (0, 0.5)$ is a tuning parameter to alleviate the overestimation problem that is referred to as the AOP parameter hereafter (see Supplementary Section S.2.3.2 for more details). Suppose that we can build a total of $J(T_n^G, \alpha) \in \mathbb{N}$ distinct indexing blocks, then the number and locations of the change points can be respectively estimated as follows:

$$\hat{q}_n = J(T_n^G, \alpha) \text{ and } \hat{k}_j^* = \underset{k \in IB_j}{\mathrm{argmax}}\, T_n^G(k), \; j = 1, \cdots, J(T_n^G, \alpha) \tag{9}$$

Our Fréchet-MOSUM method provides a theoretical critical value to control the family-wise false-positive rate at a given significance level (see (7)). This is one of the major advantages of this technique compared with some other BS-based approaches (e.g., WBS, ECP) which are difficult to obtain the asymptotic laws (of the test statistics) used for error rate control (Cho and Kirch 2021). Significance quantification for change points is extremely important for SHM applications. For instance, it can aid in automatically selecting the change points for subsequent analysis according to whether they are statistically significant. Moreover, a change detector for damage detection is desired to be capable of controlling false-positive indications at a desired level.

### *3.4. Theoretical Properties*

We next provide theoretical justifications for our method by proving the consistency of the estimators for the number and locations of change points. Establishing such consistency results is difficult. Since related theoretical results of MOSUM-type estimators in previous literature have been established based on the linear structure of Euclidean space, it is challenging to extend them to the $\mathcal{W}_2(D)$-valued data.

We start with notations and definitions. Denote with $d_0(n) := \min_{0 \leq j \leq q} |k_{j+1}^* - k_j^*|$ the shortest spacing between a pair of neighboring change points (including two boundary points $k_0^* = 0$ and $k_{q+1}^* = n$). We also define the following set:

$$B_{G,q} := \{k \in \{1, 2, \cdots, n\}: \exists k_j^* \in \{k_1^*, \cdots, k_q^*\} \text{ such that } |k - k_j^*| \leq \delta\} \tag{10}$$

where $\delta = (1 - \varepsilon)G$, with $\varepsilon$ the same as that in equation (8). We further define

$$\begin{aligned}T^G(k) := & \left(2\sigma_k^2/G\right)^{-1/2} \left|V_{[k+1,k+G]} - V_{[k-G+1,k]}\right| \\ & + \left(2\sigma_k^2/G\right)^{-1/2} \left|V_{[k+1,k+G]}^C - V_{[k+1,k+G]} + V_{[k-G+1,k]}^C - V_{[k-G+1,k]}\right|, \; k \in [G, n-G]\end{aligned} \tag{11}$$

The expressions for $V_{[k+1,k+G]}, V_{[k-G+1,k]}, V_{[k+1,k+G]}^C, V_{[k-G+1,k]}^C$, and $\sigma_k^2$ are provided in Supplementary Section S.2.4. Once the change-point model in Section 3.2 and the bandwidth $G$ are given, the $T^G$ process is deterministic, similar to the MOSUM signal presented in Figure 1 of Kirch and Reckruehm (2022). For convenience, $T^G$ is also referred to as the MOSUM signal. More precisely, under $H_A$, $T^G$ is the noise-free version of the SS process $T_n^G$ given in equation (5).

To establish the consistency results, we need the following conditions:

(C1) The shortest spacing $d_0(n)$ fulfills $\limsup\limits_{n\to\infty} d_0(n)/G = C > 2$ ($C$ is a constant).

(C2) The MOSUM signal $T^G$ fulfills $\frac{1}{\sqrt{\log(n/G)}} \min\limits_{k \in B_{G,q}} T^G(k) \to \infty$, as $n \to \infty$.

(C3) $\alpha = \alpha(n)$ (significance level) fulfills $\alpha \to 0$ and $\frac{\log\log(1/\sqrt{1-\alpha})}{\sqrt{\log(n/G)}} = O(1)$ as $n \to \infty$.

Detailed explanations for these conditions are provided in Supplementary Section S.2.5.

The following theorem states the consistency of change-point estimation:

**Theorem 2** Let $\Gamma = \{v_1, \cdots, v_n\}$ as that in Theorem 1 and follow the model defined in equation (3). Suppose that the conditions of Theorem 1 and (C1)–(C3) are satisfied. Then, under $H_A$, the change-point estimators obtained by equation (9) obey

$$\lim_{n\to\infty} P\left(\hat{q}_n = q, \max_{1\leq j \leq q}|\hat{k}_j^* - k_j^*| < G\right) = 1 \qquad (12)$$

The detailed proof of Theorem 2 is presented in Supplementary Section S.3.2.

Theorem 2 indicates that the Fréchet-MOSUM procedure can consistently estimate the number of change points, and with a probability approaching one, it can also obtain a change-point location estimate within $\left(k_j^* - G, k_j^* + G\right)$ for each $j \in \{1, \cdots, q\}$. It is noteworthy that in the change-point location estimation problem (including but not limited to using the MOSUM procedure), consistency (in the usual sense) cannot be achieved (Cho and Kirch 2022), and the convergence rate of change-point localization is $O_p(1)$ (Verzelen et al. 2020) in best-case scenarios. Equation (12) indicates that the convergence rate of our change-point location estimators is $O_p(G)$. If the change points are represented using relative locations denoted by $\{\theta_1^*, \cdots, \theta_q^*\} = \{k_1^*/n, \cdots, k_q^*/n\}$, then they can be consistently estimated as stated below.

**Corollary** Let $\theta_j^* = k_j^*/n, j = 1, \cdots, q$ and $\hat{\theta}_j^* = \hat{k}_j^*/n, j = 1, \cdots, \hat{q}_n$. Then, under the conditions of Theorem 2, we have

$$\max_{1 \leq j \leq \min(q, \hat{q}_n)} |\hat{\theta}_j^* - \theta_j^*| = o_P(1) \tag{13}$$

The above consistency results establish the asymptotic validity for our proposal.

**4. Implementation Details and Practical Considerations**

Computing the SS $T_n^G(k)$ requires calculating $\hat{V}_{[k+1,k+G]}, \hat{V}_{[k-G+1,k]}, \hat{V}_{[k+1,k+G]}^C$, and $\hat{V}_{[k-G+1,k]}^C$. We next provide the details for computing $\hat{V}_{[k+1,k+G]}$, and the remaining quantities can be computed analogously. Denoting the quantile functions of $\hat{\mu}_{[k+1,k+G]}$ and $\nu_i \in \{\nu_{k+1}, \cdots, \nu_{k+G}\}$ by $F_{\hat{\mu}_{[k+1,k+G]}}^{-1}$ and $F_i^{-1}$, respectively, then $F_{\hat{\mu}_{[k+1,k+G]}}^{-1} = \frac{1}{G}\sum_{i=k+1}^{k+G} F_i^{-1}$ holds (Lin et al. 2023). The space of quantile functions is not closed under general linear combinations but closed under convex combinations; thus, being a quantile function, $F_{\hat{\mu}_{[k+1,k+G]}}^{-1}$ can be guaranteed. Then, using equation (1), we obtain

$$\hat{V}_{[k+1,k+G]} = \frac{1}{G}\sum_{i=k+1}^{k+G} \int_0^1 \left(F_i^{-1}(t) - F_{\hat{\mu}_{[k+1,k+G]}}^{-1}(t)\right)^2 dt \tag{14}$$

The quantities $F_{\hat{\mu}_{[k+1,k+G]}}^{-1}, G \leq k \leq n - G$, can be efficiently calculated in a recursive manner as detailed in Supplementary Section S.2.6.

Computing the SS $T_n^G(k)$ also requires calculating the local AVFV estimators $\hat{\sigma}_{k,n}^2$. We propose that the estimators $\hat{\sigma}_{k,n}^2$ take the following form:

$$\hat{\sigma}_{k,n}^2 = (\hat{\sigma}_{k,n,l}^2 + \hat{\sigma}_{k,n,r}^2)/2, G \leq k \leq n - G \tag{15}$$

where $\hat{\sigma}_{k,n,l}^2 = \frac{1}{G}\sum_{i=k-G+1}^{k} d_W^4(\nu_i, \hat{\mu}_{[k-G+1,k]}) - \left(\frac{1}{G}\sum_{i=k-G+1}^{k} d_W^2(\nu_i, \hat{\mu}_{[k-G+1,k]})\right)^2$ and $\hat{\sigma}_{k,n,r}^2 = \frac{1}{G}\sum_{i=k+1}^{k+G} d_W^4(\nu_i, \hat{\mu}_{[k+1,k+G]}) - \left(\frac{1}{G}\sum_{i=k+1}^{k+G} d_W^2(\nu_i, \hat{\mu}_{[k+1,k+G]})\right)^2$.

**Proposition 1.** Under Assumption (A1), $\{\hat{\sigma}^2_{k,n}: G \leq k \leq n - G\}$ given in equation (15) fulfills $\max_{G \leq k \leq n-G} |\hat{\sigma}^2_{k,n} - \sigma^2| = o_p(1/\sqrt{\log(n/G)})$ under $H_0$.

The detailed proof of Proposition 1 is provided in Supplementary Section S.3.3.

The boundary correction for the SS process $T_n^G$ (i.e., padding the truncating values for $T_n^G$ at $1 \leq k < G$ and $(n - G) < k \leq n$) is also a matter of concern. We address this concern using a boundary extension statistic, and the computational details are presented in Supplementary Section S.2.7.

The main tuning parameter of the Fréchet-MOSUM procedure is the bandwidth $G$. Practically, $G$ can be selected based on the change-point trajectory (CPT) plot (see Supplementary Section S.2.8 for details). In Supplementary Section S.7, we also develop a multiscale detection procedure using multiple bandwidths, which is referred to as multiscale Fréchet-MOSUM. The multiscale version increases the adaptability to change-point detection, but at the cost of additional computational resources.

## 5. Simulation Results

Before analyzing real SHM data, we evaluate the finite sample performance of our proposal against alternative approaches using simulations. The competitors considered include the FPCA-ECP in Lei et al. (2023b), DSBE in Chiou et al. (2019), FMCI in Harris et al. (2022), functional BS in Rice and Zhang (2022) (hereafter FBS), and GS in Chen et al. (2023). To the best of our knowledge, FPCA-ECP is the only existing multiple change-point detector that was specifically designed for distributional data. The remaining four competitors (i.e., DSBE, FMCI, FBS, and GS) are all state-of-the-art multiple change-point detection methods designed for ordinary functional data residing in the $L^2$ space. To make these alternative methods applicable to distributional change-point detection, we employ the log quantile density (LQD) transformation (Petersen and Müller 2016) to transform the distributional data into the $L^2$ space, then apply the DSBE, FMCI, FBS, and GS detectors to the LQD-transformed data. The FPCA-ECP procedure also requires transforming the distributional data into a suitable vector space. Following Lei et al. (2023b), we consider the LQD transformation and Wasserstein-tangent-space (WassTS) mapping, and refer to the corresponding change-point detectors as FPCA-ECP(LQD) and FPCA-ECP(WassTS), respectively.

We employ two different data-generating processes (DGPs), which are referred to as DGP1 and DGP2 hereafter, to simulate the distributional sequences of DSF data. Due to space constraints, full details for data generation using the two DGPs are deferred to Supplementary Section S.4.1. Each of the synthetic DSF distributional sequences is composed of $n = 800$ PDFs and contains three change points located at $i = 200, 400,$ and $600$. To mimic the practical circumstances in which changes in the distribution of real DSF data are usually not limited to a single type, our synthetic DSF distributional sequences are allowed to undergo changes in both mean and variance structures. Representative synthetic data are visualized in Supplementary Figure S.6.

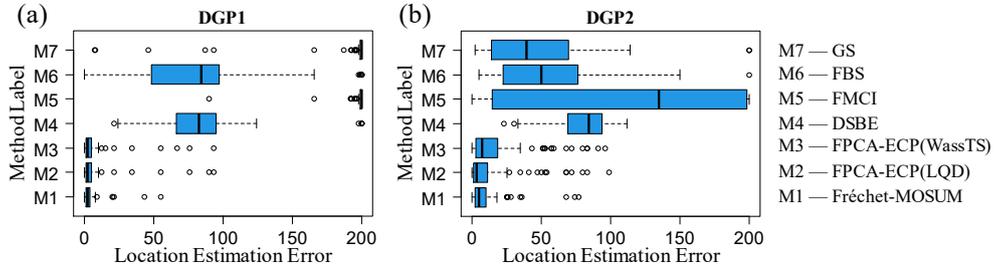

**Figure 1.** Boxplots of the estimation errors in change-point locations (measured by Hausdorff distance) for the seven considered change-point detectors. (a) The results of DGP1 and (b) the results of DGP2.

The implementation settings of the considered change-point detectors are detailed in Supplementary Section S.4.2. Based on 100 replications, the location estimation errors (measured by the Hausdorff distance (Supplementary Section S.4.3)) are compared in Figure 1, and the corresponding estimated change-point numbers are compared in Figure 2. We see that, except for the FPCA-ECP method which exhibits competitive performance with our method, the remaining competitors significantly underperform, with the FMCI being the worst. It is noteworthy that the DSBE, FBS, and GS are designed for mean change detection. Impacted by variance changes, these methods tend to overestimate the number of change points as shown in Figure 2, leading to poor performance. In contrast, the large errors of FMCI in change-point localization are attributable to underestimating the number of change points.

Although the FPCA-ECP method exhibits comparative performance with our proposal in the above simulations, several severe drawbacks for real SHM applications are evident. One major drawback is that the method is computationally intractable for large datasets due to the intensive time-consuming permutation tests of the ECP detector (Matteson and James 2014). As noted by Biau et al. (2016), the ECP detector is computationally prohibitive when the sample size exceeds a few thousand. In contrast, our Fréchet-MOSUM method is highly efficient, with a computational complexity of $O(n)$. We conduct an additional simulation study in Supplementary Section S.5.1 to compare the computational cost of our method against FPCA-ECP, revealing that our method is far more computationally efficient.

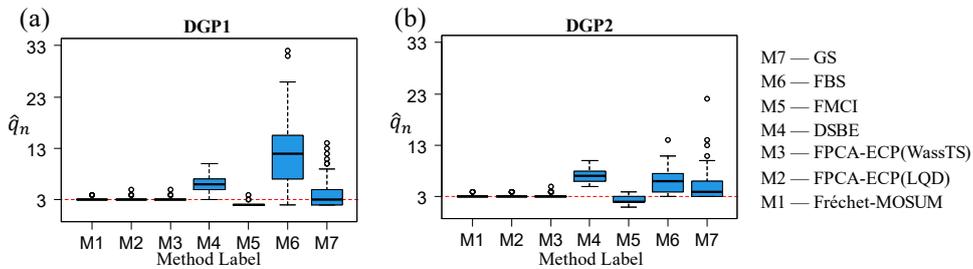

**Figure 2.** Boxplots of $\hat{q}_n$ (the estimated number of change points) for the seven considered change-point detectors. (a) The results of DGP1 and (b) the results of DGP2. The true number of change points is indicated by the dashed horizontal line.

Another major drawback of the FPCA-ECP method is related to the distributional transformation involved in the detection procedure. In contrast to our proposal, the FPCA-ECP method does not conduct change-point detection in the space where the distributional data reside, but in another space where the transformed data reside instead. This transformation approach not only sacrifices interpretability but could also produce misleading results. Figure 1 shows that the FPCA-ECP detector using the LQD transformation (FPCA-ECP(LQD)) appears to be the most competitive detector in relation to our method. However, as demonstrated in another simulation study (Supplementary Section S.5.2), LQD-transformed data are highly sensitive to changes in the slope of quantile functions, which would incur a significant shape-related magnification effect on distributional change-point detection. This magnification effect would introduce serious potential pitfalls to our structural damage detection applications such as (a) increasing the risk of producing questionable results of no practical relevance to the distributional changes of DSF data and (b) making the significance test fail to offer meaningful guidance on automatically selecting change points for potential downstream analyses (see Supplementary Section S.5.2 for an in-depth discussion).

In addition, the dimension reduction involved in the FPCA-ECP method could incur severe information loss in distributional change-point detection; see Supplementary Section S.5.3 for a demonstration of this effect). Furthermore, the FPCA-ECP method provides no analytical solution for false-positive rate control.

In the above mentioned simulations, our proposal performs sufficiently well over a broad range of data-generating mechanisms. Based on the above, we conclude that our proposal is a far more preferable technique for detecting the distributional changes of DSF data, particularly when the dataset is large. These simulation studies provide reassurance of the validity of our proposed approach for subsequent real data analysis.

**6. CTR Distributional Change-Point Analysis for Bridge Damage Detection**

We apply the Fréchet-MOSUM method to our CTR distributional sequences (Section 2) to detect potential change points for bridges' cable damage detection. As noted in the introduction, employing a statistical pattern recognition paradigm, data-driven damage detection can be cast into a framework of detecting changes in the distribution of properly extracted DSFs. In our motivating case study, the CTR data are the selected DSF data for cable damage detection.

## 6.1. Detection Results

In the Fréchet-MOSUM approach, the bandwidth is set as $G = 40$, which is chosen using the CPT plot strategy based on five randomly selected CTR distributional sequences (see Supplementary Section S.6.1 for details). The AOP parameter is set as $\varepsilon = 15/G$ following the guideline provided in Supplementary Section S.2.8. The significance level $\alpha$ is set to 0.05. Figure 3 illustrates the detection results for two representative CTR distributional sequences associated with cable pairs RCP36 and RCP41, respectively. Due to space constraints, the results of other representative CTR distributional sequences are deferred to Supplementary Section S.6.3.

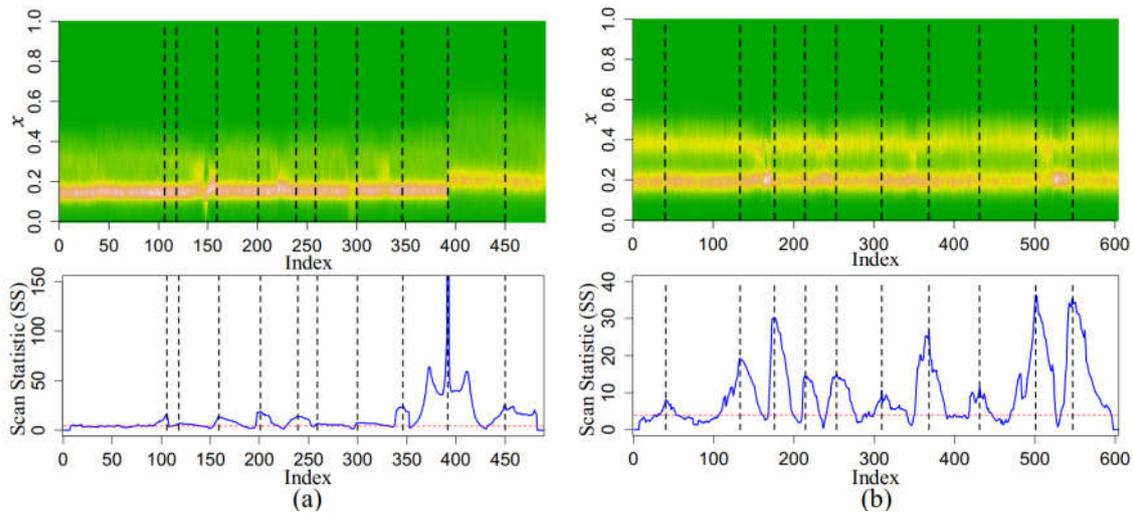

**Figure 3.** Change-point detection results obtained using the Fréchet-MOSUM method for two representative CTR distributional sequences associated with cable pairs (a) RCP36 and (b) RCP41, respectively. The first row corresponds to the heatmap of the PDF-valued sequence and the second row corresponds to the scan statistic (SS) sequence computed using the Fréchet-MOSUM procedure. Detected change points are indicated by vertical dashed lines. The horizontal dashed line in the plot of the SS sequence indicates the computed critical value $D_n(G; \alpha)$ (equation (7)).

Figure 3 reveals that most of the detected change points (indicated by vertical dashed lines) are not visually apparent in the heatmaps; hence, statistical tools that visually demonstrate detected changes are particularly helpful for examining whether the detected change points are valid. We employ the archetypal analysis (AA) technique to achieve this goal. AA builds a few archetypes to approximately represent the individual samples as convex combinations of the archetypes (Cutler and Breiman 1994). These archetypes are convex combinations of individual samples that reflect extremal or pure patterns of the data. Coefficients of individuals in archetypal representations are highly informative for exploring underlying data changes.

We use the R package *archetypes* (Eugster and Leisch 2009) to perform AA on our distributional data (see Supplementary Section S.6.2 for more details). Figure 4(a) presents the PDFs (light curves) of the CTR distributional sequence from cable pair RCP41 with the two extracted archetypes (bold curves) superimposed, revealing that the PDFs are bimodal and appear to come from two classes. The first PDF class exhibits relatively low main peaks, while those in the second class have markedly higher main peaks. The two extracted archetypes, denoted by $f_1^A$ and $f_2^A$, reflect the leading features of these two types of PDFs, respectively. The individual PDFs in the CTR distributional sequence $\{f_i\}_{i=1}^n$ can be approximated by $f_i \approx \alpha_{i1} f_1^A + (1 - \alpha_{i1}) f_2^A, i = 1, \cdots, n$, where $\alpha_{i1} \geq 0$ are called archetype mixture coefficients (AMCs). Supplementary Figure S.16 illustrates the archetypal approximations for several representative PDFs, and the results agree well with the original PDFs.

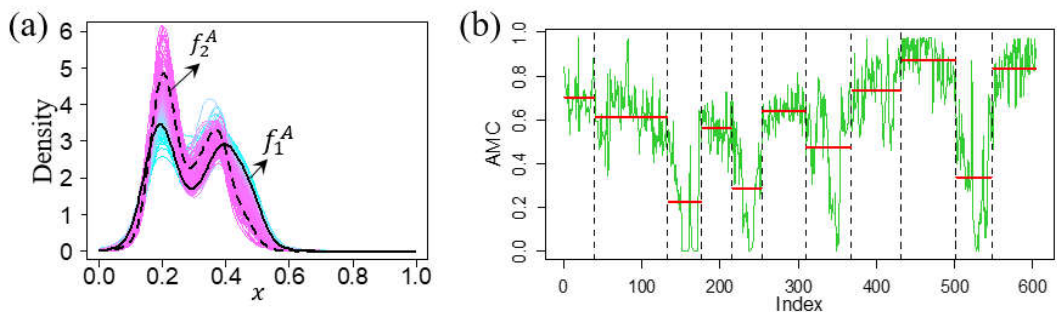

**Figure 4.** Archetypal analysis results of distributional data from cable pair RCP41. (a) The density functions (light curves) of the CTR distributional sequence along with the two estimated archetypes (bold lines), and (b) the archetype mixture coefficients (AMCs). Vertical dashed lines in the right panel indicate the detected change points, while the bold horizontal lines indicate the empirical means within segments.

The plot of the AMC sequence can be used to visually observe how individual PDFs in a distributional sequence change over time. Figure 4(b) presents the AMC sequence $\{\alpha_{i1}\}_{i=1}^n$ for the CTR distributional sequence from cable pair RCP41. The previously detected change points (i.e., those shown in Figure 3(b)) are superimposed as vertical dashed lines in Figure 4(b). We see that the AMC sequence undergoes obvious changes in mean at each of the detected change points, which confirms that all detected change points are valid. Moreover, the result in Figure 4(b) also indicates that the detected change points can provide satisfactory segmentation for the data. Given a PDF $f_i$, if its AMC $\alpha_{i1} \geq 0.5$, then the pattern of the PDF is inclined toward archetype 1, otherwise the pattern inclined toward archetype 2. The plot in Figure 4(b) reveals that the dominant pattern of the PDFs in the CTR distributional sequence transitions from archetype 1 to archetype 2 at $\hat{k}_2^*$, $\hat{k}_4^*$, $\hat{k}_6^*$, and $\hat{k}_9^*$, returning back to archetype 1 at $\hat{k}_3^*$, $\hat{k}_5^*$, $\hat{k}_7^*$, and $\hat{k}_{10}^*$, respectively. Compared with the heatmap (Figure 3), the AMC plot (Figure 4(b)) is more informative for visually confirming detected changes. In the heatmap, many of the detected changes are not revealed, and some even seem to be counterintuitive. For instance, in Figure 3(b), it seems that $\hat{k}_9^*$ should be located near $i = 520$ rather than $i = 501$; however, the corresponding AMC plot in Figure 4(b) shows that the dominant pattern of the PDFs begins to shift to archetype 2 at $i = 501$ (i.e., $\hat{k}_9^*$) and reaches an extreme state near $i = 520$. This result confirms that $\hat{k}_9^* = 501$, which is estimated using the Fréchet-MOSUM procedure, is reasonable. The detected change points shown in Figure 3(a) can be confirmed via a similar AA-based strategy. In Supplementary Section S.6.3, we provide the change-point detection results for more CTR distributional sequences, and our Fréchet-MOSUM procedure consistently provides satisfactory performance. These results verify the efficacy of our method for real SHM application, although we made some assumptions to simplify the detection problem. At this point, one may wonder why we do not use the AMC sequence to detect the change points for a CTR distributional sequence. We do not recommend such an indirect strategy because the relevant AA procedure is computationally intensive and the result is also sensitive to outliers. Therefore, the AA technique is only used as a diagnostic tool to visually examine the detected changes.

As noted previously, the CTR distributional sequence may undergo changes in mean, variance, or both. For instance, the marked shift of the heatmap of the CTR distributional sequence shown in Figure 3(a) (along the y-axis) near $i = 400$ is a clear reflection of mean change. Moreover, mean changes in a distributional sequence can also be manifested as the mean changes in AMC sequences as shown in Figure 4(b); however, variance changes are not visually apparent from the results in Figure 3 or Figure 4. To verify that the distributional sequence of real CTR data may contain variance changes, we conduct further analysis of the CTR distributional sequence shown in Figure 3(b).

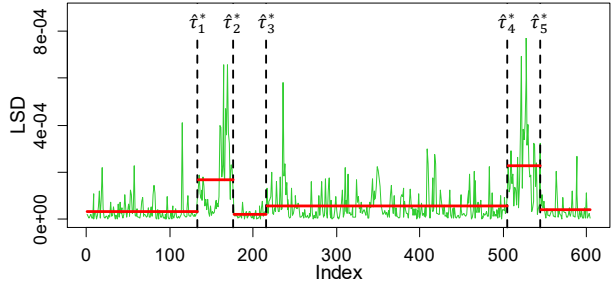

**Figure 5.** The local squared deviation (LSD) sequence of cable pair RCP41. The vertical dashed lines indicate the detected change points, while the bold horizontal lines indicate the empirical means within segments.

Specifically, let $\hat{k}_1^*, \cdots, \hat{k}_{\hat{q}_n}^*$ be the $\hat{q}_n$ detected change points, such that $0 = \hat{k}_0^* < \hat{k}_1^* < \cdots < \hat{k}_{\hat{q}_n}^* < \hat{k}_{\hat{q}_n + 1}^* = n$ that split the distributional sequence into $\hat{q}_n + 1$ segments. Empirical Fréchet means (EFM) within segments are denoted as $\hat{\mu}_{(\hat{k}_j^*, \hat{k}_{j+1}^*]} = \underset{\omega \in \mathcal{W}_2(D)}{\operatorname{argmin}} \frac{1}{\hat{k}_{j+1}^* - \hat{k}_j^*} \sum_{i=\hat{k}_j^*+1}^{\hat{k}_{j+1}^*} d_{\mathcal{W}}^2(\nu_i, \omega), j = 0, \cdots, \hat{q}_n$. We then compute the squared deviations of the distributional data around EFMs for each segment, and the results are denoted as $Y_i = d_{\mathcal{W}}^2(\nu_i, \hat{\mu}_i)$ with $\hat{\mu}_i = \sum_{j=0}^{\hat{q}_n} \hat{\mu}_{(\hat{k}_j^*, \hat{k}_{j+1}^*]} I\{\hat{k}_j^* < i \leq \hat{k}_{j+1}^*\}$ for $i = 1, \cdots, n$. We call $\mathcal{Y} = \{Y_1, \cdots, Y_n\}$ local squared deviation (LSD) sequence, since relevant EFMs are computed using local segments. If mean changes are adequately detected, the piecewise constant $\mathcal{W}_2(\mathcal{D})$-valued process $\{\hat{\mu}_i: i = 1, \cdots, n\}$ approximates the Fréchet mean process of the distributional sequence. In this situation, LSD values reveal the dispersion of the distributional data, which is closely related to the Fréchet variance defined in equation (2); therefore, the variance changes in the distributional sequence can be manifested as mean changes in the LSD sequence. To this end, we use the standard MOSUM procedure implemented in the R package *mosum* (Meier et al. 2021) to test for mean changes in the LSD sequence, and the bandwidth is also set to 40. A total of five change points ($\hat{\tau}_1^* = 133$, $\hat{\tau}_2^* = 176$, $\hat{\tau}_3^* = 216$, $\hat{\tau}_4^* = 505$, and $\hat{\tau}_5^* = 545$) are detected from the LSD sequence associated with cable pair RCP41, which are visualized in Figure 5. These five change points coincide with five of the ten change points detected earlier using the Fréchet-MOSUM procedure almost exactly ($\hat{k}_2^* = 133$, $\hat{k}_3^* = 176$, $\hat{k}_4^* = 214$, $\hat{k}_9^* = 501$, and $\hat{k}_{10}^* = 547$). The LSD sequence (Figure 5) shows marked changes in mean at the five detected change points, indicating that the CTR distributional sequence undergoes significant changes in variance. In addition, we further examine the variance changes for other CTR distributional sequences using a similar technique, revealing similar variance changes in the investigated data (see Supplementary Section S.6.4). Combined with the AA results, we conclude that a CTR distributional sequence may undergo changes in mean, variance, or both. As demonstrated in the simulation results in Section 5, the performance of a change-point detector that can only accommodate mean change may be severely impacted by variance change; therefore, change-point detectors that are used for real CTR data applications must be able to flexibly accommodate both mean and variance changes, and our Fréchet-MOSUM method is adapted well to such scenarios.

      The change points detected from the LSD sequence can also be used to refine the change-point set (Supplementary Section S.6.5 summarizes the procedure), which helps to recover some change points missed by the Fréchet-MOSUM detector. We call

this change-point compensation strategy LSD refinement. The results in Supplementary Figure S.19 reflect the enhanced performance from using LSD refinement.

We next present the results for the 50 CTR distributional sequences. The Fréchet-MOSUM algorithm (including LSD refinement processing), which is coded in R, only requires 25 seconds to accomplish detection for 50 datasets using a PC with an Intel(R) Core(TM) i-5 processor. Since we discarded missing data (Section 2), the PDF indices between different CTR distributional sequences are not matched one-to-one. Consequently, the change points detected from different sequences are not aligned in the time domain. For a better comparison, we conduct registration processing (detailed in Supplementary Section S.6.6) on the indices of the detected change points to align them in the time domain. The index of a detected change point before (after) registration is called an original (registered) time index (see Supplementary Table S.5). Change points detected from the 50 CTR distributional sequences are visualized and compared in Figure 6 using the registered indices.

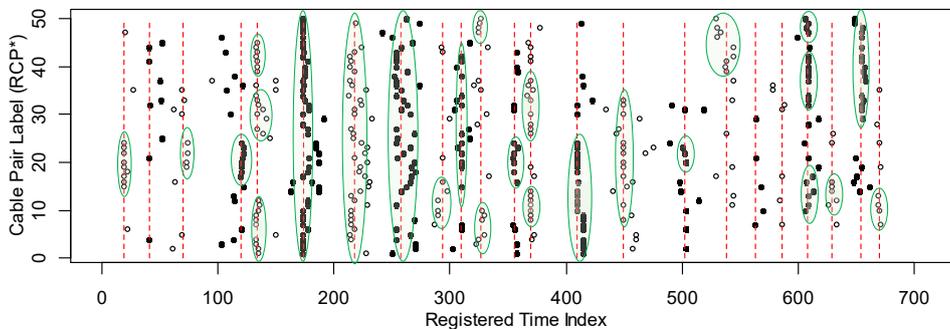

**Figure 6.** Visual representation of change points (empty/solid circles) detected from the 50 CTR distributional sequences. The change points detected from the same CTR distributional sequence are plotted along the horizontal axis at the same vertical position. The detected change points are divided into 23 different groups based on registered time indices. The data from two adjacent groups are distinguished using different symbols (i.e., empty or solid circles). The vertical dashed lines indicate the medians within groups.

Each empty (or solid) circle in Figure 6 represents a detected change point. The change points detected from the same CTR distributional sequence (labeled using the associated RCP) are plotted along the x-axis at the same vertical position. Figure 6 reveals obvious co-occurrence phenomena in the detected change points. Specifically, if a change-point event can be detected from one CTR distributional sequence, similar change-point events can usually be detected from a number of other CTR distributional sequences almost at the same time. This is not surprising, because sudden damage to a bridge's cable is usually caused by an accidental event (e.g., passing of overweight vehicles, hurricanes, impact, earthquakes) that usually incurs damage to more than one cable. Consequently, the distributional sequences of the extracted CTR data (DSFs) associated with different cable pairs may undergo changes simultaneously. Due to estimation errors, the location estimates of such simultaneous change points may not perfectly coincide, but would be closely gathered in the time domain when properly illustrated in a plot (e.g., Figure 6). We roughly divide the detected change points in Figure 6 into 23 separate groups based on registered time indices, and indicate the data from two adjacent groups using different symbols (empty or solid circles) to make them visually distinguishable. The vertical dashed lines indicate the group-wise medians of the detected change points, indicating at least 23 change-point events that widely occur in the investigated CTR distributional data. The detected change points also tend to be spatially clustered, as demonstrated by the ellipses in Figure 6. This phenomenon could occur for various reasons, including (a) a strong accidental action causing damage to multiple cables near the position of action or (b) a dead load redistribution in the cable system due to a damaged or fractured cable suddenly increasing the force carried by other nearby cables, which may also produce damage.

Since the CTR data are DSFs of stay cables (Section 2), the above detected change points can provide indications of possible damage that occurred in the corresponding cables. Finding these damage indications is a central objective of level 1 damage detection (identifying signs of damage) (Farrar and Worden 2013), which is a necessary precondition for localizing, classifying, and quantifying potential damage (levels 2–4) (Farrar and Worden 2013). Given a detected change point, it usually requires SHM practitioners to carefully assess whether the relevant change is attributed to damage or operational/environmental disturbance. Generally, if no evidence can be found to support the latter, then the detected change point was most likely induced by damage (Fugate et al. 2001). As noted in the introduction, identifying the timing of damage-induced changes is indispensable for calibrating/updating structural deterioration models that are used for remaining life prediction (level 5) (Farrar and Worden 2013) and other structural life-cycle management. The co-occurrence and spatial clustering of detected change points reflect the dependence of damage across cables. Performing change-point analysis for the DSF data from multiple cables is critical for better understanding and analyzing damage dependence in the cable system.

## *6.2. Multiscale Detection Results*

Due to space constraints, we briefly outline our multiscale Fréchet-MOSUM procedure and its detection results here, and the full details are presented in Supplementary Section S.7. Specifically, we consider multiple bandwidths on an equally-spaced grid, $G_{grid} = \{G_1, G_2, \cdots, G_L\}$. We apply the Fréchet-MOSUM procedure separately for each $G \in G_{grid}$, conducting a round of change-point detection on a given CTR distributional sequence of length $n$. We then build a matrix, $\mathcal{M} \in \mathbb{R}^{n \times L}$ with $\mathcal{M}(i,j) = 1$ if $i$ coincides with a change point detected using bandwidth $G_j$ and $\mathcal{M}(i,j) = 0$ otherwise. We call the visual representation of $\mathcal{M}$ (e.g., Figure 7(a)) a change-point indicator (CPI) diagram. In the CPI diagram, each of the nonzero entries of $\mathcal{M}$ is indicated by an empty circle, which represents the location estimate of a change point that is detected using the corresponding bandwidth. If $G_{grid}$ is not too coarse, a true change point can usually be detected repeatedly under different bandwidths, and the relevant location estimates can form a stable trajectory in the CPI diagram. This motivates us to construct an algorithm to identify the potential trajectories in the CPI diagram (Supplementary Algorithm S.4).

We then use the identified trajectories to estimate the change points (Supplementary Algorithms S.5 and S.6). This multiscale detection strategy combines the change points detected using different bandwidths to obtain the final estimates for potential change points. Using multiple bandwidths, rather than relying on a single one, could reduce the risk of parameter misspecification, but requires more computational cost. We consider 26 equally-spaced bandwidths ranging from 30 to 80 in our analysis and use default settings (Supplementary Section S.7.1.5). The multiscale detection procedure requires about 10 minutes to accomplish the detection for our 50 real datasets. Figure 7(b) illustrates the trajectories identified from the CPI diagram in Figure 7(a) using Algorithms S.4. The vertical dashed lines indicate the final change-point location estimates obtained by fusing the results associated with each of the trajectories using Algorithms S.5 and S.6. Additional results are illustrated in Supplementary Figure S.32, and Supplementary Figure S.33 presents the multiscale detection results for the 50 distributional sequences. Comparing Figure 6 with Figure S.33 reveals that the single-bandwidth version of the Fréchet-MOSUM procedure achieves comparable performance to the multiscale version, indicating that the bandwidth selected for the single-bandwidth version is valid. The single-bandwidth version is highly efficient; however, if the bandwidth choice is improper, it may be significantly inferior to the multiscale version. Practically, SHM practitioners can choose the appropriate detection strategy (single-bandwidth or multiscale) depending on individual circumstances.

**7. Discussion**

In structural health monitoring (SHM), data-driven damage detection can be achieved by detecting changes in the distribution of properly extracted damage-sensitive features (DSFs). We develop a multiple change-point detection method, named Fréchet-MOSUM, to detect the distributional changes contained in DSF data. Advantages of our method include (a) low computational complexity, (b) enabling the detection of mean and variance changes, or both, and (c) controlling the family-wise false-positive rate. We provide theoretical justifications for our proposal. Our method is general and can be easily extended for applicability to a wide variety of data residing in a metric space.

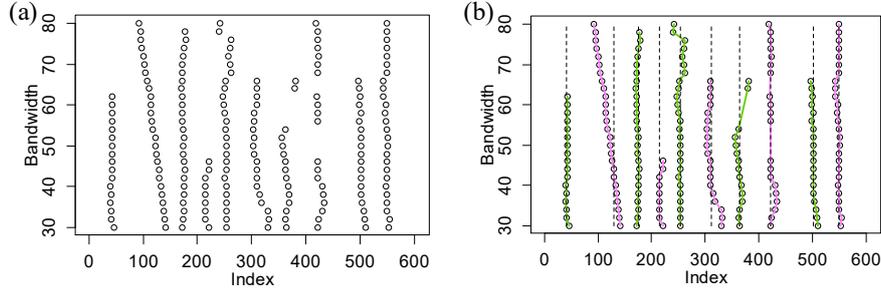

**Figure 7.** Multiscale detection results of the CTR distributional sequence associated with cable pair RCP41. (a) The CPI diagram and (b) the identified change-point trajectories. The final change-point location estimates that are obtained using Algorithms S.5 and S.6 are indicated by vertical dashed lines.

We apply our method to cable-tension monitoring data collected by a SHM system of a long-span cable-stayed bridge for cable damage detection, using cable-tension ratio (CTR) as the DSF. We conduct an extensive change-point analysis for the CTR data associated with 50 cable pairs. To the best of our knowledge, this is the first attempt to implement a systematic distributional change diagnosis for the DSF data of a long-span cable-stayed bridge over a long-term period. The detected change points exhibit notable co-occurrence tendencies across different cables and also tend to be spatially clustered, indicating that an accidental action or extreme event often produce damage to multiple cables, particularly those in nearby locations. Detecting the presence of incipient damage is crucial in SHM for early damage warnings. Incipient damage-induced DSF changes are usually visually inapparent. In our case study, most of the detected changes are subtle, and we introduce an archetypal analysis (AA)-based diagnostic tool to confirm the findings. The results demonstrate that our method can effectively detect such subtle but important changes.

Our Fréchet-MOUSM method falls into the category of functional change-point detection (FCPD) methods. At present, FCPD methods for detecting multiple change points are scarce and have primarily focused on mean change detection. Our study reveals that the investigated CTR distributional sequence can undergo changes in mean, variance, or both. Our numerical study demonstrates that a FCPD method designed specifically for mean change detection performs poorly when the data contain variance changes. To our knowledge, the FPCA-ECP procedure (Lei et al. 2023b) is the only FCPD method in multiple change-point settings that has been specifically designed for distributional data that can detect changes in mean and variance; however, serious drawbacks are found such as heavy computational burden, causing information loss, or producing questionable results in some circumstances. Our Fréchet-MOSUM method provides a highly competitive alternative FCPD method for SHM applications, particularly for large datasets. The numerical results demonstrate its superiority against competitors, and it also yields satisfactory performance in our real data application.

Our current method only provides point estimates for the change-point locations. A viable extension could construct associated confidence intervals. Another direction worth pursuing is to address the outlying PDFs that may disturb the detection result. One straightforward way to do this is to identify and remove outlying PDFs (Lei et al. 2023a, 2023b) during data preprocessing. This strategy has been incorporated into our R code for implementing the Fréchet-MOSUM procedure based on the distributional outlier detector developed by Lei, Chen and Li (2023). However, enhancing the robustness of the method is still of great future interest.

**Supplementary Materials**

Supplementary materials include the implementation and mathematical details of the Fréchet-MOSUM method, all proofs, additional simulation and comparative studies, additional figures and tables from our data analyses, full details of the multiscale Fréchet-MOSUM procedure, and the codes and data for reproducing the results of this research.

**Funding**

This work was financially supported by the National Natural Science Foundation of China (Grant Nos. 51908166 and U2239253), China Postdoctoral Science Foundation (Grant No. 2019M661287), and Postdoctoral Science Foundation of Heilong Jiang province.

# ONLINE SUPPLEMENT

# Supplementary materials for "Detecting Multiple Change Points in Distributional Sequences Derived from Structural Health Monitoring Data: An Application to Bridge Damage Detection"

Xinyi Lei[a,b,c,1] and Zhicheng Chen[a,b,c,1,*]

[a]*Key Lab of Smart Prevention and Mitigation of Civil Engineering Disasters of the Ministry of Industry and Information Technology, Harbin Institute of Technology, Harbin, 150090, China*

[b]*Key Lab of Structures Dynamic Behavior and Control of the Ministry of Education, Harbin Institute of Technology, Harbin, 150090, China*

[c]*School of Civil Engineering, Harbin Institute of Technology, Harbin, 150090, China*

---

*CONTACT Zhicheng Chen; e-mail address: zhichengchen@hit.edu.cn; Key Lab of Smart Prevention and Mitigation of Civil Engineering Disasters of the Ministry of Industry and Information Technology, Harbin Institute of Technology, Harbin, 150090, China.

[1]The two authors contributed equally to this work.

# Contents

# S.1. Front View of the Investigated Cable-Stayed Bridge and Selected Cables

The front view of the cable-stayed bridge investigated in our case study is displayed in Figure S.1. As described in Section 2 of the main text, a total of 84 pairs of stay cables are symmetrically anchored at the two edges of the bridge deck. The cable pairs are labeled as CP1 to CP84 from the south to north. A total of 50 pairs of cables are selected for investigation in this study. For convenience, these cable pairs are relabeled using regularized cable-pair labels (RCPLs). Table S.1 presents the one-to-one correspondence between the original cable-pair label and regularized cable-pair label for each of the selected cable pairs.

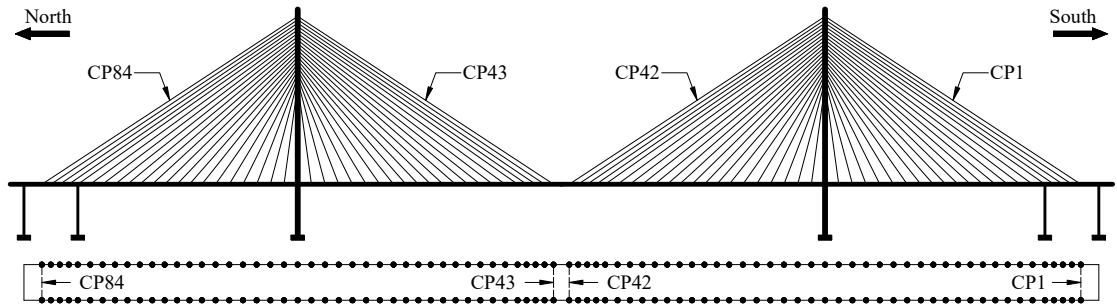

**Figure S.1.** The front view of the investigated cable-stayed bridge (Lei et al. 2023a).

**Table S.1.** Cable pairs selected for investigation. CPL stands for the cable-pair label (Figure S.1), while RCPL stands for the regularized cable-pair label.

| CPL | RCPL | CPL | RCPL | CPL | RCPL | CPL | RCPL | CPL | RCPL |
|---|---|---|---|---|---|---|---|---|---|
| CP14 | RCP1 | CP31 | RCP11 | CP48 | RCP21 | CP61 | RCP31 | CP72 | RCP41 |
| CP15 | RCP2 | CP32 | RCP12 | CP49 | RCP22 | CP62 | RCP32 | CP73 | RCP42 |
| CP17 | RCP3 | CP33 | RCP13 | CP50 | RCP23 | CP63 | RCP33 | CP74 | RCP43 |
| CP19 | RCP4 | CP35 | RCP14 | CP51 | RCP24 | CP64 | RCP34 | CP75 | RCP44 |
| CP20 | RCP5 | CP37 | RCP15 | CP53 | RCP25 | CP66 | RCP35 | CP77 | RCP45 |
| CP21 | RCP6 | CP38 | RCP16 | CP54 | RCP26 | CP67 | RCP36 | CP78 | RCP46 |
| CP24 | RCP7 | CP43 | RCP17 | CP55 | RCP27 | CP68 | RCP37 | CP80 | RCP47 |
| CP27 | RCP8 | CP45 | RCP18 | CP57 | RCP28 | CP69 | RCP38 | CP81 | RCP48 |
| CP29 | RCP9 | CP46 | RCP19 | CP59 | RCP29 | CP70 | RCP39 | CP82 | RCP49 |
| CP30 | RCP10 | CP47 | RCP20 | CP60 | RCP30 | CP71 | RCP40 | CP84 | RCP50 |

## S.2. Additional Technical Details for the Fréchet-MOUSM Method

### S.2.1. Why the Constructed Scan Statistic is a Plug-In MOSUM-Type Statistic

This subsection explains why the scan statistic (SS) $T_n^G(k)$ defined in equation (5) of the main text is a plug-in MOSUM-type statistic.

We begin with notations. Let $\Gamma = \{v_1, \cdots, v_n\}$ be a distributional ($\mathcal{W}_2(\mathcal{D})$-valued) sequence with $q$ change points located at $k_1^*, k_2^*, \cdots, k_q^*$ such that $1 < k_1^* < k_2^* < \cdots < k_q^* < n$. According to the multiple change-point setting described in Section 3.2 (of the main text), $\Gamma$ can be split into $(q+1)$ distinct segments, and the data in each segment are independently and identically distributed (i.i.d.). Specifically, let $k_0^* = 0$ and $k_{q+1}^* = n$, then we have

$$v_{k_0^*+1}, \cdots, v_{k_1^*} \sim P_1$$
$$v_{k_1^*+1}, \cdots, v_{k_2^*} \sim P_2 \tag{S.1}$$
$$\cdots \cdots \cdots$$
$$v_{k_q^*+1}, \cdots, v_{k_{q+1}^*} \sim P_{q+1}$$

where $P_j$ represents the common probability distribution, on the Wasserstein space $\mathcal{W}_2(\mathcal{D})$, that the $\mathcal{W}_2(\mathcal{D})$-valued data within the $j$th segment follow. The data from the $j$th segment have the same Fréchet mean and Fréchet variance defined as follows:

$$\mu_j = \underset{\omega \in \mathcal{W}_2(D)}{\mathrm{argmin}} E_{P_j}\left(d_\mathcal{W}^2(v,\omega)\right), v \in \mathcal{W}_2(D) \text{ and } v \sim P_j$$

$$V_j = \underset{\omega \in \mathcal{W}_2(D)}{\min} E_{P_j}\left(d_\mathcal{W}^2(v,\omega)\right), v \in \mathcal{W}_2(D) \text{ and } v \sim P_j$$

For convenience, we define the following notation:

$$E_i\left(d_\mathcal{W}^2(v_i,\omega)\right) = E_{P_j}\left(d_\mathcal{W}^2(v_i,\omega)\right), \ k_{j-1}^* < i \leq k_j^*$$

For $i = 1, 2, \cdots, n$, we can equivalently write $E_i\left(d_\mathcal{W}^2(v_i,\omega)\right)$ as

$$E_i\left(d_\mathcal{W}^2(v_i,\omega)\right) = \sum_{j=1}^{q+1} E_{P_j}\left(d_\mathcal{W}^2(v_i,\omega)\right) I\{k_{j-1}^* < i \leq k_j^*\}$$

where $I\{\cdot\}$ is the indicator function. With these notations in place, we define the following oracle versions of the sample Fréchet means $\hat{\mu}_{[k-G+1,k]} = \underset{\omega \in \mathcal{W}_2(D)}{\mathrm{argmin}} \frac{1}{G} \sum_{i=k-G+1}^{k} d_\mathcal{W}^2(v_i,\omega)$ and $\hat{\mu}_{[k+1,k+G]} = \underset{\omega \in \mathcal{W}_2(D)}{\mathrm{argmin}} \frac{1}{G} \sum_{i=k+1}^{k+G} d_\mathcal{W}^2(v_i,\omega)$:

$$\tilde{\mu}_{[k-G+1,k]} = \underset{\omega \in \mathcal{W}_2(D)}{\mathrm{argmin}} \left\{ \frac{1}{G} \sum_{i=k-G+1}^{k} E_i\left(d_\mathcal{W}^2(v_i,\omega)\right) \right\}$$

$$\tilde{\mu}_{[k+1,k+G]} = \underset{\omega \in \mathcal{W}_2(D)}{\operatorname{argmin}} \left\{ \frac{1}{G} \sum_{i=k+1}^{k+G} E_i \left( d_\mathcal{W}^2(v_i, \omega) \right) \right\}$$

Using $\tilde{\mu}_{[k-G+1,k]}$ and $\tilde{\mu}_{[k+1,k+G]}$, we further define the oracle versions of $\hat{V}_{[k-G+1,k]} = \frac{1}{G}\sum_{i=k-G+1}^{k} d_\mathcal{W}^2(v_i, \hat{\mu}_{[k-G+1,k]})$, $\hat{V}_{[k+1,k+G]} = \frac{1}{G}\sum_{i=k+1}^{k+G} d_\mathcal{W}^2(v_i, \hat{\mu}_{[k+1,k+G]})$, $\hat{V}_{[k-G+1,k]}^C = \frac{1}{G}\sum_{i=k-G+1}^{k} d_\mathcal{W}^2(v_i, \hat{\mu}_{[k+1,k+G]})$, and $\hat{V}_{[k+1,k+G]}^C = \frac{1}{G}\sum_{i=k+1}^{k+G} d_\mathcal{W}^2(v_i, \hat{\mu}_{[k-G+1,k]})$ as follows:

$$\tilde{V}_{[k-G+1,k]} = \frac{1}{G} \sum_{i=k-G+1}^{k} d_\mathcal{W}^2(v_i, \tilde{\mu}_{[k-G+1,k]})$$

$$\tilde{V}_{[k+1,k+G]} = \frac{1}{G} \sum_{i=k+1}^{k+G} d_\mathcal{W}^2(v_i, \tilde{\mu}_{[k+1,k+G]})$$

$$\tilde{V}_{[k-G+1,k]}^C = \frac{1}{G} \sum_{i=k-G+1}^{k} d_\mathcal{W}^2(v_i, \tilde{\mu}_{[k+1,k+G]})$$

$$\tilde{V}_{[k+1,k+G]}^C = \frac{1}{G} \sum_{i=k+1}^{k+G} d_\mathcal{W}^2(v_i, \tilde{\mu}_{[k-G+1,k]})$$

Given the distributional change-point model (equation (3) in the main text) and the bandwidth $G$, the oracle Fréchet means $\tilde{\mu}_{[k-G+1,k]}$ and $\tilde{\mu}_{[k+1,k+G]}$ are both deterministic elements of $\mathcal{W}_2(\mathcal{D})$. Consequently, $\tilde{V}_{[k-G+1,k]}$, $\tilde{V}_{[k+1,k+G]}$, $\tilde{V}_{[k-G+1,k]}^C$, and $\tilde{V}_{[k+1,k+G]}^C$ defined above are all moving sum (MOSUM)-type statistics. We next define the following oracle version of the scan statistic (equation (5) of the main text) with known $\sigma^2 = \operatorname{var}\{d_\mathcal{W}^2(\mu, \nu)\}$:

$$\tilde{T}_n^G(k) = (2\sigma^2/G)^{-1/2} \left| \tilde{V}_{[k+1,k+G]} - \tilde{V}_{[k-G+1,k]} \right|$$
$$+ (2\sigma^2/G)^{-1/2} \left| \tilde{V}_{[k+1,k+G]}^C - \tilde{V}_{[k+1,k+G]} + \tilde{V}_{[k-G+1,k]}^C - \tilde{V}_{[k-G+1,k]} \right|, \quad k \in [G, n-G]$$

$\tilde{T}_n^G(k)$ is a combination of the MOSUM-type statistics $\tilde{V}_{[k-G+1,k]}$, $\tilde{V}_{[k+1,k+G]}$, $\tilde{V}_{[k-G+1,k]}^C$, and $\tilde{V}_{[k+1,k+G]}^C$; thus, it is also a MOSUM-type statistic. If we replace the components $\sigma^2$, $\tilde{\mu}_{[k-G+1,k]}$, and $\tilde{\mu}_{[k+1,k+G]}$ of $\tilde{T}_n^G(k)$ with their corresponding estimators $\hat{\sigma}_{k,n}^2$, $\hat{\mu}_{[k-G+1,k]}$, and $\hat{\mu}_{[k+1,k+G]}$, then $\tilde{T}_n^G(k)$ becomes the scan statistic (equation (5) of the main text) of our Fréchet-MOSUM procedure. In this sense, the constructed scan statistic is a plug-in MOSUM-type statistic.

### S.2.2. Condition (SC1)

Let $\Gamma = \{v_1, \cdots, v_n\}$ be an distributional sequence for change-point detection. According to the setting of our change-point model specified in Section 3.2 of the main text, $v_1, \cdots, v_n$ are i.i.d. $\mathcal{W}_2(\mathcal{D})$-valued random objects under $H_0$. We denote the associated Fréchet mean under $H_0$ as $\mu = \underset{\omega \in \mathcal{W}_2(D)}{\mathrm{argmin}}\, E\left(d_\mathcal{W}^2(v, \omega)\right)$. Let $F_i^{-1}(t)$ and $F_\mu^{-1}(t)$ represent the quantile functions associated with $v_i \in \Gamma$ and $\mu$, respectively. We then define

$$e_i(t) := F_i^{-1}(t) - F_\mu^{-1}(t), \quad i = 1, \cdots, n \tag{S.2}$$

By the definition, $e_i(t)$ can be treated as the deviation of the quantile function $F_i^{-1}(t)$ around $F_\mu^{-1}(t)$.

In this study, $\{e_i\}_{i=1}^n$ is assumed to satisfy the following condition:

**(SC1)** Under $H_0$, $\{e_i\}_{i=1}^n$ fulfills

(i) $\underset{G \leq k \leq (n-G)}{\max} \int \left(\frac{1}{\sqrt{G}} \sum_{i=k-G+1}^{k} e_i(t)\right)^2 dt = O_p(1)$

(ii) $\frac{1}{\sqrt{G}} \sum_{i=k+1}^{k+G} \int e_i^2(t) dt = O_p(1)$ for each $k \in \{G, G+1, \cdots, n-G\}$

where $G = G(n)$ is the bandwidth that satisfies condition (A1) in the main text.

It is noteworthy that $e_i(t)$ resides in the Hilbert space $L^2[0,1]$ rather than the nonlinear space where the quantile functions reside. Consequently, $e_i(t)$s are Hilbert space-valued data (also called Hilbertian data). The above assumptions are mild. Condition SC1(i) has been commonly used in previous functional change-point analysis literature (e.g., Aston and Kirch, 2012). As noted in Remark 3.1 of Aston and Kirch (2012), the result of condition SC1(i) follows from functional central limit theorems for Hilbertian data. Condition SC1(ii) follows from the classical central limit theorem on noting that $Y_i = \int e_i^2(t) dt, i = 1, \cdots, n$ are real-valued data.

### S.2.3. Additional Details on the Fréchet-MOSUM Change-Point Estimator

#### S.2.3.1. Illustration of the Scan Statistic Sequence

The upper panel of Figure S.2 displays the scan statistic (SS) sequence calculated from a simulated distributional sequence with three change points ($k_1^* = 300$, $k_2^* = 700$ and $k_3^* = 1000$) by using our Fréchet-MOSUM method at a bandwidth of $G = 100$. For comparison purposes, we also present the heatmap of the distributional sequence in the lower panel of Figure S.2. In the plot of SS sequence, the calculated critical value $D_n(G; \alpha)$ at the significance level $\alpha = 0.05$ is indicated by a horizontal dashed line, which serves as a threshold. The result shows that the SS sequence exceeds the threshold within an interval around each of the change points (vertical dashed lines) and peaks near each change point.

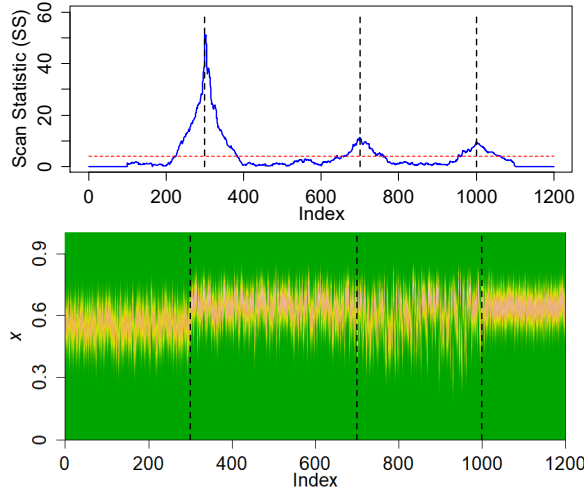

**Figure S.2.** The SS sequence (upper panel) of a distributional sequence with three change points. The heatmap of the PDF-valued sequence is visualized in the lower panel. The three true change points are indicated by vertical dashed lines. The horizontal dashed line in the upper panel indicates the critical value for the level $\alpha$ test.

*S.2.3.2. A More In-Depth Explanation of the AOP Parameter*

The AOP parameter $\varepsilon$ (equation (8) of the main text) is used to specify the minimum length of the over-threshold indexing block. Specifically, the length of a picked over-threshold indexing block, denoted by $IB_j = \{k \in \mathbb{N}: s_j \leq k \leq e_j\}$, must satisfy

$$e_j - s_j \geq \varepsilon G$$

where $G$ is the bandwidth, and $\varepsilon$ is the AOP parameter that takes a fixed value within $(0, 0.5)$. This condition plays a pivotal role in avoiding picking out an indexing block associated with a small over-threshold "bulge" of the SS sequence near the critical line, such as the one demonstrated by the dashed rectangle in the right panel of Figure S.3. Usually, such small over-threshold "bulges" are noise-induced random fluctuations in the SS sequence rather than the manifestations of underlying changes of the data (Eichinger and Kirch 2018; Kirch and Reckruehm 2022); therefore, it is necessary to exclude them from consideration to avoid overestimating the number of change points.

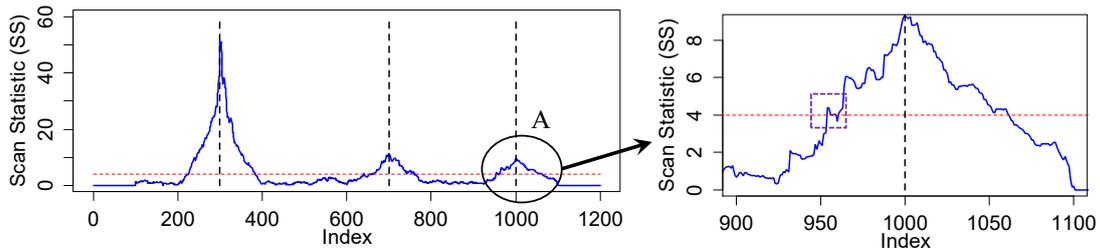

**Figure S.3.** Same as the upper panel of Figure S.2 but the local region abound the third change point is enlarged for detail.

*S.2.4. The MOSUM signal*

Recall that the scan statistic $T_n^G(k)$ (equation (5) of the main text) takes the form

$$T_n^G(k) = \left(2\hat{\sigma}_{k,n}^2/G\right)^{-1/2}\left|\hat{V}_{[k+1,k+G]} - \hat{V}_{[k-G+1,k]}\right|$$
$$+ \left(2\hat{\sigma}_{k,n}^2/G\right)^{-1/2}\left|\hat{V}_{[k+1,k+G]}^C - \hat{V}_{[k+1,k+G]} + \hat{V}_{[k-G+1,k]}^C - \hat{V}_{[k-G+1,k]}\right|, \quad k \in [G, n-G] \qquad (S.3)$$

The expression of the MOSUM signal (equation (11) of the main text), namely the noise-free version of $T_n^G$ under $H_A$, takes the form

$$T^G(k) := \left(2\sigma_k^2/G\right)^{-1/2}\left|V_{[k+1,k+G]} - V_{[k-G+1,k]}\right|$$
$$+ \left(2\sigma_k^2/G\right)^{-1/2}\left|V_{[k+1,k+G]}^C - V_{[k+1,k+G]} + V_{[k-G+1,k]}^C - V_{[k-G+1,k]}\right|, \quad k \in [G, n-G] \qquad (S.4)$$

where $G$ is the bandwidth. In this subsection, we provide the expression for each term in (S.4). Before proceeding further, we make two additional remarks.

**Remark 1.** $V_{[k+1,k+G]}$, $V_{[k-G+1,k]}$, $V_{[k+1,k+G]}^C$, $V_{[k-G+1,k]}^C$, and $\sigma_k^2$ in (S.4) are noise-free versions of $\hat{V}_{[k+1,k+G]}$, $\hat{V}_{[k-G+1,k]}$, $\hat{V}_{[k+1,k+G]}^C$, $\hat{V}_{[k-G+1,k]}^C$, and $\hat{\sigma}_{k,n}^2$ under $H_A$, respectively. Given the change-point model (equation (3) of the main text) and the bandwidth $G$, $T^G$ is a deterministic process, similar to the MOSUM signal presented in Figure 1 of Kirch and Reckruehm (2022). For convenience, $T^G$ is referred to as the MOSUM signal of our method.

**Remark 2.** The condition of the MOSUM signal is only used in proving the consistency of our Fréchet-MOSUM change-point estimator, where $T^G(k)$ must satisfy $\frac{1}{\sqrt{\log(n/G)}} \min_{k \in B_{G,q}} T^G(k) \to \infty$, as $n \to \infty$ (i.e., condition (C2) in the main text). In this sense, we only concern the asymptotic behavior of the MOSUM signal. Under condition (C1) in the main text, the window $[k-G+1, k+G]$ asymptotically does not contain more than one change point. Here we only provide the expression for the MOSUM signal for sufficiently large $n$ that the window $[k-G+1, k+G]$ only contains at most one change point.

*S.2.4.1. Notations*

We begin with notations. Given a distributional ($\mathcal{W}_2(\mathcal{D})$-valued) sequence $\Gamma = \{v_1, \cdots, v_n\}$ with $q$ unknown change points $k_1^*, k_2^*, \cdots, k_q^*$ such that $0 < k_1^* < k_2^* < \cdots < k_q^* < n$. According to our multiple change-point model described in Section 3.2 (of the main text), $\Gamma$ can be split into $(q + 1)$ distinct segments, and the data within each segment are independently and identically distributed. Specifically, let $k_0^* = 0$ and $k_{q+1}^* = n$ denote the two boundary points, we have

$$\begin{aligned} v_{k_0^*+1}, \cdots, v_{k_1^*} &\sim P_1 \\ v_{k_1^*+1}, \cdots, v_{k_2^*} &\sim P_2 \\ &\cdots \cdots \cdots \\ v_{k_q^*+1}, \cdots, v_{k_{q+1}^*} &\sim P_{q+1} \end{aligned} \quad (S.5)$$

where $P_j$ represents the common probability distribution, on the Wasserstein space $\mathcal{W}_2(\mathcal{D})$, that the data within the $j$th segment follow. Consequently, the data within each segment have the same Fréchet mean and Fréchet variance. For convenience, let $\theta_j$ denote any $\mathcal{W}_2(\mathcal{D})$-valued random object such that $\theta_j \sim P_j$. Then, the Fréchet mean and Fréchet variance of the data within the $j$th segment can be written as

$$\mu_j = \underset{\omega \in \mathcal{W}_2(D)}{\operatorname{argmin}} E_{P_j}\left(d_W^2(\theta_j, \omega)\right), \quad j = 1, 2, \cdots, (q+1) \quad (S.6a)$$

$$V_j = E_{P_j}\left(d_W^2(\theta_j, \mu_j)\right), \quad j = 1, 2, \cdots, (q+1) \quad (S.6b)$$

The first and last segments in (S.5) are referred to as the left and right boundary segments, respectively. In the following, we first present the result of $T^G(k)$ for the situation that $k$ is not located in any boundary segment, followed by the situation of boundary segment.

*S.2.4.2. Situation I: k is not Located in any Boundary Segment*

In this situation, we have $k_1^* < k \leq k_q^*$. Furthermore, due to $k_1^* < k_2^* < \cdots < k_q^*$, there exists $j \in \{1, \cdots, q-1\}$ such that

$$k \in (k_j^*, k_{j+1}^*]$$

Under condition (C1) in the main text, we have $k_{j+1}^* - k_j^* > 2G$ for sufficiently large $n$. Note that the indexing set $IDS(k) = \{j \in \{1, \cdots, n\}: k - G + 1 \leq j \leq k + G\}$ involved in the MOSUM signal $T^G$ at $k$ (see (S.4)) is of length $2G$, indicating that at most one change point can be contained in $IDS(k)$ for sufficiently large $n$. Therefore, according to Remark 2, one of the following cases must hold when $k \in (k_j^*, k_{j+1}^*]$:

**Case 1**: $IDS(k)$ contains no change point, as illustrated in Figure S.4(a);

**Case 2**: $IDS(k)$ contains the change point $k_j^*$, as illustrated in Figure S.4(b);

**Case 3**: $IDS(k)$ contains the change point $k_{j+1}^*$, as illustrated in Figure S.4(c).

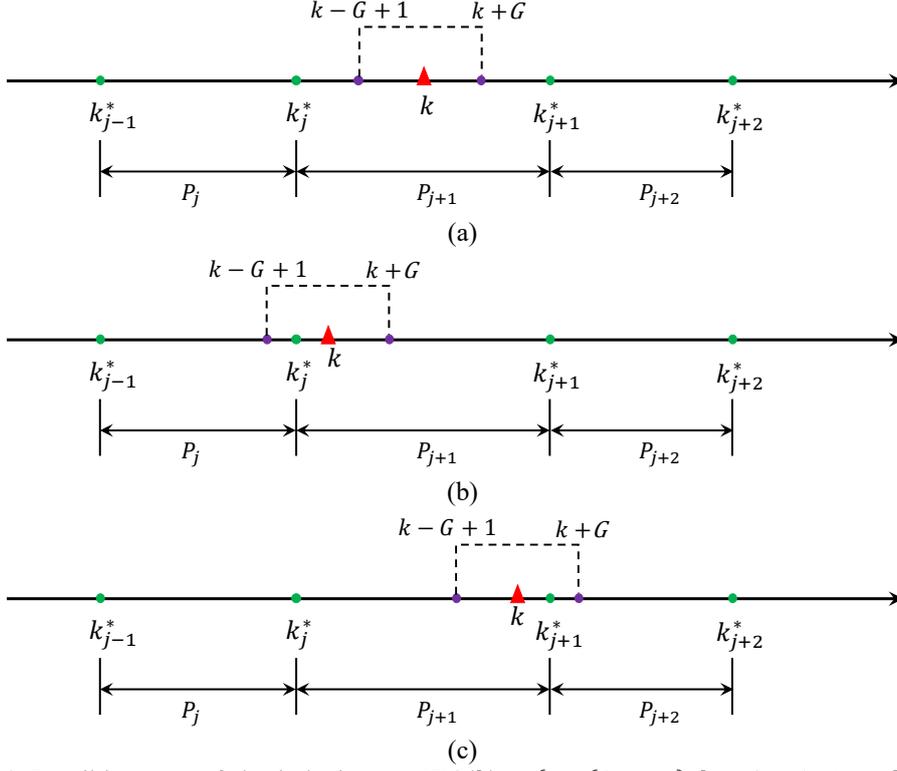

**Figure S.4.** Possible cases of the indexing set $IDS(k) = \{j \in \{1, \cdots, n\}: k - G + 1 \leq j \leq k + G\}$ when $k \in (k_j^*, k_{j+1}^*]$. (a) Case 1: $IDS(k)$ contains no change point, (b) Case 2: $IDS(k)$ contains the change point $k_j^*$, and (c) Case 3: $IDS(k)$ contains the change point $k_{j+1}^*$.

In the following, we provide the expressions for the terms $V_{[k+1,k+G]}$, $V_{[k-G+1,k]}$, $V_{[k+1,k+G]}^C$, $V_{[k-G+1,k]}^C$, and $\sigma_k^2$ of the MOSUM signal $T^G(k)$ for cases 1–3, respectively. Note that $\sigma_k^2$ is a noise-free version of $\hat{\sigma}_{k,n}^2$ (the specific form of $\hat{\sigma}_{k,n}^2$ is provided in equation (15) of the main text).

**(i)** For case 1, since $IDS(k)$ contains no change point, the distributional data within the window $[k - G + 1, k + G]$ are independently and identically distributed. More precisely, $\nu_{k-G+1}, \cdots, \nu_{k+G} \sim P_{j+1}$, as illustrated in Figure S.4(a). Consequently, the two Fréchet means $\mu_{[k+1,k+G]}$ and $\mu_{[k-G+1,k]}$ are both equal to $\mu_{j+1}$, i.e.,

$$\mu_{[k+1,k+G]} = \mu_{[k-G+1,k]} = \mu_{j+1} = \underset{\omega \in \mathcal{W}_2(D)}{\arg\min} E_{P_{j+1}}\left(d_\mathcal{W}^2(\theta_{j+1}, \omega)\right) \tag{S.7}$$

Then, $V_{[k+1,k+G]}$, $V_{[k-G+1,k]}$, $V_{[k+1,k+G]}^C$, $V_{[k-G+1,k]}^C$, and $\sigma_k^2$ in (S.4) have the following expressions:

$$V_{[k+1,k+G]} = V_{[k-G+1,k]} = V_{[k+1,k+G]}^C = V_{[k-G+1,k]}^C$$
$$= E_{P_{j+1}}\left(d_\mathcal{W}^2(\theta_{j+1}, \mu_{j+1})\right) = V_{j+1} \tag{S.8a}$$

$$\sigma_k^2 = \frac{1}{2}\left(\sigma_{k,l}^2 + \sigma_{k,r}^2\right) \tag{S.8b}$$

where $\sigma_{k,l}^2 = \sigma_{k,r}^2 = E_{P_{j+1}} d_{\mathcal{W}}^4(\theta_{j+1}, \mu_{j+1}) - \left(E_{P_{j+1}} d_{\mathcal{W}}^2(\theta_{j+1}, \mu_{j+1})\right)^2$.

**(ii)** For case 2, due to $k_j^* \in IDS(k)$ and $k_{j+1}^* \notin IDS(k)$, it holds that $k - G + 1 \leq k_j^* \leq k + G < k_{j+1}^*$. Furthermore, due to $k \in (k_j^*, k_{j+1}^*]$, we have $k_j^* < k$, which indicates $k - G + 1 \leq k_j^* < k \leq k + G < k_{j+1}^*$. Consequently, as illustrated in Figure S.4(b), we have $\nu_{k-G+1}, \cdots, \nu_{k_j^*} \sim P_j$, $\nu_{k_j^*+1}, \cdots, \nu_k \sim P_{j+1}$, and $\nu_{k+1}, \cdots, \nu_{k+G} \sim P_{j+1}$. In this circumstance, similar to Dubey and Müller (2020), the two Fréchet means $\mu_{[k-G+1,k]}$ and $\mu_{[k+1,k+G]}$ take the form

$$\mu_{[k-G+1,k]} = \underset{\omega \in \mathcal{W}_2(D)}{\mathrm{argmin}} \left\{ \frac{k_j^* - (k-G)}{G} E_{P_j} d_{\mathcal{W}}^2(\theta_j, \omega) \right. \tag{S.9a}$$
$$\left. + \frac{k - k_j^*}{G} E_{P_{j+1}} d_{\mathcal{W}}^2(\theta_{j+1}, \omega) \right\}$$

$$\mu_{[k+1,k+G]} = \underset{\omega \in \mathcal{W}_2(D)}{\mathrm{argmin}} E_{P_{j+1}} d_{\mathcal{W}}^2(\theta_{j+1}, \omega) = \mu_{j+1} \tag{S.9b}$$

respectively. Consequently, $V_{[k+1,k+G]}$, $V_{[k-G+1,k]}$, $V_{[k+1,k+G]}^C$, $V_{[k-G+1,k]}^C$, and $\sigma_k^2$ in (S.4) have the following expressions:

$$V_{[k-G+1,k]} = \frac{k_j^* - (k-G)}{G} E_{P_j} d_{\mathcal{W}}^2(\theta_j, \mu_{[k-G+1,k]}) \tag{S.10a}$$
$$+ \frac{k - k_j^*}{G} E_{P_{j+1}} d_{\mathcal{W}}^2(\theta_{j+1}, \mu_{[k-G+1,k]})$$

$$V_{[k+1,k+G]} = E_{P_{j+1}} d_{\mathcal{W}}^2(\theta_{j+1}, \mu_{j+1}) \tag{S.10b}$$

$$V_{[k-G+1,k]}^C = \frac{k_j^* - (k-G)}{G} E_{P_j} d_{\mathcal{W}}^2(\theta_j, \mu_{j+1}) \tag{S.10c}$$
$$+ \frac{k - k_j^*}{G} E_{P_{j+1}} d_{\mathcal{W}}^2(\theta_{j+1}, \mu_{j+1})$$

$$V_{[k+1,k+G]}^C = E_{P_{j+1}} d_{\mathcal{W}}^2(\theta_{j+1}, \mu_{[k-G+1,k]}) \tag{S.10d}$$

$$\sigma_k^2 = \frac{1}{2}(\sigma_{k,l}^2 + \sigma_{k,r}^2) \tag{S.10e}$$

where

$$\sigma_{k,l}^2 = \frac{k_j^* - (k-G)}{G} E_{P_j} d_{\mathcal{W}}^4(\theta_j, \mu_{[k-G+1,k]})$$
$$+ \frac{k - k_j^*}{G} E_{P_{j+1}} d_{\mathcal{W}}^4(\theta_{j+1}, \mu_{[k-G+1,k]}) - V_{[k-G+1,k]}^2$$

and

$$\sigma_{k,r}^2 = E_{P_{j+1}} d_{\mathcal{W}}^4(\theta_{j+1}, \mu_{j+1}) - \left(E_{P_{j+1}} d_{\mathcal{W}}^2(\theta_{j+1}, \mu_{j+1})\right)^2$$

**(iii)** For case 3, due to $k_{j+1}^* \in IDS(k)$ and $k_j^* \notin IDS(k)$, it holds that $k_j^* < k - G + 1 \leq k_{j+1}^* \leq k + G$. Furthermore, due to $k \in (k_j^*, k_{j+1}^*]$, we have $k \leq k_{j+1}^*$, which indicates $k_j^* < k - G + 1 \leq k \leq k_{j+1}^* \leq k + G$. Consequently, as illustrated in Figure S.4(c), we have $v_{k-G+1}, \cdots, v_k \sim P_{j+1}$, $v_{k+1}, \cdots, v_{k_{j+1}^*} \sim P_{j+1}$, and $v_{k_{j+1}^*+1}, \cdots, v_{k+G} \sim P_{j+2}$. Similar to case 2, we have

$$\mu_{[k-G+1,k]} = \underset{\omega \in \mathcal{W}_2(D)}{\operatorname{argmin}} E_{P_{j+1}} d_\mathcal{W}^2(\theta_{j+1}, \omega) = \mu_{j+1} \tag{S.11a}$$

$$\mu_{[k+1,k+G]} = \underset{\omega \in \mathcal{W}_2(D)}{\operatorname{argmin}} \left\{ \frac{k_{j+1}^* - k}{G} E_{P_{j+1}} d_\mathcal{W}^2(\theta_{j+1}, \omega) + \frac{k + G - k_{j+1}^*}{G} E_{P_{j+2}} d_\mathcal{W}^2(\theta_{j+2}, \omega) \right\} \tag{S.11b}$$

Consequently, $V_{[k+1,k+G]}$, $V_{[k-G+1,k]}$, $V_{[k+1,k+G]}^C$, $V_{[k-G+1,k]}^C$, and $\sigma_k^2$ in (S.4) have the following expressions:

$$V_{[k-G+1,k]} = E_{P_{j+1}} d_\mathcal{W}^2(\theta_{j+1}, \mu_{j+1}) \tag{S.12a}$$

$$V_{[k+1,k+G]} = \frac{k_{j+1}^* - k}{G} E_{P_{j+1}} d_\mathcal{W}^2(\theta_{j+1}, \mu_{[k+1,k+G]})$$
$$+ \frac{k + G - k_{j+1}^*}{G} E_{P_{j+2}} d_\mathcal{W}^2(\theta_{j+2}, \mu_{[k+1,k+G]}) \tag{S.12b}$$

$$V_{[k-G+1,k]}^C = E_{P_{j+1}} d_\mathcal{W}^2(\theta_{j+1}, \mu_{[k+1,k+G]}) \tag{S.12c}$$

$$V_{[k+1,k+G]}^C = \frac{k_{j+1}^* - k}{G} E_{P_{j+1}} d_\mathcal{W}^2(\theta_{j+1}, \mu_{j+1})$$
$$+ \frac{k + G - k_{j+1}^*}{G} E_{P_{j+2}} d_\mathcal{W}^2(\theta_{j+2}, \mu_{j+1}) \tag{S.12d}$$

$$\sigma_k^2 = \frac{1}{2}(\sigma_{k,l}^2 + \sigma_{k,r}^2) \tag{S.12e}$$

where

$$\sigma_{k,l}^2 = E_{P_{j+1}} d_\mathcal{W}^4(\theta_{j+1}, \mu_{j+1}) - \left( E_{P_{j+1}} d_\mathcal{W}^2(\theta_{j+1}, \mu_{j+1}) \right)^2$$

and

$$\sigma_{k,r}^2 = \frac{k_{j+1}^* - k}{G} E_{P_{j+1}} d_\mathcal{W}^4(\theta_{j+1}, \mu_{[k+1,k+G]})$$
$$+ \frac{k + G - k_{j+1}^*}{G} E_{P_{j+2}} d_\mathcal{W}^4(\theta_{j+2}, \mu_{[k+1,k+G]}) - V_{[k+1,k+G]}^2$$

*S.2.4.3. Situation II: k is Located in a Boundary Segment*

If $k$ is within the left boundary segment (i.e., $G \leq k < k_1^*$), then only the case 1 or case 3 illustrated in Figure S.4 can happen. Consequently, $V_{[k+1,k+G]}$, $V_{[k-G+1,k]}$, $V^C_{[k+1,k+G]}$, $V^C_{[k-G+1,k]}$, and $\sigma_k^2$ have similar expressions with their counterparts in equation (S.8) or (S.12) with $j = 0$. If $k$ is within the right boundary segment (i.e., $k_q^* < k \leq n - G$), then only the case 1 or case 2 illustrated in Figure S.4 can happen. Consequently, $V_{[k+1,k+G]}$, $V_{[k-G+1,k]}$, $V^C_{[k+1,k+G]}$, $V^C_{[k-G+1,k]}$, and $\sigma_k^2$ have similar expressions with their counterparts in equation (S.8) or (S.10) with $j = q$.

*S.2.5. Explanations of Conditions (C1)–(C3)*

Condition (C1) and (C3) have been commonly used in previous MOSUM literature (e.g., Muhsal 2013; Eichinger and Kirch 2018; Kirch and Reckruehm 2022; Kim et al. 2023). Condition (C1) requires that the spacing between any two adjacent change points is asymptotically longer than $2G$, where $G$ is the bandwidth of the MOSUM procedure. Under condition (C1), at most one change point is permitted to be contained in a window of length $2G$. Condition (C3) requires that the sequence of significance levels $\alpha_n$ converges to zero but not too fast, as $n \to \infty$. Condition (C2) imposes a constraint on the strength of the MOSUM signal, which requires that $\min_{k \in B_{G,q}} T^G(k)$ is of larger order than $\sqrt{\log(n/G)}$. Similar assumptions for signal strength have been commonly used in previous MOSUM literature, such as condition (C2) of Kim et al. (2023) and assumption 3.1 (b) of Kirch and Reckruehm (2022).

*S.2.6. Recursive Formulas for Calculating the Mean Quantile Function*

This subsection provides recursive formulas for rapidly calculating $F^{-1}_{\hat{\mu}_{[k+1,k+G]}}$ and $F^{-1}_{\hat{\mu}_{[k-G+1,k]}}$ involved in computing Wasserstein metrics (see equation (14) of the main text).

We denote with $\Gamma^G_k = \{\nu_{k-G+1}, \cdots, \nu_k, \nu_{k+1}, \cdots, \nu_{k+G}\}$ the sub-sequence of the distributional sequence $\Gamma = \{\nu_1, \cdots, \nu_n\}$ isolated by a sliding window at $k$ of length $2G$, and denote with $\{F^{-1}_{k-G+1}, \cdots, F^{-1}_k, F^{-1}_{k+1}, \cdots, F^{-1}_{k+G}\}$ the corresponding quantile functions. Define

$$\overline{F^{-1}_{right}(k)} := F^{-1}_{\hat{\mu}_{[k+1,k+G]}} = \frac{1}{G} \sum_{i=k+1}^{k+G} F^{-1}_i$$

$$\overline{F^{-1}_{right}(k+1)} := F^{-1}_{\hat{\mu}_{[k+2,k+G+1]}} = \frac{1}{G}\sum_{i=k+2}^{k+G+1} F_i^{-1}$$

$$\overline{F^{-1}_{left}(k)} := F^{-1}_{\hat{\mu}_{[k-G+1,k]}} = \frac{1}{G}\sum_{i=k-G+1}^{k} F_i^{-1}$$

$$\overline{F^{-1}_{left}(k+1)} := F^{-1}_{\hat{\mu}_{[k-G+2,k+1]}} = \frac{1}{G}\sum_{i=k-G+2}^{k+1} F_i^{-1}$$

Then, we have the following recursive formulas for calculating $F^{-1}_{\hat{\mu}_{[k+1,k+G]}}$ and $F^{-1}_{\hat{\mu}_{[k-G+1,k]}}$ with a computational complexity of order $O(n)$:

$$\overline{F^{-1}_{right}(k+1)} = \overline{F^{-1}_{right}(k)} + \frac{1}{G}(F^{-1}_{k+G+1} - F^{-1}_{k+1}),\ k = G, G+1, \cdots, n-G-1$$

$$\overline{F^{-1}_{left}(k+1)} = \overline{F^{-1}_{left}(k)} + \frac{1}{G}(F^{-1}_{k+1} - F^{-1}_{k-G+1}),\ k = G, G+1, \cdots, n-G-1$$

The derivations of the above recursive formulas are straightforward, we omit the details.

### S.2.7. Boundary Correction for the Scan Statistic (SS) Sequence

In the Fréchet-MOSUM procedure, the scan statistic (SS) $T_n^G$ at $k$ is calculated from the local distributional data $\{v_{k-G+1}, \cdots, v_{k+G}\}$ (isolated by a symmetric window with bandwidth $G$). Consequently, the SS sequence is only computable within the interval $n - G \leq k \leq n$, as the number of distributional samples within $1 \leq k < G$ (left boundary) or $n - G < k \leq n$ (right boundary) is smaller than $G$. In practical applications, the issue of boundary correction, namely how to pad the unavailable values of $T_n^G$ at the two boundaries with substituted values, is of great concern. We address this issue by following the boundary extension strategy described in Meier et al. (2021), which are originally designed for the standard MOSUM procedure. This strategy pads the boundary values based on a CUSUM-type boundary extension statistic that is modified from the MOSUM-type scan statistic. To pad the values within the left boundary, we tailor the scan statistic given in equation (5) of the main text to a CUSUM-type version as follows:

$$T^G_{Left}(k) = \sqrt{\frac{k(2G-k)}{2G\hat{\sigma}_k^2}}\left(\left|\hat{V}_{[1,k]} - \hat{V}_{[k+1,2G]}\right| + \left|\hat{V}^C_{[1,k]} - \hat{V}_{[1,k]} + \hat{V}^C_{[k+1,2G]} - \hat{V}_{[k+1,2G]}\right|\right),$$

$$k = 1, \cdots, G-1$$

where

$$\hat{\sigma}_k^2 = (\hat{\sigma}^2_{k,l} + \hat{\sigma}^2_{k,r})/2, \quad k = 1, \cdots, G-1$$

in which

$$\hat{\sigma}_{k,l}^2 = \frac{1}{k}\sum_{i=1}^{k} d_{\mathcal{W}}^4(v_i, \hat{\mu}_{[1,k]}) - \left(\frac{1}{k}\sum_{i=1}^{k} d_{\mathcal{W}}^2(v_i, \hat{\mu}_{[1,k]})\right)^2$$

and

$$\hat{\sigma}_{k,r}^2 = \frac{1}{2G-k}\sum_{i=k+1}^{2G} d_{\mathcal{W}}^4(v_i, \hat{\mu}_{[k+1,2G]}) - \left(\frac{1}{2G-k}\sum_{i=k+1}^{2G} d_{\mathcal{W}}^2(v_i, \hat{\mu}_{[k+1,2G]})\right)^2$$

For convenience, we call $T_{Left}^G(k)$ the left boundary extension (LBE) statistic. One can easily verify that the LBE statistic $T_{Left}^G(k)$ is equivalent to the Fréchet-MOSUM scan statistic $T_n^G(k)$ at $k = G$, meaning that the LBE statistic coincides with the scan statistic at its left cutoff point. Values within the right boundary can be corrected in a similar way.

In practical applications, distributional samples used for computing the empirical Fréchet means or variances involved in the above CUSUM-type statistic should not be too few, otherwise the uncertainty of the results would be substantially increased. Therefore, we only compute the boundary extension statistic for the points within $\lceil 2cG \rceil \le k < G$ or $n - G \le k < n - \lceil 2cG \rceil$ for some positive $c$. In our analysis, the default value of $c$ is set to 0.1.

### S.2.8. Selections for Tuning Parameters

Our Fréchet-MOSUM procedure involves two important tuning parameters. One is the bandwidth $G$, and the other is the AOP parameter $\varepsilon \in (0, 0.5)$ (see equation (8) of the main text). Practically, $G$ can be selected using a change-point trajectory (CPT) plot-based strategy, which will be detailed later in this subsection. The AOP parameter $\varepsilon$, combined with the bandwidth $G$, is used to specify the minimum length of the over-threshold indexing block (OTIB) that can be picked out for change-point estimation. Specifically, the length of a valid OTIB, denoted by $IB_j = \{k \in \mathbb{N}: s_j \le k \le e_j\}$, must satisfy

$$e_j - s_j \ge \varepsilon G$$

As detailed in Section S.2.3.2, this constraint is used to alleviate the overestimation issue. In practical applications, the choice of such a minimum block length is somewhat subjective. Generally, the larger $\varepsilon G$ is the lower risk of overestimating the number of change points; however, a too large $\varepsilon G$ can also significantly reduce the detection sensitivity of the change-point detector. We recommend the user to first specify the minimum length of the OTIB, denoted as $L_m = \varepsilon G$, then $\varepsilon$ can be automatically determined as follows:

$$\varepsilon = \min(0.5, L_m/G)$$

Our experience suggests that choosing $L_m$ from $[10, 20]$ can strike a good balance between overestimation and detection sensitivity.

We next discuss how to use the CPT plot to make a proper choice for the bandwidth $G$. We first introduce the CPT plot. Let $\Gamma = \{v_1,\cdots,v_n\}$ be an investigated CTR distributional sequence. In order to select an appropriate bandwidth for the Fréchet-MOSUM detector, we consider a collection of candidate bandwidths on an equally-spaced grid that is denoted as $G_{grid} = \{G_1, G_2, \cdots, G_L\}$. For each $G_i \in G_{grid}$, we separately apply the Fréchet-MOSUM detector to the distributional sequence $\Gamma$ by setting $G = G_i$ and $\varepsilon = \min(0.5, L_m/G_i)$, and we store the detection result for each of the bandwidth candidates. We then plot all of the detected change points in one plot with the $x$-axis representing the indices (of the CTR distributions) and the $y$-axis bandwidth, as shown in Figure S.5. In Figure S.5, each of the detected change points is indicated by an empty circle. If $G_{grid}$ is not too coarse, a true change point can usually be detected repeatedly under different bandwidth candidates, and the relevant location estimates can form a trajectory in the plot. Therefore, we call such a plot the CPT plot. A change point in the CPT plot is called a stable change point if it can be detected several times. In our analysis, if a change point can be detected more than 3 times, then it is treated as stable.

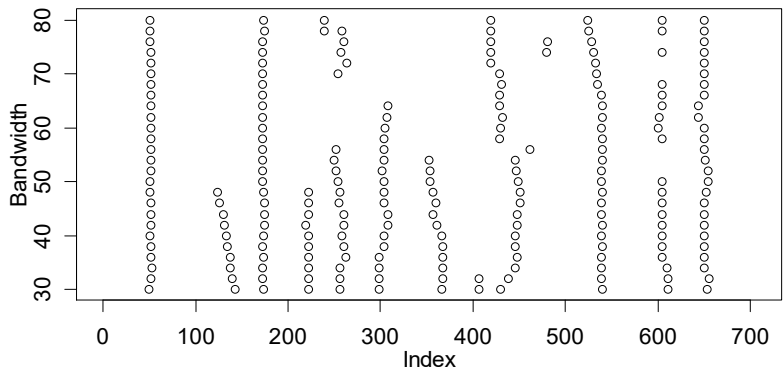

**Figure S.5.** The change-point trajectory (CPT) plot of a real CTR distributional sequence selected from our CTR datasets. Each empty circle indicates a detected change point.

Now, we are ready to select the bandwidth based on the CPT plot. Practically, the bandwidth should not be too large, otherwise the location estimate of a change point has high risk to be disturbed by its adjacent change points. For instance, the location estimates of two nearby change points may merge into one estimate when the bandwidth is too large. Therefore, too large a bandwidth can generally lead to underestimating the number of change points. On the other hand, the bandwidth should also not be too small, otherwise the functional samples that can be used to compute the scan statistic in the Fréchet-MOSUM procedure would be scarce. This can incur significant efficiency loss in estimating the change points. Therefore, the bandwidth is recommended to select from the candidates that can detect as many stable change points as possible but do not take too small values. For instance, the CPT plot shown in Figure S.5 reveals that a bandwidth between 35 and 45 would be a relatively ideal choice.

**Remark 3.** In structural health monitoring, our application of interest, there are usually many distributional sequences (associated with different monitoring sites) that are required to perform change-point detection. To save the computational time spent on selecting the bandwidth, we recommend to randomly select a few of distributional sequences from the investigated dataset to create their CPT plots, and choose a reasonable common bandwidth accordingly and then use it to perform change-point detection on each of the investigated distributional sequences.

**Remark 4.** In addition to the single-bandwidth detection version, we also develop a multiscale detection version of our Fréchet-MOSUM method that uses a collection of bandwidths rather than a single choice to perform change-point detection. The full details of this multiscale Fréchet-MOSUM procedure is provided later in Section S.7. The main idea of this approach is to properly fuse the change-point detection results obtained by using different bandwidths to produce the final change-point estimates. Using multiple bandwidths, rather than relying on a single one, could reduce the risk of parameter misspecification, but requires more computational cost.

## S.3. Details of Proofs

### S.3.1. Proof of Theorem 1

*Proof of Theorem 1.* Recall that the expressions for $\hat{V}_{[k-G+1,k]}$, $\hat{V}_{[k+1,k+G]}$, $\hat{V}^C_{[k-G+1,k]}$, and $\hat{V}^C_{[k+1,k+G]}$ in the test statistic $T_n(G) = \max_{G \leq k \leq n-G} T_n^G(k)$ (equation (4) of the main text) are as follows:

$$\hat{V}_{[k-G+1,k]} = \frac{1}{G} \sum_{i=k-G+1}^{k} d_W^2(v_i, \hat{\mu}_{[k-G+1,k]}) \qquad (S.13a)$$

$$\hat{V}_{[k+1,k+G]} = \frac{1}{G} \sum_{i=k+1}^{k+G} d_W^2(v_i, \hat{\mu}_{[k+1,k+G]}) \qquad (S.13b)$$

$$\hat{V}^C_{[k-G+1,k]} = \frac{1}{G}\sum_{i=k-G+1}^{k} d^2_{\mathcal{W}}(\nu_i, \hat{\mu}_{[k+1,k+G]}) \tag{S.13c}$$

$$\hat{V}^C_{[k+1,k+G]} = \frac{1}{G}\sum_{i=k+1}^{k+G} d^2_{\mathcal{W}}(\nu_i, \hat{\mu}_{[k-G+1,k]}) \tag{S.13d}$$

where $\hat{\mu}_{[k-G+1,k]}$ and $\hat{\mu}_{[k+1,k+G]}$ are the empirical Fréchet means given as follows:

$$\hat{\mu}_{[k-G+1,k]} = \underset{\omega \in \mathcal{W}_2(D)}{\operatorname{argmin}} \frac{1}{G}\sum_{i=k-G+1}^{k} d^2_{\mathcal{W}}(\nu_i, \omega)$$

$$\hat{\mu}_{[k+1,k+G]} = \underset{\omega \in \mathcal{W}_2(D)}{\operatorname{argmin}} \frac{1}{G}\sum_{i=k+1}^{k+G} d^2_{\mathcal{W}}(\nu_i, \omega)$$

For $k = G, G+1, \cdots, n-G$, we define the following oracle versions of $\hat{V}_{[k-G+1,k]}$, $\hat{V}_{[k+1,k+G]}$, $\hat{V}^C_{[k-G+1,k]}$, and $\hat{V}^C_{[k+1,k+G]}$ under $H_0$:

$$\tilde{V}_{[k-G+1,k]} = \frac{1}{G}\sum_{i=k-G+1}^{k} d^2_{\mathcal{W}}(\nu_i, \mu) \tag{S.14a}$$

$$\tilde{V}_{[k+1,k+G]} = \frac{1}{G}\sum_{i=k+1}^{k+G} d^2_{\mathcal{W}}(\nu_i, \mu) \tag{S.14b}$$

$$\tilde{V}^C_{[k-G+1,k]} = \frac{1}{G}\sum_{i=k-G+1}^{k} d^2_{\mathcal{W}}(\nu_i, \mu) \tag{S.14c}$$

$$\tilde{V}^C_{[k+1,k+G]} = \frac{1}{G}\sum_{i=k+1}^{k+G} d^2_{\mathcal{W}}(\nu_i, \mu) \tag{S.14d}$$

where $\mu = \underset{\omega \in \mathcal{W}_2(D)}{\operatorname{argmin}} E\left(d^2_{\mathcal{W}}(\nu, \omega)\right)$ is the Fréchet mean under $H_0$.

Recall that $\sigma^2 = \operatorname{var}\{d^2_{\mathcal{W}}(\mu, \nu)\}$ is the asymptotic variance of the Fréchet variance (AVFV) (Dubey and Müller 2019) under $H_0$, in which $\mu$ is the Fréchet mean. By replacing $\hat{\sigma}^2_{k,n}$ in the scan statistic $T^G_n(k)$ (equation (5) of the main text) with $\sigma^2$, we further define a new statistic as follows:

$$T^{\#,G}_n(k) = (2\sigma^2/G)^{-1/2}\left|\hat{V}_{[k+1,k+G]} - \hat{V}_{[k-G+1,k]}\right|$$
$$+ (2\sigma^2/G)^{-1/2}\left|\hat{V}^C_{[k+1,k+G]} - \hat{V}_{[k+1,k+G]} + \hat{V}^C_{[k-G+1,k]} - \hat{V}_{[k-G+1,k]}\right|, \quad k \in [G, n-G] \tag{S.15}$$

In order to prove Theorem 1, we need two additional lemmas:

**Lemma S.1** Under the conditions of Theorem 1, we have

(i) Under $H_0$, it holds that

$$\max_{G \leq k \leq n-G} \sqrt{G}\left|\frac{1}{G}\sum_{i=k-G+1}^{k}\left(d^2_{\mathcal{W}}(\nu_i, \hat{\mu}_{[k-G+1,k]}) - d^2_{\mathcal{W}}(\nu_i, \mu)\right)\right| = o_p\left(\frac{1}{\sqrt{\log(n/G)}}\right)$$

(ii) Under $H_0$, it holds that

$$\max_{G \leq k \leq n-G} \sqrt{G} \left| \frac{1}{G} \sum_{i=k-G+1}^{k} \left( d_W^2(\nu_i, \hat{\mu}_{[k+1,k+G]}) - d_W^2(\nu_i, \mu) \right) \right| = o_p\left( \frac{1}{\sqrt{\log(n/G)}} \right)$$

***Proof of Lemma S.1.*** Here, we only provide details for the proof of the first assertion of Lemma S.1, the second assertion can be proved by adopting similar arguments.

Recall that $\hat{\mu}_{[k-G+1,k]} = \underset{\omega \in \mathcal{W}_2(D)}{\operatorname{argmin}} \frac{1}{G} \sum_{i=k-G+1}^{k} d_W^2(\nu_i, \omega)$ is the sample Fréchet mean, which is also referred to as the sample Wasserstein barycenter in the literature (e.g., Fan and Müller 2021). Let $F_{\hat{\mu}_{[k-G+1,k]}}^{-1}$ denote the quantile function associated with $\hat{\mu}_{[k-G+1,k]}$. From the definition of the Wasserstein metric given in equation (1) of the main text, we have

$$d_W^2(\nu_i, \mu) = \int_0^1 \left( F_i^{-1}(t) - F_\mu^{-1}(t) \right)^2 dt$$

$$d_W^2(\nu_i, \hat{\mu}_{[k-G+1,k]}) = \int_0^1 \left( F_i^{-1}(t) - F_{\hat{\mu}_{[k-G+1,k]}}^{-1}(t) \right)^2 dt$$

According to the property of Wasserstein barycenter provided in Lin et al. (2023), it holds that $F_{\hat{\mu}_{[k-G+1,k]}}^{-1} = \frac{1}{G} \sum_{i=k-G+1}^{k} F_i^{-1}$, where $F_i^{-1}$ is the quantile function associated with $\nu_i \in \Gamma$. Furthermore, using equation (S.2), the above two metrics can be equivalently written as

$$d_W^2(\nu_i, \mu) = \int_0^1 e_i^2(t) dt$$

$$d_W^2(\nu_i, \hat{\mu}_{[k-G+1,k]}) = \int_0^1 \left( e_i(t) - \bar{e}_{[k-G+1,k]}(t) \right)^2 dt$$

where $e_i(t) = F_i^{-1}(t) - F_\mu^{-1}(t)$ (see (S.2)), and $\bar{e}_{[k-G+1,k]} = \frac{1}{G} \sum_{i=k-G+1}^{k} e_i(t)$.

By some straight-forward calculations, we have

$$\frac{1}{G} \sum_{i=k-G+1}^{k} \left( d_W^2(\nu_i, \hat{\mu}_{[k-G+1,k]}) - d_W^2(\nu_i, \mu) \right)$$

$$= \int_0^1 \frac{1}{G} \sum_{i=k-G+1}^{k} \left( -2 e_i(t) \bar{e}_{[k-G+1,k]}(t) + \left( \bar{e}_{[k-G+1,k]}(t) \right)^2 \right) dt$$

$$= \int_0^1 \left( -2 \bar{e}_{[k-G+1,k]}(t) \left( \frac{1}{G} \sum_{i=k-G+1}^{k} e_i(t) \right) + \left( \bar{e}_{[k-G+1,k]}(t) \right)^2 \right) dt$$

$$= - \int_0^1 \left( \bar{e}_{[k-G+1,k]}(t) \right)^2 dt$$

Consequently, the term $\max_{G \leq k \leq n-G} \sqrt{G} \left| \frac{1}{G} \sum_{i=k-G+1}^{k} \left( d_W^2(v_i, \hat{\mu}_{[k-G+1,k]}) - d_W^2(v_i, \mu) \right) \right|$ in assertion (i) of Lemma S.1 can be equivalently written as

$$\max_{G \leq k \leq n-G} \sqrt{G} \left| \frac{1}{G} \sum_{i=k-G+1}^{k} \left( d_W^2(v_i, \hat{\mu}_{[k-G+1,k]}) - d_W^2(v_i, \mu) \right) \right|$$

$$= \max_{G \leq k \leq n-G} \sqrt{G} \int_0^1 \left( \bar{e}_{[k-G+1,k]}(t) \right)^2 dt$$

$$= \sqrt{G} \max_{G \leq k \leq n-G} \int_0^1 \left( \frac{1}{G} \sum_{i=k-G+1}^{k} e_i(t) \right)^2 dt \qquad \text{(S.16)}$$

$$= \frac{1}{\sqrt{G}} \max_{G \leq k \leq n-G} \int_0^1 \left( \frac{1}{\sqrt{G}} \sum_{i=k-G+1}^{k} e_i(t) \right)^2 dt$$

Furthermore, assumption (A3) in the main text implies that the functional sequence $\{e_i(t)\}$ fulfills condition (SC1) in Section S.2.2 (of the supplement). Under condition (SC1(i)), we have

$$\max_{G \leq k \leq (n-G)} \int_0^1 \left( \frac{1}{\sqrt{G}} \sum_{i=k-G+1}^{k} e_i(t) \right)^2 dt = O_p(1)$$

Consequently,

$$\max_{G \leq k \leq n-G} \sqrt{G} \left| \frac{1}{G} \sum_{i=k-G+1}^{k} \left( d_W^2(v_i, \hat{\mu}_{[k-G+1,k]}) - d_W^2(v_i, \mu) \right) \right| = O_p(\frac{1}{\sqrt{G}}) \qquad \text{(S.17)}$$

Observe that

$$\frac{\frac{1}{\sqrt{G}}}{\frac{1}{\sqrt{\log(n/G)}}} = \sqrt{\frac{\log(n/G)}{G}} = \sqrt{\frac{\log n - \log G}{G}} \leq \sqrt{\frac{\log n}{G}} \leq \sqrt{\frac{n^{\frac{2}{2+\Delta}} \log n}{G}}$$

Under assumption (A1) in the main text, it follows that $\frac{\frac{1}{\sqrt{G}}}{\frac{1}{\sqrt{\log(n/G)}}} = o(1)$, thus

$$\frac{1}{\sqrt{G}} = o\left( \frac{1}{\sqrt{\log(n/G)}} \right)$$

Combination with (S.17) yields

$$\max_{G \leq k \leq n-G} \sqrt{G} \left| \frac{1}{G} \sum_{i=k-G+1}^{k} \left( d_W^2(v_i, \hat{\mu}_{[k-G+1,k]}) - d_W^2(v_i, \mu) \right) \right| = o_p(\frac{1}{\sqrt{\log(n/G)}})$$

This completes the proof for the first assertion of Lemma S.1. The proof for the second assertion can be obtained by following similar arguments, and we omit it here for the sake of brevity.

**Lemma S.2** Under the conditions of Theorem 1, it holds that
$$\sqrt{G}\left|\tilde{V}_{[k+1,k+G]} - \tilde{V}_{[k-G+1,k]}\right| = O_p(1), \text{ for each } k \in \{G, G+1, \cdots, n-G\}$$
under $H_0$.

***Proof of Lemma S.2.*** From equation (S.14), we have
$$\tilde{V}_{[k+1,k+G]} = \frac{1}{G}\sum_{i=k+1}^{k+G} d_W^2(v_i, \mu) \text{ and } \tilde{V}_{[k-G+1,k]} = \frac{1}{G}\sum_{i=k-G+1}^{k} d_W^2(v_i, \mu)$$

Then, using the triangle inequality and equation (S.2), we obtain

$$\sqrt{G}\left|\tilde{V}_{[k+1,k+G]} - \tilde{V}_{[k-G+1,k]}\right| = \sqrt{G}\left|\frac{1}{G}\sum_{i=k+1}^{k+G} d_W^2(v_i, \mu) - \frac{1}{G}\sum_{i=k-G+1}^{k} d_W^2(v_i, \mu)\right|$$

$$\leq \frac{1}{\sqrt{G}}\sum_{i=k+1}^{k+G} d_W^2(v_i, \mu) + \frac{1}{\sqrt{G}}\sum_{i=k-G+1}^{k} d_W^2(v_i, \mu)$$

$$\leq \frac{1}{\sqrt{G}}\sum_{i=k+1}^{k+G}\int_0^1 \left(F_i^{-1}(t) - F_\mu^{-1}(t)\right)^2 dt$$
$$+ \frac{1}{\sqrt{G}}\sum_{i=k-G+1}^{k}\int_0^1 \left(F_i^{-1}(t) - F_\mu^{-1}(t)\right)^2 dt$$

$$\leq \frac{1}{\sqrt{G}}\sum_{i=k+1}^{k+G}\int_0^1 e_i^2(t)dt + \frac{1}{\sqrt{G}}\sum_{i=k-G+1}^{k}\int_0^1 e_i^2(t)dt$$

Assumption (A3) in the main text implies that condition (SC1) in Section S.2.2 (of the supplement) holds. Then, under condition (SC1) (ii), we have

$$\sqrt{G}\left|\tilde{V}_{[k-G+1,k]} - \tilde{V}_{[k+1,k+G]}\right| = O_p(1)$$

This completes the proof.

Next, we resume the proof of Theorem 1. The rest of the proof can be divided into three steps.

***Step 1:*** We first show that $T_n^G(k) = T_n^{\#,G}(k) + o_p\left(\frac{1}{\sqrt{\log(n/G)}}\right)$ under $H_0$, for each $k \in \{G, G+1, \cdots, n-G\}$. Here, $T_n^{\#,G}(k)$ is the statistic defined in equation (S.15), and $T_n^G(k)$ is the scan statistic defined in equation (5) of the main text.

Given $k \in \{G, \cdots, n-G\}$, we first observe that

$$|T_n^G(k) - T_n^{\#,G}(k)| = \frac{1}{\sqrt{2}}\left|\frac{1}{\hat{\sigma}_{k,n}} - \frac{1}{\sigma}\right|\left(\sqrt{G}|\hat{V}_{[k+1,k+G]} - \hat{V}_{[k-G+1,k]}|\right.$$
$$\left. + \sqrt{G}|\hat{V}^C_{[k+1,k+G]} - \hat{V}_{[k+1,k+G]} + \hat{V}^C_{[k-G+1,k]} - \hat{V}_{[k-G+1,k]}|\right) \quad \text{(S.18)}$$
$$:= \frac{1}{\sqrt{2}}\left|\frac{1}{\hat{\sigma}_{k,n}} - \frac{1}{\sigma}\right|(|J_n^{G,I}(k)| + |J_n^{G,II}(k)|)$$

where

$$J_n^{G,I}(k) := \sqrt{G}(\hat{V}_{[k+1,k+G]} - \hat{V}_{[k-G+1,k]}) \quad \text{(S.19a)}$$
$$J_n^{G,II}(k) := \sqrt{G}(\hat{V}^C_{[k+1,k+G]} - \hat{V}_{[k+1,k+G]} + \hat{V}^C_{[k-G+1,k]} - \hat{V}_{[k-G+1,k]}) \quad \text{(S.19b)}$$

We first handle the term $\left|\frac{1}{\hat{\sigma}_{k,n}} - \frac{1}{\sigma}\right|$ in (S.18). Under assumption (A2) in the main text, we have $|\hat{\sigma}_{k,n}^2 - \sigma^2| \leq \max_{G \leq j \leq n-G}|\hat{\sigma}_{j,n}^2 - \sigma^2| = o_p\left(\frac{1}{\sqrt{\log(n/G)}}\right) = o_p(1)$. By the continuous mapping theorem, we obtain $|\hat{\sigma}_{k,n} - \sigma| = o_p(1)$. Consequently, we have $\hat{\sigma}_{k,n} = \sigma + o_p(1)$ and $\hat{\sigma}_{k,n}\sigma = \sigma^2 + o_p(1)$. By some straight-forward calculations, we obtain that for $k = G, G+1, \cdots, n-G$

$$\left|\frac{1}{\hat{\sigma}_{k,n}} - \frac{1}{\sigma}\right| = \frac{|\hat{\sigma}_{k,n} - \sigma||\hat{\sigma}_{k,n} + \sigma|}{\hat{\sigma}_{k,n}\sigma|\hat{\sigma}_{k,n} + \sigma|} = \frac{1}{(\sigma^2 + o_p(1))(2\sigma + o_p(1))}|\hat{\sigma}_{k,n}^2 - \sigma^2|$$

Note that $\sigma$ is a constant such that $0 < \sigma < \infty$, then $\frac{1}{(\sigma^2+o_p(1))(2\sigma+o_p(1))} = O_p(1)$. Furthermore, assumption (A2) in the main text implies $|\hat{\sigma}_{k,n}^2 - \sigma^2| = o_p\left(\frac{1}{\sqrt{\log(n/G)}}\right)$. Consequently, it holds that

$$\left|\frac{1}{\hat{\sigma}_{k,n}} - \frac{1}{\sigma}\right| = O_p(1)o_p\left(\frac{1}{\sqrt{\log(n/G)}}\right) = o_p\left(\frac{1}{\sqrt{\log(n/G)}}\right) \quad \text{(S.20)}$$

Next, we handle the term $|J_n^{G,I}(k)|$ in (S.18). We first observe that

$$|J_n^{G,I}(k)| = \sqrt{G}|\hat{V}_{[k+1,k+G]} - \hat{V}_{[k-G+1,k]}|$$
$$= \sqrt{G}|\hat{V}_{[k+1,k+G]} - \tilde{V}_{[k+1,k+G]} + \tilde{V}_{[k+1,k+G]} - \tilde{V}_{[k-G+1,k]} + \tilde{V}_{[k-G+1,k]} - \hat{V}_{[k-G+1,k]}|$$
$$\leq U_n^{G,I}(k) + U_n^{G,II}(k) + U_n^{G,III}(k)$$

where $U_n^{G,I}(k) = \sqrt{G}|\hat{V}_{[k+1,k+G]} - \tilde{V}_{[k+1,k+G]}|$, $U_n^{G,II}(k) = \sqrt{G}|\tilde{V}_{[k+1,k+G]} - \tilde{V}_{[k-G+1,k]}|$ and $U_n^{G,III}(k) = \sqrt{G}|\hat{V}_{[k-G+1,k]} - \tilde{V}_{[k-G+1,k]}|$. From (S.17) in the proof Lemma S.1, it follows that

$$U_n^{G,III}(k) = \sqrt{G}|\hat{V}_{[k-G+1,k]} - \tilde{V}_{[k-G+1,k]}| = O_p\left(\frac{1}{\sqrt{G}}\right) = O_p(1)$$

By following similar arguments, we can also obtain

$$U_n^{G,I}(k) = \sqrt{G}|\hat{V}_{[k+1,k+G]} - \tilde{V}_{[k+1,k+G]}| = O_p(1)$$

For $U_n^{G,II}(k)$, by Lemma S.2, we have

$$U_n^{G,II}(k) = \sqrt{G}|\tilde{V}_{[k+1,k+G]} - \tilde{V}_{[k-G+1,k]}| = O_p(1)$$

Consequently,
$$|J_n^{G,I}(k)| \leq U_n^{G,I}(k) + U_n^{G,II}(k) + U_n^{G,III}(k) = O_p(1) \tag{S.21}$$

Now, we turn to handle the term $|J_n^{G,II}(k)|$ in (S.18). We first observe that
$$\begin{aligned}|J_n^{G,II}(k)| &= \sqrt{G}|\hat{V}_{[k+1,k+G]}^C - \hat{V}_{[k+1,k+G]} + \hat{V}_{[k-G+1,k]}^C - \hat{V}_{[k-G+1,k]}| \\ &\leq \sqrt{G}|\hat{V}_{[k-G+1,k]} - \hat{V}_{[k-G+1,k]}^C| + \sqrt{G}|\hat{V}_{[k+1,k+G]} - \hat{V}_{[k+1,k+G]}^C|\end{aligned} \tag{S.22}$$

In view of (S.14a) and (S.14c), we see under $H_0$ that
$$\tilde{V}_{[k-G+1,k]} = \tilde{V}_{[k-G+1,k]}^C = \frac{1}{G}\sum_{i=k-G+1}^{k} d_W^2(v_i, \mu)$$

Then, it follows that
$$\sqrt{G}|\hat{V}_{[k-G+1,k]} - \hat{V}_{[k-G+1,k]}^C|$$
$$= \sqrt{G}|\hat{V}_{[k-G+1,k]} - \tilde{V}_{[k-G+1,k]} + \tilde{V}_{[k-G+1,k]}^C - \hat{V}_{[k-G+1,k]}^C|$$
$$\leq \sqrt{G}|\hat{V}_{[k-G+1,k]} - \tilde{V}_{[k-G+1,k]}| + \sqrt{G}|\hat{V}_{[k-G+1,k]}^C - \tilde{V}_{[k-G+1,k]}^C|$$

By Lemma S.1, we have
$$\sqrt{G}|\hat{V}_{[k-G+1,k]} - \tilde{V}_{[k-G+1,k]}| = o_p\left(\frac{1}{\sqrt{\log(n/G)}}\right)$$
$$\sqrt{G}|\hat{V}_{[k-G+1,k]}^C - \tilde{V}_{[k-G+1,k]}^C| = o_p\left(\frac{1}{\sqrt{\log(n/G)}}\right)$$

Consequently,
$$\sqrt{G}|\hat{V}_{[k-G+1,k]} - \hat{V}_{[k-G+1,k]}^C| = o_p\left(\frac{1}{\sqrt{\log(n/G)}}\right)$$

By following similar arguments, we can also obtain
$$\sqrt{G}|\hat{V}_{[k+1,k+G]} - \hat{V}_{[k+1,k+G]}^C| = o_p\left(\frac{1}{\sqrt{\log(n/G)}}\right)$$

Combining the above results with (S.22), we get
$$|J_n^{G,II}(k)| = o_p\left(\frac{1}{\sqrt{\log(n/G)}}\right) \tag{S.23}$$

Then, combining (S.18), (S.20), (S.21), and (S.23) together yields
$$|T_n^G(k) - T_n^{\#,G}(k)| = o_p\left(\frac{1}{\sqrt{\log(n/G)}}\right)\left(O_p(1) + o_p\left(\frac{1}{\sqrt{\log(n/G)}}\right)\right)$$
$$= o_p\left(\frac{1}{\sqrt{\log(n/G)}}\right)$$

We can therefore conclude for each $k \in \{G, \cdots, n - G\}$ that

$$T_n^G(k) = T_n^{\#,G}(k) + o_p\left(\frac{1}{\sqrt{\log(n/G)}}\right) \tag{S.24}$$

under $H_0$.

**Step 2:** We next show that $T_n^{\#,G}(k) = \frac{1}{\sqrt{2}\sigma}|\tilde{J}_n^{G,I}(k)| + o_p\left(\frac{1}{\sqrt{\log(n/G)}}\right)$, where $\tilde{J}_n^{G,I}(k)$ is the following oracle version of $J_n^{G,I}(k)$ given in equation (S.19a):

$$\tilde{J}_n^{G,I}(k) = \sqrt{G}\left(\tilde{V}_{[k+1,k+G]} - \tilde{V}_{[k-G+1,k]}\right) \tag{S.25}$$

From (S.15) and (S.19), it follows that

$$T_n^{\#,G}(k) = \frac{1}{\sqrt{2}\sigma}\left(|J_n^{G,I}(k)| + |J_n^{G,II}(k)|\right) \tag{S.26}$$

For $|J_n^{G,II}(k)|$, we have obtained in (S.23) that

$$|J_n^{G,II}(k)| = o_p\left(\frac{1}{\sqrt{\log(n/G)}}\right) \tag{S.27}$$

For $|J_n^{G,I}(k)|$, observe that

$$|J_n^{G,I}(k) - \tilde{J}_n^{G,I}(k)| = |\sqrt{G}(\hat{V}_{[k+1,k+G]} - \hat{V}_{[k-G+1,k]}) - \sqrt{G}(\tilde{V}_{[k+1,k+G]} - \tilde{V}_{[k-G+1,k]})|$$

$$\leq \sqrt{G}|\hat{V}_{[k-G+1,k]} - \tilde{V}_{[k-G+1,k]}| + \sqrt{G}|\hat{V}_{[k+1,k+G]} - \tilde{V}_{[k+1,k+G]}|$$

By Lemma S.1, we have

$$\sqrt{G}|\hat{V}_{[k-G+1,k]} - \tilde{V}_{[k-G+1,k]}| = o_p\left(\frac{1}{\sqrt{\log(n/G)}}\right)$$

$$\sqrt{G}|\hat{V}_{[k+1,k+G]} - \tilde{V}_{[k+1,k+G]}| = o_p\left(\frac{1}{\sqrt{\log(n/G)}}\right)$$

It follows that $|J_n^{G,I}(k) - \tilde{J}_n^{G,I}(k)| = o_p\left(\frac{1}{\sqrt{\log(n/G)}}\right)$. Consequently, $J_n^{G,I}(k)$ can be written as $J_n^{G,I}(k) = \tilde{J}_n^{G,I}(k) + o_p\left(\frac{1}{\sqrt{\log(n/G)}}\right)$. By the continuous mapping theorem, we obtain

$$|J_n^{G,I}(k)| = |\tilde{J}_n^{G,I}(k)| + o_p\left(\frac{1}{\sqrt{\log(n/G)}}\right) \tag{S.28}$$

Note that $\sigma$ is a constant such that $0 < \sigma < \infty$, then from equations (S.26), (S.27), and (S.28), we can therefore conclude that

$$T_n^{\#,G}(k) = \frac{1}{\sqrt{2}\sigma}|\tilde{J}_n^{G,I}(k)| + o_p\left(\frac{1}{\sqrt{\log(n/G)}}\right) \tag{S.29}$$

**Step 3:** Now we are ready to prove the final result of Theorem 1. First, combining the conclusions in the former two steps (i.e., equations (S.24) and (S.29)), we obtain for $k = G, G + 1, \cdots, n - G$ that

$$T_n^G(k) = T_n^{\#,G}(k) + o_p\left(\frac{1}{\sqrt{\log(n/G)}}\right)$$

$$= \frac{1}{\sqrt{2}\sigma}|\tilde{J}_n^{G,I}(k)| + o_p\left(\frac{1}{\sqrt{\log(n/G)}}\right) + o_p\left(\frac{1}{\sqrt{\log(n/G)}}\right)$$

$$= \frac{1}{\sqrt{2}\sigma}|\tilde{J}_n^{G,I}(k)| + o_p\left(\frac{1}{\sqrt{\log(n/G)}}\right)$$

Recall that the test statistic is $T_n(G) = \max_{G \leq k \leq n-G} T_n^G(k)$, see equation (4) in the main text. Since $\gamma_1(n/G)$ and $\gamma_2(n/G)$ in Theorem 1 satisfy $\gamma_1(n/G) \geq 0$ and $\gamma_2(n/G) \geq 0$, then we have

$$\gamma_1(n/G)T_n(G) - \gamma_2(n/G) = \gamma_1(n/G)\max_{G \leq k \leq n-G} T_n^G(k) - \gamma_2(n/G)$$

$$= \max_{G \leq k \leq n-G}\{\gamma_1(n/G)T_n^G(k) - \gamma_2(n/G)\}$$

$$= \max_{G \leq k \leq n-G}\left\{\gamma_1(n/G)\left(\frac{1}{\sqrt{2}\sigma}|\tilde{J}_n^{G,I}(k)| + o_p\left(\frac{1}{\sqrt{\log(n/G)}}\right)\right) - \gamma_2(n/G)\right\}$$

$$= \max_{G \leq k \leq n-G}\left\{\frac{\gamma_1(n/G)}{\sqrt{2}\sigma}|\tilde{J}_n^{G,I}(k)| - \gamma_2(n/G) + \gamma_1(n/G)o_p\left(\frac{1}{\sqrt{\log(n/G)}}\right)\right\}$$

Recall that the expression of $\gamma_1(n/G)$ is $\gamma_1(n/G) = \sqrt{2\log(n/G)}$, thus the above result becomes

$$\gamma_1(n/G)T_n(G) - \gamma_2(n/G) = \max_{G \leq k \leq n-G}\left\{\frac{\gamma_1(n/G)}{\sqrt{2}\sigma}|\tilde{J}_n^{G,I}(k)| - \gamma_2(n/G) + o_p(1)\right\} \quad (S.30)$$

For $k \in \{G, G+1, \cdots, n-G\}$, define $u = k/n$. Denote with $c_G = G/n$, then $u \in I_c^G = [c_G, 1-c_G] \subset [0,1]$. We further define $\beta(u) = \frac{\gamma_1(n/G)}{\sqrt{2}\sigma}|\tilde{J}_n^{G,I}(\lfloor nu \rfloor)| - \gamma_2(n/G)$ for $u \in I_c^G$. Then $\{\beta(u): u \in I_c^G\}$ is a stochastic process indexed by $I_c^G = [c_G, 1-c_G]$. By following similar arguments as in the proof of Theorem 1 of Dubey and Müller (2020), we can prove that $\{\beta(u): u \in I_c^G\}$ is asymptotically equicontinuous in $u$. This implies the weak convergence of the stochastic process $\{\beta(u) - (\beta(u) + o_p(1)): u \in I_c^G\}$ to zero. Consequently, $\{\beta(u) - (\beta(u) + o_p(1)): u \in I_c^G\}$ convergence to zero in probability, which implies that

$$\{(\beta(u) + o_p(1)): u \in I_c^G\} \xrightarrow{P} \{\beta(u): u \in I_c^G\}$$

where $\xrightarrow{P}$ denotes convergence in probability. By the continuous mapping theorem, we have $\sup_{u \in I_c^G}(\beta(u) + o_p(1)) \xrightarrow{P} \sup_{u \in I_c^G} \beta(u)$. In view of $\beta(u) = \frac{\gamma_1(n/G)}{\sqrt{2}\sigma}|\tilde{J}_n^{G,I}(\lfloor nu \rfloor)| - \gamma_2(n/G)$ and $I_c^G = [c_G, 1-c_G]$ with $c_G = G/n$, the above result implies that

$$\max_{G \leq k \leq n-G}\left\{\frac{\gamma_1(n/G)}{\sqrt{2}\sigma}|\tilde{J}_n^{G,I}(k)| - \gamma_2(n/G) + o_p(1)\right\} \xrightarrow{P} \max_{G \leq k \leq n-G}\left\{\frac{\gamma_1(n/G)}{\sqrt{2}\sigma}|\tilde{J}_n^{G,I}(k)| - \gamma_2(n/G)\right\}$$

Combination with (S.30) yields

$$\gamma_1(n/G)T_n(G) - \gamma_2(n/G) \xrightarrow{P} \max_{G \leq k \leq n-G}\left\{\frac{\gamma_1(n/G)}{\sqrt{2}\sigma}\left|\tilde{J}_n^{G,I}(k)\right| - \gamma_2(n/G)\right\}$$

Also due to $\gamma_1(n/G), \gamma_2(n/G) \geq 0$, the above result can be equivalently written as

$$\gamma_1(n/G)T_n(G) - \gamma_2(n/G) \xrightarrow{P} \gamma_1(n/G)\max_{G \leq k \leq n-G}\left\{\frac{1}{\sqrt{2}\sigma}\left|\tilde{J}_n^{G,I}(k)\right|\right\} - \gamma_2(n/G) \quad (S.31)$$

For $\frac{1}{\sqrt{2}\sigma}\left|\tilde{J}_n^{G,I}(k)\right|$, from equations (S.14a), (S.14b), and (S.25), it follows that

$$\frac{1}{\sqrt{2}\sigma}\left|\tilde{J}_n^{G,I}(k)\right| = \frac{1}{\sqrt{2}\sigma}\left|\sqrt{G}\left(\tilde{V}_{[k+1,k+G]} - \tilde{V}_{[k-G+1,k]}\right)\right|$$

$$= \frac{1}{\sigma}\frac{1}{\sqrt{2G}}\left|\sum_{i=k+1}^{k+G}d_\mathcal{W}^2(\nu_i,\mu) - \sum_{i=k-G+1}^{k}d_\mathcal{W}^2(\nu_i,\mu)\right|$$

$$:= \frac{1}{\sigma}\frac{1}{\sqrt{2G}}\left|\sum_{i=k+1}^{k+G}Z_i - \sum_{i=k-G+1}^{k}Z_i\right|, \quad k = G,\cdots,(n-G)$$

where $Z_i := d_\mathcal{W}^2(\nu_i,\mu)$, $i = G,\cdots,n-G$. Under $H_0$, $\nu_i$s are i.i.d. $\mathcal{W}_2(\mathcal{D})$-valued random objects and $\mu = \underset{\omega \in \mathcal{W}_2(D)}{\arg\min} E\left(d_\mathcal{W}^2(\nu_1,\omega)\right)$ is a deterministic but unknown $\mathcal{W}_2(\mathcal{D})$-valued object. Consequently, $Z_i$s are i.i.d. real-valued random variables. Under the boundedness assumption for $\mathcal{W}_2(\mathcal{D})$, namely $\underset{\nu_1,\nu_2 \in \mathcal{W}_2(D)}{\sup} d_\mathcal{W}(\nu_1,\nu_2) < \infty$ (Section 3.1 of the main text), we have that, for any $\Delta > 0$,

$$E|Z_i|^{2+\Delta} = E|d_\mathcal{W}^2(\nu_i,\mu)|^{2+\Delta} < \infty \quad (S.32)$$

Furthermore, under assumption (A1) in the main text, we have $n^{\frac{2}{2+\Delta}}\log n/G \to 0$ and $n/G \to \infty$, as $n \to \infty$. Then, it follows that

$$\frac{G}{n}\log(n/G) = \frac{-\log(G/n)}{n/G} \to 0 \quad (S.33a)$$

$$\frac{G}{n^{2/(2+\Delta)}} \geq \frac{G}{n^{2/(2+\Delta)}\log n} = \frac{1}{n^{\frac{2}{2+\Delta}}\log n/G} \to \infty \quad (S.33b)$$

as $n \to \infty$.

The results in (S.32) and (S.33) imply that the i.i.d. real-valued random variables $Z_i = d_\mathcal{W}^2(\nu_i,\mu)$, $i = G,\cdots,n-G$, satisfy the conditions of Theorem 2.1 of Hušková and Slabý (2001), under $H_0$. By Theorem 2.1 of Hušková and Slabý (2001), we have

$$\gamma_1(n/G)\max_{G \leq k \leq n-G}\left\{\frac{1}{\sqrt{2}\sigma}\left|\tilde{J}_n^{G,I}(k)\right|\right\} - \gamma_2(n/G) \xrightarrow{d} \Gamma_{Gum} \quad (S.34)$$

where $\xrightarrow{d}$ denotes convergence in distribution, and $\Gamma_{Gum}$ denotes a random variable that possesses the distribution function $P(\Gamma_{Gum} \leq x) = \exp(-2\exp(-x))$. Combining (S.34) with (S.31), we can therefore conclude that

$$\gamma_1(n/G)T_n(G) - \gamma_2(n/G) \xrightarrow{d} \Gamma_{Gum}$$

Consequently,
$$\lim_{n\to\infty} P(\gamma_1(n/G)T_n(G) - \gamma_2(n/G) \leq x) = \exp(-2\exp(-x))$$

This completes the proof.

### S.3.2. Proof of Theorem 2

***Proof of Theorem 2.*** In view of the well-known probability inequality $P(AB) \geq 1 - P(A^C) - P(B^C)$ (Boole inequality) (Lin and Bai, 2010), we see that

$$P\left(\hat{q}_n = q, \max_{1\leq j\leq q}|\hat{k}_j^* - k_j^*| < G\right) \\ \geq 1 - P(\hat{q}_n \neq q) - P\left(\max_{1\leq j\leq q}|\hat{k}_j^* - k_j^*| \geq G\right) \quad \text{(S.35)}$$

Consequently, to prove $P\left(\hat{q}_n = q, \max_{1\leq j\leq q}|\hat{k}_j^* - k_j^*| < G\right) \to 1$, as $n \to \infty$, it is sufficient to show that

$$P(\hat{q}_n \neq q) \to 0 \text{ and } P\left(\max_{1\leq j\leq q}|\hat{k}_j^* - k_j^*| \geq G\right) \to 0 \quad \text{(S.36)}$$

as $n \to \infty$.

We next present a sufficient condition for $P(\hat{q}_n \neq q) \to 0$. To this end, similar to the proof of Theorem 6.1 of Muhsal (2013), we define the following two sets:

$$B_{G,q} := \{k \in \{1,2,\cdots,n\}: \exists k_j^* \in \{k_1^*,\cdots,k_q^*\} \text{ such that } |k - k_j^*| \leq (1-\varepsilon)G\} \quad \text{(S.37a)}$$

$$A_{G,q} := \{k \in \{1,2,\cdots,n\}: |k - k^*| \geq G \text{ for any } k^* \in \{k_0^*, k_1^*, \cdots, k_q^*, k_{q+1}^*\}\} \quad \text{(S.37b)}$$

where $k_1^*,\cdots,k_q^*$ are the $q$ different change points such that $0 = k_0^* < k_1^* < k_2^* < \cdots < k_q^* < k_{q+1}^* = n$, $G$ is the bandwidth, and $\varepsilon$ is the AOP parameter given in (8) of the main text. Here, $k_0^* = 0$ and $k_{q+1}^* = n$ denote the two boundary points of the distributional sequence. $B_{G,q}$ in (S.37a) is the same as that defined in equation (10) of the main text. Let $T_n^G(k)$ be the scan statistic given in equation (5) of the main text, and $D_n(G, \alpha_n)$ be the critical value (see (8) of the main text) of the Fréchet-MOSUM change-point estimator for a given significance level $\alpha_n$ that fulfills condition (C3) in the main text. Then, similar to that in the proof of Theorem 6.1 of Muhsal (2013), $\min_{k \in B_{G,q}} T_n^G(k) \geq D_n(G, \alpha_n)$ ensures that the Fréchet-MOSUM change-point estimator can detect at least $q$ change points, while $\max_{k \in A_{G,q}} T_n^G(k) < D_n(G, \alpha_n)$ ensures the detection of at most $q$ change points. Consequently,

$$\left\{\min_{k \in B_{G,q}} T_n^G(k) \geq D_n(G, \alpha_n), \max_{k \in A_{G,q}} T_n^G(k) < D_n(G, \alpha_n)\right\} \text{ implies } \{\hat{q}_n = q\}$$

Thus
$$P\left(\min_{k\in B_{G,q}} T_n^G(k) \geq D_n(G,\alpha_n), \max_{k\in A_{G,q}} T_n^G(k) < D_n(G,\alpha_n)\right) \leq P(\hat{q}_n = q)$$

Also, in view of the Boole inequality $P(AB) \geq 1 - P(A^C) - P(B^C)$, we have

$$P(\hat{q}_n = q) \geq 1 - P\left(\min_{k\in B_{G,q}} T_n^G(k) < D_n(G,\alpha_n)\right) - P\left(\max_{k\in A_{G,q}} T_n^G(k) \geq D_n(G,\alpha_n)\right)$$

It follows that

$$P(\hat{q}_n \neq q) = 1 - P(\hat{q}_n = q)$$
$$\leq P\left(\min_{k\in B_{G,q}} T_n^G(k) < D_n(G,\alpha_n)\right) + P\left(\max_{k\in A_{G,q}} T_n^G(k) \geq D_n(G,\alpha_n)\right) \quad \text{(S.38)}$$

Therefore, a sufficient condition for $P(\hat{q}_n \neq q) \to 0$ is

$$P\left(\min_{k\in B_{G,q}} T_n^G(k) < D_n(G,\alpha_n)\right) \to 0 \text{ and } P\left(\max_{k\in A_{G,q}} T_n^G(k) \geq D_n(G,\alpha_n)\right) \to 0, \text{ as } n \to \infty$$

Combining this with (S.36), we see, to prove the result of Theorem 2, it is sufficient to show under $H_A$ that

$$P\left(\min_{k\in B_{G,q}} T_n^G(k) < D_n(G,\alpha_n)\right) \to 0 \quad \text{(S.39a)}$$

$$P\left(\max_{k\in A_{G,q}} T_n^G(k) \geq D_n(G,\alpha_n)\right) \to 0 \quad \text{(S.39b)}$$

$$P\left(\max_{1\leq j\leq q}|\hat{k}_j^* - k_j^*| \geq G\right) \to 0 \quad \text{(S.39c)}$$

as $n \to \infty$.

Consequently, the proof of Theorem 2 can be split into three steps, and each step proves one condition of (S.39).

***Step 1:*** We first prove under $H_A$ that

$$P\left(\min_{k\in B_{G,q}} T_n^G(k) < D_n(G,\alpha_n)\right) \to 0, \text{ as } n \to \infty \quad \text{(S.40)}$$

For $k = G, G+1, \cdots, n-G$, we define

$$S_n^G(k) = (1/\hat{\sigma}_{k,n})|\hat{V}_{[k+1,k+G]} - \hat{V}_{[k-G+1,k]}| \\ + (1/\hat{\sigma}_{k,n})|\hat{V}_{[k+1,k+G]}^C - \hat{V}_{[k+1,k+G]} + \hat{V}_{[k-G+1,k]}^C - \hat{V}_{[k-G+1,k]}| \quad \text{(S.41a)}$$

$$S^G(k) = (1/\sigma_k)|V_{[k+1,k+G]} - V_{[k-G+1,k]}| \\ + (1/\sigma_k)|V_{[k+1,k+G]}^C - V_{[k+1,k+G]} + V_{[k-G+1,k]}^C - V_{[k-G+1,k]}| \quad \text{(S.41b)}$$

where $\hat{V}_{[k+1,k+G]}$, $\hat{V}_{[k-G+1,k]}$, $\hat{V}^C_{[k+1,k+G]}$, $\hat{V}^C_{[k-G+1,k]}$, and $\hat{\sigma}_{k,n}$ are the same as their counterparts in equation (5) of the main text, and $V_{[k+1,k+G]}$, $V_{[k-G+1,k]}$, $V^C_{[k+1,k+G]}$, $V^C_{[k-G+1,k]}$, and $\sigma_k$ are the same as their counterparts in equation (11) of the main text. Then, $T_n^G(k)$ and $T^G(k)$ given in (5) and (11) in the main text can be rewritten as

$$T_n^G(k) = \sqrt{G/2}\, S_n^G(k), \quad k = G, G+1, \cdots, n-G \qquad (S.42a)$$
$$T^G(k) = \sqrt{G/2}\, S^G(k), \quad k = G, G+1, \cdots, n-G \qquad (S.42b)$$

respectively.

For notational convenience, we further define

$$\begin{aligned}\hat{\psi}_G(k) &= \hat{V}_{[k+1,k+G]} - \hat{V}_{[k-G+1,k]} \\ \hat{\lambda}_G(k) &= \hat{V}^C_{[k+1,k+G]} - \hat{V}_{[k+1,k+G]} + \hat{V}^C_{[k-G+1,k]} - \hat{V}_{[k-G+1,k]}\end{aligned} \qquad (S.43a)$$

$$\begin{aligned}\psi_G(k) &= V_{[k+1,k+G]} - V_{[k-G+1,k]} \\ \lambda_G(k) &= V^C_{[k+1,k+G]} - V_{[k+1,k+G]} + V^C_{[k-G+1,k]} - V_{[k-G+1,k]}\end{aligned} \qquad (S.43b)$$

In order to prove (S.40), we need an additional lemma:

**Lemma S.3** Under the conditions of Theorem 2, it holds that

$$\max_{G \le k \le n-G} |\hat{\psi}_G(k) - \psi_G(k)| = O_p\left(G^{-\frac{1}{2}}\right), \quad \text{under } H_A \qquad (S.44a)$$

$$\max_{G \le k \le n-G} |\hat{\lambda}_G(k) - \lambda_G(k)| = O_p\left(G^{-\frac{1}{2+\beta}}\right), \quad \text{under } H_A \qquad (S.44b)$$

for some $0 < \beta < 2$.

***Proof of Lemma S.3.*** The proof proceeds analogously to those of Lemmas 4–5 of Dubey and Müller (2020).

Given the distributional sequence $\Gamma = \{v_1, \cdots, v_n\}$ with $q$ unknown change points $k_1^*, k_2^*, \cdots, k_q^*$ such that $1 < k_1^* < k_2^* < \cdots < k_q^* < n$. As detailed in Section S.2.4, the forms of $V_{[k-G+1,k]}$, $V_{[k+1,k+G]}$, $V^C_{[k-G+1,k]}$, and $V^C_{[k+1,k+G]}$ in situation I ($k_1^* < k \le k_q^*$) are slightly different from their counterparts in situation II ($G \le k < k_1^*$ or $k_q^* < k \le n - G$). Here, we first address the case of $k_1^* < k \le k_q^*$ (i.e., situation I), the result of $G \le k < k_1^*$ or $k_q^* < k \le n - G$ (i.e., situation II) can be obtained by using the similar tactics.

For situation I, due to $k_1^* < k \le k_q^*$ and $k_1^* < k_2^* < \cdots < k_q^*$

there exists $j \in \{1, \cdots, q-1\}$ such that $k \in (k_j^*, k_{j+1}^*]$ \qquad (S.45)

Let $IDS(k) = \{j \in \{1, \cdots, n\}: k - G + 1 \le j \le k + G\}$ be the indexing set defined in Section S.2.4.2. According to the analysis in Section S.2.4.2, $IDS(k)$ contains at most one change point ($k_j^*$ or $k_{j+1}^*$) for $n$ sufficiently large. Then, from (S.45), it follows that one of the following cases must hold for $n$ sufficiently large:

**Case 1**: $IDS(k)$ contains no change point;

**Case 2**: $IDS(k)$ contains the change point $k_j^*$;

**Case 3**: $IDS(k)$ contains the change point $k_{j+1}^*$.

Accordingly, we define the following three events:

$$A_1 = \{(k, G) \in (k_j^*, k_{j+1}^*] \times \mathbb{Z}^+ \text{ such that case 1 happen}\}$$

$$A_2 = \{(k, G) \in (k_j^*, k_{j+1}^*] \times \mathbb{Z}^+ \text{ such that case 2 happen}\}$$

$$A_3 = \{(k, G) \in (k_j^*, k_{j+1}^*] \times \mathbb{Z}^+ \text{ such that case 3 happen}\}$$

where $\mathbb{Z}^+$ denotes the set of positive integers. From the analysis in Section S.2.4.2, we can deduce that $A_1$, $A_2$, and $A_3$ are asymptotically mutually exclusive, and the associated indicator functions satisfy

$$I\{(k, G) \in A_1\} + I\{(k, G) \in A_2\} + I\{(k, G) \in A_3\} = 1 \tag{S.46}$$

Then, according to the expressions for $\mu_{[k-G+1,k]}$ in cases 1–3 provided in Section S.2.4.2, we can write $\mu_{[k-G+1,k]}$ as

$$\mu_{[k-G+1,k]} = \mu_{j+1} I\{(k, G) \in A_1\}$$

$$+ \underset{\omega \in \mathcal{W}_2(D)}{\arg\min} \left\{ \frac{k_j^* - (k-G)}{G} E_{P_j} d_\mathcal{W}^2(\theta_j, \omega) + \frac{k - k_j^*}{G} E_{P_{j+1}} d_\mathcal{W}^2(\theta_{j+1}, \omega) \right\} I\{(k, G) \in A_2\}$$

$$+ \mu_{j+1} I\{(k, G) \in A_3\}$$

(S.47)

Similarly, $\mu_{[k+1,k+G]}$, $V_{[k-G+1,k]}$, $V_{[k+1,k+G]}$, $V_{[k-G+1,k]}^C$, and $V_{[k+1,k+G]}^C$ can be equivalently written as

$$\mu_{[k+1,k+G]} = \mu_{j+1} I\{(k, G) \in A_1\}$$

$$+ \mu_{j+1} I\{(k, G) \in A_2\}$$

$$+ \underset{\omega \in \mathcal{W}_2(D)}{\arg\min} \left\{ \frac{k_{j+1}^* - k}{G} E_{P_{j+1}} d_\mathcal{W}^2(\theta_{j+1}, \omega) + \frac{k + G - k_{j+1}^*}{G} E_{P_{j+2}} d_\mathcal{W}^2(\theta_{j+2}, \omega) \right\} I\{(k, G) \in A_3\}$$

(S.48)

$$V_{[k-G+1,k]} = E_{P_{j+1}} \left( d_\mathcal{W}^2(\theta_{j+1}, \mu_{j+1}) \right) I\{(k, G) \in A_1\}$$

$$+ \left\{ \frac{k_j^* - (k-G)}{G} E_{P_j} d_\mathcal{W}^2(\theta_j, \mu_{[k-G+1,k]}) + \frac{k - k_j^*}{G} E_{P_{j+1}} d_\mathcal{W}^2(\theta_{j+1}, \mu_{[k-G+1,k]}) \right\} I\{(k, G) \in A_2\}$$

$$+ E_{P_{j+1}} \left( d_\mathcal{W}^2(\theta_{j+1}, \mu_{j+1}) \right) I\{(k, G) \in A_3\}$$

(S.49)

$$V_{[k+1,k+G]} = E_{P_{j+1}} \left( d_\mathcal{W}^2(\theta_{j+1}, \mu_{j+1}) \right) I\{(k, G) \in A_1\}$$

$$+ E_{P_{j+1}} \left( d_\mathcal{W}^2(\theta_{j+1}, \mu_{j+1}) \right) I\{(k, G) \in A_2\}$$

$$+\left\{\frac{k_{j+1}^* - k}{G}E_{P_{j+1}}d_\mathcal{W}^2(\theta_{j+1}, \mu_{[k+1,k+G]}) + \frac{k+G-k_{j+1}^*}{G}E_{P_{j+2}}d_\mathcal{W}^2(\theta_{j+2}, \mu_{[k+1,k+G]})\right\}I\{(k,G) \in A_3\}$$
(S.50)

$$V_{[k-G+1,k]}^C = E_{P_{j+1}}\left(d_\mathcal{W}^2(\theta_{j+1}, \mu_{j+1})\right)I\{(k,G) \in A_1\}$$
$$+\left\{\frac{k_j^* - (k-G)}{G}E_{P_j}d_\mathcal{W}^2(\theta_j, \mu_{j+1}) + \frac{k-k_j^*}{G}E_{P_{j+1}}d_\mathcal{W}^2(\theta_{j+1}, \mu_{j+1})\right\}I\{(k,G) \in A_2\}$$
$$+E_{P_{j+1}}\left(d_\mathcal{W}^2(\theta_{j+1}, \mu_{[k+1,k+G]})\right)I\{(k,G) \in A_3\}$$
(S.51)

$$V_{[k+1,k+G]}^C = E_{P_{j+1}}\left(d_\mathcal{W}^2(\theta_{j+1}, \mu_{j+1})\right)I\{(k,G) \in A_1\}$$
$$+E_{P_{j+1}}\left(d_\mathcal{W}^2(\theta_{j+1}, \mu_{[k-G+1,k]})\right)I\{(k,G) \in A_2\}$$
$$+\left\{\frac{k_{j+1}^* - k}{G}E_{P_{j+1}}d_\mathcal{W}^2(\theta_{j+1}, \mu_{j+1}) \frac{k+G-k_{j+1}^*}{G}E_{P_{j+2}}d_\mathcal{W}^2(\theta_{j+2}, \mu_{j+1})\right\}I\{(k,G) \in A_3\}$$
(S.52)

On the other hand, it is noteworthy that no matter whether it is under $H_0$ or under $H_A$, the empirical versions of $\mu_{[k-G+1,k]}$ and $\mu_{[k+1,k+G]}$ are obtained as follows:

$$\hat{\mu}_{[k-G+1,k]} = \underset{\omega \in \mathcal{W}_2(D)}{\operatorname{argmin}} \frac{1}{G}\sum_{i=k-G+1}^{k} d_\mathcal{W}^2(\nu_i, \omega) \tag{S.53a}$$

$$\hat{\mu}_{[k+1,k+G]} = \underset{\omega \in \mathcal{W}_2(D)}{\operatorname{argmin}} \frac{1}{G}\sum_{i=k+1}^{k+G} d_\mathcal{W}^2(\nu_i, \omega) \tag{S.53b}$$

To prove the result of Lemma S.3, we first prove the following two important assertions:

$$\max_{G \le k \le n-G} d_\mathcal{W}(\hat{\mu}_{[k-G+1,k]}, \mu_{[k-G+1,k]}) = O_p(G^{-\frac{1}{2+\beta}}), \text{ under } H_A \tag{S.54a}$$

$$\max_{G \le k \le n-G} d_\mathcal{W}(\hat{\mu}_{[k+1,k+G]}, \mu_{[k+1,k+G]}) = O_p(G^{-\frac{1}{2+\beta}}), \text{ under } H_A \tag{S.54b}$$

for some $0 < \beta < 2$.

Here, we only provide details for the proof of assertion (S.54a), assertion (S.54b) can be proved in a similar way.

For $k_1^* < k \le k_q^*$, we define

$$M_G(\omega, k) = \frac{1}{G}\sum_{i=k-G+1}^{k} d_\mathcal{W}^2(\nu_i, \omega)$$

and

$$\widetilde{M}_G(\omega, k) = E_{P_{j+1}}\left(d_\mathcal{W}^2(\theta_{j+1}, \omega)\right)I\{(k,G) \in A_1\}$$

$$+\left\{\frac{k_j^* - (k-G)}{G}E_{P_j}d_{\mathcal{W}}^2(\theta_j, \omega) + \frac{k - k_j^*}{G}E_{P_{j+1}}d_{\mathcal{W}}^2(\theta_{j+1}, \omega)\right\}I\{(k,G) \in A_2\}$$

$$+E_{P_{j+1}}\left(d_{\mathcal{W}}^2(\theta_{j+1}, \omega)\right)I\{(k,G) \in A_3\}$$

From equation (S.53a), we have $\hat{\mu}_{[k-G+1,k]} = \underset{\omega \in \mathcal{W}_2(D)}{\operatorname{argmin}} M_G(\omega, k)$. Recall that $\mu_{j+1} = \underset{\omega \in \mathcal{W}_2(D)}{\operatorname{argmin}} E_{P_{j+1}}\left(d_{\mathcal{W}}^2(\theta_{j+1}, \omega)\right)$, see equation (S.7); then, from (S.47) we have $\mu_{[k-G+1,k]} = \underset{\omega \in \mathcal{W}_2(D)}{\operatorname{argmin}} \widetilde{M}_G(\omega, k)$.

In view of (S.46), the expression for $M_G(\omega, k)$ can be decomposed as

$$M_G(\omega, k) = \frac{1}{G}\sum_{i=k-G+1}^{k} d_{\mathcal{W}}^2(\nu_i, \omega)$$

$$= \frac{1}{G}\sum_{i=k-G+1}^{k} d_{\mathcal{W}}^2(\nu_i, \omega)\, I\{(k,G) \in A_1\}$$

$$+ \frac{1}{G}\sum_{i=k-G+1}^{k} d_{\mathcal{W}}^2(\nu_i, \omega)\, I\{(k,G) \in A_2\}$$

$$+ \frac{1}{G}\sum_{i=k-G+1}^{k} d_{\mathcal{W}}^2(\nu_i, \omega)\, I\{(k,G) \in A_3\}$$

Then, we have

$$\left|M_G(\omega, k) - \widetilde{M}_G(\omega, k)\right|$$

$$\leq \left|\frac{1}{G}\sum_{i=k-G+1}^{k} d_{\mathcal{W}}^2(\nu_i, \omega) - E_{P_{j+1}}\left(d_{\mathcal{W}}^2(\theta_{j+1}, \omega)\right)\right| I\{(k,G) \in A_1\}$$

$$+ \left|\frac{1}{G}\sum_{i=k-G+1}^{k} d_{\mathcal{W}}^2(\nu_i, \omega) - \left\{\frac{k_j^* - (k-G)}{G}E_{P_j}d_{\mathcal{W}}^2(\theta_j, \omega) + \frac{k - k_j^*}{G}E_{P_{j+1}}d_{\mathcal{W}}^2(\theta_{j+1}, \omega)\right\}\right| I\{(k,G) \in A_2\}$$

$$+ \left|\frac{1}{G}\sum_{i=k-G+1}^{k} d_{\mathcal{W}}^2(\nu_i, \omega) - E_{P_{j+1}}\left(d_{\mathcal{W}}^2(\theta_{j+1}, \omega)\right)\right| I\{(k,G) \in A_3\}$$

$$:= H_G^{I}(\omega, k) + H_G^{II}(\omega, k) + H_G^{III}(\omega, k)$$

From the analysis in Section S.2.4, we have $\nu_{k-G+1}, \cdots, \nu_k \overset{i.i.d}{\sim} P_{j+1}$ when event $A_1$ or $A_3$ happens. Consequently, by the weak law of large numbers, it holds for each $\omega \in \mathcal{W}_2(D)$

$$H_G^{I}(\omega, k) = \left|\frac{1}{G}\sum_{i=k-G+1}^{k} d_{\mathcal{W}}^2(\nu_i, \omega) - E_{P_{j+1}}\left(d_{\mathcal{W}}^2(\theta_{j+1}, \omega)\right)\right| I\{(k,G) \in A_1\}$$

$$\leq \left|\frac{1}{G}\sum_{i=k-G+1}^{k} d_{\mathcal{W}}^2(\nu_i, \omega) - E_{P_{j+1}}\left(d_{\mathcal{W}}^2(\theta_{j+1}, \omega)\right)\right| = o_p(1)$$

and

$$H_G^{III}(\omega, k) = \left|\frac{1}{G}\sum_{i=k-G+1}^{k} d_{\mathcal{W}}^2(\nu_i, \omega) - E_{P_{j+1}}\left(d_{\mathcal{W}}^2(\theta_{j+1}, \omega)\right)\right| I\{(k,G) \in A_3\} = o_p(1)$$

For $H_G^{II}(\omega, k)$, we first observe that

$$\frac{1}{G}\sum_{i=k-G+1}^{k}(\cdot) = \frac{1}{G}\sum_{i=k-G+1}^{k_j^*}(\cdot) + \frac{1}{G}\sum_{i=k_j^*+1}^{k}(\cdot)$$

$$= \frac{k_j^* - (k-G)}{G}\frac{1}{k_j^* - (k-G)}\sum_{i=k-G+1}^{k_j^*}(\cdot) + \frac{k - k_j^*}{G}\frac{1}{k - k_j^*}\sum_{i=k_j^*+1}^{k}(\cdot)$$

Additionally, on noting that $\frac{|k_j^* - (k-G)|}{G} \leq 1$ and $\frac{|k - k_j^*|}{G} \leq 1$, we have

$$H_G^{II}(\omega, k) = \left|\frac{1}{G}\sum_{i=k-G+1}^{k} d_{\mathcal{W}}^2(\nu_i, \omega) - \right.$$

$$\left.\left\{\frac{k_j^* - (k-G)}{G} E_{P_j} d_{\mathcal{W}}^2(\theta_j, \omega) + \frac{k - k_j^*}{G} E_{P_{j+1}} d_{\mathcal{W}}^2(\theta_{j+1}, \omega)\right\}\right| I\{(k, G) \in A_2\}$$

$$\leq \left|\frac{1}{k_j^* - (k-G)}\sum_{i=k-G+1}^{k_j^*} d_{\mathcal{W}}^2(\nu_i, \omega) - E_{P_j} d_{\mathcal{W}}^2(\theta_j, \omega)\right| I\{(k, G) \in A_2\}$$

$$+ \left|\frac{1}{k - k_j^*}\sum_{i=k_j^*+1}^{k} d_{\mathcal{W}}^2(\nu_i, \omega) - E_{P_{j+1}} d_{\mathcal{W}}^2(\theta_{j+1}, \omega)\right| I\{(k, G) \in A_2\}$$

From the analysis in Section S.2.4, we have $\nu_{k-G+1}, \cdots, \nu_{k_j^*} \stackrel{i.i.d}{\sim} P_j$ and $\nu_{k_j^*+1}, \cdots, \nu_k \stackrel{i.i.d}{\sim} P_{j+1}$ when event $A_2$ happens. Then, by the weak law of large numbers, we have for each $\omega \in \mathcal{W}_2(D)$, $H_G^{II}(\omega, k) = o_p(1)$. Combining the results obtained above, we get

$$\left|M_G(\omega, k) - \widetilde{M}_G(\omega, k)\right| = o_p(1) \text{ for each } \omega \in \mathcal{W}_2(D) \tag{S.55}$$

Under the boundedness assumption for $\mathcal{W}_2(\mathcal{D})$ (see Section 3.1 of the main text), we can show that the stochastic process $\{M_G(\omega, k): \omega \in \mathcal{W}_2(D)\}$ indexed by $\mathcal{W}_2(D)$ is asymptotically equicontinuous in $\omega$ by adopting similar arguments as in the proof of Lemma 1 of Dubey and Müller (2020). Combining this with the pointwise convergence conclusion given in (S.55), we can conclude that the stochastic process $\{M_G(\omega, k) - \widetilde{M}_G(\omega, k): \omega \in \mathcal{W}_2(D)\}$ converges weakly to zero. This implies that $\{M_G(\omega, k) - \widetilde{M}_G(\omega, k): \omega \in \mathcal{W}_2(D)\}$ converges to zero in probability. Consequently, by the continuous mapping theorem, we have

$$\sup_{\omega \in \mathcal{W}_2(D)} \left|M_G(\omega, k) - \widetilde{M}_G(\omega, k)\right| = o_p(1) \tag{S.56}$$

We next use $\sup_{\omega \in \mathcal{W}_2(D)} \left|M_G(\omega, k) - \widetilde{M}_G(\omega, k)\right| = o_p(1)$ to prove, for each $k \in (k_1^*, k_q^*]$, $d_{\mathcal{W}}(\hat{\mu}_{[k-G+1,k]}, \mu_{[k-G+1,k]}) = o_p(1)$. For this purpose, we first show that $\mu_{[k-G+1,k]} = \underset{\omega \in \mathcal{W}_2(D)}{\mathrm{argmin}}\, \widetilde{M}_G(\omega, k)$ is a well-separated minimizer of the process $\omega \mapsto \widetilde{M}_G(\omega, k)$, namely it holds for any $\delta > 0$ that

$$\widetilde{M}_G\big(\mu_{[k-G+1,k]}, k\big) < \inf_{\omega \in \mathcal{W}_2(D): d_{\mathcal{W}}(\omega, \mu_{[k-G+1,k]}) \geq \delta} \widetilde{M}_G(\omega, k) \tag{S.57}$$

Such a well-separated condition is similar to that in Lemma 3.2.1 of Van der Vaart and Wellner (1996) for the argmax problem.

Recall that

$$\mu_{[k-G+1,k]} = \underset{\omega \in \mathcal{W}_2(D)}{\operatorname{argmin}} \widetilde{M}_G(\omega, k)$$

$$= \underset{\omega \in \mathcal{W}_2(D)}{\operatorname{argmin}} E_{P_{j+1}}\left(d_{\mathcal{W}}^2(\theta_{j+1}, \omega)\right) I\{(k, G) \in A_1 \cup A_3\} \tag{S.58}$$

$$+ \underset{\omega \in \mathcal{W}_2(D)}{\operatorname{argmin}} \left\{ \frac{k_j^* - (k-G)}{G} E_{P_j} d_{\mathcal{W}}^2(\theta_j, \omega) + \frac{k - k_j^*}{G} E_{P_{j+1}} d_{\mathcal{W}}^2(\theta_{j+1}, \omega) \right\} I\{(k, G) \in A_2\}$$

For the case of $(k, G) \in A_1 \cup A_3$, the result in (S.58) can be equivalently written as

$$\mu_{[k-G+1,k]} = \underset{\omega \in \mathcal{W}_2(D)}{\operatorname{argmin}} E_{P_{j+1}}\left(d_{\mathcal{W}}^2(\theta_{j+1}, \omega)\right) = \underset{\omega \in \mathcal{W}_2(D)}{\operatorname{argmin}} \widetilde{U}_G(\omega, k)$$

where $\widetilde{U}_G(\omega, k) = E_{P_{j+1}}\left(d_{\mathcal{W}}^2(\theta_{j+1}, \omega)\right) = \widetilde{M}_G(\omega, k) I\{(k, G) \in A_1 \cup A_3\}$. On the other hand, the Wasserstein space $(\mathcal{W}_2(D), d_{\mathcal{W}})$ equipped with the Wasserstein metric $d_{\mathcal{W}}$ fulfills assumptions (A1)–(A3) in Dubey and Müller (2020). Assumption (A1) implies that for any $\delta > 0$, we have

$$\widetilde{U}_G\big(\mu_{[k-G+1,k]}, k\big) < \inf_{\omega \in \mathcal{W}_2(D): d_{\mathcal{W}}(\omega, \mu_{[k-G+1,k]}) \geq \delta} \widetilde{U}_G(\omega, k)$$

Using $\widetilde{U}_G(\omega, k) = \widetilde{M}_G(\omega, k) I\{(k, G) \in A_1 \cup A_3\}$, the above result can be equivalently written as

$$\widetilde{M}_G\big(\mu_{[k-G+1,k]}, k\big) I\{(k, G) \in A_1 \cup A_3\}$$
$$< \inf_{\omega \in \mathcal{W}_2(D): d_{\mathcal{W}}(\omega, \mu_{[k-G+1,k]}) \geq \delta} \widetilde{M}_G(\omega, k) I\{(k, G) \in A_1 \cup A_3\} \tag{S.59}$$

For the case of $(k, G) \in A_2$, the result in (S.58) can be equivalently written as

$$\mu_{[k-G+1,k]} = \underset{\omega \in \mathcal{W}_2(D)}{\operatorname{argmin}} \left\{ \frac{k_j^* - (k-G)}{G} E_{P_j} d_{\mathcal{W}}^2(\theta_j, \omega) + \frac{k - k_j^*}{G} E_{P_{j+1}} d_{\mathcal{W}}^2(\theta_{j+1}, \omega) \right\}$$

$$= \underset{\omega \in \mathcal{W}_2(D)}{\operatorname{argmin}} \widetilde{R}_G(\omega, k)$$

where $\widetilde{R}_G(\omega, k) = \frac{k_j^* - (k-G)}{G} E_{P_j} d_{\mathcal{W}}^2(\theta_j, \omega) + \frac{k - k_j^*}{G} E_{P_{j+1}} d_{\mathcal{W}}^2(\theta_{j+1}, \omega) = \widetilde{M}_G(\omega, k) I\{(k, G) \in A_2\}$. Similarly, applying assumption (A1) in Dubey and Müller (2020) leads to

$$\widetilde{R}_G\big(\mu_{[k-G+1,k]}, k\big) < \inf_{\omega \in \mathcal{W}_2(D): d_{\mathcal{W}}(\omega, \mu_{[k-G+1,k]}) \geq \delta} \widetilde{R}_G(\omega, k), \text{ for any } \delta > 0$$

Using $\widetilde{R}_G(\omega, k) = \widetilde{M}_G(\omega, k) I\{(k, G) \in A_2\}$, the above result can be equivalently written as

$$\widetilde{M}_G(\mu_{[k-G+1,k]}, k) I\{(k, G) \in A_2\}$$
$$< \inf_{\omega \in \mathcal{W}_2(D):\, d_\mathcal{W}(\omega, \mu_{[k-G+1,k]}) \geq \delta} \widetilde{M}_G(\omega, k) I\{(k, G) \in A_2\} \quad (S.60)$$

Combining (S.59) and (S.60), we can conclude that $\mu_{[k-G+1,k]} = \operatorname{argmin}_{\omega \in \mathcal{W}_2(D)} \widetilde{M}_G(\omega, k)$ satisfies the well-separated condition stated in (S.57).

Fix an arbitrary $\delta > 0$, we define
$$\varphi(\delta) = \inf_{\omega \in \mathcal{W}_2(D):\, d_\mathcal{W}(\omega, \mu_{[k-G+1,k]}) \geq \delta} \widetilde{M}_G(\omega, k) - \widetilde{M}_G(\mu_{[k-G+1,k]}, k) \quad (S.61)$$

If $d_\mathcal{W}(\hat{\mu}_{[k-G+1,k]}, \mu_{[k-G+1,k]}) \geq \delta$, then it implies $\hat{\mu}_{[k-G+1,k]} \in B_\delta^C(\mu_{[k-G+1,k]}) = \{\omega \in \mathcal{W}_2(D):\, d_\mathcal{W}(\omega, \mu_{[k-G+1,k]}) \geq \delta\}$. Thus, we have
$$\widetilde{M}_G(\hat{\mu}_{[k-G+1,k]}, k) \geq \inf_{\omega \in B_\delta^C(\mu_{[k-G+1,k]})} \widetilde{M}_G(\omega, k) = \inf_{\omega \in \mathcal{W}_2(D):\, d_\mathcal{W}(\omega, \mu_{[k-G+1,k]}) \geq \delta} \widetilde{M}_G(\omega, k)$$

It follows that
$$\{d_\mathcal{W}(\hat{\mu}_{[k-G+1,k]}, \mu_{[k-G+1,k]}) \geq \delta\}$$
$$\Rightarrow \widetilde{M}_G(\hat{\mu}_{[k-G+1,k]}, k) \geq \inf_{\omega \in \mathcal{W}_2(D):\, d_\mathcal{W}(\omega, \mu_{[k-G+1,k]}) \geq \delta} \widetilde{M}_G(\omega, k)$$

where $A \Rightarrow B$ denotes $A$ implies $B$. Using equation (S.61), the above result becomes
$$\{d_\mathcal{W}(\hat{\mu}_{[k-G+1,k]}, \mu_{[k-G+1,k]}) \geq \delta\}$$
$$\Rightarrow \widetilde{M}_G(\hat{\mu}_{[k-G+1,k]}, k) \geq \varphi(\delta) + \widetilde{M}_G(\mu_{[k-G+1,k]}, k) \quad (S.62)$$
$$\Rightarrow \widetilde{M}_G(\hat{\mu}_{[k-G+1,k]}, k) - \widetilde{M}_G(\mu_{[k-G+1,k]}, k) \geq \varphi(\delta)$$

On the other hand, recall that $M_G(\omega, k) = \frac{1}{G} \sum_{i=k-G+1}^{i=k} d_\mathcal{W}^2(\nu_i, \omega)$ and $\hat{\mu}_{[k-G+1,k]} = \operatorname{argmin}_{\omega \in \mathcal{W}_2(D)} M_G(\omega, k)$; then, we have
$$M_G(\mu_{[k-G+1,k]}, k) - M_G(\hat{\mu}_{[k-G+1,k]}, k) = M_G(\mu_{[k-G+1,k]}, k) - \min_{\omega \in \mathcal{W}_2(D)} M_G(\omega, k) \geq 0$$

It follows that
$$\widetilde{M}_G(\hat{\mu}_{[k-G+1,k]}, k) - \widetilde{M}_G(\mu_{[k-G+1,k]}, k)$$
$$\leq \left(\widetilde{M}_G(\hat{\mu}_{[k-G+1,k]}, k) - \widetilde{M}_G(\mu_{[k-G+1,k]}, k)\right) + \left(M_G(\mu_{[k-G+1,k]}, k) - M_G(\hat{\mu}_{[k-G+1,k]}, k)\right)$$
$$= \left(\widetilde{M}_G(\hat{\mu}_{[k-G+1,k]}, k) - M_G(\hat{\mu}_{[k-G+1,k]}, k)\right) - \left(\widetilde{M}_G(\mu_{[k-G+1,k]}, k) - M_G(\mu_{[k-G+1,k]}, k)\right)$$
$$\leq \left|\widetilde{M}_G(\hat{\mu}_{[k-G+1,k]}, k) - M_G(\hat{\mu}_{[k-G+1,k]}, k)\right| + \left|\widetilde{M}_G(\mu_{[k-G+1,k]}, k) - M_G(\mu_{[k-G+1,k]}, k)\right|$$
$$\leq 2 \sup_{\omega \in \mathcal{W}_2(D)} |\widetilde{M}_G(\omega, k) - M_G(\omega, k)|$$

Combining this with (S.62), we get
$$\{d_\mathcal{W}(\hat{\mu}_{[k-G+1,k]}, \mu_{[k-G+1,k]}) \geq \delta\} \Rightarrow \sup_{\omega \in \mathcal{W}_2(D)} |\widetilde{M}_G(\omega, k) - M_G(\omega, k)| \geq \frac{\varphi(\delta)}{2}$$

Thus,

$$P\big(d_{\mathcal{W}}(\hat{\mu}_{[k-G+1,k]}, \mu_{[k-G+1,k]}) \geq \delta\big) \leq P\left(\sup_{\omega \in \mathcal{W}_2(D)} |\widetilde{M}_G(\omega, k) - M_G(\omega, k)| \geq \frac{\varphi(\delta)}{2}\right)$$

Furthermore, in view of (S.57) and (S.61), we see that $\varphi(\delta) > 0$. Finally, using the result $\sup_{\omega \in \mathcal{W}_2(D)} |\widetilde{M}_G(\omega, k) - M_G(\omega, k)| = o_p(1)$ obtained in (S.56), it holds for any $\delta > 0$ that

$$P\big(d_{\mathcal{W}}(\hat{\mu}_{[k-G+1,k]}, \mu_{[k-G+1,k]}) \geq \delta\big) = o(1) \text{ as } G = G(n) \to \infty$$

From the above analysis, we can see that the above conclusion holds for each $k \in (k_1^*, k_q^*]$. Thus,

$$d_{\mathcal{W}}(\hat{\mu}_{[k-G+1,k]}, \mu_{[k-G+1,k]}) = o_p(1) \text{ for each } k \in (k_1^*, k_q^*]$$

With this pointwise convergence conclusion, we can show for some $0 < \beta < 2$ that

$$\max_{k_1^* < k \leq k_q^*} d_{\mathcal{W}}(\hat{\mu}_{[k-G+1,k]}, \mu_{[k-G+1,k]}) = O_p\left(G^{-\frac{1}{2+\beta}}\right)$$

by adopting similar arguments as in the proof of Lemma 4 in Dubey and Müller (2020).

By following similar theoretical analysis, we can show that the above uniform convergence rate can also be achieved for the cases of $G \leq k \leq k_1^*$ (corresponds to the case when $k$ is within the left boundary segment in situation II described in Section S.2.4.3) and $k_q^* < k \leq n - G$ (corresponds to the case that $k$ is within the right boundary segment in situation II described in Section S.2.4.3). Combining the above results, we can conclude that

$$\max_{G \leq k \leq n-G} d_{\mathcal{W}}(\hat{\mu}_{[k-G+1,k]}, \mu_{[k-G+1,k]}) = O_p\left(G^{-\frac{1}{2+\beta}}\right) \tag{S.63}$$

for some $0 < \beta < 2$.

Using the similar tactics, we can show that

$$\max_{G \leq k \leq n-G} d_{\mathcal{W}}(\hat{\mu}_{[k+1,k+G]}, \mu_{[k+1,k+G]}) = O_p\left(G^{-\frac{1}{2+\beta}}\right) \tag{S.64}$$

for some $0 < \beta < 2$.

The rest of the proof follows from the same arguments as in the proof of Lemma 5 in Dubey and Müller (2020). Specifically, by similar arguments as in the proof of Lemma 5 (Dubey and Müller, 2020), we can show for some $0 < \beta < 2$ that

$$\max_{G \leq k \leq n-G} |\hat{V}_{[k+1,k+G]} - V_{[k+1,k+G]}| = O_p\left(G^{-\frac{1}{2}}\right)$$

$$\max_{G \leq k \leq n-G} |\hat{V}_{[k-G+1,k]} - V_{[k-G+1,k]}| = O_p\left(G^{-\frac{1}{2}}\right)$$

$$\max_{G \leq k \leq n-G} |\hat{V}^C_{[k+1,k+G]} - V^C_{[k+1,k+G]}| = O_p\left(G^{-\frac{1}{2+\beta}}\right)$$

$$\max_{G\leq k\leq n-G}|\widehat{V}^C_{[k-G+1,k]} - V^C_{[k-G+1,k]}| = O_p\left(G^{-\frac{1}{2+\beta}}\right)$$

under $H_A$. By the triangle inequality and the results obtained above, we have

$$\max_{G\leq k\leq n-G}|\widehat{\psi}_G(k) - \psi_G(k)| = O_p\left(G^{-\frac{1}{2}}\right), \quad \text{under } H_A$$

$$\max_{G\leq k\leq n-G}|\widehat{\lambda}_G(k) - \lambda_G(k)| = O_p\left(G^{-\frac{1}{2+\beta}}\right), \quad \text{under } H_A$$

for some $0 < \beta < 2$. This completes the proof.

Next, we resume the proof of step 1. From (S.41a) and (S.43a), we have

$$\frac{\widehat{\sigma}_{k,n}S_n^G(k)}{\sigma_k} = \frac{1}{\sigma_k}(|\widehat{\psi}_G(k)| + |\widehat{\lambda}_G(k)|)$$

$$= \frac{|\widehat{\psi}_G(k)| - |\psi_G(k)| + |\widehat{\lambda}_G(k)| - |\lambda_G(k)|}{\sigma_k} + \frac{|\psi_G(k)| + |\lambda_G(k)|}{\sigma_k} \quad \text{(S.65)}$$

$$:= \Delta^G(k) + S^G(k)$$

where $\Delta^G(k) = (1/\sigma_k)(|\widehat{\psi}_G(k)| - |\psi_G(k)| + |\widehat{\lambda}_G(k)| - |\lambda_G(k)|)$ and $S^G(k)$ is that given in equation (S.41b). For $\Delta^G(k)$, we first observe that

$$\max_{G\leq k\leq n-G}|\Delta^G(k)| = \max_{G\leq k\leq n-G}\frac{\left||\widehat{\psi}_G(k)| - |\psi_G(k)| + |\widehat{\lambda}_G(k)| - |\lambda_G(k)|\right|}{\sigma_k}$$

$$\leq \max_{G\leq k\leq n-G}\frac{\left||\widehat{\psi}_G(k)| - |\psi_G(k)|\right| + \left||\widehat{\lambda}_G(k)| - |\lambda_G(k)|\right|}{\sigma_k}$$

It can be easily verified that the inequality $(|a| - |b|)^2 \leq (a - b)^2$ holds for any $a, b \in \mathbb{R}$. This implies that $||a| - |b|| \leq |a - b|$. Then, the above result becomes

$$\max_{G\leq k\leq n-G}|\Delta^G(k)| \leq \max_{G\leq k\leq n-G}\frac{|\widehat{\psi}_G(k) - \psi_G(k)| + |\widehat{\lambda}_G(k) - \lambda_G(k)|}{\sigma_k}$$

$$\leq \frac{1}{\min_{G\leq k\leq n-G}\{\sigma_k\}}\max_{G\leq k\leq n-G}(|\widehat{\psi}_G(k)-\psi_G(k)| + |\widehat{\lambda}_G(k) - \lambda_G(k)|)$$

$$\leq \frac{1}{\min_{G\leq k\leq n-G}\{\sigma_k\}}\left(\max_{G\leq k\leq n-G}|\widehat{\psi}_G(k)-\psi_G(k)| + \max_{G\leq k\leq n-G}|\widehat{\lambda}_G(k) - \lambda_G(k)|\right)$$

Due to $0 < \sigma_k < \infty$ for each $G \leq k \leq n - G$, there exists a universal positive constant $C_0$ such that $\frac{1}{\min_{G\leq k\leq n-G}\{\sigma_k\}} \leq C_0$. Consequently,

$$\max_{G\leq k\leq n-G}|\Delta^G(k)| \leq C_0\left(\max_{G\leq k\leq n-G}|\widehat{\psi}_G(k)-\psi_G(k)| + \max_{G\leq k\leq n-G}|\widehat{\lambda}_G(k) - \lambda_G(k)|\right)$$

Therefore, for any fixed $L > 0$, if $\max_{G\leq k\leq n-G}|\widehat{\psi}_G(k) - \psi_G(k)| \leq LG^{-\frac{1}{2}}$ and $\max_{G\leq k\leq n-G}|\widehat{\lambda}_G(k) - \lambda_G(k)| \leq LG^{-\frac{1}{2+\beta}}$, then

$$\max_{G\leq k\leq n-G}|\Delta^G(k)| \leq 2C_0 LG^{-\frac{1}{2+\beta}} \quad \text{(S.66)}$$

Recall that the set $B_{G,q}$ defined in equation (10) of the main text takes the form

$$B_{G,q} := \{k \in \{1,2,\cdots,n\}: \exists k_j^* \in \{k_1^*,\cdots,k_q^*\} \text{ such that } |k - k_j^*| \leq (1-\varepsilon)G\}$$

where $k_1^*, \cdots, k_q^*$ are the $q$ different change points such that $1 < k_1^* < k_2^* < \cdots < k_q^* < n$, and $\varepsilon \in (0, 0.5)$ is the AOP parameter given in (8) of the main text. Under condition (C1) in the main text, it holds that $B_{G,q} \subset \{G, G+1, \cdots, n-G\}$ for sufficiently large $n$. Thus,

$$\max_{k \in B_{G,q}} |\Delta^G(k)| \leq \max_{G \leq k \leq n-G} |\Delta^G(k)| \leq 2C_0 L G^{-\frac{1}{2+\beta}}, \text{ for sufficiently large } n \quad (S.67)$$

This implies that there exists a finite positive constant $C_1$ such that

$$\min_{k \in B_{G,q}} \Delta^G(k) \geq -C_1 G^{-\frac{1}{2+\beta}} \quad (S.68)$$

Observe that for $\Delta > 0$, it holds that $n^{\frac{2}{2+\Delta}} \log n \to \infty$, as $n \to \infty$. Under assumption (A1) in the main text, we have $n^{\frac{2}{2+\Delta}} \log n / G \to 0$, as $n \to \infty$. These results imply that $G \to \infty$, as $n \to \infty$. Consequently, we have

$$-C_1 G^{-\frac{1}{2+\beta}} \to 0^-, \text{ as } n \to \infty \quad (S.69)$$

due to $0 < C_1 < \infty$. Since $S^G(k)$ given in (S.41b) fulfills $\min_{k \in B_{G,q}} S^G(k) \geq 0$, it follows from (S.69) and (S.68) that

$$\min_{k \in B_{G,q}} \Delta^G(k) \geq -C_1 G^{-\frac{1}{2+\beta}} \geq -\frac{\min_{k \in B_{G,q}} S^G(k)}{2}, \text{ for sufficiently large } n \quad (S.70)$$

Combining (S.65) and (S.70), we have for $k \in B_{G,q}$ that

$$\frac{\hat{\sigma}_{k,n} S_n^G(k)}{\sigma_k} = \Delta^G(k) + S^G(k) \geq \min_{k \in B_{G,q}} \Delta^G(k) + \min_{k \in B_{G,q}} S^G(k) \geq \frac{1}{2} \min_{k \in B_{G,q}} S^G(k)$$

when $n$ is sufficiently large. Consequently, it holds that

$$\min_{k \in B_{G,q}} \frac{\hat{\sigma}_{k,n} S_n^G(k)}{\sigma_k} \geq \frac{1}{2} \min_{k \in B_{G,q}} S^G(k), \quad \text{for sufficiently large } n \quad (S.71)$$

On the other hand, from (S.42a), we have

$$P\left\{\min_{k \in B_{G,q}} T_n^G(k) \geq D_n(G, \alpha_n)\right\} = P\left\{\min_{k \in B_{G,q}} \sqrt{G/2}\, S_n^G(k) \geq D_n(G, \alpha_n)\right\}$$

$$= P\left\{\sqrt{G/2} \min_{k \in B_{G,q}} \frac{\hat{\sigma}_{k,n} S_n^G(k)}{\sigma_k} \geq \frac{\hat{\sigma}_{k,n}}{\sigma_k} D_n(G, \alpha_n)\right\} \quad (S.72)$$

For notational convenience, we define the following events:

$$A = \left\{\max_{G \leq k \leq n-G} |\hat{\psi}_G(k) - \psi_G(k)| \leq LG^{-\frac{1}{2}}, \max_{G \leq k \leq n-G} |\hat{\lambda}_G(k) - \lambda_G(k)| \leq LG^{-\frac{1}{2+\beta}}\right\} \quad (S.73a)$$

$$B = \left\{\min_{k \in B_{G,q}} \frac{\hat{\sigma}_{k,n} S_n^G(k)}{\sigma_k} \geq \frac{1}{2} \min_{k \in B_{G,q}} S^G(k), \text{for sufficiently large } n\right\} \quad (S.73b)$$

$$C = \left\{\sqrt{G/2}\, \frac{1}{2} \min_{k \in B_{G,q}} S^G(k) \geq \frac{\hat{\sigma}_{k,n}}{\sigma_k} D_n(G, \alpha_n)\right\} \tag{S.73c}$$

First, it is noteworthy that the result in (S.71) is derived under the conditions of $\max_{G \leq k \leq n-G} |\hat{\psi}_G(k) - \psi_G(k)| \leq LG^{-\frac{1}{2}}$ and $\max_{G \leq k \leq n-G} |\hat{\lambda}_G(k) - \lambda_G(k)| \leq LG^{-\frac{1}{2+\beta}}$. Thus, we have $A \subset B$. It follows that $A \cap C \subset B \cap C$. Furthermore, we have

$$B \cap C = \left\{\min_{k \in B_{G,q}} \frac{\hat{\sigma}_{k,n} S_n^G(k)}{\sigma_k} \geq \frac{1}{2} \min_{k \in B_{G,q}} S^G(k),\; \frac{1}{2}\sqrt{\frac{G}{2}} \min_{k \in B_{G,q}} S^G(k) \geq \frac{\hat{\sigma}_{k,n}}{\sigma_k} D_n(G, \alpha_n)\right\}$$

$$\subset \left\{\sqrt{\frac{G}{2}} \min_{k \in B_{G,q}} \frac{\hat{\sigma}_{k,n} S_n^G(k)}{\sigma_k} \geq \frac{1}{2}\sqrt{\frac{G}{2}} \min_{k \in B_{G,q}} S^G(k),\; \frac{1}{2}\sqrt{\frac{G}{2}} \min_{k \in B_{G,q}} S^G(k) \geq \frac{\hat{\sigma}_{k,n}}{\sigma_k} D_n(G, \alpha_n)\right\}$$

$$\subset \left\{\sqrt{\frac{G}{2}} \min_{k \in B_{G,q}} \frac{\hat{\sigma}_{k,n} S_n^G(k)}{\sigma_k} \geq \frac{\hat{\sigma}_{k,n}}{\sigma_k} D_n(G, \alpha_n)\right\}$$

for sufficiently large $n$. Combining the above results, we get

$$A \cap C \subset B \cap C \subset \left\{\sqrt{\frac{G}{2}} \min_{k \in B_{G,q}} \frac{\hat{\sigma}_{k,n} S_n^G(k)}{\sigma_k} \geq \frac{\hat{\sigma}_{k,n}}{\sigma_k} D_n(G, \alpha_n)\right\}$$

for sufficiently large $n$. This implies

$$P\left(\sqrt{G/2} \min_{k \in B_{G,q}} \frac{\hat{\sigma}_{k,n} S_n^G(k)}{\sigma_k} \geq \frac{\hat{\sigma}_{k,n}}{\sigma_k} D_n(G, \alpha_n)\right) \geq P(A \cap C)$$

for sufficiently large $n$. Using the property $P(C) = P(C \cap A) + P(C \cap A^C) \leq P(C \cap A) + P(A^C)$, we have $P(C \cap A) \geq P(C) - P(A^C)$. Consequently, the above result becomes

$$P\left(\sqrt{G/2} \min_{k \in B_{G,q}} \frac{\hat{\sigma}_{k,n} S_n^G(k)}{\sigma_k} \geq \frac{\hat{\sigma}_{k,n}}{\sigma_k} D_n(G, \alpha_n)\right) \geq P(C) - P(A^C) \tag{S.74}$$

for sufficiently large $n$. Combining (S.72) and (S.74), we get

$$P\left\{\min_{k \in B_{G,q}} T_n^G(k) \geq D_n(G, \alpha_n)\right\} \geq P(C) - P(A^C) \tag{S.75}$$

for sufficiently large $n$.

Let $A_1 = \left\{\max_{G \leq k \leq n-G} |\hat{\psi}_G(k) - \psi_G(k)| \leq LG^{-\frac{1}{2}}\right\}$ and $A_2 = \left\{\max_{G \leq k \leq n-G} |\hat{\lambda}_G(k) - \lambda_G(k)| \leq LG^{-\frac{1}{2+\beta}}\right\}$. From (S.73a), we have $A = A_1 \cap A_2$. Consequently, the term $P(A^C)$ in (S.75) becomes

$$P(A^C) = P((A_1 \cap A_2)^C) \leq P(A_1^C) + P(A_2^C)$$
$$= P\left(\max_{G \leq k \leq n-G} |\hat{\psi}_G(k) - \psi_G(k)| > LG^{-\frac{1}{2}}\right)$$
$$+ P\left(\max_{G \leq k \leq n-G} |\hat{\lambda}_G(k) - \lambda_G(k)| > LG^{-\frac{1}{2+\beta}}\right)$$

By Lemma S.3, it follows that one can choose a sufficiently large $L$ to make $P(A^C) = o(1)$. Consequently, the result in (S.75) becomes

$$\begin{aligned}
P\left(\min_{k \in B_{G,q}} T_n^G(k) \geq D_n(G, \alpha_n)\right) &\geq P(C) + o(1) \\
&= P\left(\sqrt{G/2}\, \frac{1}{2} \min_{k \in B_{G,q}} S^G(k) \geq \frac{\hat{\sigma}_{k,n}}{\sigma_k} D_n(G, \alpha_n)\right) + o(1) \\
&= P\left(\frac{1}{2} \min_{k \in B_{G,q}} \left\{\sqrt{G/2}\, S^G(k)\right\} \geq \frac{\hat{\sigma}_{k,n}}{\sigma_k} D_n(G, \alpha_n)\right) + o(1) \\
&= P\left(\min_{k \in B_{G,q}} T^G(k) \geq 2\frac{\hat{\sigma}_{k,n}}{\sigma_k} D_n(G, \alpha_n)\right) + o(1)
\end{aligned} \quad (S.76)$$

The third equality follows by $T^G(k) = \sqrt{G/2}\, S^G(k)$ given in (S.42b).

In the following, we first handle the term $D_n(G, \alpha_n)$ in (S.76). From equation (7) in the main text, the expression of $D_n(G, \alpha_n)$ can be written as

$$D_n(G, \alpha_n) = \frac{-\log\log(1/\sqrt{1-\alpha_n})}{\gamma_1(n/G)} + \frac{\gamma_2(n/G)}{\gamma_1(n/G)} \quad (S.77)$$

where $\gamma_1(x) = \sqrt{2\log x}$ and $\gamma_2(x) = 2\log x + 0.5\log\log x + \log(1.5) - 0.5\log\pi$. Under condition (C3) in the main text, it follows that

$$\frac{-\log\log(1/\sqrt{1-\alpha_n})}{\gamma_1(n/G)} = -\frac{1}{\sqrt{2}} \frac{\log\log(1/\sqrt{1-\alpha_n})}{\sqrt{\log(n/G)}} = O(1) \quad (S.78)$$

On the other hand, by some straight-forward calculations, we have

$$\lim_{x \to \infty} \frac{\gamma_2(x)/\gamma_1(x)}{\sqrt{2\log x}} = 1$$

which implies $\gamma_2(x)/\gamma_1(x) = O(\sqrt{\log x})$, as $x \to \infty$. Under assumption (A1) in the main text, we have $(n/G) \to \infty$. Consequently, we obtain

$$\frac{\gamma_2(n/G)}{\gamma_1(n/G)} = O\left(\sqrt{\log(n/G)}\right) \quad (S.79)$$

Combining these arguments, we have

$$D_n(G, \alpha_n) = O(1) + O\left(\sqrt{\log(n/G)}\right) = O\left(\sqrt{\log(n/G)}\right) \quad (S.80)$$

We next turn to handle the term $\frac{\hat{\sigma}_{k,n}}{\sigma_k}$ in (S.76). First, using the results in (S.63) and (S.64), we can show for each $k \in [G, n-G]$ that

$$|\hat{\sigma}_{k,n} - \sigma_k| = o_p(1), \text{ under } H_A$$

by following similar arguments as in the proof of Lemma 4 of Dubey and Müller (2020). By the boundedness of $\sigma_k$, we further have $\hat{\sigma}_{k,n} = O_p(1)$. On the other hand, due to $0 < \sigma_k < \infty$ for each $G \leq k \leq n - G$, there exists a universal positive constant $C_0$ such that $\frac{1}{\min_{G \leq k \leq n-G}\{\sigma_k\}} \leq C_0$, thus $\frac{1}{\sigma_k} \leq \frac{1}{\min_{G \leq k \leq n-G}\{\sigma_k\}} \leq C_0$. Consequently, we have

$$\frac{\hat{\sigma}_{k,n}}{\sigma_k} = O_p(1) \tag{S.81}$$

Substituting the results in (S.80) and (S.81) into (S.76) yields

$$\begin{aligned} P\left(\min_{k \in B_{G,q}} T_n^G(k) \geq D_n(G, \alpha_n)\right) \\ \geq P\left(\min_{k \in B_{G,q}} T^G(k) \geq \sqrt{\log(n/G)} O_p(1)\right) + o(1) \\ = P\left(\frac{1}{\sqrt{\log(n/G)}} \min_{k \in B_{G,q}} T^G(k) \geq O_p(1)\right) + o(1) \\ = P\left(O_p(1) \leq \frac{1}{\sqrt{\log(n/G)}} \min_{k \in B_{G,q}} T^G(k)\right) + o(1) \end{aligned} \tag{S.82}$$

Under condition (C2) in the main text, we have $\frac{1}{\sqrt{\log(n/G)}} \min_{k \in B_{G,q}} T^G(k) \to \infty$, as $n \to \infty$.

Consequently, it follows that $P\left(O_p(1) \leq \frac{1}{\sqrt{\log(n/G)}} \min_{k \in B_{G,q}} T^G(k)\right) \to 1$, as $n \to \infty$.

Combining this with (S.82), we get

$$P\left(\min_{k \in B_{G,q}} T_n^G(k) \geq D_n(G, \alpha_n)\right) \to 1 \text{ as } n \to \infty$$

We can therefore conclude under $H_A$ that

$$P\left\{\min_{k \in B_{G,q}} T_n^G(k) < D_n(G, \alpha_n)\right\} \to 0, \text{ as } n \to \infty$$

This completes step 1.

***Step 2:*** We next show under $H_A$ that

$$P\left(\max_{k \in A_{G,q}} T_n^G(k) \geq D_n(G, \alpha_n)\right) \to 0, \text{ as } n \to \infty \tag{S.83}$$

Recall that the set $A_{G,q}$ defined in (S.37b) takes the form

$$A_{G,q} := \left\{k \in \{1, 2, \cdots, n\} : |k - k^*| \geq G \text{ for any } k^* \in \{k_0^*, k_1^*, \cdots, k_q^*, k_{q+1}^*\}\right\}$$

The condition $k \in A_{G,q}$ involved in (S.83) indicates that we only need to consider the $k$ such that

$$k \in A_{G,q} = \left\{k \in \{1, 2, \cdots, n\} : |k - k^*| \geq G \text{ for any } k^* \in \{k_0^*, k_1^*, \cdots, k_q^*, k_{q+1}^*\}\right\} \tag{S.84}$$

Also, recall that the $q$ different but unknown change points $k_1^*, \cdots, k_q^*$ contained in the distributional sequence $\Gamma = \{\nu_1, \cdots, \nu_n\}$ satisfy

$$0 = k_0^* < k_1^* < k_2^* < \cdots < k_q^* < k_{q+1}^* = n$$

where $k_0^* = 0$ and $k_{q+1}^* = n$ denote the left and right boundary points, respectively. Then, for any $k \in \{1, 2, \cdots, n\}$, there exists $j \in \{0, 1, \cdots, q\}$ such that

$$k \in (k_j^*, k_{j+1}^*] \tag{S.85}$$

Combining (S.84) and (S.85) yields

$$k - k_j^* \geq G \text{ and } k_{j+1}^* - k \geq G$$

Thus,

$$k - G + 1 \geq k_j^* + 1 \text{ and } k + G \leq k_{j+1}^*$$

This implies that the sub-sequence $\Gamma_k^G = \{v_{k-G+1}, \cdots, v_k, \cdots, v_{k+G}\}$ satisfies

$$\Gamma_k^G = \{v_{k-G+1}, \cdots, v_k, \cdots, v_{k+G}\} \subset \{v_{k_j^*+1}, \cdots, v_k, \cdots, v_{k_{j+1}^*}\} = \Gamma_{j+1}$$

where $\Gamma_{j+1}$ is the $(j+1)$th segment of the distributional sequence, see equation (3) of the main text. From the change-point model given in equation (3) of the main text, we see that the distributional data within each segment separated by two adjacent change points (including boundary points) are i.i.d. $\mathcal{W}_2(\mathcal{D})$-valued random objects.

Consequently, if $k \in A_{G,q}$, the distributional data contained in $\Gamma_k^G = \{v_{k-G+1}, \cdots, v_k, \cdots, v_{k+G}\}$ are also i.i.d. under $H_A$. Therefore, the statistic $\max_{k \in A_{G,q}} T_n^G(k)$ under $H_A$ has the same asymptotic behavior as its $H_0$ counterpart. By Theorem 1 in the main text, we have under $H_0$ that

$$\gamma_1(n/G) \max_{G \leq k \leq n-G} T_n^G(k) - \gamma_2(n/G) \xrightarrow{d} \Gamma_{Gum} \tag{S.86}$$

where $\xrightarrow{d}$ denotes convergence in distribution, and $\Gamma_{Gum}$ denotes a random variable that possesses the distribution function $P(\Gamma_{Gum} \leq x) = \exp(-2\exp(-x))$. The above analysis indicates that we can use the asymptotic distribution of $\max_{k \in A_{G,q}} T_n^G(k)$ under $H_0$ to bound the probability $P(\max_{k \in A_{G,q}} T_n^G(k) \geq D_n(G, \alpha_n))$ under $H_A$, similar to that in the proof of Theorem 6.1 of Muhsal (2013). Consequently, we have

$$P\left(\max_{k \in A_{G,q}} T_n^G(k) \geq D_n(G, \alpha_n)\right) \leq P\left(\max_{G \leq k \leq n-G} T_n^G(k) \geq D_n(G, \alpha_n)\right)$$

$$= P\left(\gamma_1(n/G) \max_{G \leq k \leq n-G} T_n^G(k) - \gamma_2(n/G) \geq \gamma_1(n/G) D_n(G, \alpha_n) - \gamma_2(n/G)\right)$$

$$= P\left(\gamma_1(n/G) \max_{G \leq k \leq n-G} T_n^G(k) - \gamma_2(n/G) \geq -\log\log(1/\sqrt{1-\alpha_n})\right)$$

The last equality follows by $D_n(G; \alpha) = \left(-\text{loglog}(1/\sqrt{1-\alpha}) + \gamma_2(n/G)\right)/\gamma_1(n/G)$ given in equation (7) of the main text. Combining this with (S.86), we have for sufficiently large $n$ that

$$P\left(\max_{k \in A_{G,q}} T_n^G(k) \geq D_n(G, \alpha_n)\right) \leq P(\Gamma_{Gum} \geq -\text{loglog}(1/\sqrt{1-\alpha_n}))$$

Recall that $P(\Gamma_{Gum} \leq x) = \exp(-2\exp(-x))$, it follows that

$$P\left(\max_{k \in A_{G,q}} T_n^G(k) \geq D_n(G, \alpha_n)\right) \leq \alpha_n$$

for sufficiently large $n$. Furthermore, under condition (C3) in the main text, we have $\alpha_n \to 0$, as $n \to \infty$. This implies that

$$P\left(\max_{k \in A_{G,q}} T_n^G(k) \geq D_n(G, \alpha_n)\right) \to 0, \text{ as } n \to \infty$$

This completes step 2.

***Step 3:*** Finally, we prove the last condition given in (S.39), that is

$$P\left(\max_{1 \leq j \leq q} |\hat{k}_j^* - k_j^*| \geq G\right) \to 0, \text{ as } n \to \infty, \text{ under } H_A$$

First, we observe that

$$P\left(\max_{1 \leq j \leq q} |\hat{k}_j^* - k_j^*| \geq G\right)$$
$$= P\left(\max_{1 \leq j \leq q} |\hat{k}_j^* - k_j^*| \geq G, q = \hat{q}_n\right) + P\left(\max_{1 \leq j \leq q} |\hat{k}_j^* - k_j^*| \geq G, q \neq \hat{q}_n\right) \quad \text{(S.87)}$$
$$:= I + II$$

Combining the conclusions in steps 1–2 with (S.38), we get $P(q \neq \hat{q}_n) = o(1)$. Consequently, for the second term $II$ in (S.87), we have

$$II = P\left(\max_{1 \leq j \leq q} |\hat{k}_j^* - k_j^*| \geq G, q \neq \hat{q}_n\right) \leq P(q \neq \hat{q}_n) = o(1)$$

On the other hand, by combining the conclusions in steps 1–2 together and adopting similar arguments as in the proof of Corollary 6.3 of Muhsal (2013), we can show that

$$I = P\left(\max_{1 \leq j \leq q} |\hat{k}_j^* - k_j^*| \geq G, q = \hat{q}_n\right) = o(1)$$

We can therefore conclude that

$$P\left(\max_{1 \leq j \leq q} |\hat{k}_j^* - k_j^*| \geq G\right) \to 0 \text{ as } n \to \infty$$

This completes step 3.

The above three steps prove that the three conditions given in (S.39) are all satisfied under the conditions of Theorem 2. Since (S.39) is a sufficient condition for the result of Theorem 2, this completes the proof of Theorem 2.

*S.3.3. Proof of Proposition 1*

**Proof of Proposition 1.** Recall that $\hat{\sigma}_{k,n}^2$ given in equation (15) of the main text takes the form

$$\hat{\sigma}_{k,n}^2 = (\hat{\sigma}_{k,n,l}^2 + \hat{\sigma}_{k,n,r}^2)/2, \qquad G \leq k \leq n - G$$

where

$$\hat{\sigma}_{k,n,l}^2 = \frac{1}{G}\sum_{i=k-G+1}^{k} d_{\mathcal{W}}^4(v_i, \hat{\mu}_{[k-G+1,k]}) - \left(\frac{1}{G}\sum_{i=k-G+1}^{k} d_{\mathcal{W}}^2(v_i, \hat{\mu}_{[k-G+1,k]})\right)^2 \quad \text{(S.88a)}$$

$$\hat{\sigma}_{k,n,r}^2 = \frac{1}{G}\sum_{i=k+1}^{k+G} d_{\mathcal{W}}^4(v_i, \hat{\mu}_{[k+1,k+G]}) - \left(\frac{1}{G}\sum_{i=k+1}^{k+G} d_{\mathcal{W}}^2(v_i, \hat{\mu}_{[k+1,k+G]})\right)^2 \quad \text{(S.88b)}$$

Observe that

$$\begin{aligned}\max_{G \leq k \leq n-G} |\hat{\sigma}_{k,n}^2 - \sigma^2| &= \max_{G \leq k \leq n-G} \left|\frac{\hat{\sigma}_{k,n,l}^2 + \hat{\sigma}_{k,n,r}^2}{2} - \frac{\sigma^2}{2} - \frac{\sigma^2}{2}\right| \\ &\leq \frac{1}{2}\max_{G \leq k \leq n-G}|\hat{\sigma}_{k,n,l}^2 - \sigma^2| + \frac{1}{2}\max_{G \leq k \leq n-G}|\hat{\sigma}_{k,n,r}^2 - \sigma^2|\end{aligned} \quad \text{(S.89)}$$

We first prove that $\max_{G \leq k \leq n-G}|\hat{\sigma}_{k,n,l}^2 - \sigma^2| = o_p\left(\frac{1}{\sqrt{\log(n/G)}}\right)$. Using the notation defined in equation (S.13a), we rewrite $\hat{\sigma}_{k,n,l}^2$ given in (S.88a) as

$$\hat{\sigma}_{k,n,l}^2 = \frac{1}{G}\sum_{i=k-G+1}^{k} d_{\mathcal{W}}^4(v_i, \hat{\mu}_{[k-G+1,k]}) - \hat{V}_{[k-G+1,k]}^2 \quad \text{(S.90)}$$

where $\hat{V}_{[k-G+1,k]} = \frac{1}{G}\sum_{i=k-G+1}^{k} d_{\mathcal{W}}^2(v_i, \hat{\mu}_{[k-G+1,k]})$. Furthermore, we define the following oracle version of $\hat{\sigma}_{k,n,l}^2$:

$$\begin{aligned}\tilde{\sigma}_{k,n,l}^2 &= \frac{1}{G}\sum_{i=k-G+1}^{k} d_{\mathcal{W}}^4(v_i, \mu) - \left(\frac{1}{G}\sum_{i=k-G+1}^{k} d_{\mathcal{W}}^2(v_i, \mu)\right)^2 \\ &= \frac{1}{G}\sum_{i=k-G+1}^{k} d_{\mathcal{W}}^4(v_i, \mu) - \tilde{V}_{[k-G+1,k]}^2\end{aligned} \quad \text{(S.91)}$$

where $\mu = \underset{\omega \in \mathcal{W}_2(D)}{\operatorname{argmin}} E\left(d_{\mathcal{W}}^2(v_1, \omega)\right)$ is the Fréchet mean under $H_0$, and $\tilde{V}_{[k-G+1,k]}$ is the oracle version of $\hat{V}_{[k-G+1,k]}$ defined in equation (S.14a). Then, we have

$$\begin{aligned}\max_{G \leq k \leq n-G}|\hat{\sigma}_{k,n,l}^2 - \sigma^2| &= \max_{G \leq k \leq n-G}|\hat{\sigma}_{k,n,l}^2 - \tilde{\sigma}_{k,n,l}^2 + \tilde{\sigma}_{k,n,l}^2 - \sigma^2| \\ &\leq \max_{G \leq k \leq n-G}|\hat{\sigma}_{k,n,l}^2 - \tilde{\sigma}_{k,n,l}^2| + \max_{G \leq k \leq n-G}|\tilde{\sigma}_{k,n,l}^2 - \sigma^2|\end{aligned} \quad \text{(S.92)}$$

where $S_I = \max_{G \leq k \leq n-G} |\hat{\sigma}^2_{k,n,l} - \tilde{\sigma}^2_{k,n,l}|$, and $S_{II} = \max_{G \leq k \leq n-G} |\tilde{\sigma}^2_{k,n,l} - \sigma^2|$.

For $S_I = \max_{G \leq k \leq n-G} |\hat{\sigma}^2_{k,n,l} - \tilde{\sigma}^2_{k,n,l}|$, by some straight-forward calculations, we obtain

$$\begin{aligned} S_I &= \max_{G \leq k \leq n-G} |\hat{\sigma}^2_{k,n,l} - \tilde{\sigma}^2_{k,n,l}| \\ &\leq \frac{1}{G} \max_{G \leq k \leq n-G} \left| \sum_{i=k-G+1}^{k} \left( d_{\mathcal{W}}^4(\nu_i, \hat{\mu}_{[k-G+1,k]}) - d_{\mathcal{W}}^4(\nu_i, \mu) \right) \right| \\ &\quad + \max_{G \leq k \leq n-G} |\hat{V}^2_{[k-G+1,k]} - \tilde{V}^2_{[k-G+1,k]}| \end{aligned} \quad (S.93)$$

$$:= T_I + T_{II}$$

where

$$T_I = \frac{1}{G} \max_{G \leq k \leq n-G} \left| \sum_{i=k-G+1}^{k} \left( d_{\mathcal{W}}^4(\nu_i, \hat{\mu}_{[k-G+1,k]}) - d_{\mathcal{W}}^4(\nu_i, \mu) \right) \right|$$

$$T_{II} = \max_{G \leq k \leq n-G} |\hat{V}^2_{[k-G+1,k]} - \tilde{V}^2_{[k-G+1,k]}|$$

For $T_I$, some algebra shows that

$$\begin{aligned} T_I &= \frac{1}{G} \max_{G \leq k \leq n-G} \left| \sum_{i=k-G+1}^{k} \left( d_{\mathcal{W}}^4(\nu_i, \hat{\mu}_{[k-G+1,k]}) - d_{\mathcal{W}}^4(\nu_i, \mu) \right) \right| \\ &\leq \frac{1}{G} \max_{G \leq k \leq n-G} \sum_{i=k-G+1}^{k} |d_{\mathcal{W}}(\nu_i, \hat{\mu}_{[k-G+1,k]}) - d_{\mathcal{W}}(\nu_i, \mu)| h_1(k,i) h_2(k,i) \\ &\leq \frac{1}{G} \max_{G \leq k \leq n-G} \sum_{i=k-G+1}^{k} d_{\mathcal{W}}(\hat{\mu}_{[k-G+1,k]}, \mu) h_1(k,i) h_2(k,i) \end{aligned} \quad (S.94)$$

where $h_1(k,i) = d_{\mathcal{W}}(\nu_i, \hat{\mu}_{[k-G+1,k]}) + d_{\mathcal{W}}(\nu_i, \mu)$, and $h_2(k,i) = d_{\mathcal{W}}^2(\nu_i, \hat{\mu}_{[k-G+1,k]}) + d_{\mathcal{W}}^2(\nu_i, \mu)$. The last inequality is obtained by the triangle inequality. Under the boundedness assumption for $\mathcal{W}_2(\mathcal{D})$, namely $\sup_{\nu_1, \nu_2 \in \mathcal{W}_2(D)} d_{\mathcal{W}}(\nu_1, \nu_2) < \infty$ (see Section 3.1 of the main text), it holds that there exists a universal constant $C_d$ such that

$$C_d = \text{diam}(\mathcal{W}_2(D)) = \sup_{\nu_1, \nu_2 \in \mathcal{W}_2(D)} d_{\mathcal{W}}(\nu_1, \nu_2) < \infty$$

where $\text{diam}(\mathcal{W}_2(D))$ denotes the diameter of $\mathcal{W}_2(D)$. Consequently, $h_1(k,i) h_2(k,i) \leq 2C_d \cdot 2C_d^2 = 4C_d^3 < \infty$. Let $C_0 = 4C_d^3$. Then, the result in (S.94) becomes

$$\begin{aligned} T_I &\leq C_0 \frac{1}{G} \max_{G \leq k \leq n-G} \sum_{i=k-G+1}^{k} d_{\mathcal{W}}(\hat{\mu}_{[k-G+1,k]}, \mu) \\ &\leq C_0 \max_{G \leq k \leq n-G} d_{\mathcal{W}}(\hat{\mu}_{[k-G+1,k]}, \mu) \end{aligned} \quad (S.95)$$

It is noteworthy that the Wasserstein space $(\mathcal{W}_2(D), d_{\mathcal{W}})$ equipped with the Wasserstein metric $d_{\mathcal{W}}$ fulfills assumptions (A1)–(A3) in Dubey and Müller (2020). Therefore, under $H_0$, the empirical Fréchet mean $\hat{\mu}_{[k-G+1,k]}$ computed from $\{\nu_{k-G+1}, \cdots,$ $v_k\}$ satisfy the conditions of Lemma 1 of Dubey and Müller (2020). This implies that $\max_{G \leq k \leq n-G} d_{\mathcal{W}}(\hat{\mu}_{[k-G+1,k]}, \mu) = O_p\left(\frac{1}{\sqrt{G}}\right)$. Combination with (S.95) yields

$$T_I = O_p\left(\frac{1}{\sqrt{G}}\right) \tag{S.96}$$

Furthermore, observe that

$$\frac{\frac{1}{\sqrt{G}}}{\frac{1}{\sqrt{\log(n/G)}}} = \sqrt{\frac{\log(n/G)}{G}} = \sqrt{\frac{\log n - \log G}{G}} \leq \sqrt{\frac{\log n}{G}} \leq \sqrt{\frac{n^{\frac{2}{2+\Delta}}\log n}{G}}$$

Under assumption (A1) in the main text, it follows that $\frac{\frac{1}{\sqrt{G}}}{\frac{1}{\sqrt{\log(n/G)}}} = o(1)$, thus

$$\frac{1}{\sqrt{G}} = o\left(\frac{1}{\sqrt{\log(n/G)}}\right) \tag{S.97}$$

Combination with (S.96) yields

$$T_I = o_p\left(\frac{1}{\sqrt{\log(n/G)}}\right) \tag{S.98}$$

By following similar arguments, we can also show that

$$T_{II} = \max_{G \leq k \leq n-G} |\hat{V}^2_{[k-G+1,k]} - \tilde{V}^2_{[k-G+1,k]}| = o_p\left(\frac{1}{\sqrt{\log(n/G)}}\right) \tag{S.99}$$

Combining (S.93), (S.98) and (S.99), we get

$$S_I = \max_{G \leq k \leq n-G} |\hat{\sigma}^2_{k,n,l} - \tilde{\sigma}^2_{k,n,l}| = T_I + T_{II} = o_p\left(\frac{1}{\sqrt{\log(n/G)}}\right) \tag{S.100}$$

We next turn to handle the second term $S_{II} = \max_{G \leq k \leq n-G} |\tilde{\sigma}^2_{k,n,l} - \sigma^2|$ given in (S.92). Substituting expression (S.91) for $\tilde{\sigma}^2_{k,n,l}$ into $S_{II}$ yields

$$S_{II} = \max_{G \leq k \leq n-G} \left| \frac{1}{G} \sum_{i=k-G+1}^{k} d_{\mathcal{W}}^4(v_i, \mu) - \left(\frac{1}{G}\sum_{i=k-G+1}^{k} d_{\mathcal{W}}^2(v_i, \mu)\right)^2 - \sigma^2 \right|$$

$$:= \max_{G \leq k \leq n-G} \left| \frac{1}{G} \sum_{i=k-G+1}^{k} Z_i^2 - \left(\frac{1}{G}\sum_{i=k-G+1}^{k} Z_i\right)^2 - \sigma^2 \right| \tag{S.101}$$

$$:= \max_{G \leq k \leq n-G} |\hat{\sigma}^{\#,2}_{k,n,l} - \sigma^2|$$

where $Z_i = d_{\mathcal{W}}^2(v_i, \mu)$, $i = 1, \cdots, n$, and $\hat{\sigma}^{\#,2}_{k,n,l} = \frac{1}{G}\sum_{i=k-G+1}^{k} Z_i^2 - \left(\frac{1}{G}\sum_{i=k-G+1}^{k} Z_i\right)^2$. It is noteworthy that, under $H_0$, $v_i$s are i.i.d. $\mathcal{W}_2(\mathcal{D})$-valued random objects, while $\mu = \underset{\omega \in \mathcal{W}_2(D)}{\mathrm{argmin}} E\left(d_{\mathcal{W}}^2(v_1, \omega)\right)$ is a deterministic but unknown $\mathcal{W}_2(\mathcal{D})$-valued object.

Consequently, $Z_i$s are i.i.d. real-valued random variables. Furthermore, by the definition of $\sigma^2$, we have $\sigma^2 = \text{var}\{d_{\mathcal{W}}^2(\mu, \nu)\} = \text{var}(Z) = E(Z^2) - \left(E(Z)\right)^2$. Consequently, $\hat{\sigma}_{k,n,l}^{\#,2}$ is an estimator of $\sigma^2 = \text{var}(Z)$. Under the boundedness assumption for $\mathcal{W}_2(\mathcal{D})$ (see Section 3.1 of the main text), the above variance estimator $\hat{\sigma}_{k,n,l}^{\#,2}$ computed from the i.i.d. real-valued random variables $Z_i = d_{\mathcal{W}}^2(\nu_i, \mu)$, $i = k - G + 1, \cdots, k$, fulfills the conditions of Corollary 6.16 of Muhsal (2013) under $H_0$. By Corollary 6.16 in Muhsal (2013), it follows that $\max_{G \leq k \leq n-G} |\hat{\sigma}_{k,n,l}^{\#,2} - \sigma^2| = o_p\left(\frac{1}{\log(n/G)}\right)$. Combination with (S.101) yields

$$S_{II} = o_p\left(\frac{1}{\log(n/G)}\right) = o_p\left(\frac{1}{\sqrt{\log(n/G)}}\right) \quad \text{(S.102)}$$

We can therefore conclude that

$$\max_{G \leq k \leq n-G} |\hat{\sigma}_{k,n,l}^2 - \sigma^2| = o_p\left(\frac{1}{\sqrt{\log(n/G)}}\right) \quad \text{(S.103)}$$

by combining (S.92), (S.100), and (S.102) together.

By adopting similar arguments as before, we can also obtain

$$\max_{G \leq k \leq n-G} |\hat{\sigma}_{k,n,r}^2 - \sigma^2| = o_p\left(\frac{1}{\sqrt{\log(n/G)}}\right) \quad \text{(S.104)}$$

Finally, from (S.89), it follows that

$$\max_{G \leq k \leq n-G} |\hat{\sigma}_{k,n}^2 - \sigma^2| \leq \frac{1}{2} \max_{G \leq k \leq n-G} |\hat{\sigma}_{k,n,l}^2 - \sigma^2| + \frac{1}{2} \max_{G \leq k \leq n-G} |\hat{\sigma}_{k,n,r}^2 - \sigma^2|$$

$$= o_p\left(\frac{1}{\sqrt{\log(n/G)}}\right) + o_p\left(\frac{1}{\sqrt{\log(n/G)}}\right)$$

$$= o_p\left(\frac{1}{\sqrt{\log(n/G)}}\right)$$

This completes the proof.

## S.4. Technical Details of the Simulation Study in the Main Text

This section provides additional technical details of the simulation study presented in Section 5 of the main text.

### S.4.1. Data-Generating Processes

We consider a DSF distributional sequence consisting of $n = 800$ PDFs, which is denoted as $\Gamma = \{f_1, \cdots, f_n\}$. We assume that the sequence contains three equally-spaced change points located at $k_1^* = 200$, $k_2^* = 400$, and $k_3^* = 600$. For convenience, we denote with $k_0^* = 0$ and $k_4^* = n$ the two boundary points. Consequently, $\Gamma$ can be divided into the following four segments separated by the three change points:

$$\Gamma_j^{seg} = \{f_i : i \in \mathbb{N} \text{ and } k_{j-1}^* < i \leq k_j^*\}, \quad j = 1, 2, 3, 4 \tag{S.105}$$

We consider two different data-generating processes (DGPs), referred to as DGP1 and DGP2, to generate DSF distributional sequences.

**(a) DGP1**

The DSF distributional data are generated based on a normal distribution model. We begin with notations. Let $f_{norm}(x|m, \sigma)$ denote the density function of a normal distribution with mean $m$ and standard deviation $\sigma$, and let $f_{norm}^T(x|m, \sigma)$ denote the density function after truncating $f_{norm}(x|m, \sigma)$ to the interval $[0,1]$ as follows:

$$f_{norm}^T(x|m, \sigma) = \frac{f_{norm}(x|m, \sigma)}{\int_0^1 f_{norm}(u|m, \sigma) du}, x \in [0,1]$$

We use the following piecewise stationary distributional process to generate the distributional sequence $\Gamma = \{\Gamma_1^{seg}, \Gamma_2^{seg}, \Gamma_3^{seg}, \Gamma_4^{seg}\}$:

$$\Gamma_j^{seg} = \{f_{norm,i}^T(x|m_i, 0.02) : i \in \mathbb{N} \text{ and } k_{j-1}^* < i \leq k_j^*\}, \quad j = 1, 2, 3, 4$$

with $m_i$ generated separately from the following four different models:

$$\begin{cases} m_i \stackrel{iid}{\sim} U(c_1 - \Delta_1, c_1 + \Delta_1) \text{ with } c_1 = 0.44 \text{ and } \Delta_1 = 0.005, & k_0^* < i \leq k_1^* \\ m_i \stackrel{iid}{\sim} U(c_2 - \Delta_2, c_2 + \Delta_2) \text{ with } c_2 = 0.44 \text{ and } \Delta_2 = 0.050, & k_1^* < i \leq k_2^* \\ m_i \stackrel{iid}{\sim} U(c_3 - \Delta_3, c_3 + \Delta_3) \text{ with } c_3 = 0.48 \text{ and } \Delta_3 = 0.050, & k_2^* < i \leq k_3^* \\ m_i \stackrel{iid}{\sim} U(c_4 - \Delta_4, c_4 + \Delta_4) \text{ with } c_4 = 0.40 \text{ and } \Delta_4 = 0.100, & k_3^* < i \leq k_4^* \end{cases}$$

in which $U(c_j - \Delta_j, c_j + \Delta_j)$ stands for a uniform distribution on $[c_j - \Delta_j, c_j + \Delta_j]$. $c_j$ and $\Delta_j$ control the central tendency and dispersion of the generated random samples $\{m_i : k_{j-1}^* < i \leq k_j^*\}$ associated with $\Gamma_j^{seg}$, respectively. In this setting, due to the fact that the parameter $\sigma$ of the truncated normal distribution is fixed, a change in $c_j$ induces a change in the Fréchet mean of the generated DSF distributional data, while a change in $\Delta_j$ induces a change in the Fréchet variance. This is to mimic the practical situation that a DSF distributional sequence extracted from real SHM data usually undergoes changes in both mean and variance structures.

The distributional sequence simulated using DGP1 is illustrated in Figure S.6 (a).

**(b) DGP 2**

The DSF distributional data are generated based on the log quantile density (LQD) transformation proposed by Petersen and Müller (2016). A brief introduction to this transformation is presented in Appendix S.1. Let $T = [0,1] \subset \mathbb{R}$ denote a compact interval, and let $L^2(T)$ denote the Hilbert space of square integrable real functions on $T$. Taking advantage of the linear structure of $L^2(T)$, we first generate a $L^2(T)$-valued functional sequence denoted as $\Psi = \{\psi_1(t), \cdots, \psi_n(t)\}$; we then map each element of $\Psi$ to the density space using the inverse LQD transformation (equation (S.128) in Appendix S.1), producing a DSF distributional sequence represented as

$$\Gamma = \{f_1(x), \cdots, f_n(x)\} = \{\text{LQD}^{-1}[\psi_1](x), \cdots, \text{LQD}^{-1}[\psi_n](x)\}$$

where $\text{LQD}^{-1}[\psi_i]$ stands for the inverse LQD transformation of $\psi_i \in \Psi$.

With the three pre-specified change points, the $L^2(T)$-valued functional sequence $\Psi$ can be divided into the following four segments (analogous to (S.105)):

$$\Psi_j^{seg} = \{\psi_i : i \in \mathbb{N} \text{ and } k_{j-1}^* < i \leq k_j^*\}, \quad j = 1, 2, 3, 4$$

We then employ the following $L^2(T)$-valued piecewise stationary process to generate $\Psi$:

$$\psi_i(t) = \sum_{j=1}^{4} \delta_j(t) I\{k_{j-1}^* < i \leq k_j^*\} + e_i(t), \quad i = 1, \cdots, n \quad \text{(S.106)}$$

where $\delta_j(t), j \in \{1,2,3,4\}$ is the mean function associated with the $j$th segment $\Psi_j^{seg}$, $\{e_i(t): i = 1, \cdots, n\}$ is the sequence of error functions with zero mean and piecewise constant covariance function (indexed by $i$). In this setting, the generated $L^2(T)$-valued functional sequence has both piecewise constant mean and covariance structures. In other words, the functional sequence $\Psi$ is allowed to undergo changes in mean, variance, or both at each change point.

For $j \in \{1,2,3,4\}$, the mean function $\delta_j(t)$ in (S.106) is generated by using the LQD transformation (equation (S.127) in Appendix S.1) to map a PDF $(f_{m,j})$ to $L^2(T)$ as follows:

$$\delta_j(t) = \text{LQD}[f_{m,j}](x), \quad j = 1, 2, 3, 4$$

with

$$f_{m,j}(x) = 0.9\big(0.8 f_{Beta}(x|28, 22 + 2j) + 0.2 f_{Beta}(x|14, 31 + j)\big) + 0.1$$

in which $f_{Beta}(x|a, b)$ stands for the density function of the Beta distribution with parameters $a$ and $b$.

The sequence of error functions (i.e., $\{e_i(t): i = 1, \cdots, n\}$) is generated using the following zero-mean Gaussian process (GP) on $T = [0,1]$:

$$e_i \sim \mathcal{GP}(0, c_i), \quad i = 1, \cdots, n$$

where $c_i = c_i(s, t)$ with $s, t \in T \times T$ is the covariance function of the Gaussian process at the location $i \in \{1, 2, \cdots, n\}$. The covariance structure of the error functions is allowed to change between segments separated by a change point. To this end, the process $\{c_i(s,t): i = 1,2,\cdots,n\}$ is produced based on the following piecewise constant model:

$$c_i(s,t) = \kappa(s,t|\eta,\varrho)\sum_{j=1}^{4}\theta_j I\{k_{j-1}^* < i \leq k_j^*\}, \qquad i = 1,2,\cdots,n$$

where $\kappa(s,t|\eta,\varrho) = \exp\{-|s-t|^\eta\}/(2\varrho^2)$ is the exponential covariance kernel with parameters $\eta$ and $\varrho$, and $\theta_j, j \in \{1,2,3,4\}$ is a factor for adjusting the amplitude of the covariance surface $c_i(s,t)$ associated with the $j$th segment. We set $\eta = 1.99999$, $\varrho = 0.2$, and $\theta = (0.003, 0.030, 0.001, 0.035)$ in this simulation. Consequently, the simulated sequence of error functions undergoes a change in covariance structure in each of the change points $k_1^*$, $k_2^*$, and $k_3^*$.

The distributional sequence simulated using DGP2 is illustrated in Figure S.6 (b).

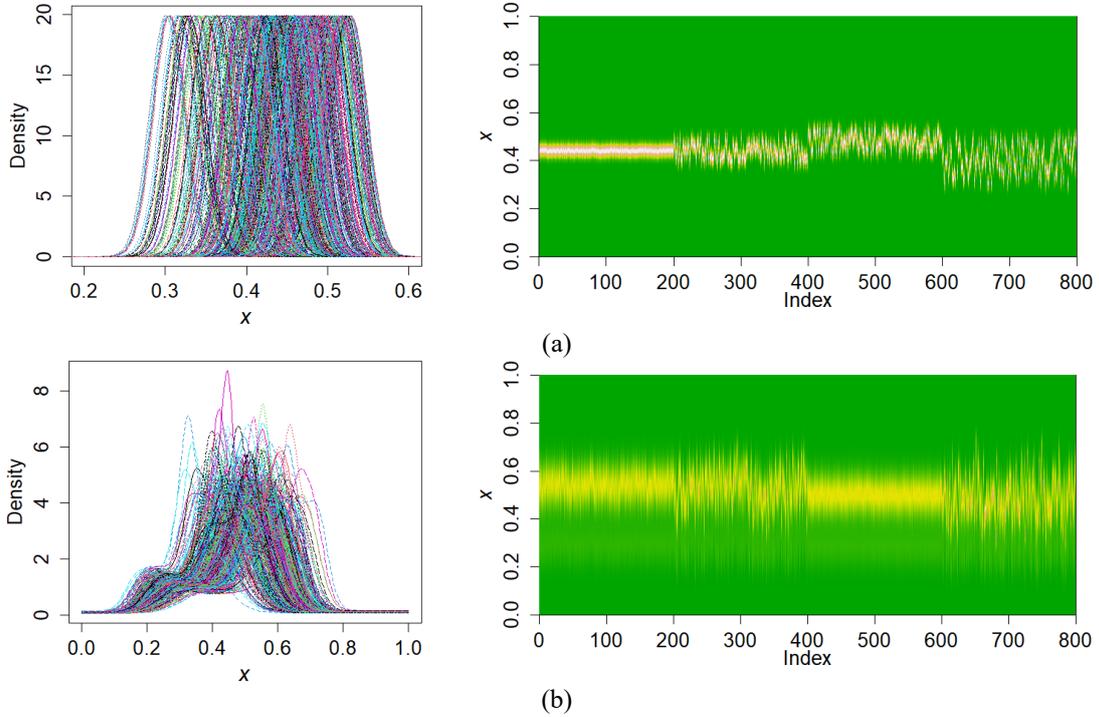

**Figure S.6.** Representative distributional data generated by (a) the first data-generating process (DGP1) and (b) the second data-generating process (DGP2), respectively. The left column corresponds to PDF curves, while the right column corresponds to the heatmap of the PDF-valued sequence.

### *S.4.2. Implementation Settings*

This subsection presents implementation settings for our Fréchet-MOSUM method and the considered competitors. Throughout the rest of this study, significance levels of related hypothesis testing-based change-point detectors (e.g., Fréchet-MOSUM, FPCA-ECP, GS) are all set to $\alpha = 0.05$, unless otherwise stated. The other settings are detailed as follows:

For the Fréchet-MOSUM method, the bandwidth $G$ and the AOP parameter $\varepsilon$ are set as $G = 80$ and $\varepsilon = 0.2$, respectively, on the basis of the recommended settings described in Section S.2.8. The parameter $c$ involved in the boundary correction procedure (Section S.2.7) is set to 0.1 (the default value).

For the FPCA-ECP method (Lei et al. 2023b), we use the recommended settings in the paper, except that the dimensionality $m$ of the reduced functional data (see (13) in Lei et al. (2023b)) is selected adaptively from the data, such that, 99% of the total variation can be explained by the resulting truncated Karhunen–Loève representation. In other words, $m$ is determined by setting the fraction of variance explained (FVE) in the FPCA to be 99%.

For the DSBE method (Chiou et al. 2019), the parameter $K$ (i.e., the number of equally-spaced segments in the initial segmentation) involved in step (D1) of Algorithm DS in Chiou et al. (2019) is set to 10. Given a functional sequence indexed by the set $\{1,2,\cdots,n\}$, the $j$th change point in our analysis is represented by $k_j^* \in \{1,2,\cdots,n\}$ rather than $\theta_j^* = (k_j^*/n) \in (0,1)$ (the relative position) as that in Chiou et al. (2019). Consequently, the searching interval $\mathfrak{I}_j^{(r)}(h) = \left(\theta_{j-1}^{(r+1)} + h, \ \theta_{j+1}^{(r)} - h\right] \subset (0,1)$ (involved in step (D3) of Algorithm DS in Chiou et al. (2019)) needs to be changed to $\mathfrak{I}_j^{(r)}(\tilde{h}) = \left(k_{j-1}^{(r+1)} + \tilde{h}, \ k_{j+1}^{(r)} - \tilde{h}\right] \subset (1,n)$ with $\tilde{h} = \lfloor nh \rfloor$, and we set $\tilde{h} = 5$. On the other hand, the DSBE procedure requires projecting the functional data onto a low dimensional subspace denoted by $V_m = \text{span}\{e_1,\cdots,e_m\}$, where $\{e_l\}_{l=1}^m$ is a collection of $m$ orthonormal basis functions of $L^2(T)$. Following Chiou et al. (2019), $V_m$ is built based on the eigenspace of the empirical covariance operator associated with the functional samples. Consequently, the orthonormal system $\{e_l\}_{l=1}^m$ is composed of $m$ FPCs associated with the first $m$ largest eigenvalues. Similar to the setting of FPCA-ECP, $m$ is also selected adaptively from the data by setting FVE=99%. Moreover, we choose the thresholding criterion, detailed in Appendix S.2 of the supplement, as the stopping rule for the backward elimination (BE) procedure.

For the FMCI method (Harris et al. 2022), we implement the change-point detection procedure using the R package *fmci* downloaded from the authors' website (*https://trevor-harris.github.io/code/*), and adopt the default parameter settings.

For the FBS method (Rice and Zhang 2022), we use the default parameter settings recommended by the authors.

For the GS method (Chen et al. 2023), the trimming parameter $h$ involved in step (G2) of the GS estimator is set to 0.02 (the recommended value provided by the authors), and the empirical quantile of the asymptotic distribution of the test statistic is computed based on 200 Monte Carlo samples of Brownian bridges. Similar to DSBE, the GS procedure also requires projecting the functional data onto a subspace denoted by $V_m = \text{span}\{e_1, \cdots, e_m\}$. Following the authors' recommendation, $V_m$ is determined based on the FPCA technique with FVE=99%.

Recall that five of the considered competing methods (i.e., FPCA-ECP(LQD), DSBE, FMCI, FBS, and GS) require performing transformations on the distributional data using the LQD transformation, see Section 5 of the main text for details. As pointed out for instance by Lei, Chen and Li (2023) and Lei et al. (2023b), the LQD-transformed data are insensitive to the horizontal shift of the distributional data in the density space. Practically, this issue can be alleviated by performing a pretreatment on the distributional data using Eq. (10) in Lei et al. (2023b). In our analysis, we also apply this pretreatment to the distributional data before performing the LQD transformation, and adopt the default setting in Lei et al. (2023b).

### S.4.3. Quality Measure (Hausdorff Distance)

Suppose that the distributional sequence contains $q$ different but unknown change points denoted as $k_1^*, k_2^*, \cdots, k_q^*$. The error of a multiple change-point detector in estimating the locations of these change points can be measured by the Hausdorff distance (a metric for quantifying the dis-similarity between two sets (Brault et al. 2018)). The true change points and their estimators constitute two different sets denoted as $S_{CP} = \{k_1^*, k_2^*, \cdots, k_q^*\}$ and $\hat{S}_{CP} = \{\hat{k}_1^*, \hat{k}_2^*, \cdots, \hat{k}_{\hat{q}}^*\}$, respectively. The Hausdorff distance between $S_{CP}$ and $\hat{S}_{CP}$, denoted as $d_H(S_{CP}, \hat{S}_{CP})$, is defined as (Brault et al. 2018)

$$d_H(S_{CP}, \hat{S}_{CP}) = \begin{cases} \max\{d_1(S_{CP}, \hat{S}_{CP}), d_2(\hat{S}_{CP}, S_{CP})\}, & \text{if } \hat{S}_{CP} \neq \emptyset \\ \max_{k \in S_{CP}} |k|, & \text{if } \hat{S}_{CP} = \emptyset \end{cases} \quad (S.107)$$

where $d_1(S_{CP}, \hat{S}_{CP}) = \max_{k \in S_{CP}} \min_{\hat{k} \in \hat{S}_{CP}} |k - \hat{k}|$ and $d_2(\hat{S}_{CP}, S_{CP}) = \max_{\hat{k} \in \hat{S}_{CP}} \min_{k \in S_{CP}} |\hat{k} - k|$.

## S.5. Additional Simulation Studies

### *S.5.1. Additional Simulation Study 1*

In this simulation study, we compare the computational efficiency between our Fréchet-MOSUM method and the FPCA-ECP method. In FPCA-ECP, the detection of functional change points is achieved by detecting change points in the time series of FPC scores using the E-Divisive (ECP) method proposed by Matteson and James (2014). A major drawback of the E-Divisive method is computationally intensive due to the inefficient permutation tests that are involved (each permutation test requires re-calculating the energy divergence measure with a computational complexity of $O(n^2)$) (Fryzlewicz 2014; Cleynen and Lebarbier 2017; Arlot et al. 2019). Consequently, the run time of the E-Divisive detector will rapidly increase with the sample size $n$ (length of the data sequence). As pointed out by Biau et al. (2016), the E-Divisive method is computationally prohibitive when $n$ exceeds a few thousand. In contrast, our Fréchet-MOSUM method is highly efficient with a computational complexity of $O(n)$. In the following, we use simulated data to compare the computational efficiency between the two methods.

### *S.5.1.1. Data Generation and Implementation Settings*

The data-generating procedure used in this simulation study is summarized in Algorithm S.1.

---
**Algorithm S.1**: Algorithm for generating $m$ distributional sequences of lengths $L_1, \cdots, L_m$, such that $L_1 < \cdots < L_m$.

---
**Input**: The number of duplications $N_{dup}$ and a vector $S_L = (L_1, \cdots, L_m)$ for specifying the lengths of the $m$ distributional sequences to be generated

**Output**: The generated $m$ distributional sequences of lengths $L_1, \cdots, L_m$

1: Generate a baseline distributional sequence of length 500 with one change point using the Beta distribution model
$$\Gamma_{bl} = \{f_i(x) = \text{BetaPdf}(x; a_i, 32): i = 1, 2, \cdots, 500\}$$
in which $a_i \overset{i.i.d}{\sim} U(13, 17)$ for $i = 1, 2, \cdots, 250$ and $a_i \overset{iid}{\sim} U(16, 20)$ for $i = 251, 252, \cdots, 500$. Here, $U(\alpha, \beta)$ stands for the uniform distribution on $[\alpha, \beta]$.

2: Duplicate $\Gamma_{bl}$ for $N_{dup}$ times to produce a long distributional sequence
$$\Gamma_{long} = \{\Lambda_1, \Lambda_2, \cdots, \Lambda_{N_{dup}}\} \text{ with } \Lambda_j = \Gamma_{bl} \text{ for } j = 1, 2, \cdots, N_{dup}$$

3: Extract $m$ sub-sequences of lengths $L_1, \cdots, L_m$ from $\Gamma_{long}$
$$\Gamma_s^{sub} = \{\Gamma_{long}(1), \Gamma_{long}(2), \cdots, \Gamma_{long}(L_s)\} \text{ for s} = 1, 2, \cdots, m.$$

4: Output $\Gamma_1^{sub}, \Gamma_2^{sub}, \cdots,$ and $\Gamma_m^{sub}$.

---

First, we set $N_{dup} = 4$ and $S_L = (250, 500, 750, 1000, 1250, 1500, 1750, 2000)$ in Algorithm S.1 to generate 8 distributional sequences of lengths ranging from $n = 250$ to $n = 2000$ with the increments of 250. We then employ the Fréchet-MOSUM and FPCA-ECP methods to perform change-point detections on the 8 distributional sequences and record the run time (of handling each sequence) for the two methods. For the FPCA-ECP method, we only consider the FPCA-ECP(LQD) detector. The other detector (i.e., FPCA-ECP(WassTS)) has similar computational complexity with its LQD counterpart, thus we do not consider it in this comparison. The implementation settings for Fréchet-MOSUM and FPCA-ECP are the same as those described in Section S.4.2 of the supplement.

*S.5.1.2. Results*

All the numerical experiments are performed on a PC with an Intel(R) Core(TM) i-5 processor, and we run the code in a R software of version 4.2.3. The run time of the two methods with respect to the sample size $n$ is shown in Figure S.7, revealing that the run time of our Fréchet-MOSUM method is far shorter than the FPCA-ECP method. Increasing $n$ from 1000 to 2000 has no significant impact on the run time of our method, but imposes a substantial burden to the FPCA-ECP method.

To demonstrate that our Fréchet-MOSUM method has a linear computational complexity, we conduct an additional experiment. We use Algorithm S.1 to generate 10 distributional sequences of lengths ranging from $n = 2000$ to $n = 20000$ with the increments of 2000, which can be achieved by setting $N_{dup} = 40$ and $S_L = (2000, 4000, 6000, 8000, 10000, 12000, 14000, 16000, 18000, 20000)$. Figure S.8 displays the time spent of the Fréchet-MOSUM method in performing change-point detection on each of the simulated distributional sequences. As expected, the run time increases linearly with the sample size $n$, revealing that our method has a linear computational complexity. From Figure S.8, we see that our method only requires about 20 seconds to accomplish the detection for a long distributional sequence of 20000. In contrast, the FPCA-ECP method requires more than 180 seconds (3 minutes) to accomplish the detection for a much shorter sequence that only contains 2000 functional samples (see Figure S.7). Therefore, our method is of immense computational advantage in analyzing large datasets.

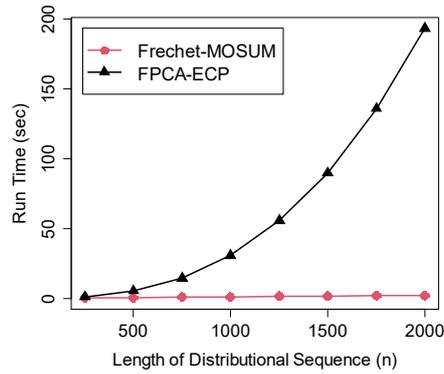

**Figure S.7.** Comparison of time spent of the Fréchet-MOSUM and PFCA-ECP methods to accomplish change-point detection for distributional sequences with different lengths.

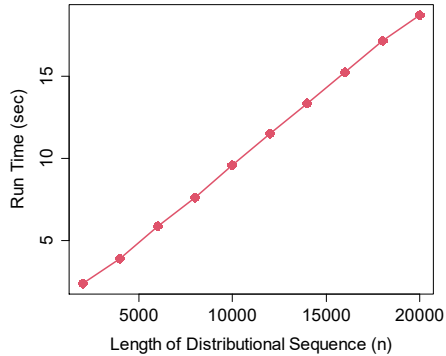

**Figure S.8.** The time spent of the Fréchet-MOSUM method to accomplish change-point detection for each of the long distributional sequences.

### *S.5.2. Additional Simulation Study 2*

This simulation study is designed to demonstrate a major drawback of the LQD transformation in distributional change-point detection.

Our Fréchet-MOSUM method is general and can be easily extended for applicability to data residing in a general metric space $(\Omega, d)$, and we only need to replace the Wasserstein distance $d_{\mathcal{W}}$ with the distance $d$ (endowed with the metric space $\Omega$). The LQD-transformed data are ordinary functional data residing in the $L^2([0,1])$ space (see Appendix S.1 of the supplement), which is a metric space endowed with the $L^2$ distance

$$d_{L^2}(\psi_1, \psi_2) = \left(\int \big(\psi_1(\tau) - \psi_2(\tau)\big)^2 d\tau\right)^{1/2}, \forall \psi_1, \psi_2 \in L^2([0,1])$$

Therefore, our Fréchet-MOSUM detector is also applicable to the LQD-transformed data. However, we do not adopt such a LQD transformation strategy in our distributional change-point analyses. The main reason is that the LQD transformation has a crucial drawback that it can magnify some minor shape changes of no close relation to the distributional changes, which may lead to serious issues in both interpretability and practicability; more detailed discussions and demonstrations will be provided later in this subsection. Due to this, all of the LQD transformation-based competitors (e.g., the FPCA-ECP (LQD) detector) considered in the simulation study of the main text suffer from a similar limitation.

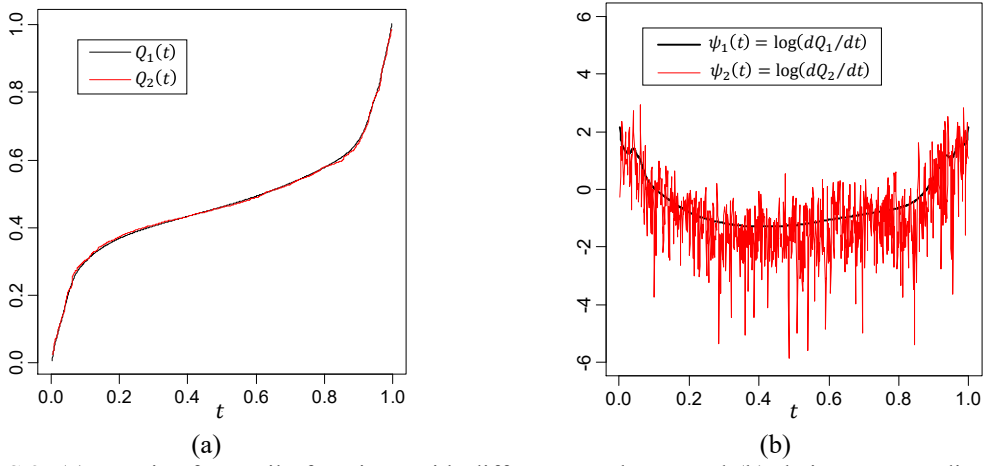

**Figure S.9.** (a) A pair of quantile functions with different roughness and (b) their corresponding LQD-transformed curves.

Before proceeding, we provide an in-depth discussion on the LQD transformation. Given a PDF $f(x)$, let $Q(t)$ be the associated quantile function, then the LQD-transformation of $f(x)$ is the logarithm of the derivative of the quantile function $Q(t)$ defined as follows (Petersen and Müller 2016) (see Appendix S.1 for more details):

$$\psi(t) = \text{LQD}[f](t) = \log\left(\frac{dQ(t)}{dt}\right) \qquad (S.108)$$

Due to the derivative operation, the LQD-transformed result of a distributional sequence is highly sensitive to the changes in slope of quantile functions. Consequently, the LQD transformation may magnify some minor shape changes of quantile functions that are not closely related to the changes in the distribution of the raw observations (e.g., the DSF data in our analysis). For instance, the two quantile functions $Q_1$ and $Q_2$ shown in Figure S.9 (a) are very close together in amplitude, indicating that there is only a small difference between the two associated probability distributions; however, the corresponding LQD-transformed curves (Figure S.9 (b)) are substantially different with each other, primarily because the curves of the two quantile functions are significant different in roughness. In this situation, a significant change in roughness of the quantile functions does not mean that the associated distributions are significantly changed, but it can produce a significant change to the LQD-transformed data. Therefore, a statistically significant change detected from LQD-transformed data may not be a manifestation of a significant change of the associated probability distributions. If we use the LQD-transformed data to perform change-point detection for the distributional data, the interpretability of the detected changes might be lost. Moreover, the aforementioned shape-related magnification effect of the LQD transformation can also bring some other serious issues in real applications, and a detailed discussion is deferred to the end of this simulation study.

*S.5.2.1. Data Generation and Implementation Settings*

In the following, we use synthetic data to investigate the impact of the shape-related amplification effect of the LQD transformation on distributional change-point detection. We first generate a sequence of quantile functions associated with an identical probability distribution, that undergoes a change in roughness of the curves. Specifically, let $\Gamma_Q = \{Q_1, \cdots, Q_{500}\}$ denote the functional sequence consisting of 500 quantile functions to be generated in this simulation, and we divide it into two equi-length segments denoted as $\Gamma_Q^{left} = \{Q_1, \cdots, Q_{250}\}$ and $\Gamma_Q^{right} = \{Q_{251}, \cdots, Q_{500}\}$, respectively. All the quantile functions are estimated from samples generated from an identical Beta distribution, but the estimation strategies are different between $\Gamma_Q^{left}$ and $\Gamma_Q^{right}$. The data-generating procedure is detailed as follows:

We independently generate 500 different groups of i.i.d. samples from an identical Beta distribution Beta(35,24), and denote the results as

$$\{X_k^1\}_{k=1}^N \overset{iid}{\sim} \text{Beta}(35,24), \cdots, \{X_k^{500}\}_{k=1}^N \overset{iid}{\sim} \text{Beta}(35,24) \quad \quad \text{(S.109)}$$

where $N$ is the sample size, which we set to 400 in this simulation. These random samples correspond to the raw DSF data in our analysis.

The quantile functions in $\Gamma_Q^{left}$ are calculated from the kernel density functions estimated from the first 250 groups of Beta distributed samples, respectively. Let $\hat{f}_i^{kerl}(x)$ ($i \in \{=1, \cdots, 250\}$) denote the density function estimated from $\{X_k^i\}_{k=1}^N$ using a kernel density estimator. For $i = 1, 2, \cdots, 250$, the quantile function $Q_i(t)$ is computed as $Q_i(t) = F_i^{-1}(t)$ with $F_i(x) = \int_{-\infty}^x \hat{f}_i^{kerl}(u) du$. For convenience, this estimation strategy for quantile functions is referred to as the kernel smoothing estimation (KSE) strategy throughout the rest of this simulation study.

The quantile functions in $\Gamma_Q^{right}$ are obtained based on sample quantiles computed from the last 250 groups of Beta distributed samples, respectively. Let $T_{grid} = \{t_1, \cdots, t_T\} \subset [0,1]$ be an equally-spaced grid for computing the sample quantiles, and we set $T = 201$, $t_1 = 0$, and $t_T = 1$ in this simulation. For $i = 251, 252, \cdots, 500$, the values of $Q_i(t)$ at the grid are estimated as the sample quantiles computed using the R function "quantile(x, probs)" with x= $\{X_k^i\}_{k=1}^N$ and probs= $\{t_1, \cdots, t_T\}$. The full function of $Q_i(t)$ is obtained by linear interpolation between the estimated values on the grid. For convenience, this estimation strategy for quantile functions is referred to as the sample quantile interpolation (SQI) strategy.

The functional samples of a representative quantile-function sequence produced by the above data-generating procedure are visualized on the left panel of Figure S.10. For comparison, the functional samples associated with $\Gamma_Q^{left}$ and $\Gamma_Q^{right}$ are visualized on the middle and right panels of Figure S.10, respectively. The corresponding LQD-transformed data are visualized in Figure S.11. Since all of the 500 groups of samples given in equation (S.109) are generated from an identical distribution, no change occurs in the underlying distribution of the raw data. The minor differences of the quantile functions between $\Gamma_Q^{left}$ and $\Gamma_Q^{right}$ are primarily due to the fact that the functions are estimated using different strategies. For a better comparison, Figure S.12 displays two quantile functions that are estimated from an identical group of samples using the KSE and SQI strategies, respectively. Except for the lower and upper tails, the two quantile functions agree well in amplitude with each other. The significant discrepancies near the tails may be caused by the scarcity of samples, indicating the uncertainties of estimates near the tails are much larger than those in the central part of the curve.

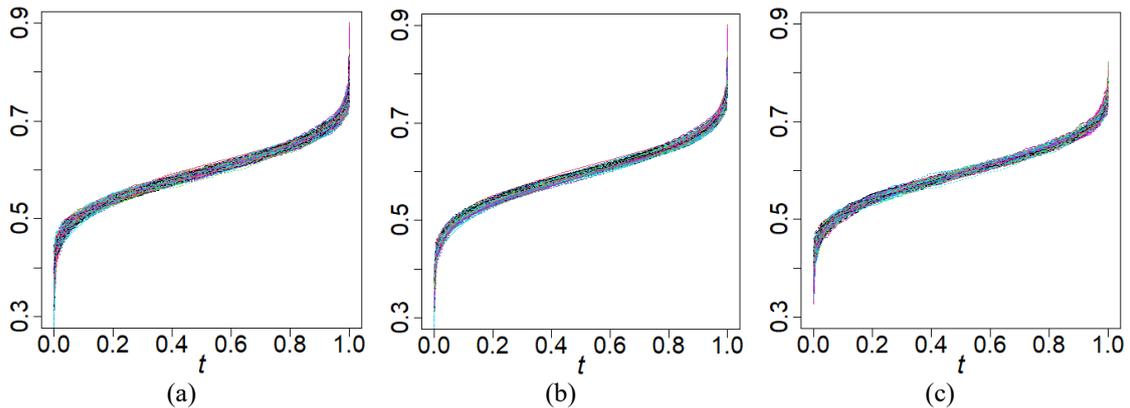

(a)          (b)          (c)

**Figure S.10.** Functional samples of a representative generated sequence of quantile functions. (a) Functional samples of the whole sequence $\Gamma_Q$, (b) functional samples of the left half segment $\Gamma_Q^{left}$, and (c) functional samples of the right half segment $\Gamma_Q^{right}$.

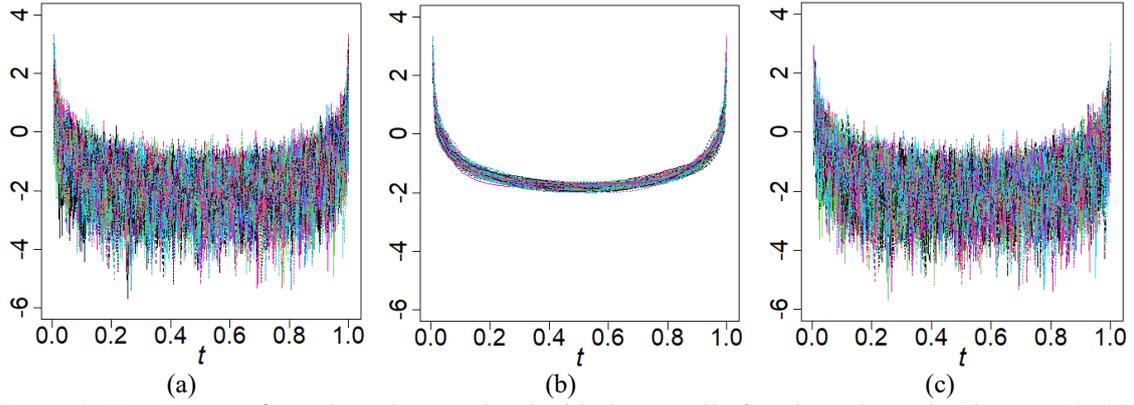

(a)                        (b)                      (c)

**Figure S.11.** LQD-transformed results associated with the quantile functions shown in Figure S.10. (a) Results correspond to the whole sequence $\Gamma_Q$, (b) results correspond to the left half segment $\Gamma_Q^{left}$, and (c) results correspond to the right half segment $\Gamma_Q^{right}$.

The primary objective of this simulation study is to investigate the amplification effect on local minor differences of curves induced by the LQD transformation; therefore, we can exclude the tail parts of the curves in subsequent functional change-point analysis. Specifically, all the quantile functions are postprocessed by truncating them to the interval $A = [0.1, 0.9]$, and the resulting functional sequence is denoted as

$$\Gamma_Q^T = \{Q_1(t)I_A(t), \cdots, Q_{500}(t)I_A(t)\} \tag{S.110}$$

where $I_A(\cdot)$ is the indicator function. We then recompute the LQD transformation by substituting $Q(t)$ in equation (S.108) with the truncated version $Q(t)I_A(t)$, and the resulting LQD-transformed sequence is denoted as

$$\Gamma_\psi^T = \left\{\log\left(\frac{d(Q_1(t)I_A(t))}{dt}\right), \cdots, \log\left(\frac{d(Q_{500}(t)I_A(t))}{dt}\right)\right\} \tag{S.111}$$

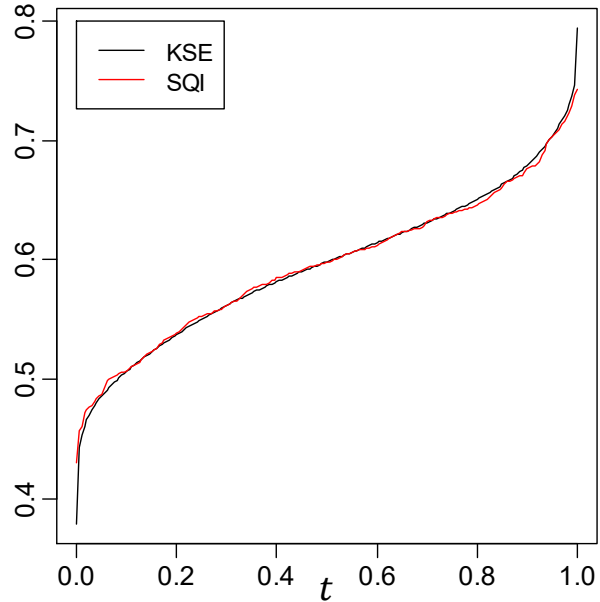

**Figure S.12.** Comparison of two representative quantile functions estimated from the same group of samples using the KSE and SQI strategies, respectively.

To investigate the impact of the shape-related amplification effect of the LQD transformation on distributional change-point detection, we conduct a performance comparison between the following two different distributional change-point detection schemes:

Scheme I: Perform change-point detection on the quantile-function sequence $\Gamma_Q^T$ using the Fréchet-MOSUM detector with $L^2$ distance.

Scheme II: Perform change-point detection on the LQD-transformed sequence $\Gamma_\psi^T$ using the Fréchet-MOSUM detector with $L^2$ distance.

**Remark 5.** Scheme I corresponds to our proposed distributional change-point detection strategy. Mathematically, a distributional sequence $\Gamma = \{\nu_1, \cdots, \nu_n\}$ can be equivalently represented by the corresponding quantile-function sequence $\Gamma_Q = \{Q_1, \cdots, Q_n\} = \{F_1^{-1}, \cdots, F_n^{-1}\}$. The Wasserstein distance between two probability measures $\nu_i, \nu_j \in \mathcal{W}_2(\mathcal{D})$ is equivalent to the $L^2$ distance between the associated quantile functions, namely $d_\mathcal{W}(\nu_i, \nu_j) = d_{L^2}(F_{\nu_i}^{-1}, F_{\nu_j}^{-1}) = d_{L^2}(Q_i, Q_j)$ (see equation (1) of the main test). From the implementation details of the Fréchet-MOSUM method, described in Section 4 of the main text, we see that performing change-point detection on the distributional sequence $\Gamma = \{\nu_1, \cdots, \nu_n\}$ using the Wasserstein distance is equivalent to performing change-point detection on the associated quantile-function sequence $\Gamma_Q = \{F_1^{-1}, \cdots, F_n^{-1}\}$ using the $L^2$ distance. However, if we perform change-point detection on the LQD-transformed sequence using the Fréchet-MOSUM detector with $L^2$ distance (scheme II), then it is no longer equivalent to our proposed distributional change-point detection strategy using the Wasserstein distance. This is because the Wasserstein distance between two probability measures is not equivalent to the $L^2$ distance between the associated LQD-transformed functions, namely $d_\mathcal{W}(\nu_i, \nu_j) \neq d_{L^2}(\psi_i, \psi_j)$.

We set $G = 100$, $\varepsilon = 0.2$, and $\alpha = 0.05$ for the Fréchet-MOSUM detector, and do not consider the boundary correction for the SS sequence.

*S.5.2.2. Results*

For a fair comparison, we repeat the change-point detection experiments for $N = 200$ times. In each experiment, the data are re-generated using the same data-generating procedure described above.

Let $N_{QF}^+$ and $N_{LQD}^+$ be the total number of experiments that at least one change point is identified using scheme I and scheme II, respectively. We define the positive rates as $r_{QF}^+ = N_{QF}^+/N$ and $r_{LQD}^+ = N_{LQD}^+/N$ for the two schemes, where $N$ is the total number of experiments. The higher the positive rate, the higher the chance that the change-point detector identifies statistically significant changes in the repeated experiments. As noted earlier, the generated raw data contain no distributional change, and there is only a minor difference between the estimated distributional data (represented by quantile functions) associated with the left and right half segments. Therefore, we do not expect the positive rate to take a high value.

Based on the detection results of the $N = 200$ experiments, the calculated positive rates of scheme I and scheme II are $r_{QF}^+ = 0.405$ and $r_{LQD}^+ = 1.000$, respectively. The computed SS sequences in three representative experiments are depicted in Figure S.13, where the left and right columns correspond to the results of scheme I and scheme II, respectively. Here we see that, near the middle of the SS sequence, the results of scheme II are far larger than those of scheme I. Combining with the calculated positive rate, we see that, in scheme II, the Fréchet-MOSUM detector identifies changes in all of the 200 experiments; however, in scheme I (the recommend strategy), the calculated positive rate is much lower. The above results indicate that the LQD transformation-based distributional change-point detection strategy is highly sensitive to changes in slope of quantile functions, while our recommended strategy is far less sensitive to such changes.

In this simulation, although the change in roughness of the curves is primarily induced by the switching of curve estimation strategy, it provides some profound insights into the shape-related magnification effect of the LQD transformation. One may argue that similar estimation-induced changes can be avoided by using the same estimation strategy; however, in practical applications, due to the high sensitivity to changes in slope (of quantile functions), the LQD transformation might produce some other unexpected magnification phenomena, leading to unforeseen negative impacts on distributional change-point analyses.

It is noteworthy that a significant change in slope of quantile functions (manifested as a statistically significant change in the LQD-transformed data) cannot offer us sufficient confidence to conclude that the underlying distribution of the raw observations (e.g., the data in equation (S.109)) is changed, such as the case in this simulation. In structural damage detection, our application of interest, if a change-point detector employed for detecting changes in distribution of DSF data is highly sensitive to changes in slope of the associated quantile functions, then it has a high risk of producing questionable results of no practical relevance to the distributional changes of DSF data. On the other hand, we only expect to keep the change points that have statistically significant changes in distribution of the raw DSF data for potential downstream analyses (e.g., ascertain whether the distributional change is caused by operational/environmental disturbance or damage). However, the shape-related magnification effect of the LQD transformation might make the significance test fail to offer a meaningful guidance on automatically screening change points that fulfill this practical requirement.

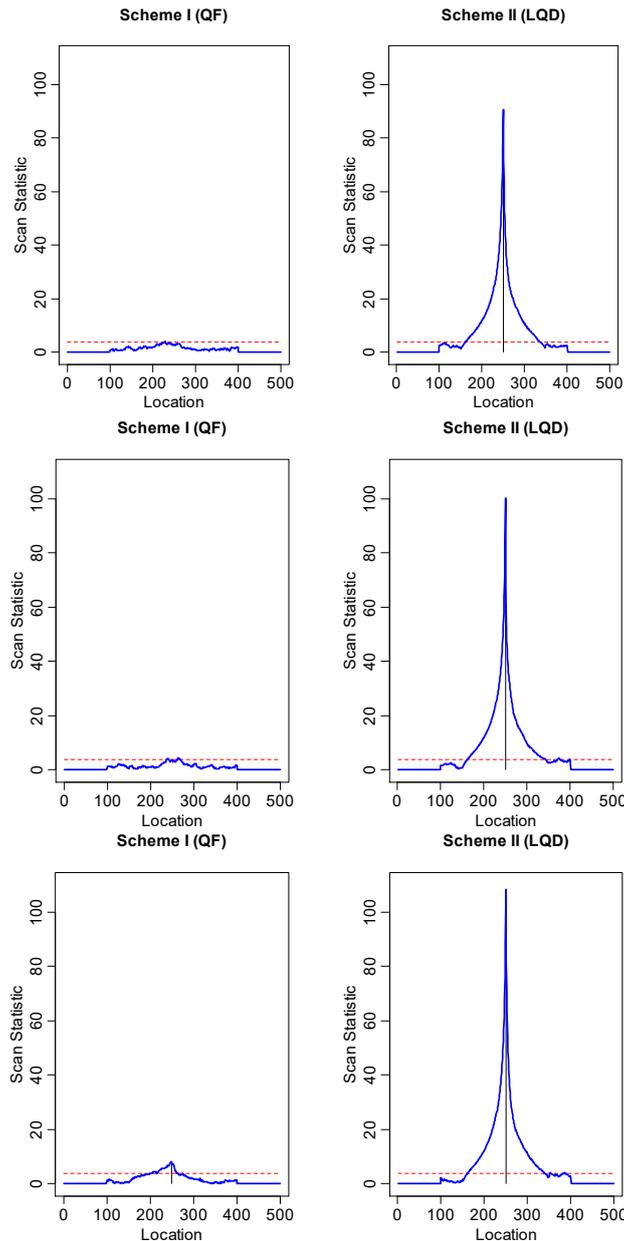

**Figure S.13.** SS sequences (blue lines) with the detected change points (marked by vertical solid lines) in three representative experiments. The left column corresponds to the result of scheme I, while the right column corresponds to the result of scheme II. The critical value in the significance test is marked by the horizontal dashed line, and the location of a detected change point is marked by the vertical solid line. A plot without a vertical solid line indicates that no change point is identified using the change-point detector.

*S.5.3. Additional Simulation Study 3*

The competing methods DSBE (Chiou et al. 2019), GS (Chen et al. 2023), and FPCA-ECP (Lei et al. 2023b) considered in the simulation study of the main text (Section 5) involve a dimension reduction processing by projecting the functional data onto a low dimensional subspace spanned by dominant principal curves. If the function that characterizes a certain type of change (e.g., the mean change) of the data is orthogonal to the projection subspace, then the subspace projection can incur a substantial loss of information, as pointed out for instance in Aue et al. (2018). In contrast, our method does not rely on a low dimensional approximation of the functional data, thus it does not suffer from a similar information loss problem. To demonstrate this, we take FPCA-ECP as the representative competitor to conduct an in-depth comparison with our proposal.

*S.5.3.1. Data Generation and Implementation Settings*

Similar to DGP2 in Section S.4.1, the functional data are first generated in the $L^2(T)$ space with $T = [0,1]$, then they are transformed into the density space by the inverse LQD transformation (Petersen and Müller 2016) to obtain the distributional data. The Hilbert structure of the $L^2(T)$ space allows us to conveniently generate the functional data based on the orthogonal basis expansion.

We consider an orthogonal system formed by the first forty Fourier basis functions on the interval $T = [0,1]$, denoted by $(\eta_1(t), \cdots, \eta_{40}(t))$, to generate the functional data in the $L^2(T)$ space. Specifically,

$$\eta_1(t) = 1, \ \eta_2(t) = \sqrt{2}\sin(2\pi t), \ \eta_3(t) = \sqrt{2}\cos(2\pi t), \cdots, \eta_{40}(t) = \sqrt{2}\sin(40\pi t)$$

We consider a functional sequence, denoted as $\Gamma_\psi = (\psi_1(t), \cdots, \psi_n(t))$, with $q$ change points $k_1^*, \cdots, k_q^*$ such that $0 = k_0^* < k_1^* < k_2^* < \cdots < k_q^* < k_{q+1}^* = n$ ( $k_0^* = 0$ and $k_{q+1}^* = n$ are the two boundary points). We set $n = 800$ and $q = 9$ in this simulation. The $q = 9$ change-points are supposed to be equally spaced, which implies $k_j^* = 80j$ for $j = 1, 2, \cdots, 9$. We employ the following model to generate the functional sequence $\Gamma_\psi$:

$$\psi_i(t) = \psi_{base}(t) + \sum_{j=0}^{9} \delta_j(t) I\{k_j^* < i \leq k_{j+1}^*\} + e_i(t), \quad i = 1, 2, \cdots, 800 \quad \text{(S.112)}$$

where $\psi_{base}(t)$ is the baseline mean function, $\delta_j(t)$ is the break function associated with $k_j^*$, $I\{\cdot\}$ is the indicator function, and $e_i(t)$ is the zero mean error function. At each change point $k_j^* \in \{k_0^*, k_1^*, \cdots, k_9^*\}$, the change of $\Gamma_\psi$ is achieved by adding the break function $\delta_j(t)$ to the baseline mean function $\psi_{base}(t)$ for the data indexed by $k_j^* + 1, k_j^* + 2, \cdots, k_{j+1}^*$. In the following, we present data generating mechanisms for the components $\psi_{base}(t)$, $\delta_j(t)$ and $e_i(t)$ in (S.112).

The baseline mean function $\psi_{base}(t)$ in (S.112) is generated by transforming the following density function into the $L^2(T)$ space using the LQD transformation (equation (S.127) in Appendix S.1):

$$f_{base}(x) = 0.9\big(0.8 f_{beta}(x; 28, 24) + 0.2 f_{beta}(x; 14, 32)\big) + 0.1, \quad x \in [0,1]$$

where $f_{beta}(x; a, b)$ denotes the density function of the Beta distribution with parameters $a$ and $b$. Thus, $\psi_{base} = \text{LQD}[f_{beta}]$, where $\text{LQD}[\cdot]$ denotes the LQD transformation.

The break functions $\delta_j(t)$, $j = 1, \cdots, 9$, in (S.112) are generated based on the following basis expansions:

$$\delta_j(t) = \sum_{l=1}^{40} \xi_{jl}^\delta \eta_l(t), \quad j = 1, \cdots, 9 \qquad (S.113)$$

where $\xi_{jl}^\delta = \langle \delta_j, \eta_l \rangle$, $l = 1, \cdots, 40$, are Fourier coefficients of $\delta_j$ with respect to the Fourier basis functions $\eta_l: l = 1, \cdots, 40$. As noted above, the break function $\delta_j(t)$ plays the role of introducing a change to the mean structure of the functional sequence $\Gamma_\psi$ at $k = k_j^*$. If $\xi_{jl}^\delta \neq 0$, then it means that the functional sequence undergoes a mean change at $k_j^*$ on the $l$th dimension (with respect to the orthogonal system $\{\eta_l: l = 1, 2, \cdots, 40\}$). At each change point, we only allow a few dimensions of the functional data to undergo a change. Consequently, only a few Fourier coefficients of $\delta_j(t)$ have non-zero values. Let $S_{CD}(k_j^*) = \{l \in \{1, 2, \cdots, 40\}: \xi_{jl}^\delta \neq 0\}$ denote the indexing set of non-zero Fourier coefficients of $\delta_j$, and the settings for $S_{CD}(k_j^*)$ are listed in Table S.2 for $j = 0, 1, \cdots, 9$. Then, based on a similar data-generating mechanism described in Aue et al. (2018), the break functions $\delta_j(t)$ for $j = 1, \cdots, 9$ in this simulation are generated by:

$$\delta_j(t) = e^{-j/2} \frac{1}{\sqrt{\text{Card}\big(S_{CD}(k_j^*)\big)}} \sum_{l \in S_{CD}(k_j^*)} \eta_l(t), \quad j = 1, \cdots, 9 \qquad (S.114)$$

where $\text{Card}\big(S_{CD}(k_j^*)\big)$ represents the cardinality of the set $S_{CD}(k_j^*)$. For the case of $j = 0$, due to $S_{CD}(k_0^*) = \emptyset$, we set $\delta_0(t) = 0$.

The error functions $e_i(t)$, $i = 1, \cdots, 800$, in (S.112) are generated based on the following basis expansions:

$$e_i(t) = 0.15 \sum_{l=1}^{40} \omega_{il}^e \eta_l(t), \qquad i = 1, \cdots, 800 \tag{S.115}$$

where $\omega_{il}^e = \langle e_i, \eta_l \rangle$ ($l = 1, \cdots, 40$) are the Fourier coefficients of $e_i$ with respect to Fourier basis functions $\eta_l: l = 1, \cdots, 40$. $(\omega_{i1}^e, \cdots, \omega_{i40}^e)$ are independently generated from the following multivariate normal distribution model for each $i \in \{1, \cdots, 800\}$:

$$(\omega_{i1}^e, \cdots, \omega_{i40}^e)^{\mathrm{T}} \sim N(\boldsymbol{\mu}_\omega, \boldsymbol{\Sigma}_\omega), i = 1, \cdots, 800 \tag{S.116}$$

with $\boldsymbol{\mu}_\omega = (0, \cdots, 0, \cdots, 0)^{\mathrm{T}}$ and $\boldsymbol{\Sigma}_\omega = \mathrm{diag}\left((20^{-1}, \cdots, 20^{-l}, \cdots, 20^{-40})\right)$. The rapid decay of the diagonal elements of $\boldsymbol{\Sigma}_\omega$ leads to a rapid decay of eigenvalues in the FPCA, similar to that in Aue et al. (2018).

**Table S.2.** The indexing set of the non-zero Fourier coefficients for simulating the break function using equation (S.114) for each of the change points.

| Change point | $S_{CD}(k_j^*)$ | Change point | $S_{CD}(k_j^*)$ |
|---|---|---|---|
| $k_0^* = 0$ | $S_{CD}(k_0^*) = \emptyset$ | $k_1^* = 80$ | $S_{CD}(k_1^*) = \{1,2,3\}$ |
| $k_2^* = 160$ | $S_{CD}(k_2^*) = \{4,5,\cdots,8\}$ | $k_3^* = 240$ | $S_{CD}(k_3^*) = \{11,12,13\}$ |
| $k_4^* = 320$ | $S_{CD}(k_4^*) = \{12,13,14,15\}$ | $k_5^* = 400$ | $S_{CD}(k_5^*) = \{16,17,18\}$ |
| $k_6^* = 480$ | $S_{CD}(k_6^*) = \{21,22,23\}$ | $k_7^* = 540$ | $S_{CD}(k_7^*) = \{24,25,\cdots,28\}$ |
| $k_8^* = 640$ | $S_{CD}(k_8^*) = \{29,30,\cdots,35\}$ | $k_9^* = 720$ | $S_{CD}(k_9^*) = \{35,36,\cdots,40\}$ |

Finally, we use the inverse LQD transformation (equation (S.128) in Appendix S.1) to transform the generated $L^2(T)$-valued data $\psi_1(t), \cdots, \psi_n(t)$ into the density space, yielding a distributional sequence denoted as

$$\Gamma = (f_1(t), \cdots, f_n(t)) = ((\mathrm{LQD}^{-1}[\psi_1])(t), \cdots, (\mathrm{LQD}^{-1}[\psi_n])(t)) \tag{S.117}$$

For performance comparison, we apply the Fréchet-MOSUM method and the FPCA-ECP method to the generated distributional sequence $\Gamma = (f_1(t), \cdots, f_n(t))$. For the Fréchet-MOSUM method, the bandwidth $G$ is set to 30 and the AOP parameter $\varepsilon$ is set to 0.4. Moreover, we do not consider the boundary correction processing in computing the SS sequence in this simulation study. For the FPCA-ECP method, we directly apply the functional change-point detector to the generated $L^2(T)$-valued functional sequence $\Gamma_\psi$, and we choose the same implementation settings as those described in Section S.4.2 (of the supplement).

*S.5.3.2. Results*

We repeat the above change-point detection experiment for 20 times, and the distributional data are generated separately using the same data-generating process for each replicate. The estimated numbers of change points in the 20 replicates are compared in Table S.3 for the two methods. We see that our method correctly estimates the number of change points in each of the 20 replicates, while the FPCA-ECP method exhibits a severe underestimation issue. The location estimation errors of change points, which are quantified by the Hausdorff distance, are compared in Table S.4. We see that the location estimation errors of the FPCA-ECP method are considerably larger than those of our method.

**Table S.3.** Numbers of estimated change points in 20 replicates.

| Experiment index | Number of estimated change points | | Experiment index | Number of estimated change points | |
|---|---|---|---|---|---|
| | Fréchet-MOSUM | FPCA-ECP | | Fréchet-MOSUM | FPCA-ECP |
| 1 | 9 | 6 | 11 | 9 | 7 |
| 2 | 9 | 9 | 12 | 9 | 6 |
| 3 | 9 | 6 | 13 | 9 | 7 |
| 4 | 9 | 6 | 14 | 9 | 6 |
| 5 | 9 | 7 | 15 | 9 | 7 |
| 6 | 9 | 7 | 16 | 9 | 6 |
| 7 | 9 | 6 | 17 | 9 | 6 |
| 8 | 9 | 6 | 18 | 9 | 7 |
| 9 | 9 | 7 | 19 | 9 | 6 |
| 10 | 9 | 8 | 20 | 9 | 6 |

**Table S.4.** Location estimation errors (quantified by Hausdorff distance) of change points in 20 replicates.

| Experiment index | Location estimation error | | Experiment index | Location estimation error | |
|---|---|---|---|---|---|
| | Fréchet-MOSUM | FPCA-ECP | | Fréchet-MOSUM | FPCA-ECP |
| 1 | 13 | 160 | 11 | 1 | 80 |
| 2 | 0 | 29 | 12 | 1 | 160 |
| 3 | 1 | 160 | 13 | 2 | 85 |
| 4 | 1 | 160 | 14 | 1 | 160 |
| 5 | 1 | 80 | 15 | 9 | 80 |
| 6 | 1 | 80 | 16 | 1 | 160 |
| 7 | 3 | 160 | 17 | 1 | 160 |
| 8 | 1 | 160 | 18 | 1 | 80 |
| 9 | 1 | 80 | 19 | 0 | 160 |
| 10 | 4 | 71 | 20 | 0 | 160 |

The poor performance of the FPCA-ECP method is primarily attributed to the information loss incurred by the subspace projection during the FPCA-based dimension reduction. In the FPCA-ECP method (Lei et al. 2023b), the functional data are required to be projected onto an orthogonal system formed by dominant FPCs, so as to reduce them to a low dimensional vector-valued data composed of the FPC scores; then, the ECP change-point detector (Matteson and James 2014) is applied to detect the changes in the resulting time series of score vectors. As pointed out by Aue et al. (2018), if functions representing the changes (e.g., the break functions in this simulation) are orthogonal to the projection subspace, then the subspace projection can lead to a loss of information. Such a limitation makes the FPCA-ECP method significantly underestimate the number of change points in this simulation.

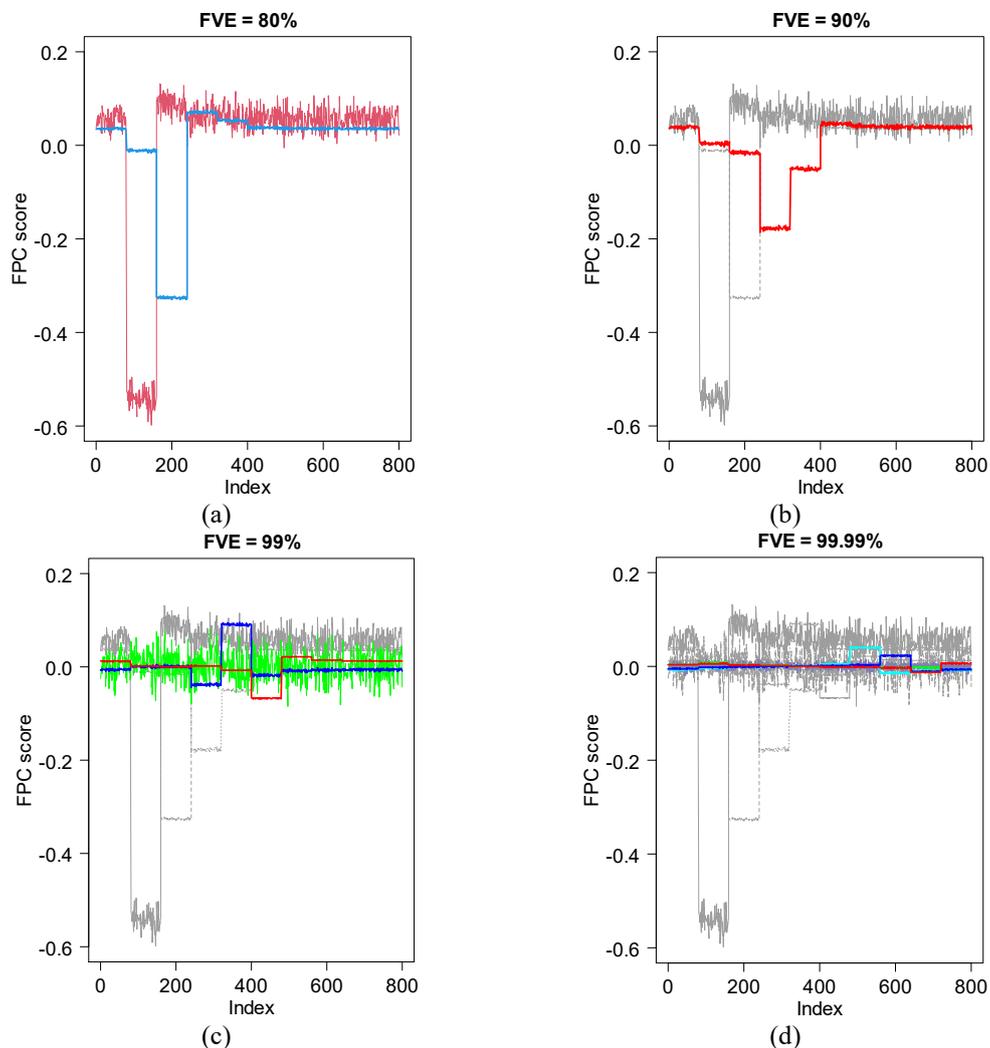

**Figure S.14.** FPC-score time series calculated from a simulated $L^2(T)$-valued functional sequence $\Gamma_\psi$ under different truncation levels (controlled by the fraction of variance explained (FVE)). (a) FVE=80%, (b) FVE=90%, (c) FVE=99%, and (d) FVE=99.99%. The dimension of the FPC-score time series is non-decreasing with increasing FVE, and the FPC scores associated with the new added dimensions under the new FVE are represented by colored lines.

In order to gain more insight into this FPC truncation-induced information loss problem, Figure S.14 presents the FPC-score time series calculated from an identical simulated $L^2(T)$-valued functional sequence $\Gamma_\psi$ under different truncation levels (controlled by the FVE, see Section S.4.2 of the supplement for more details). It is noteworthy that the dimension of the FPC-score time series is non-decreasing with increasing FVE. In Figure S.14, FPC scores associated with the new added dimensions under the new FVE are represented by colored lines. We see from Figure S.14 that the change information of some change points is contained in higher dimensions of the FPC scores. Therefore, the truncation processing in the FPCA-based dimension reduction can incur a loss of such change information. In contrast, our method does not rely on a low-dimensional subspace projection processing; therefore, as expected, it substantially outperforms the FPCA-ECP method in this simulation study.

## S.6. Supplemental Materials for Real Data Study

### S.6.1. Bandwidth Selection

The bandwidth $G$ is selected using the CPT plot-based strategy described in Section S.2.8. Following the recommendation in Remark 3, we randomly select five CTR distributional sequences from our datasets to create CPT plots, which are presented in Figure S.15. Following the CPT plot-based bandwidth selection principle (Section S.2.8), we see from Figure S.15 that a bandwidth of $G = 40$ is a relatively ideal choice.

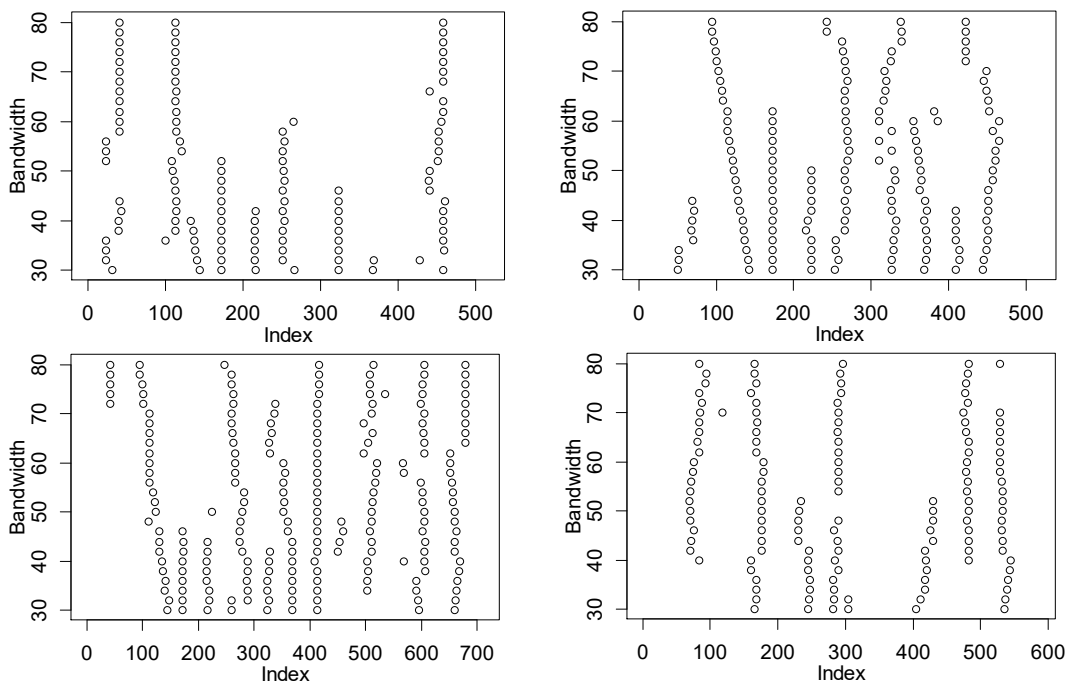

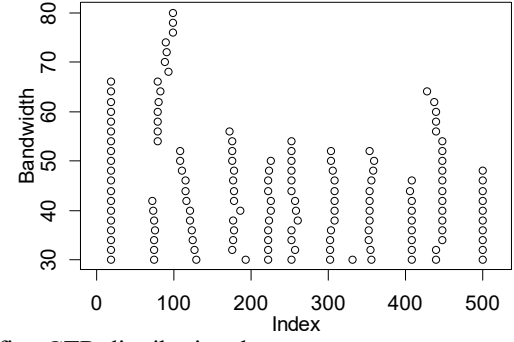

**Figure S.15.** CPT plots of five CTR distributional sequences.

### S.6.2. Archetypal Analysis for Distributional Data

This subsection describes how to use the archetype analysis (AA) algorithm in the R package *archetypes* (Eugster and Leisch 2009) to perform AA on our CTR distributional data. The AA algorithm in the package *archetypes* is designed for multivariate data; therefore, we first convent our PDFs into multivariate data by discretizing them at an equally-spaced grid. It is noteworthy that (a) both the archetype extraction and archetype representation (for the data) involved in the AA are achieved by performing convex combinations on the data; and (b) the PDF space is closed under convex combinations. Therefore, if the grid is not too sparse, the above discretization has no significant impact on the AA result of a PDF-valued dataset, compared with treating the PDFs as continuous functions. Let $\Gamma = \{f_1(x), \cdots, f_n(x)\}$ be an arbitrary investigated CTR distributional sequence, and let $\mathcal{D}$ be the common support of the distributions. In our analysis, each of the PDFs in $\Gamma$ is discretized on a regular grid $\{x_1, \cdots, x_{250}\} \subset \mathcal{D}$, and the results are arranged in a matrix of size $n \times 250$ as follows:

$$X = \begin{bmatrix} f_1(x_1) & \cdots & f_1(x_{250}) \\ \vdots & \vdots & \vdots \\ f_n(x_1) & \cdots & f_n(x_{250}) \end{bmatrix}$$

After this discretization processing, the AA algorithm in the package *archetypes* can be directly applied to $X$, and the resulting archetypes are convex combinations of the discretized PDFs as follows:

$$Z = X^T \beta$$

where $\beta \in \mathbb{R}^{n \times k}$ is the coefficient matrix (solved by the AA algorithm) for determining the $k$ (pre-specified by users) different archetypes. The elements in $\beta$ obey the following constraints:

$$\sum_{i=1}^{n} \beta_{ji} = 1 \text{ and } \beta_{j1}, \beta_{j2}, \cdots, \beta_{jn} \geq 0, \quad j = 1, \cdots, k$$

The archetypes obtained above are archetypes of the multivariate data (i.e., the discretized PDFs). Since the PDF space is closed under convex combinations, the continuous versions of the archetypes (for the PDF-valued data) can be obtained as

$$f_j^A(x) = \sum_{i=1}^{n} \beta_{ji}\, f_i(x), \ x \in \mathcal{D} \text{ and } j = 1, \cdots, k$$

Figure S.16 compares the curves of the original PDFs with their archetypal approximations (which are the mixtures of two archetypes) for six representative distributions associated with cable pair RCP41.

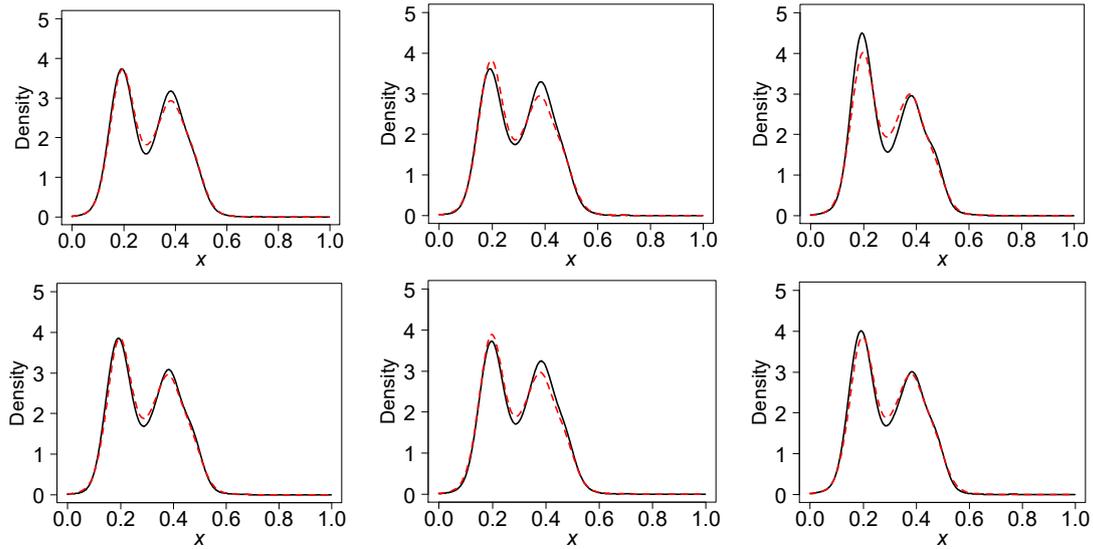

**Figure S.16.** Archetypal approximations (red dashed lines) for several representative PDFs (black solid lines).

*S.6.3. More Detection Results Obtained by the Fréchet-MOSUM Method*

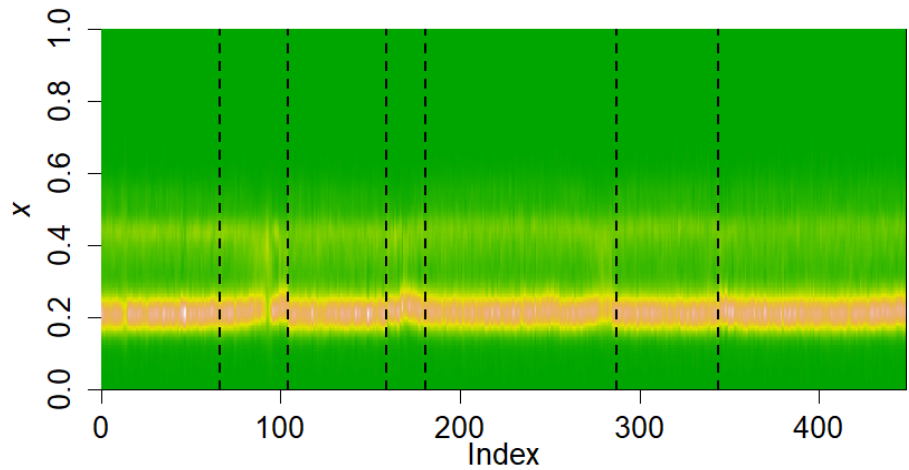

The heatmap of the PDF-valued sequence

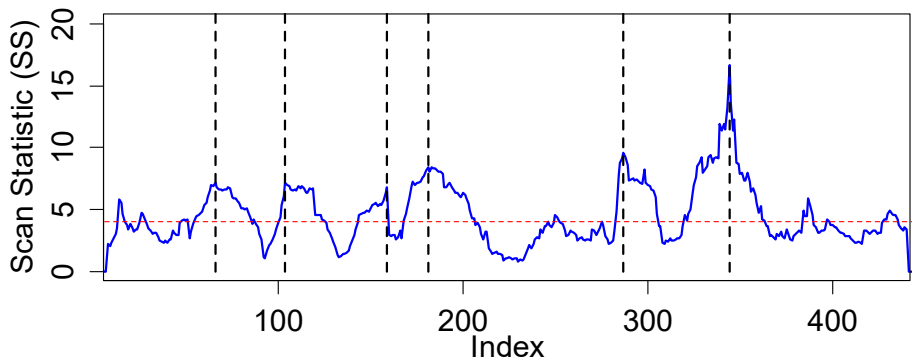

The scan statistic (SS)

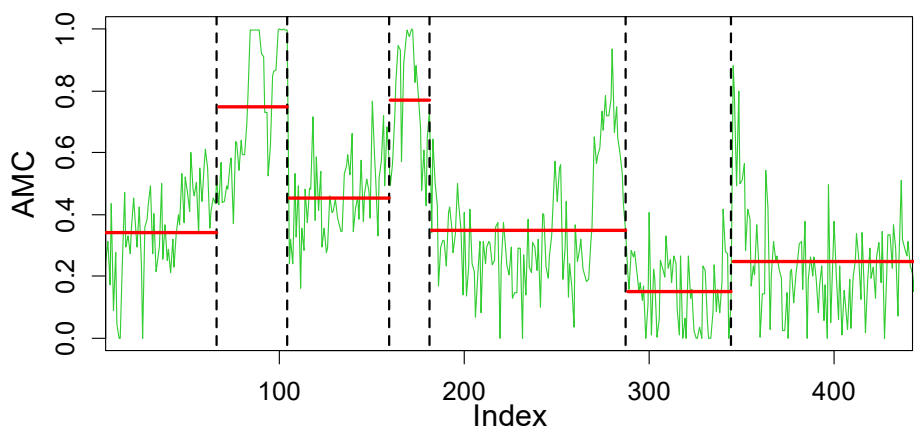

Archetype mixture coefficients (AMCs)

(a) Detection results of the distributional sequence associated with cable pair RCP1

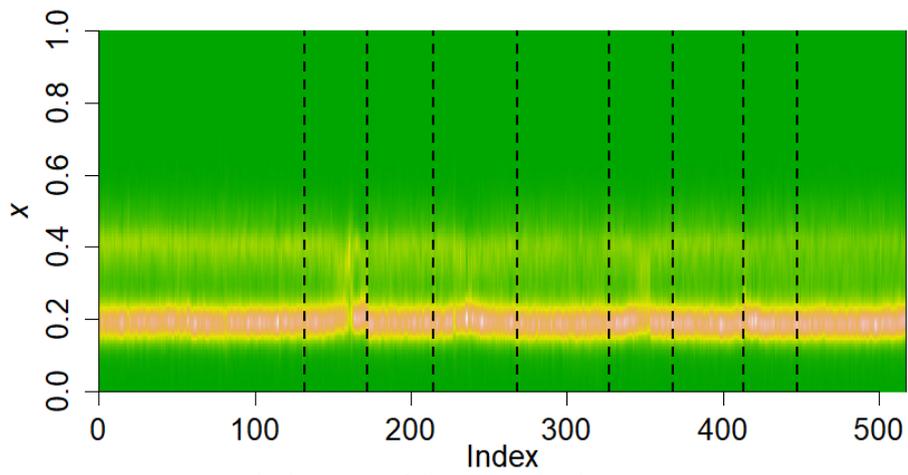

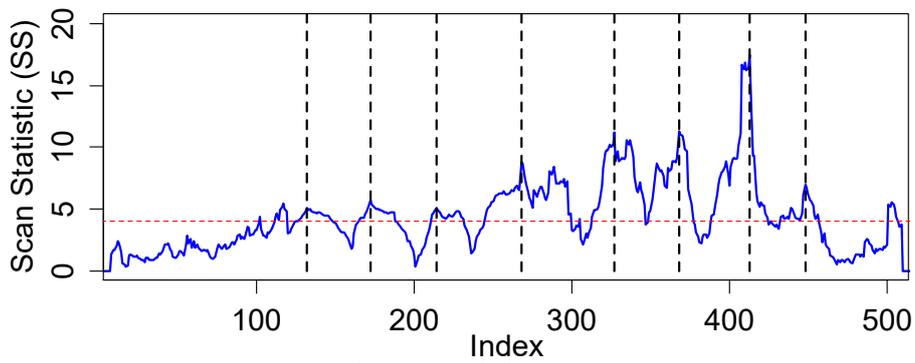

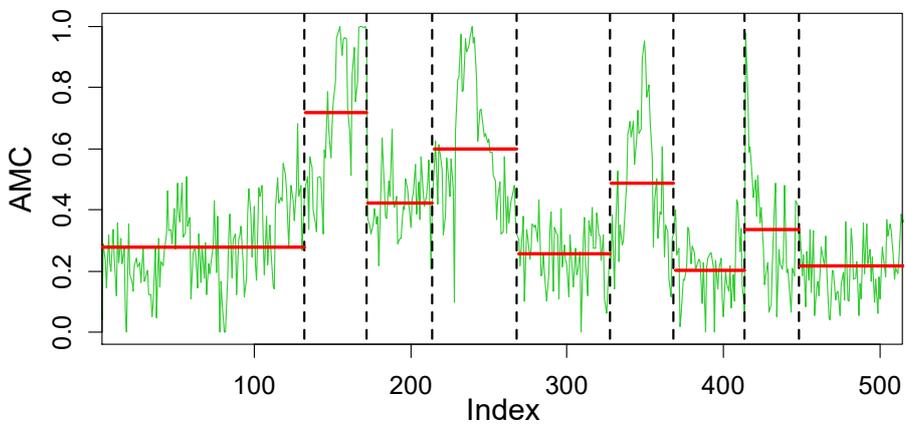

(b) Detection results of the distributional sequence associated with cable pair RCP8

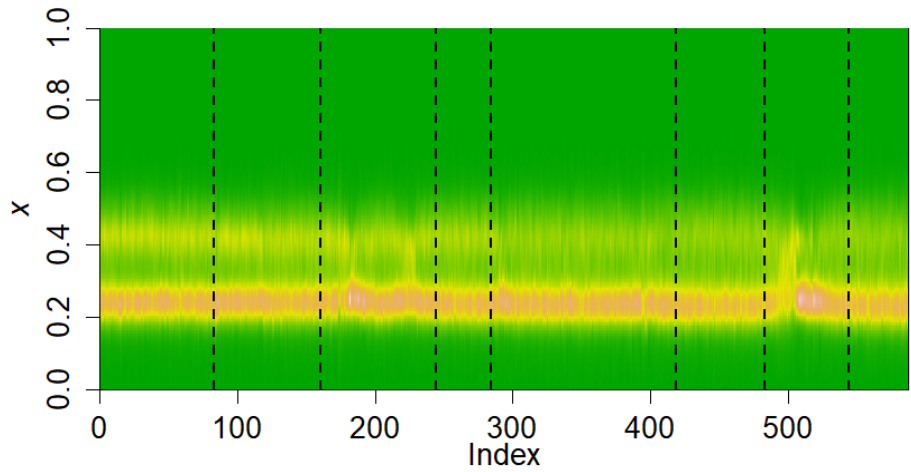

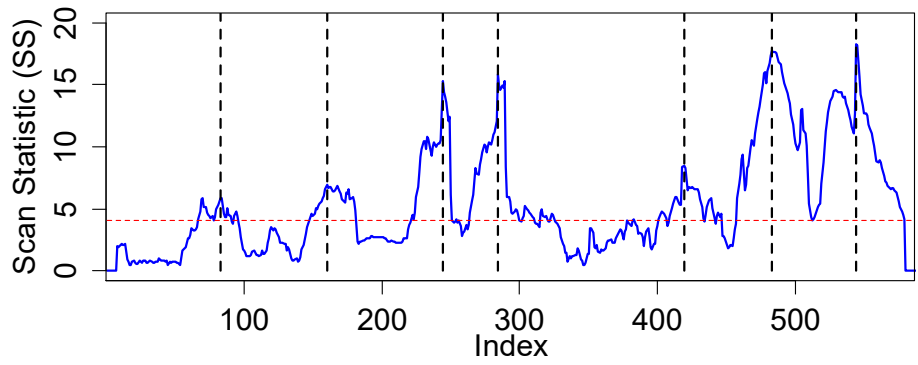

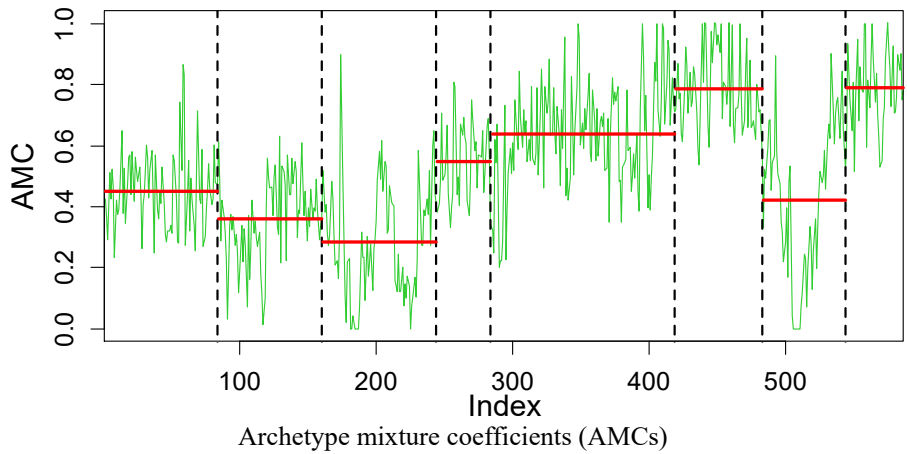

(c) Detection results of the distributional sequence associated with cable pair RCP13

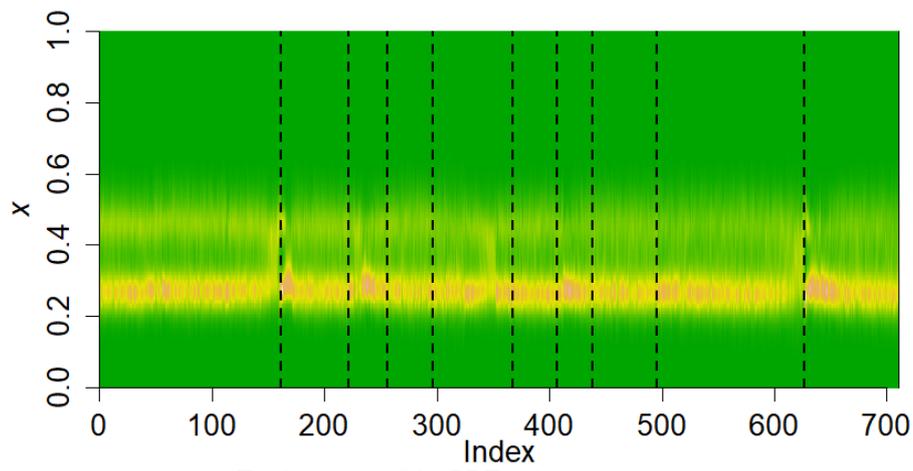
The heatmap of the PDF-valued sequence

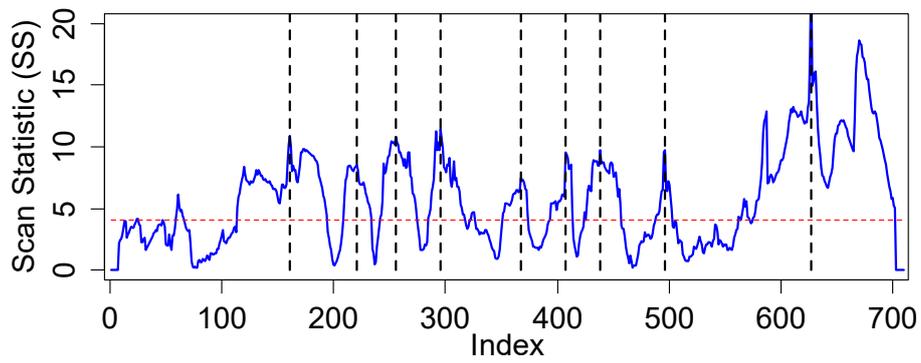
The scan statistic (SS)

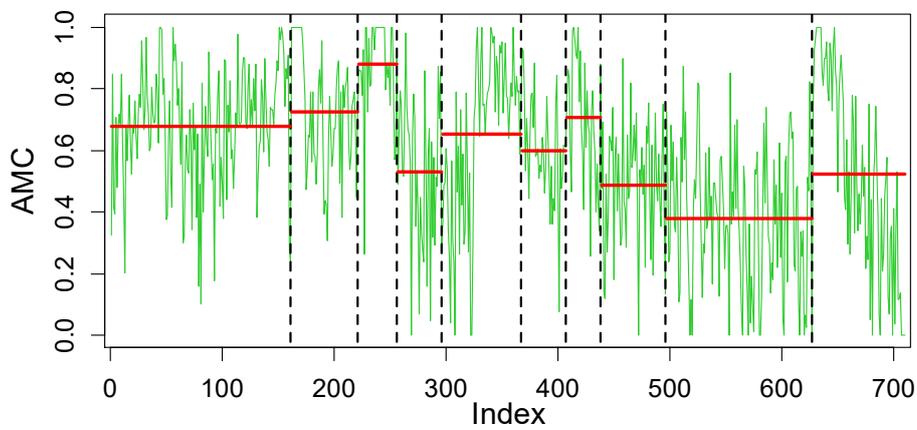
Archetype mixture coefficients (AMCs)

(d) Detection results of the distributional sequence associated with cable pair RCP14

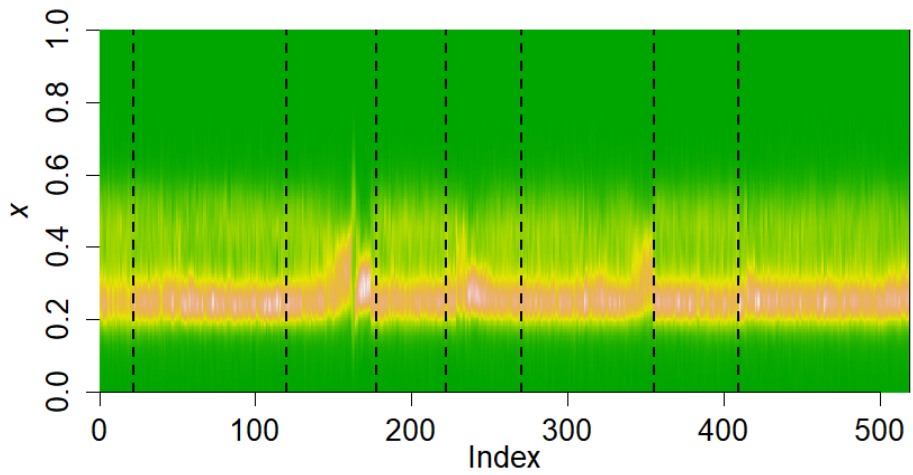

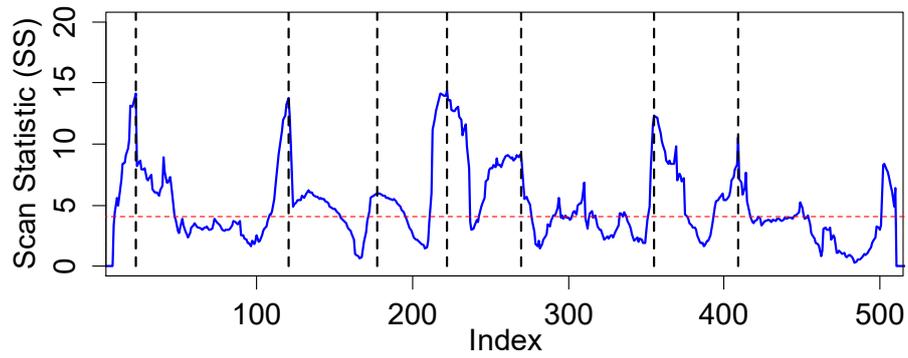

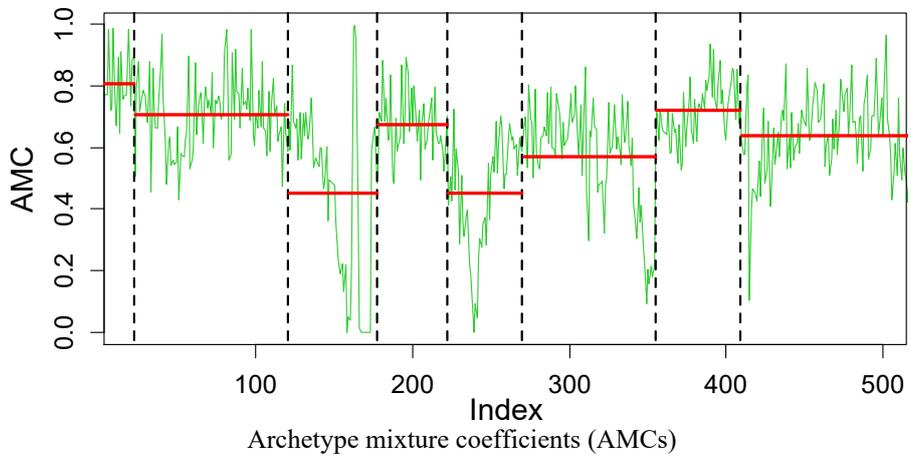

(e) Detection results of the distributional sequence associated with cable pair RCP18

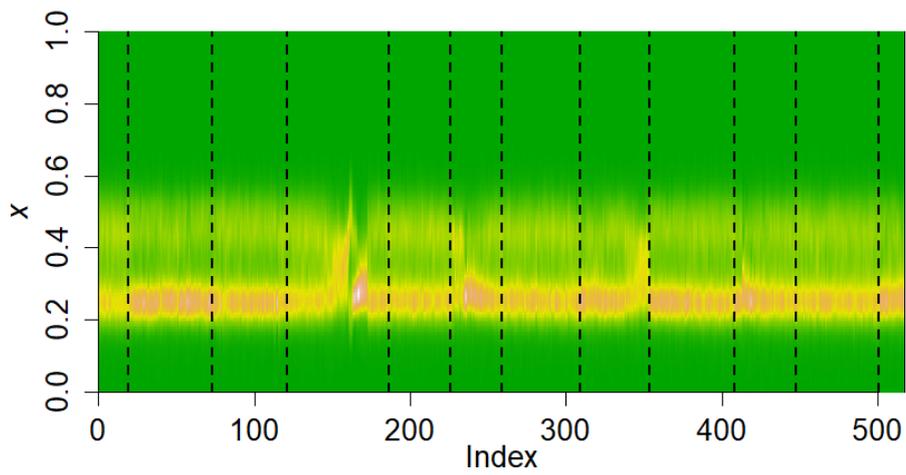

The heatmap of the PDF-valued sequence

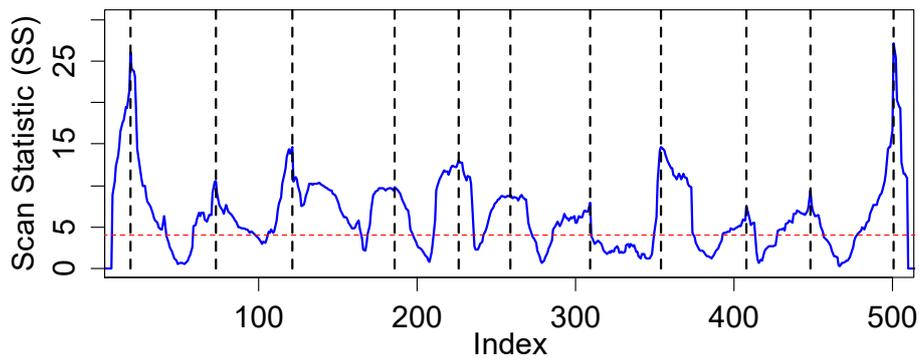

The scan statistic (SS)

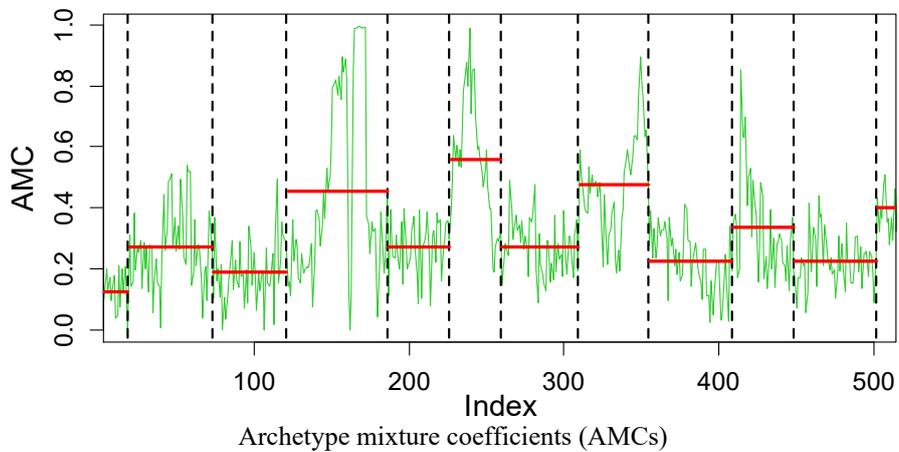

Archetype mixture coefficients (AMCs)

(f) Detection results of the distributional sequence associated with cable pair RCP22

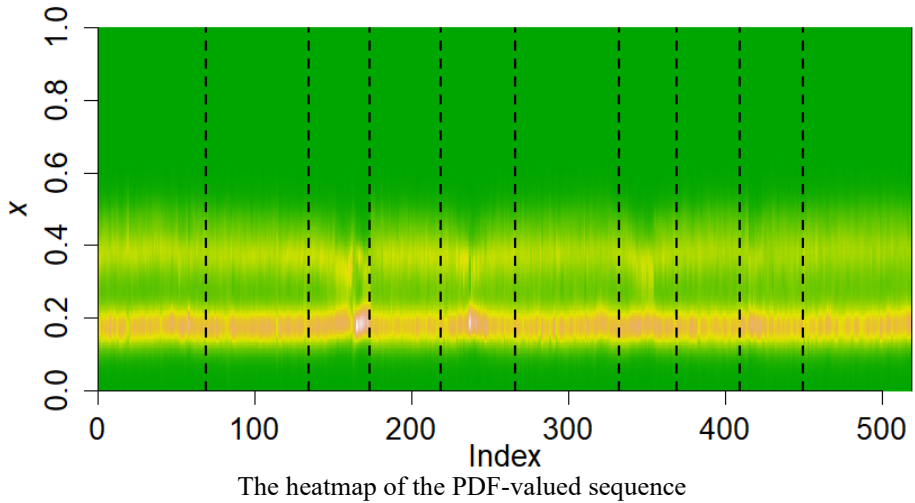
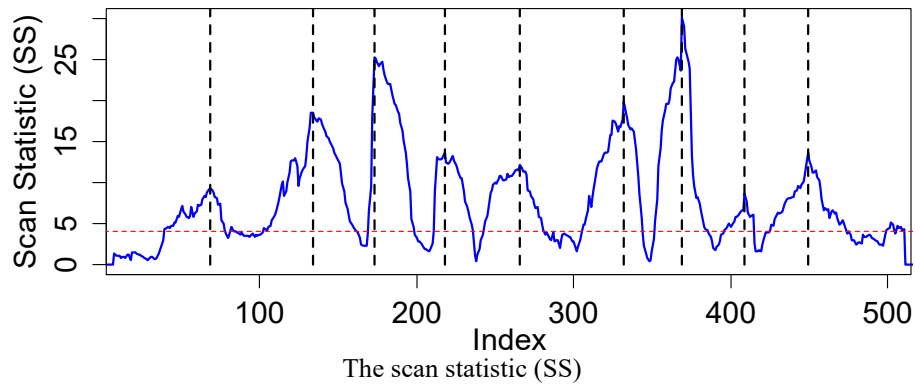
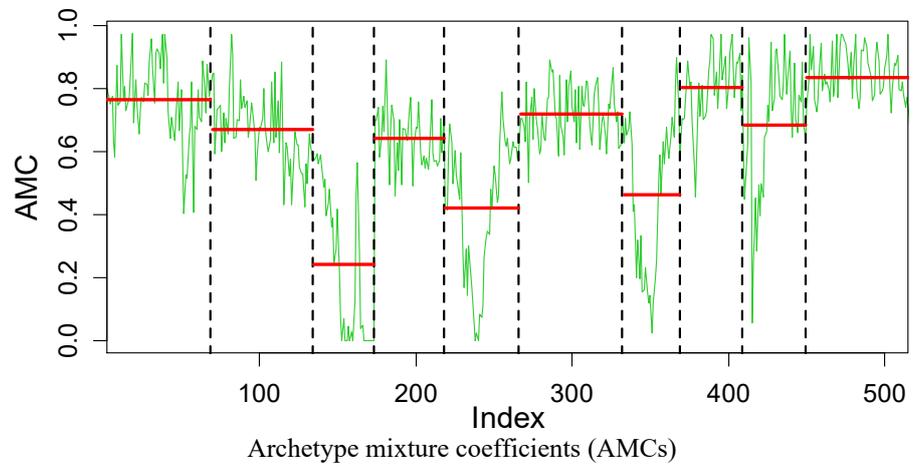

(g) Detection results of the distributional sequence associated with cable pair RCP30

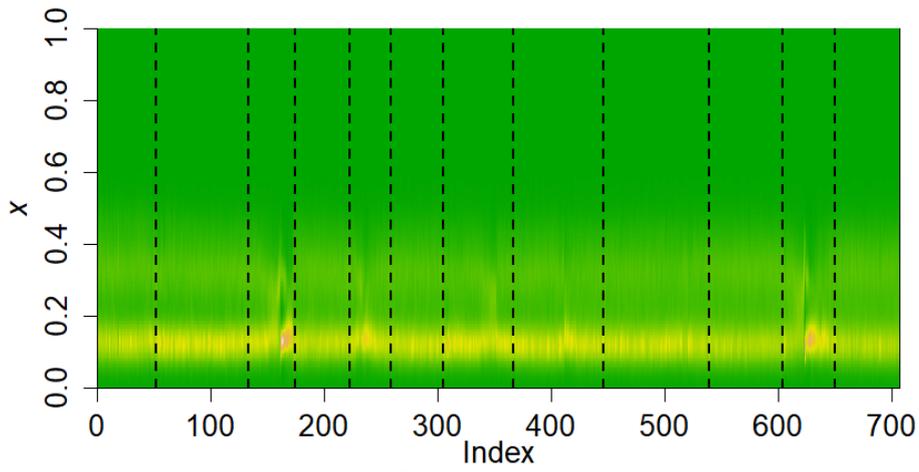

The heatmap of the PDF-valued sequence

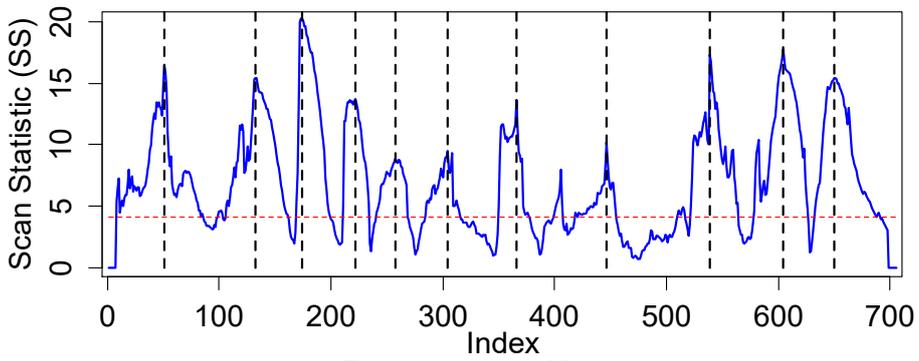

The scan statistic (SS)

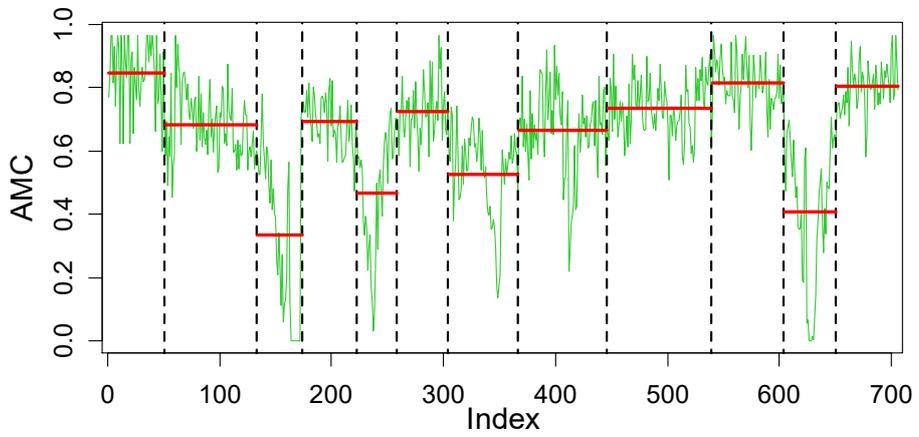

Archetype mixture coefficients (AMCs)

(h) Detection results of the distributional sequence associated with cable pair RCP33

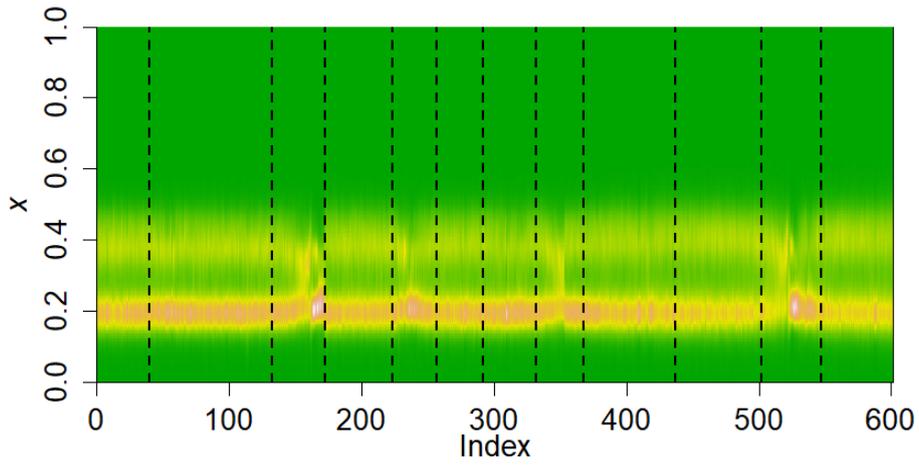

The heatmap of the PDF-valued sequence

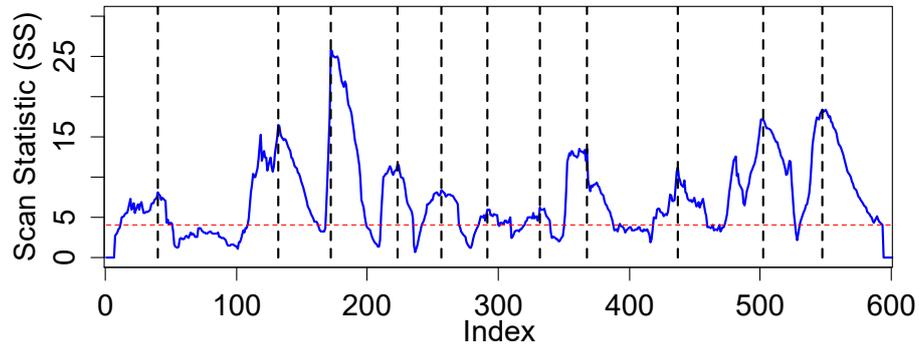

The scan statistic (SS)

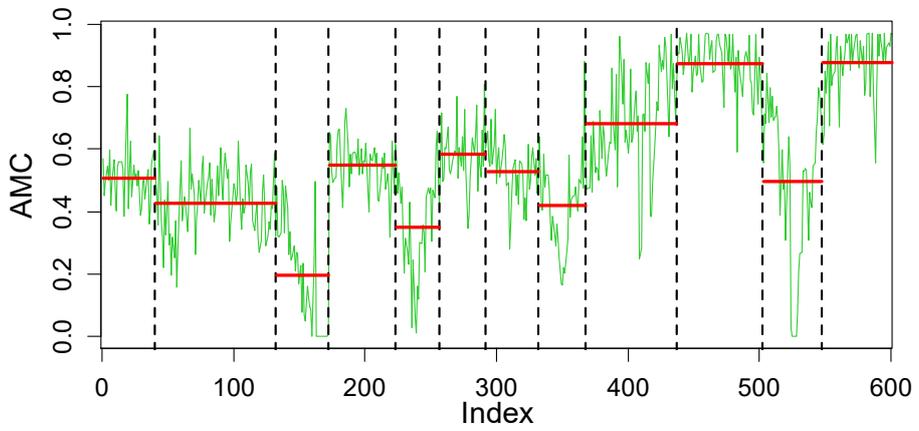

Archetype mixture coefficients (AMCs)

(i) Detection results of the distributional sequence associated with cable pair RCP44

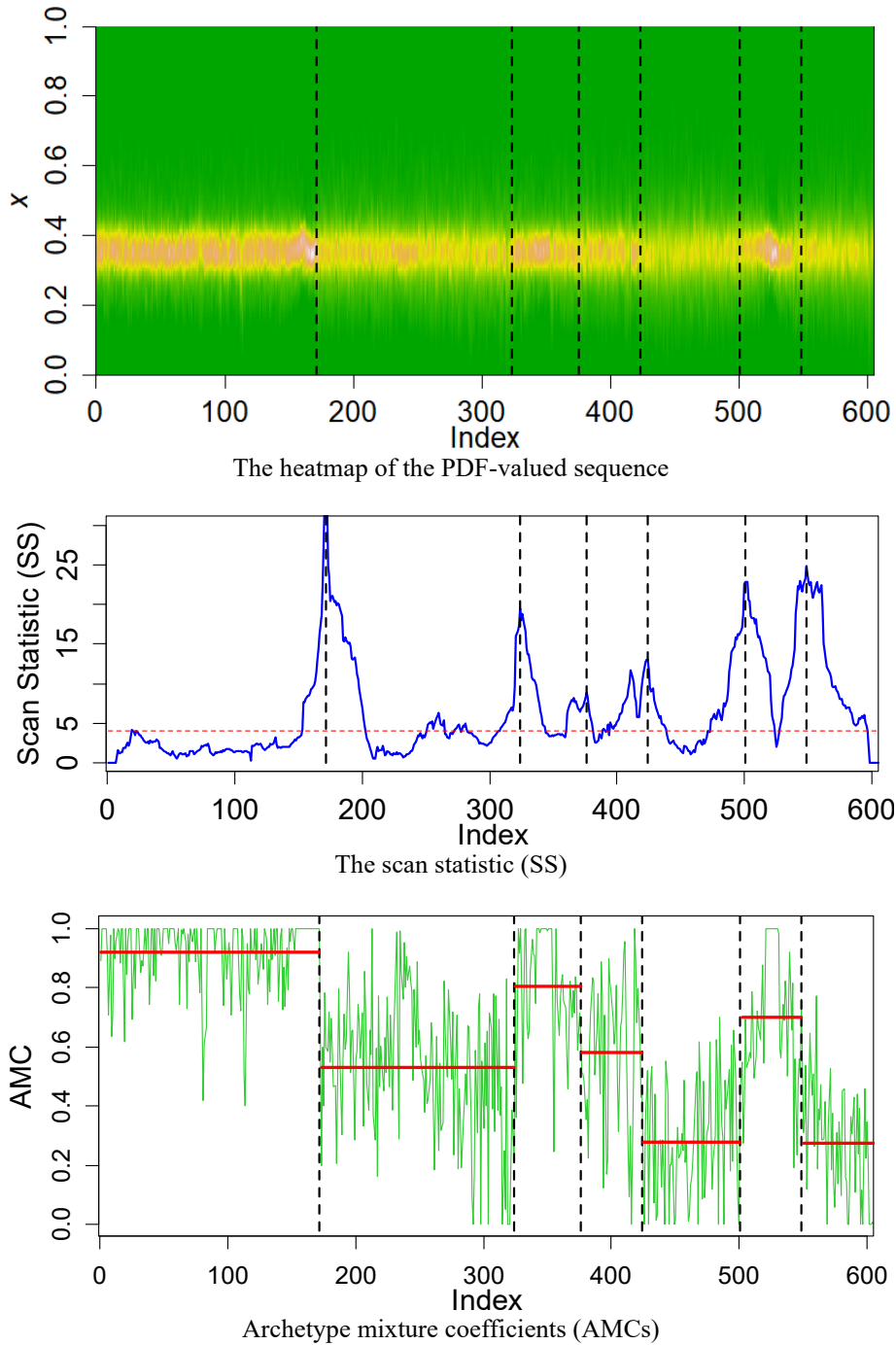

(j) Detection results of the distributional sequence associated with cable pair RCP48

**Figure S.17.** Change-point detection results obtained by the Fréchet-MOSUM method for ten representative CTR distributional sequences associated with cable pairs (a) RCP1, (b) RCP8, (c) RCP13, (d) RCP14, (e) RCP18, (f) RCP22, (g) RCP30, (h) RCP33, (i) RCP44 and (j) RCP48, respectively. The first row corresponds to the heatmap of the PDF-valued sequence, the second row corresponds to the SS sequence computed using the Fréchet-MOSUM procedure, and the third row corresponds to the archetype mixture coefficients (obtained by the archetypal analysis) used for visually confirming the detected changes (similar to the right panel of Figure 4 of the main text). The detected change points are indicated by vertical dashed lines. The horizontal dashed line in the plot of SS sequence (second row) indicates the computed critical value $D_n(G; \alpha)$. The bold horizontal lines in the plot of archetype mixture coefficients (third row) indicate the empirical means within segments.

*S.6.4. Additional Results of Variance-Change Examination*

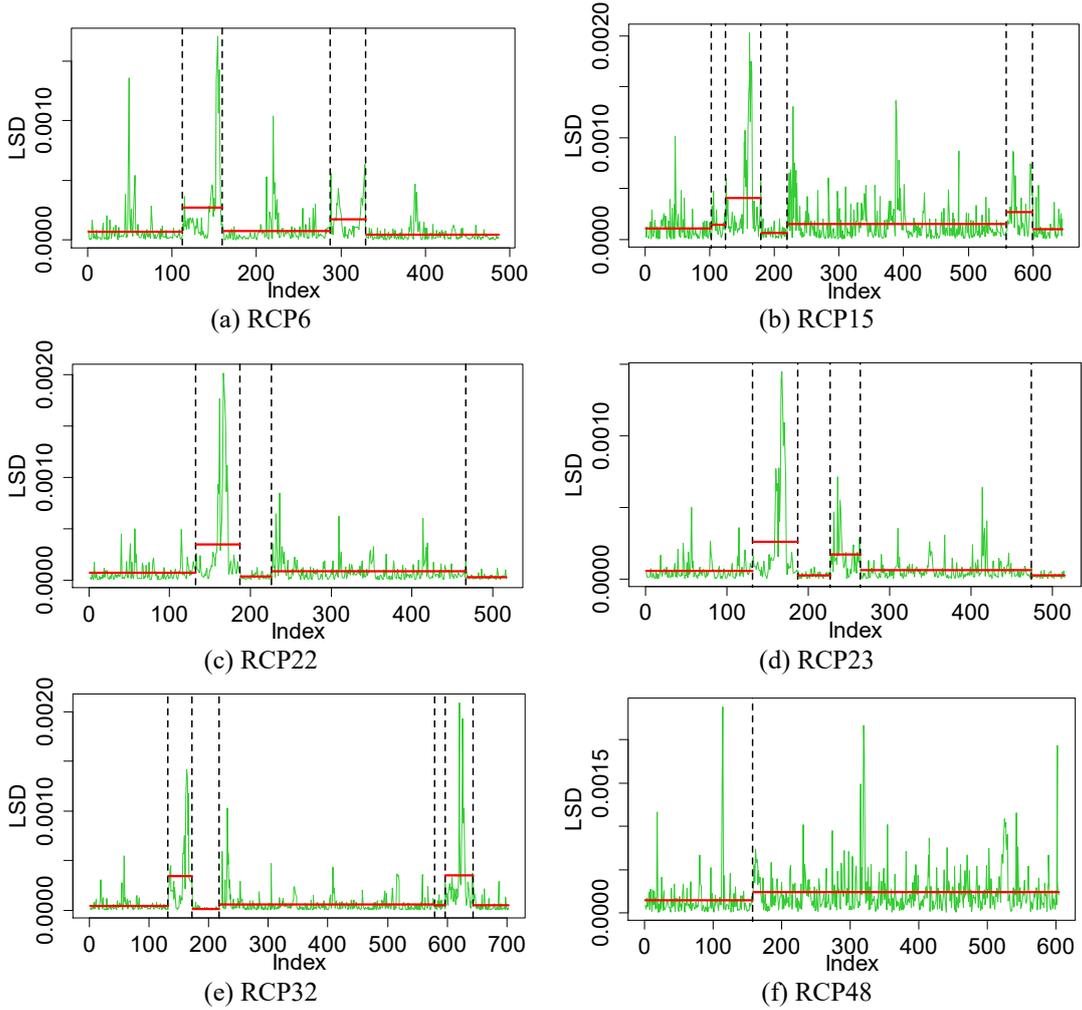

**Figure S.18.** Same as Figure 5 in the main text but the CTR distributional sequence is from cable pair (a) RCP6, (b) RCP15, (c) RCP22, (d) RCP23, (e) RCP32, and (f) RCP48, respectively.

*S.6.5. LSD Refinement*

Given a CTR distributional sequence, the change points detected from the associated LSD sequence can be used to refine the change-point set, which can help to recover some change points missed by our Fréchet-MOSUM detector. This change-point compensation strategy is referred to as the LSD refinement. Let $S_{cp}^{FM} = \{\hat{k}_1^*, \cdots, \hat{k}_{\hat{q}_n}^*\}$ be the set of change points detected by the Fréchet-MOSUM procedure and let $S_{cp}^{LSD} = \{\hat{\tau}_1^*, \cdots, \hat{\tau}_m^*\}$ be the set of change points detected from the LSD sequence. We start with $\hat{\tau}_1^*$, if $\min_{k \in S_{cp}^{FM}} |\hat{\tau}_1^* - k| > \varepsilon G$ ($\varepsilon$ is the AOP parameter and $G$ is the bandwidth), then $S_{cp}^{FM}$ is updated as $S_{cp}^{FM} = S_{cp}^{FM} \cup \{\hat{\tau}_1^*\}$. The above procedure proceeds recursively until all of the elements in $S_{cp}^{LSD}$ have been visited, and the final form of $S_{cp}^{FM}$ is treated as its refined version.

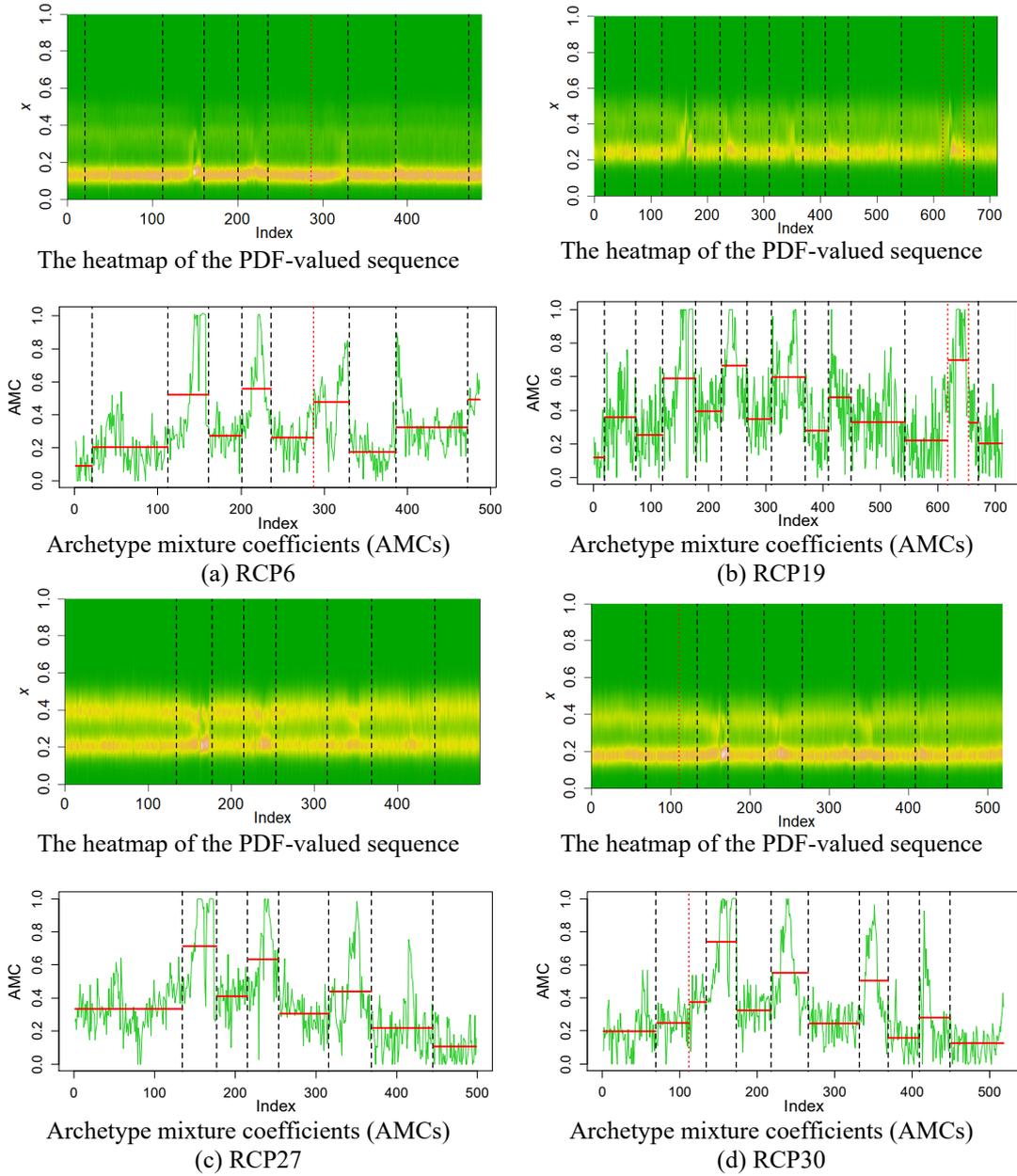

**Figure S.19.** Change-point detection results of four CTR distributional sequences after implementing the LSD refinement. The first row corresponds to the heatmap of the PDF-valued sequence, while the second row corresponds to the archetype mixture coefficients obtained by the archetypal analysis. The original change points detected by the Fréchet-MOSUM detector are indicated by black vertical dashed lines, while the new added change points after the LSD refinement are indicated by red vertical dotted lines. The bold horizontal lines in the plot of archetype mixture coefficients indicate the empirical means within segments.

### S.6.6. Registration of Change-Point Indices

Given a CTR distributional sequence $\Gamma = \{f_1(x), \cdots, f_n(x)\}$ with $f_j(x)$ representing the PDF estimated from the CTR data of the $j$th day, it can be arranged in the following structured form:

$$\left\{\begin{matrix} 1 \\ t_1 \\ f_1(x) \end{matrix}\right\}, \left\{\begin{matrix} 2 \\ t_2 \\ f_2(x) \end{matrix}\right\}, \cdots, \left\{\begin{matrix} n \\ t_n \\ f_n(x) \end{matrix}\right\}$$

where the first row corresponds to the time indices of the distributions, and the second row corresponds to the time. For instance, if the density function $f_j(x)$ is estimated from the CTR data extracted from the measurements collected on 20 May 2008, then $t_j = 20080520$. For convenience, $T = \{t_1, \cdots, t_n\}$ is referred to as the time grid of the distributional sequence Γ. As noted in the main text, due to the fact that we discarded missing data, the time indices of PDFs between different CTR distributional sequences are not matched one-to-one. Consequently, the change points detected from different sequences are not aligned in the time domain. For a better comparison, we conduct a registration processing on the indices of the detected change points to align them in the time domain.

**Table S.5.** Illustration of change-point detection results for partial of the investigated CTR distributional sequences.

| Cable pair | | Detected change points |
|---|---|---|
| RCP1 | Original time index | 66, 104, 159, 181, 287, 344 |
| | Registered time index | 135, 173, 228, 250, 357, 414 |
| | Time location | 20070121, 20070228, 20071231, 20080306, 20090227, 20090506 |
| RCP2 | Original time index | 61, 135, 173, 270, 302, 342, 385, 431 |
| | Registered time index | 61, 135, 173, 270, 302, 414, 457, 503 |
| | Time location | 20061101, 20070121, 20070228, 20080327, 20080428, 20090506, 20100422, 20100615 |
| RCP3 | Original time index | 103, 133, 173, 270, 322, 356, 413 |
| | Registered time index | 103, 133, 173, 270, 323, 357, 414 |
| | Time location | 20061214, 20070119, 20070228, 20080327, 20080707, 20090227, 20090506 |

Let $\Gamma_1, \Gamma_2, \cdots, \Gamma_{50}$ denote the 50 CTR distributional sequences extracted for investigation in this study, and let $T_1, T_2, \cdots, T_{50}$ be the corresponding time grids. We first create a larger time grid as a union of individual time grids

$$T_{UTG} = T_1 \cup T_2 \cup \cdots \cup T_{50}$$

The elements in $T_{UTG}$ are arranged in increasing order. Given an individual CTR distributional sequence $\Gamma \in \{\Gamma_1, \Gamma_2, \cdots, \Gamma_{50}\}$ with $T = \{t_1, \cdots, t_n\}$ representing the associated time grid, let $\hat{k}_1^*, \cdots, \hat{k}_{\hat{q}_n}^*$ be the $\hat{q}_n$ change points detected from Γ. In the registration processing, each of the detected change points is re-expressed by a new time index determined by

$$\hat{k}_j^\# = \{i \in \{1,2,\cdots,|T_{UTG}|\}: T_{UTG}(i) = T(\hat{k}_j^*)\}, \quad j = 1, \cdots, \hat{q}_n$$

where $|T_{UTG}|$ denotes the cardinality of $T_{UTG}$. $\hat{k}_j^{\#}$ is referred to as the registered time index of $\hat{k}_j^*$. After the registration, the indices of change points detected from different CTR distributional sequences are aligned in the time domain, as illustrated in Table S.5.

## S.7. Multiscale Fréchet-MOSUM Procedure

The implementation procedure of our multiscale Fréchet-MOSUM method is briefly outlined in Section 6.2 of the main text. This section presents the full details on our multiscale Fréchet-MOSUM method.

Let $G_{grid} = \{G_1, G_2, \cdots, G_L\}$ be the grid of bandwidths considered in the multiscale detection procedure, which is referred to as the G-grid hereafter. Given $G_l \in G_{grid}$, let $\hat{q}_n(G_l)$ be the number of change points estimated by the Fréchet-MOSUM detector at the bandwidth $G_l$, and let $S_{cp}(G_l) = \{\hat{k}_1^*(G_l), \hat{k}_2^*(G_l), \cdots, \hat{k}_{\hat{q}_n(G_l)}^*(G_l)\} \subset \{1, \cdots, n\}$ be the corresponding location estimates of the $\hat{q}_n(G_l)$ change points. We then define the following binary vector:

$$\lambda(G_l) = (\lambda_{l1}, \cdots, \lambda_{li}, \cdots, \lambda_{ln}) \text{ with } \lambda_{li} = I\{i \in S_{cp}(G_l)\}, i = 1, \cdots, n \qquad (S.118)$$

where $I\{\cdot\}$ is the indicator function. By definition, $\lambda_{li} = 1$ if $i$ coincides with the location estimate of a detected change point stored in $S_{cp}(G_l)$, and $\lambda_{li} = 0$ otherwise. Recall that $G_{grid} = \{G_1, G_2, \cdots, G_L\}$, the relevant binary vectors defined above can be arranged into the following matrix:

$$\mathcal{M} = \begin{bmatrix} \lambda(G_L) \\ \vdots \\ \lambda(G_l) \\ \vdots \\ \lambda(G_1) \end{bmatrix} \qquad (S.119)$$

With a slight abuse of terminology, we call $\mathcal{M}$ the change-point indicator (CPI) matrix. The visual representation of $\mathcal{M}$ (as illustrated in Figure S.20) is called the CPI diagram. Throughout the rest of this study, the term "CPI matrix" will be used interchangeably with "CPI diagram" for $\mathcal{M}$.

In the CPI diagram, any of the non-zero elements of the CPI matrix $\mathcal{M}$ is marked by an empty circle, which represents the location estimate of a change point detected using the corresponding bandwidth. If $G_{grid}$ is not too coarse, a true change point can usually be detected repeatedly under different bandwidths, and the relevant location estimates can form a stable trajectory in the CPI diagram, as shown in Figure S.20. In the following, we develop an algorithm to automatically identify potential trajectories from the CPI diagram, and estimate the change points based on the identified trajectories.

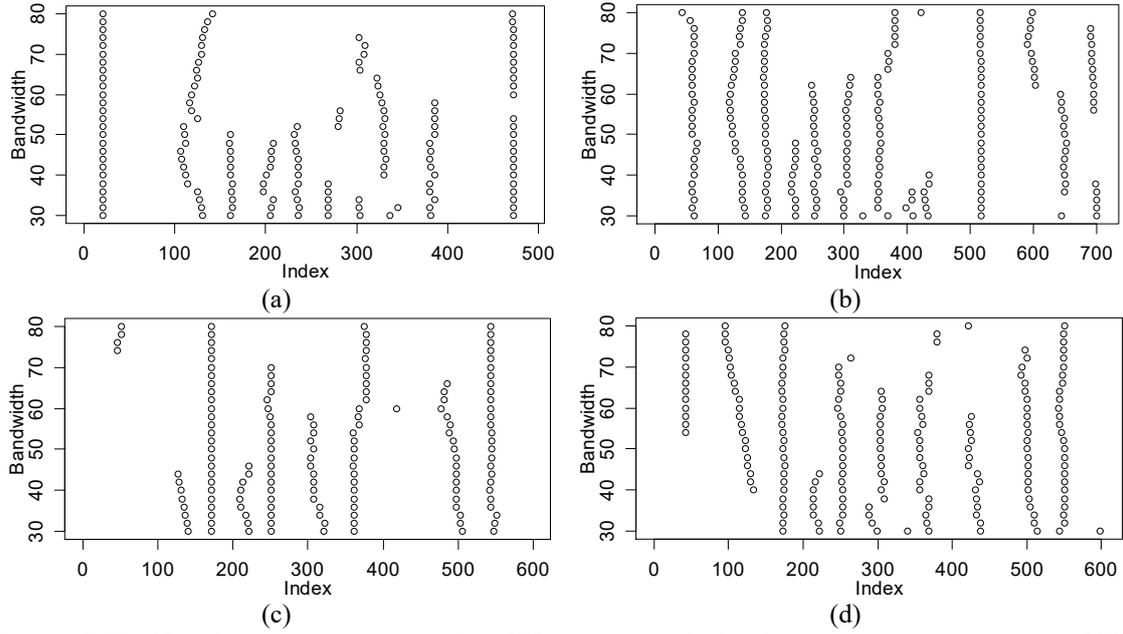

**Figure S.20.** Visual representations of the CPI matrices calculated from four representative CTR distributional sequences associated with cable pairs (a) RCP6, (b) RCP31, (c) RCP34, and (d) RCP40, respectively. The non-zero elements of each CPI matrix $\mathcal{M}$ are marked by empty circles, while the zero elements are represented by background pixels.

As noted above, in the multiscale detection the location estimates of a true change point usually tend to form a stable trajectory consisting of multiple points in the CPI diagram. If an identified trajectory contains only a few points (e.g., the one in Figure S.20 (c) that contains only a single point near $i = 400$), it is likely to be an "unreliable" trajectory that is composed of falsely detected change points (e.g., those resulted from the type I error). Once the trajectories are identified, we can eliminate the extremely short trajectories, which is beneficial for filtering out some falsely detected change points. This is the main advantage of our proposal over an alternative strategy named bottom-up merging described in Meier et al. (2021). In the bottom-up merging approach, a detected change point is added to the final change-point (FCP) set if its distance from the current FCP set exceeds a prescribed threshold, no matter whether the underlying change point is detected repeatedly (under different bandwidths) or just once.

### S.7.1. Trajectory Identification

Before proceeding, we provide a formal definition of the change-point trajectory. Given an estimated location $\hat{k}^*$ of a change point detected under the bandwidth $G_l$, it is represented by a marked point at $(\hat{k}^*, G_l)$ in the CPI diagram. Therefore, any marked point in the CPI diagram corresponds to a detected change point. The collection of the marked points in the CPI diagram that correspond to an identical underlying change point is called a change-point trajectory (hereafter trajectory) in this study. In this sense, a trajectory reflects the information of the location estimates of an identical change point varying with the bandwidths.

Identifying trajectories from the CPI diagram of real data is not straightforward. We see from Figure S.20 that not all of the potential trajectories are well-behaved straight lines. Instead, some trajectories are markedly curved, and some are fragmented. Moreover, in some CPI diagrams, there may exist inverted-"Y"-shaped composite trajectories, such as the one shown in Figure S.21 (demonstrated by red line). In the MOSUM procedure, the location estimate of a change point can be disturbed by other nearby change points if the bandwidth is too large. Consequently, the estimates of two change points at nearby locations might merge into one under a comparatively large bandwidth, resulting in an inverted-"Y"-shaped composite trajectory in the CPI diagram. Such composite trajectories increase the difficulty of trajectory identification. In our proposal, we employ the upward search and trajectory pruning operations (introduced later in this section) to cope with the negative impacts of composite trajectories.

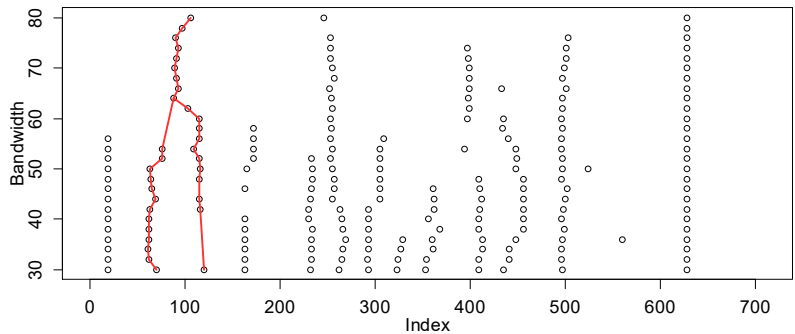

**Figure S.21.** A representative inverted-"Y"-shaped composite trajectory (indicated by red line). The CPI diagram is calculated from the CTR distributional sequence associated with cable pair RCP16.

*S.7.1.1. Basic Operations for Trajectory Search*

Our trajectory identification algorithm is developed based on four basic trajectory search operations, namely seed point selection, initial search, upward search, and trajectory pruning. This subsection presents technical details of these basic operations.

Before proceeding, we introduce some additional definitions. A change point that has not been assigned to any trajectory in the current CPI diagram is called the "free" change point. A "free" change point becomes a "frozen" change point once it is assigned to a trajectory. In each trajectory search operation, only the "free" change points can be assigned to a trajectory. In some particular circumstances, such as the trajectory pruning operation, the change points that are pruned from a trajectory will be set to "free" change points again.

(1) Seed Point Selection

This operation is used to select the reference point for initializing a trajectory. Once a reference point is determined, new points can be added into the collection of the relevant trajectory by some subsequent operations introduced later on. In such a strategy, it looks like that the trajectory grows from the reference point, see Figure S.22 for an illustration. Hence, the reference point is referred to as the seed point.

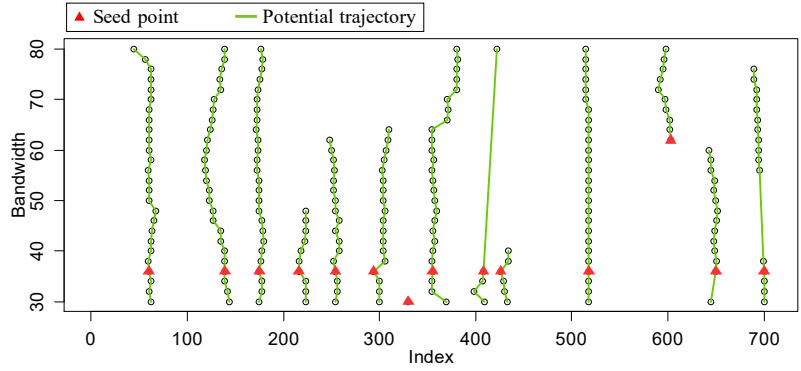

**Figure S.22.** Illustration of seed points (solid triangles) along with their corresponding potential trajectories.

We next present the details on seed point selection. Given a CPI diagram denoted by $\mathcal{M}$, we first count the number of "free" change points associated with each bandwidth $G_l$ ($l \in \{1, \cdots, L\}$), yielding a frequency vector as follows:

$$\boldsymbol{\omega} = \{\omega_1, \cdots, \omega_l, \cdots, \omega_L\} \text{ with } \boldsymbol{\omega}(l) = \omega_l = \sum_{i=1}^{n} \mathcal{M}(i,l) I\{\mathcal{M}(i,l) = 1\} \quad (S.120)$$

where $\mathcal{M}(i,l)$ denotes the element of $\mathcal{M}$ located at $(i,l)$. Suppose that we want to select seed points within the G-interval $\mathcal{B}_G = [G_{Low}, G_{Up}]$, we first find the lowest maximum point of the frequency vector $\boldsymbol{\omega}$ as follows:

$$l_{sp} = \min(\underset{l \in \{1,\cdots,L\}}{\operatorname{argmax}}(\boldsymbol{\omega}(l) I\{G_l \in \mathcal{B}_G\})) \quad (S.121)$$

we then choose the change points in $\mathcal{M}$ associated with the bandwidth $G_{l_{sp}}$ as the seed points, i.e.,

$$SP = \{(j, G_{l_{sp}}) : 1 \leq j \leq n \text{ and } \mathcal{M}(j,l) = 1\} \quad (S.122)$$

For convenience, if $SP \neq \emptyset$, it is equivalently represented as

$$SP = \{(j_*^1, G_{l_{sp}}), (j_*^2, G_{l_{sp}}), \cdots, (j_*^T, G_{l_{sp}})\} \quad (S.123)$$

where $T$ denotes the total number of selected seed points.

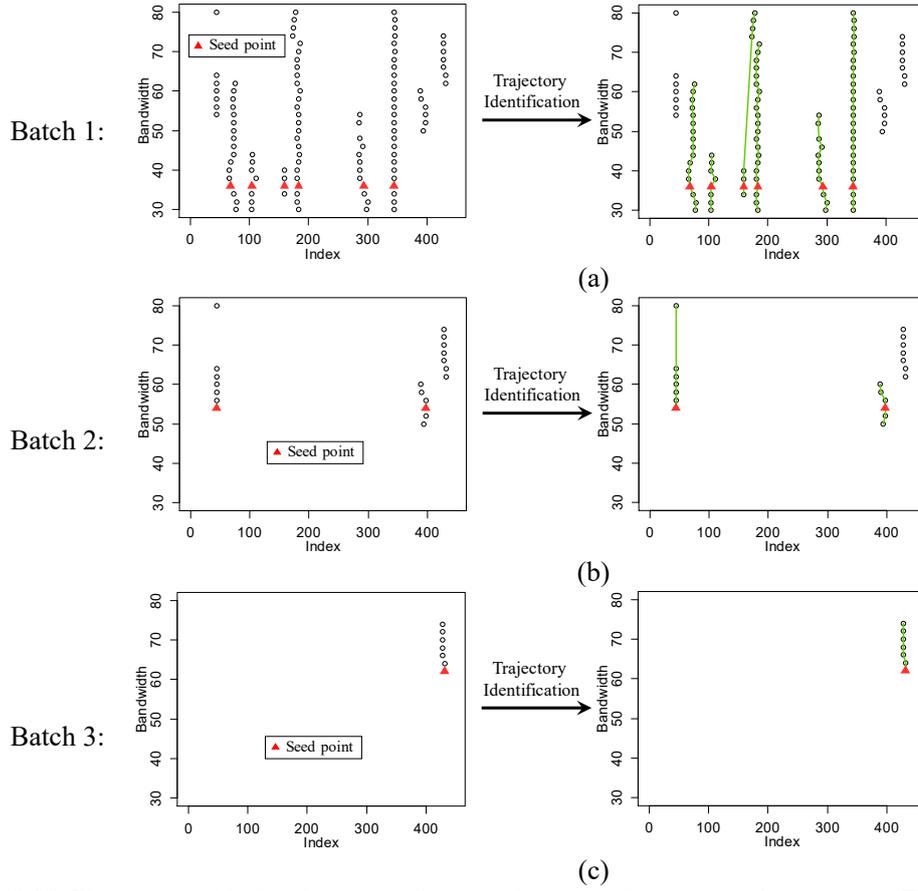

**Figure S.23.** Illustration of the batch-wise seed point selection and trajectory identification. The left column corresponds to the seed point selection in each batch, while the right column corresponds to the trajectory identification. (a) Batch 1, (b) Batch 2, and (c) Batch 3.

Each of the seed points in $SP$ represents an initial point of a trajectory that is required to be identified. If $SP$ contains $T$ different elements, it means that we have $T$ different trajectories to be searched at this round. The other potential trajectories that have no intersection with the current $SP$ set will be postponed into the next round of search using new selected seed points. Consequently, the trajectories are identified in batches, as illustrated in Figure S.23. Specifically, whenever a seed point selection operation is executed, we conduct a round of trajectory identification. Such a process proceeds recursively until the CPI diagram contains no "free" change point.

(2) Initial Search

Given a seed point $\left(j_*, G_{l_{sp}}\right) \in SP$, the initial search is the first attempt to find candidate points in $\mathcal{M}$ that can be assigned to the target trajectory initialized at $\left(j_*, G_{l_{sp}}\right)$. In the initial search operation, we find new candidate points from the following band-type region:

$$\mathrm{BD}\left(j_*, G_{l_{sp}} \middle| \Delta_h\right) \qquad (\text{S.}124)$$
$$= \{(j, G_l) \in \mathcal{M} : \max(1, j_* - \Delta_h) \leq j \leq \min(n, j_* + \Delta_h) \text{ and } 1 \leq l \leq L\}$$

where $\Delta_h$ represents the half-band width. Specifically, any "free" change point in $\mathcal{M}$ that falls into $\mathrm{BD}\left(j_*, G_{l_{sp}} \middle| \Delta_h\right)$ is assigned to the target trajectory.

(3) Upward Search

The upward search operation searches upward from an initial point to find new candidate points in $\mathcal{M}$ that can be assigned to the target trajectory. In the upward search operation, given an initial point denoted by $\left(j_\#, G_{l_\#}\right)$, we find new candidate points from the following upper-half band region:

$$\mathrm{BD}^{UH}\left(j_\#, G_{l_\#} \middle| \Delta_h\right) \qquad (\text{S.}125)$$
$$= \{(j, G_l) \in \mathcal{M} : \max(1, j_\# - \Delta_h) \leq j \leq \min(n, j_\# + \Delta_h) \text{ and } l_\# \leq l \leq L\}$$

where $\Delta_h$ represents the half-band width. Specifically, any "free" change point in $\mathcal{M}$ that falls into $\mathrm{BD}^{UH}\left(j_\#, G_{l_\#} \middle| \Delta_h\right)$ is assigned to the target trajectory. In the CPI diagram, the search domain $\mathrm{BD}^{UH}\left(j_\#, G_{l_\#} \middle| \Delta_h\right)$ is located on the top of the initial point, as illustrated in Figure S.24 (a); thus, this trajectory search operation is called the upward search.

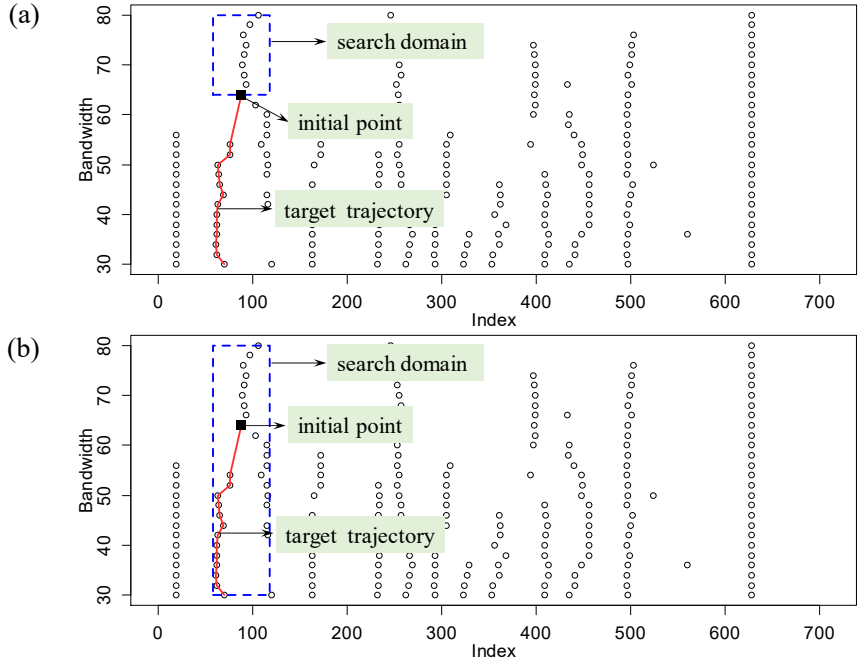

**Figure S.24.** Illustration of two different search domains. (a) $\mathrm{BD}^{UH}\left(j_\#, G_{l_\#} \middle| \Delta_h\right)$ (upper half band) and (b) $\mathrm{BD}\left(j_\#, G_{l_\#} \middle| \Delta_h\right)$ (full band).

Using the upper-half band as the search domain is less risky in falsely merging the "free" change points that belong to nearby trajectories, particularly for those associated with an inverted-"Y"-shaped composite trajectory (Figure S.21). As illustrated in Figure S.24, suppose that the initial point is the one represented by the solid square, if we use the following full band (equation (S.126)) as the search domain, then some points belonging to the other "branch" of the inverted-"Y"-shaped composite trajectory would be falsely assigned to the target trajectory (see Figure S.24 (b) for an illustration):

$$\text{BD}(j_\#, G_{l_\#}|\Delta_h) = \{(j, G_l) \in \mathcal{M}: \max(1, j_\# - \Delta_h) \leq j \leq \min(n, j_\# + \Delta_h) \text{ and } 1 \leq l \leq L\} \quad (S.126)$$

### (4) Trajectory Pruning

The trajectory pruning operation is designed for removing redundant points from a trajectory. Recall that the trajectory to be identified is a collection of estimated locations of an identical underlying change point obtained by the Fréchet-MOSUM procedure using different bandwidths. Location estimates of an identical change point under different bandwidths can differ due to estimation errors; however, the same change point is not allowed to have more than one different location estimates under the same bandwidth. In other words, the desired trajectory is only allowed to contain at most one point at the same vertical position in the CPI diagram. Therefore, the redundant candidate points (as demonstrated by the dashed ellipse in Figure S.25) are required to be removed from the trajectory.

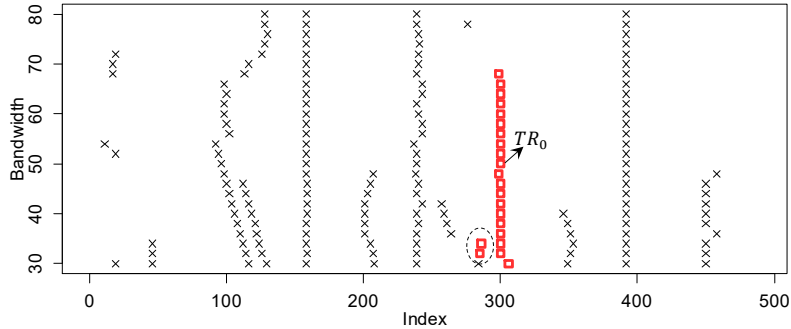

**Figure S.25.** Illustration of potential redundant points (demonstrated by the dashed ellipse) of an identified trajectory $TR_0$ (indicated by red squares).

We employ an elimination and backward insertion strategy to achieve the desired trajectory pruning. Let $TR_0 = \{(i'_1, G_{l'_1}), (i'_2, G_{l'_2}), \cdots, (i'_m, G_{l'_m})\} \subset \mathcal{M}$ denote the input trajectory for prune processing. We first find and eliminate the points from $TR_0$ that have duplicated $G$-values, and the collection of the remaining points is denoted as

$$TR_r = \{(i'_{r,1}, G_{l'_{r,1}}), (i'_{r,2}, G_{l'_{r,2}}), \cdots, (i'_{r,m_r}, G_{l'_{r,m_r}})\}$$

where $m_r$ is the number of points contained in the remaining trajectory. Suppose that the eliminated points contain $K$ different $G$-values denoted as $\{\tilde{G}_1, \cdots, \tilde{G}_K\}$, which are classified into the following $K$ different groups with the points in the same group have the same $G$-value:

Group 1: $(i_{1,1}, \tilde{G}_1), \cdots, (i_{1,m_1}, \tilde{G}_1)$

Group 2: $(i_{2,1}, \tilde{G}_2), \cdots, (i_{2,m_2}, \tilde{G}_2)$

$\vdots \qquad \vdots \qquad \vdots$

Group $K$: $(i_{K,1}, \tilde{G}_K), \cdots, (i_{K,m_K}, \tilde{G}_K)$

where $m_k$, $k \in \{1, \cdots, K\}$ denotes the number of points that are classified into the $k$th group. Among the $m_k$ points, the one that is closest to the remaining trajectory is put back into the trajectory, whereas the rest are all pruned. This processing corresponds to the backward insertion procedure. If there exists more than one point in a group that are closest to the remaining trajectory, we only retain one via random selection.

The backward insertion procedure can be implemented in a sequential manner. Given the points of group 1 and the remaining trajectory $TR_r$, we first estimate the intersection point (IP) between $TR_r$ and the horizontal line $G = \tilde{G}_1$, see Figure S.26 for an illustration. We employ a locally weighted averaging strategy that is outlined in Algorithm S.2 to estimate the IP, and the result is denoted as $(i_1^{IP}, \tilde{G}_1)$. We then find one point from group 1 that is closest to $(i_1^{IP}, \tilde{G}_1)$ to be reallocated to $TR_r$. We next use the updated remaining trajectory to find the new closest point from group 2 to update $TR_r$. This process proceeds sequentially until all of the $K$ groups have been processed, and the updated $TR_r$ at the last step is set as the after pruning trajectory. The pruning result of the trajectory $TR_0$ in Figure S.25 is illustrated in Figure S.27.

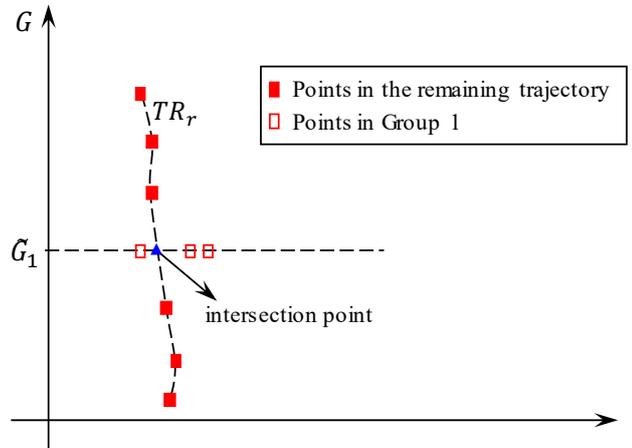

**Figure S.26.** Illustration of the intersection point between the remaining trajectory and the horizontal line $G = \tilde{G}_1$.

**Algorithm S.2**: Algorithm for estimating the intersection point between a given trajectory and the horizontal line $G = \tilde{G}$.

**Input**: The trajectory $TR_r = \{(i'_{r,1}, G'_{l_{r,1}}), (i'_{r,2}, G'_{l_{r,2}}), \cdots, (i'_{r,m_r}, G'_{l_{r,m_r}})\}$ and $\tilde{G}$ (the vertical intercept of the line $G = \tilde{G}$) such that $\tilde{G} \notin \{G'_{l_{r,1}}, G'_{l_{r,2}}, \cdots, G'_{l_{r,m_r}}\}$

**Output**: The estimated intersection point $(i^{IP}, \tilde{G})$

1: For each point in $TR_r$, calculate its distance from the horizontal line $G = \tilde{G}$, and denote the result as
$$\boldsymbol{\delta_G} = \{\delta_{G,1}, \cdots, \delta_{G,k}, \cdots, \delta_{G,m_r}\} \text{ with } \delta_{G,k} = \left|G'_{l_{r,k}} - \tilde{G}\right|.$$

2: Sort the distance vector $\boldsymbol{\delta_G}$ into ascending order and the result is written as
$$\boldsymbol{\delta^*_G} = \{\delta_{G,(1)}, \cdots, \delta_{G,(k)}, \cdots, \delta_{G,(m_r)}\} \text{ such that } \delta_{G,(1)} \leq \cdots \leq \delta_{G,(k)} \leq \cdots \leq \delta_{G,(m_r)}.$$

3: Calculate the threshold (used for subsequent locally weighted averaging)
$$\theta = \delta_{G,(p^*)} \in \boldsymbol{\delta^*_G} \text{ with } p^* = \max\{\lceil 0.5 m_r \rceil, \min\{4, m_r\}\}.$$

4: Calculate the weight vector
$$\boldsymbol{w} = \{w_1, \cdots, w_k, \cdots, w_{m_r}\} \text{ with } w_k = \frac{(1-\delta_{G,k}/\theta)I\{\delta_{G,k}\leq\theta\}}{\sum_{j=1}^{m_r}(1-\delta_{G,j}/\theta)I\{\delta_{G,j}\leq\theta\}},$$
where $I\{\cdot\}$ is the indicator function.

5: Estimate the intersection point through locally weighted averaging
$$(i^{IP}, \tilde{G}) = \left(\sum_{k=1}^{m_r} i'_{r,k} w_k, \tilde{G}\right).$$

6: Output $(i^{IP}, \tilde{G})$.

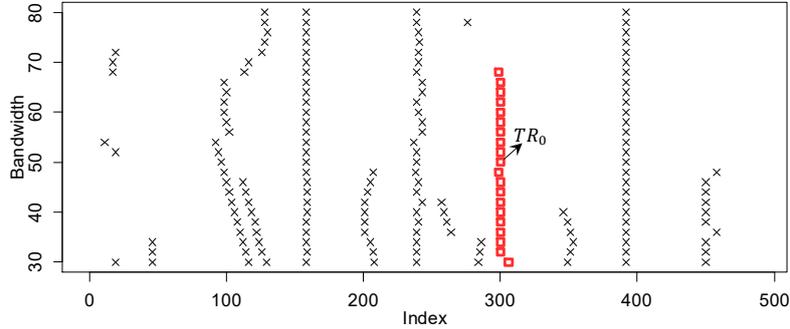

**Figure S.27.** The updated result of the trajectory $TR_0$ in Figure S.25 after conducting the trajectory pruning.

## S.7.1.2. Coarse-Grained Trajectory Search

Given the selected seed points, we first combine the initial search, upward search, and trajectory pruning operations to search for the corresponding trajectories at a coarse-grained level; we then use a technique called close neighboring point tracking (will be presented later on) to refine the identified trajectories. The implementation details of the coarse-grained search procedure are summarized in Algorithm S.3.

**Algorithm S.3**: Coarse-grained trajectory search

**Input**: The CPI diagram $\mathcal{M}$, the seed points $SP = \left\{\left(j_*^1, G_{l_{sp}}\right), \cdots, \left(j_*^T, G_{l_{sp}}\right)\right\}$, and the half-band width $\Delta_h$ of the search domain

**Output**: A total of $T$ different identified trajectories and the updated CPI diagram $\mathcal{M}$

1: **for** $i=1$ to $T$ **do**

    a: Perform the initial search operation using $SP(i) = \left(j_*^i, G_{l_{sp}}\right)$ as the seed point and $\text{BD}\left(j_*, G_{l_{sp}} \big| \Delta_h\right)$ as the search domain to find the first part of the $i$th trajectory, and the result is denoted as $TR_i$. Also, "freeze" the points in $\mathcal{M}$ that have been assigned to $TR_i$.

    b: Perform the trajectory pruning on $TR_i$ and update $\mathcal{M}$ accordingly.

    c: **repeat**

        i): Find the "topmost" point (i.e., the one with the largest G-value) of $TR_i$ and denote the result as $\left(j_\#, G_{l_\#}\right)$.

        ii): Perform the upward search operation for $TR_i$ using $\left(j_\#, G_{l_\#}\right)$ as the initial point and $\text{BD}^{UH}\left(j_\#, G_{l_\#} \big| \Delta_h\right)$ as the search domain, then update $TR_i$ and $\mathcal{M}$ accordingly.

        iii): Perform the trajectory pruning on $TR_i$ and update $\mathcal{M}$ accordingly.

    **until** there is no "free" change point in the current search domain $\text{BD}^{UH}\left(j_\#, G_{l_\#} \big| \Delta_h\right)$.

**end for**

2: Output the $T$ different identified trajectories denoted as $TR_1, \cdots, TR_T$ and the updated $\mathcal{M}$.

### S.7.1.3. Trajectory Refinement

After implementing the coarse-grained trajectory search by using Algorithm S.3, some trajectories can be completely identified; however, in some circumstances, there still exist some "free" change points in $\mathcal{M}$ that obviously belong to an identified trajectory, such as those demonstrated by the dashed ellipse in Figure S.28. This motivates us to design a trajectory refinement procedure to search for and add such points to an identified trajectory to which they may belong.

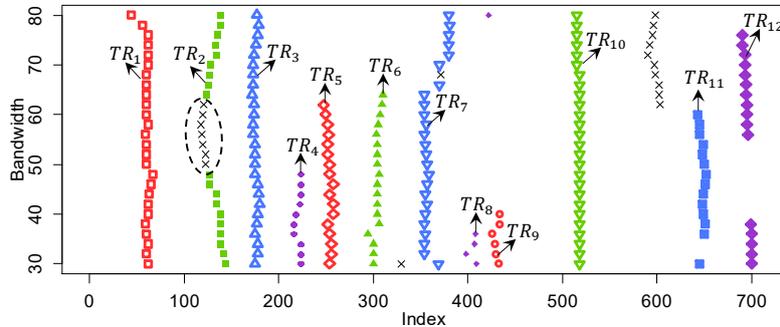

**Figure S.28.** The trajectories identified by using the coarse-grained trajectory search algorithm (Algorithm S.3) based on the seed points selected in the first batch. The distributional sequence is from cable pair RCP31. The identified trajectories are marked by colored symbols, while the points marked by "×" represent the "free" change points that have not been assigned to any trajectory. A trajectory is distinguished from its neighbouring trajectories by using different color and symbol.

We employ a close neighbor tracking strategy to achieve the desired trajectory refinement. Let $TR_i$ denote the target trajectory (identified by the coarse-grained search procedure) to be processed. Given a point $(j_\#, G_{l_\#}) \in TR_i$, a "free" change point $(j, G_l) \in \mathcal{M}$ is called the horizontal $\delta$-neighbor of $(j_\#, G_{l_\#})$ if $|j - j_\#| \leq \delta$. In the close neighbor tracking, we use a relatively small $\delta$ to search for and add such neighboring points to $TR_i$ under certain conditions. In fact, the neighboring points defined above are the location estimates (obtained in the multiscale detection) that fall into the $\delta$-neighborhood of $j_\#$. To reduce the risk of falsely merging "free" change points belonging to other nearby trajectories, we use the upward search strategy to search for potential $\delta$-neighbors that can be added to the target trajectory $TR_i$. Consequently, the close neighbor tracking can be achieved by implementing a series of upward search operations. Specifically, we sequentially perform the upward search operation on the target trajectory $TR_i$ using each point in $TR_i$ as the initial point, and the search domain $\text{BD}^{UH}(j_\#, G_{l_\#} | \Delta_h)$ (equation (S.125)) is set to $\text{BD}^{UH}(j_\#, G_{l_\#} | \delta)$. After each upward search operation, the new found "free" change points are added to $TR_i$, and the upward search operation proceeds sequentially until all of the points in $TR_i$ have served as the initial points. The parameter $\delta$ is recommended to be set as $\delta = \lceil \Delta_h/2 \rceil$, where $\Delta_h$ is the half-band width of the search domain involved in the coarse-grained trajectory search procedure.

After implementing the close neighbor tracking procedure, we also perform a trajectory pruning processing on each of the updated trajectories to remove potential duplicated points. The above close neighbor tracking together with the trajectory pruning processing constitutes the trajectory refinement procedure.

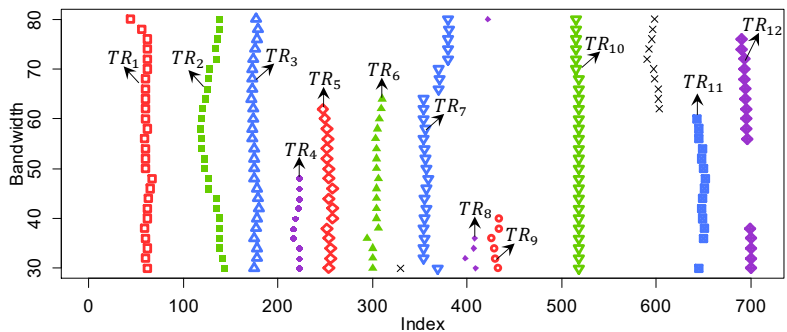

**Figure S.29**. The updated results of the trajectories in Figure S.28 after implementing the trajectory refinement processing. The meanings of the color and symbols are the same as those in Figure S.28.

After implementing the trajectory refinement procedure, the trajectories in Figure S.28 are updated as those shown in Figure S.29. We see that the "free" change points that are demonstrated by the dashed ellipse in Figure S.28 have been added to the trajectory $TR_2$.

*S.7.1.4. Trajectory Identification Algorithm*

The implementation details of our proposed trajectory identification algorithm are outlined in Algorithm S.4.

---

**Algorithm S.4**: The final trajectory identification procedure

---

**Input**: The CPI diagram $\mathcal{M}$, the half-band width $\Delta_h$ of the search domain, and the G-interval $\mathcal{B}_G = [G_{Low}, G_{Up}]$

**Output**: The identified trajectories arranged in batches

1: Initialize the batch label by setting $p = 1$.
2: **repeat**
    a: Initialize the set of identified trajectories in the $k$th batch by setting $S_p^{batch} = \emptyset$.
    b: Initialize the set of seed points by setting $SP = \emptyset$.
    c: **if** $p = 1$ **then**
        Perform the seed point selection operation on $\mathcal{M}$ within the G-interval $\mathcal{B}_G = [G_{Low}, G_{Up}]$, and the selected seed points are stored in $SP$. Then, update $\mathcal{M}$ accordingly.
    **else**
        Perform the seed point selection operation on $\mathcal{M}$ within the G-interval $\mathcal{B}_G = [\min\{G_{grid}\}, \max\{G_{grid}\}]$, and the selected seed points are stored in $SP$. Then, update $\mathcal{M}$ accordingly.
    **end if**
    d: Perform the coarse-grained trajectory search using Algorithm S.3 with $\mathcal{M}$, $SP$ and $\Delta_h$ as the inputs.
    e: Use the output $\mathcal{M}$ of Algorithm S.3 implemented in the former step to update $\mathcal{M}$.
    f: Perform trajectory refinement on all of the trajectories identified in step d by setting the parameter $\delta = \lceil \Delta_h/2 \rceil$ (used in the close neighbor tracking procedure), and update $\mathcal{M}$ accordingly. The refined trajectories are denoted as
$$\{TR_1^p, TR_2^p, \cdots, TR_{N_p}^p\}$$
    where $N_p$ denote the number of trajectories identified in step d.
    g: Set $S_p^{batch} \leftarrow \{TR_1^p, TR_2^p, \cdots, TR_{N_p}^p\}$
    h: Set $p = p + 1$ (update the batch label to enter the next iteration)
  **until** there is no "free" change point in $\mathcal{M}$.
3: Compute the total number of batches (i.e., the iterations)
$$m_b = p - 1$$
4: Output the identified trajectories in each batch, namely $S_1^{batch}, S_2^{batch}, \cdots, S_{m_b}^{batch}$

---

We see from Algorithm S.4 that the trajectories are identified in batches. In our trajectory identification strategy, a potential trajectory can only be identified once one of its points has been selected as the seed point. Seed points are adaptively selected by the seed point selection operation introduced earlier. Although executing the seed point selection operation once can accomplish the seed point selection for multiple trajectories, it does not guarantee to cover all of the potential trajectories. Therefore, the seed point selection operation together with the trajectory search procedure may be required to be repeated several times to accomplish the trajectory identification for a CPI diagram. Consequently, the trajectories are identified in batches. Specifically, whenever a seed point selection operation is executed, we conduct a round of trajectory search operations via steps (d)–(g) of Algorithm S.4, and the corresponding identified trajectories are classified into the same batch. For instance, if we use Algorithm S.4 to identify the potential trajectories in the CPI diagram shown in Figure S.30 (a), it requires a total of three different batches to accomplish the identification, and the results are shown in Figure S.30.

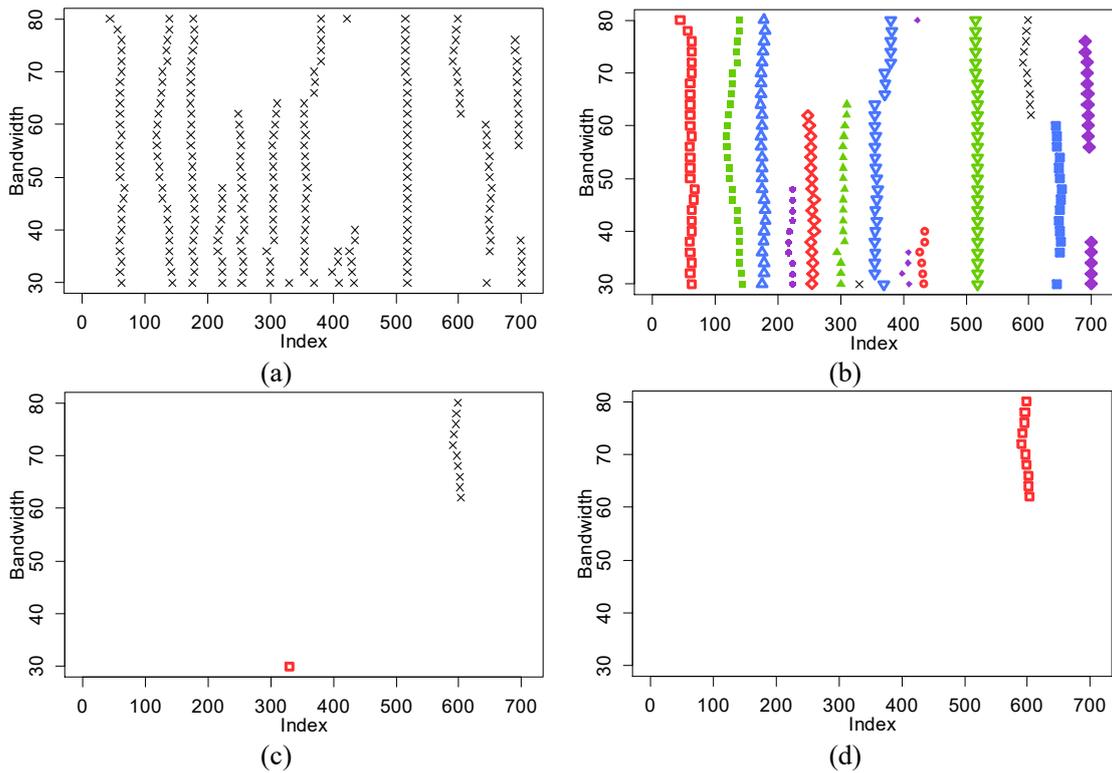

**Figure S.30**. Trajectory identification results of the CTR distributional sequence associated with cable pair RCP31. (a) The original CPI diagram before trajectory identification, (b) the identified trajectories (colored symbols) in the first batch, (c) the identified trajectory (colored square) in the second batch, and (d) the identified trajectory (colored squares) in the third batch. The points marked by "×" represent the "free" change points at the current CPI diagram.

*S.7.1.5. Recommended Implementation Settings*

This subsection provides our recommended implementation settings for the trajectory identification algorithm. The trajectory identification algorithm (outlined in Algorithm S.4) has two parameters to be set, namely the half-band width $\Delta_h$ and the G-interval $\mathcal{B}_G = [G_{Low}, G_{Up}]$.

The parameter $\Delta_h$ is used to construct the band-type search domains given in (S.124) and (S.125), and we recommend to set it as $\Delta_h = \varepsilon G_r$, where $\varepsilon$ is the AOP parameter (see (8) of the main text) and $G_r$ stands for the G-value associated with the reference point. Recall that the reference point of the initial search is the seed point denoted as $(j_*, G_{l_{sp}})$ (see (S.124)); for the upward search, the reference point is the initial point denoted as $(j_\#, G_{l_\#})$ (see (S.125)). Consequently, the recommended values of $\Delta_h$ for the initial search and upward search are $\varepsilon G_{l_{sp}}$ and $\varepsilon G_{l_\#}$, respectively. In our real data analysis, the value of $\varepsilon G$ is fixed at 15 (based on the recommended settings for $\varepsilon$ and $G$ described in Section S.2.8). Consequently, $\Delta_h$ takes the constant value 15.

The G-interval $\mathcal{B}_G = [G_{Low}, G_{Up}]$ plays the role of specifying a feasible region (see (S.121)) for the first round of seed point selection. For a MOUSM detector, the detection results of a change point are more likely to be disturbed by nearby change points when the bandwidth takes a relatively large value. Hence, the seed point of a trajectory is recommended to be selected from the lower half part of the trajectory, as the corresponding location estimates of the change point are obtained using relatively small bandwidths. On the other hand, if the bandwidth is too small, the detection result might also become less reliable, as the samples used for computing the SS sequence in the MOSUM procedure are scarce. Therefore, in the first round of seed point selection, the upper half and the bottom 10% of the G-grid are recommended to be excluded from the feasible region for seed point selection. Consequently, the recommended setting for $\mathcal{B}_G = [G_{Low}, G_{Up}]$ is as follows:

$$\mathcal{B}_G = [G_{Low}, G_{Up}] = [q_{0.1}(\boldsymbol{G}_{grid}), q_{0.5}(\boldsymbol{G}_{grid})]$$

where $q_{0.1}(\boldsymbol{G}_{grid})$ and $q_{0.5}(\boldsymbol{G}_{grid})$ denote the 10th and 50th percentiles of the G-grid $\boldsymbol{G}_{grid} = \{G_1, \cdots, G_L\}$.

*S.7.2. Change-Point Aggregation*

Recall that a trajectory is defined as a collection of location estimates (obtained by the Fréchet-MOSUM procedure using different bandwidths) that correspond to an identical underlying change point. Therefore, we aggregate the estimated change points from an identical identified trajectory to produce an estimate for the underlying change point. Practically, the following two treatments are considered in our change-point aggregation procedure:

(a) In the CPI diagram, each point of a trajectory stands for a location estimate of a change point. The change-point estimates situated at the upper part of a trajectory have a higher risk to be disturbed by nearby change points as they are estimated by using relatively large bandwidths. Therefore, in the change-point aggregation, we downweight the contributions of such change-point estimates if the trajectory is relatively long.

(b) Generally, if a true change point can be detected by a certain bandwidth, it can also be detected by other nearby bandwidths in the G-grid, if the G-grid is not too coarse. In this sense, a stable trajectory tends to contain a series of change points detected under different bandwidths. In contrast, if a trajectory contains only a few change points (e.g., one or two), it is more likely to be an unreliable trajectory that is composed of falsely detected change points (e.g., those resulted from the type I error). Therefore, if a trajectory is too short, we recommend to delete it.

Based on the above considerations, we design a change-point aggregation procedure, whose implementation details are summarized in Algorithm S.5. Before the final change-point aggregation (i.e., step (e)), step (d) of Algorithm S.5 uses a data-driven threshold $\theta$ (computed in step (c)) to eliminate some change points that are detected by using relatively large bandwidths, so as to exclude their contributions to the aggregated result. A trajectory of length less than 4 is treated as an unstable trajectory, and it is prevented from entering the aggregation module (i.e., steps (a)~(e)); the corresponding aggregated change point is set to be "NaN". The aggregated change points obtained by Algorithm S.5 are illustrated in Figure S.31 (indicated by vertical dashed lines) for the identified trajectories shown in Figure S.30 (b) and Figure S.30 (d), respectively.

**Algorithm S.5**: Change point aggregation for a single trajectory

**Input**: The trajectory $TR = \{(i'_1, G_{l'_1}), (i'_2, G_{l'_2}), \cdots, (i'_m, G_{l'_m})\}$ consisting of $m$ points

**Output**: The aggregated change-point estimate $\hat{k}^*$

1: **If** $m \geq 4$ **then**

  a: Extract the bandwidths from $TR$
  $$\boldsymbol{G}_{TR} = \{G_{l'_1}, \cdots, G_{l'_k}, \cdots, G_{l'_m}\}$$

  b: Sort the bandwidths in $\boldsymbol{G}_{TR}$ into ascending order and the result is written as
  $$\boldsymbol{G}^{\#}_{TR} = \{G_{l'_{(1)}}, \cdots, G_{l'_{(k)}}, \cdots, G_{l'_{(m)}}\} \text{ such that } G_{l'_{(1)}} < \cdots < G_{l'_{(k)}} < \cdots < G_{l'_{(m)}}.$$

  c: Calculate the threshold
  $$\theta = G_{l'_{(p^{\#})}} \in \boldsymbol{G}^{\#}_{TR} \text{ with } p^{\#} = \max\{[0.5m], 4\}.$$

  d: Extract the following sub-trajectory from $TR$ based on the threshold $\theta$:
  $$TR_{sub} = \{(i', G_{l'}) \in TR: G_{l'} \leq \theta\}$$
  $$= \{(i'_{*,1}, G_{l'_{*,1}}), (i'_{*,2}, G_{l'_{*,2}}), \cdots, (i'_{*,m_*}, G_{l'_{*,m_*}})\}$$

  e: Take the median of the $m_*$ change points $\{i'_{*,1}, \cdots, i'_{*,m_*}\}$ associated with $TR_{sub}$ as the aggregated change point
  $$\hat{k}^* = [\text{median}\{i'_{*,1}, \cdots, i'_{*,m_*}\}] \text{ ([x] stands for rounding the value of } x \text{ to be an integer)}$$

  **else**

  f: Set $\hat{k}^* = \text{NaN}$ (represents no change-point estimate)

  **end if**

2: Output $\hat{k}^*$.

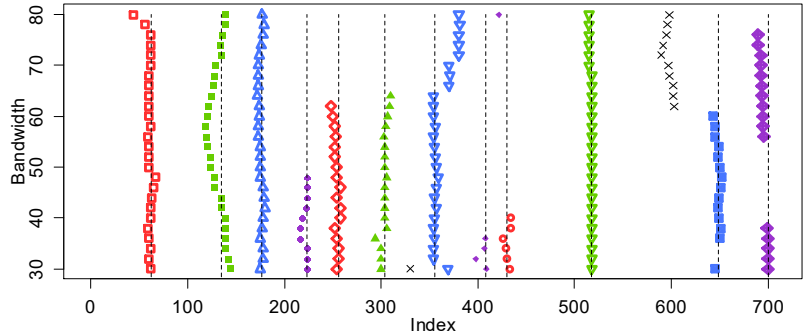

(a)

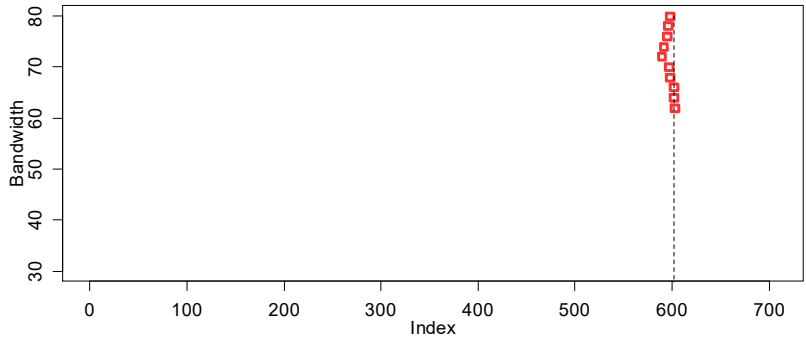

(b)

**Figure S.31**. Aggregated change points (marked by vertical dashed lines) of the identified trajectories in Figure S.30. (a) The result of the first batch and (b) the result of the third batch.

### S.7.3. Final Change-Point Estimation

After implementing the trajectory identification (Algorithm S.4) and change-point aggregation (Algorithm S.5) procedures, we can obtain the aggregated change-point estimate associated with each of the identified trajectories. After eliminating "NaN" values, the aggregated change points that are arranged in batches are illustrated in Table S.6.

**Table S.6.** Illustration of aggregated change points associated with different batches after removing "NaN" values.

| Batch label ($p$) | Aggregated change points |
|---|---|
| 1 | $K_1^b = \{\hat{k}_{1,1}^*, \hat{k}_{1,2}^*, \cdots, \hat{k}_{1,N_1}^*\}$ |
| 2 | $K_2^b = \{\hat{k}_{2,1}^*, \hat{k}_{2,2}^*, \cdots, \hat{k}_{2,N_2}^*\}$ |
| $\vdots$ | $\vdots$ |
| $m_b$ | $K_{m_b}^b = \{\hat{k}_{m_b,1}^*, \hat{k}_{m_b,2}^*, \cdots, \hat{k}_{m_b,N_{m_b}}^*\}$ |

It is noteworthy that an aggregated change point from a later batch may be a duplicated version of the one obtained in the former batch. This is because a trajectory identified in the later batch may be formed by the missed "free" change points that should belong to an existing trajectory identified in the former batch. To reduce the risk of overestimating the number of change points, we use Algorithm S.6 to merge the aggregated change points from different batches to obtain the final location estimates of the change points potentially contained in the distributional sequence.

In Algorithm S.6, Card($K_{cand}$) stands for the cardinality of the set $K_{cand}$, $G(\hat{k}_c^*)$ stands for the bandwidth by which the change point $\hat{k}_c^*$ is detected in the multiscale detection procedure, and $\varepsilon$ is the AOP parameter (see (8) of the main text). In our analysis, the value of $\varepsilon G$ is fixed at 15; thus, the condition $\min_{k \in K_{merge}} |\hat{k}_c^* - k| > \varepsilon G(\hat{k}_c^*)$ in step 2 (b) of Algorithm S.6 is equivalent to $\min_{k \in K_{merge}} |\hat{k}_c^* - k| > 15$. Consequently, given $\hat{k}_c^* \in K_{cand}$ (the union of the aggregated change points obtained in the late batches), we only add it to the set $K_{merge}$ if its distance from the current $K_{merge}$ exceeds $\varepsilon G$. Actually, the $\varepsilon G$ here is the allowed minimum length of an over-threshold indexing block that can be used to locate a potential change point in the MOSUM procedure (see (8) in the main text). Therefore, if two change points from different batches are within the $\varepsilon G$-neighbourhood of each other, then they are treated as duplicated estimates of an identical underlying change point and we delete the one coming from the later batch.

---

**Algorithm S.6**: Algorithm for merging the aggregated change points from different batches

**Input**: The aggregated change points arranged in batches $K_1^b, K_2^b, \cdots, K_{m_b}^b$ as listed in Table S.6

**Output**: The set of the merged change points denoted by $K_{merge}$

1: Set $K_{merge} = K_1^b$ and set $K_{cand} = \bigcup_{p=2}^{m_b} K_p^b$.

2: **if** $K_{cand} \neq \emptyset$ **then**

    **for** $j = 1$ **to** $\text{Card}(K_{cand})$ **do**

        a): Set $\hat{k}_c^* = K_{cand}(j)$.

        b): **if** $\min_{k \in K_{merge}} |\hat{k}_c^* - k| > \varepsilon G(\hat{k}_c^*)$ **then**

            Set $K_{merge} \leftarrow \hat{k}_c^*$.

        **end if**

    **end for**

  **end if**

3: Sort $K_{merge}$ into ascending order.

4: Output $K_{merge}$.

---

We take the output $K_{merge}$ of Algorithm S.6 as the final change-point detection result of our multiscale Fréchet-MOSUM method. Suppose $K_{merge}$ contains $N_{merge}$ elements denoted by $K_{merge} = \{\hat{k}_1^*, \hat{k}_2^*, \cdots, \hat{k}_{N_{merge}}^*\}$ such that $0 < \hat{k}_1^* < \hat{k}_2^* < \cdots < \hat{k}_{N_{merge}}^*$, then the number and locations of the change points contained in the distributional sequence are estimated as follows:

    Number of change points: $\hat{q}_n = N_{merge} = \text{Card}(K_{merge})$

    Locations of change points: $\hat{k}_1^*, \hat{k}_2^*, \cdots, \hat{k}_{\hat{q}_n}^*$

*S.7.4. Results*

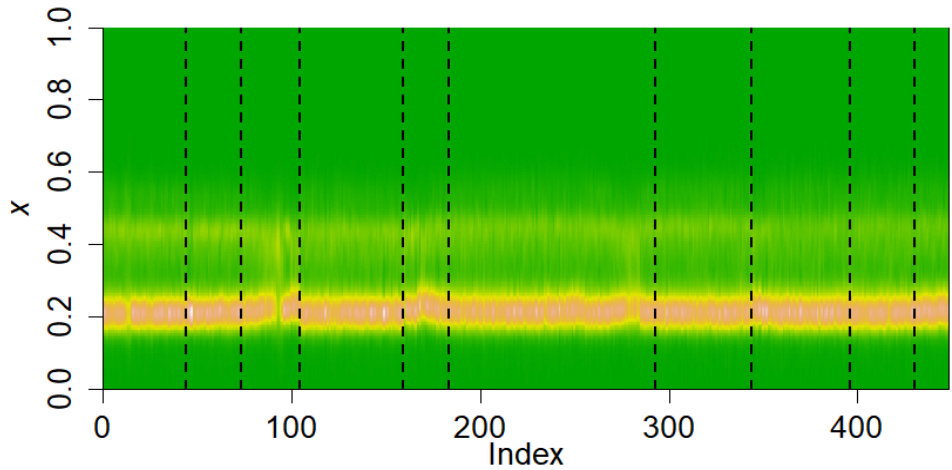
The heatmap of the PDF-valued sequence

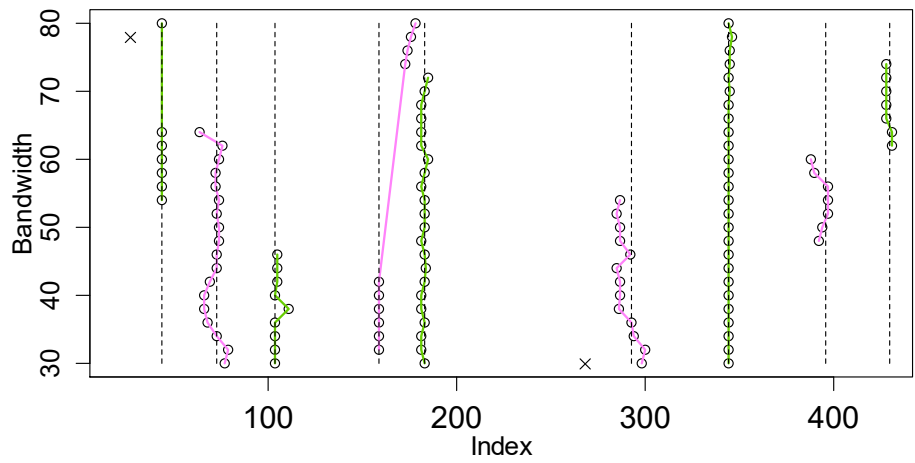
The identified change-point trajectories

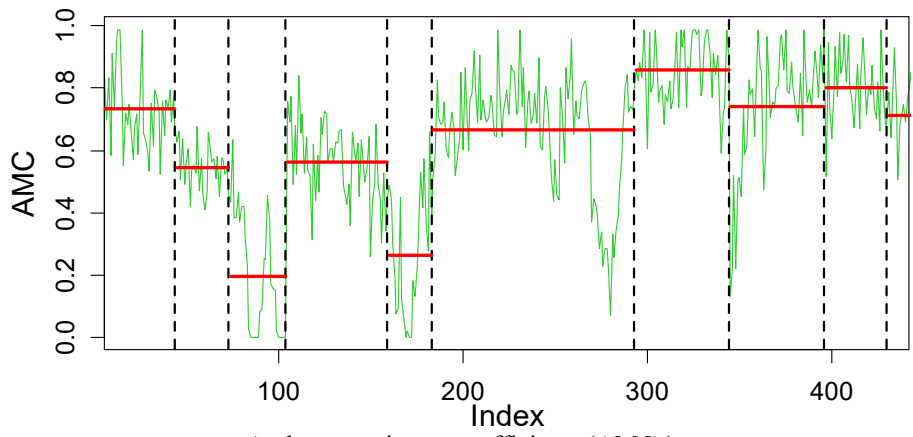
Archetype mixture coefficients (AMCs)

(a) Multiscale detection result of the distributional sequence associated with cable pair RCP1

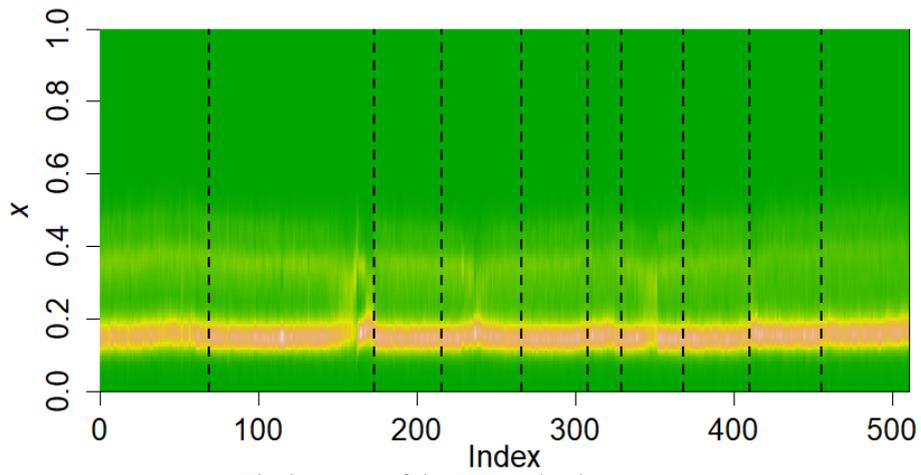
The heatmap of the PDF-valued sequence

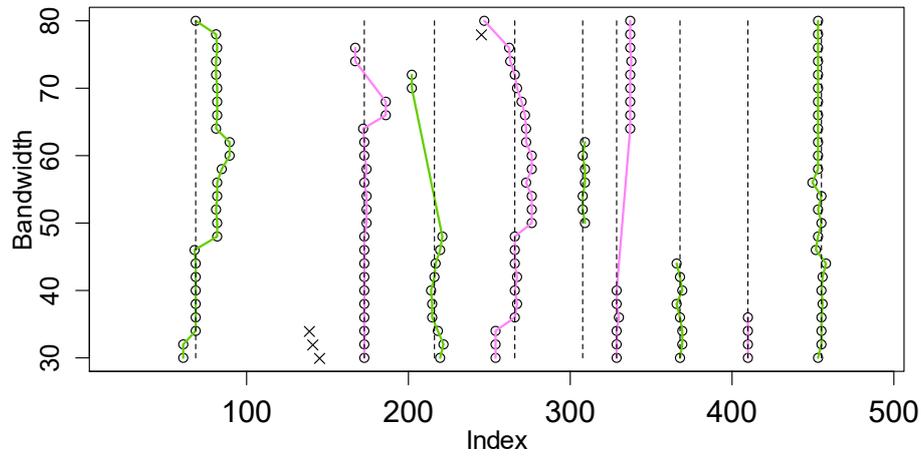
The identified change-point trajectories

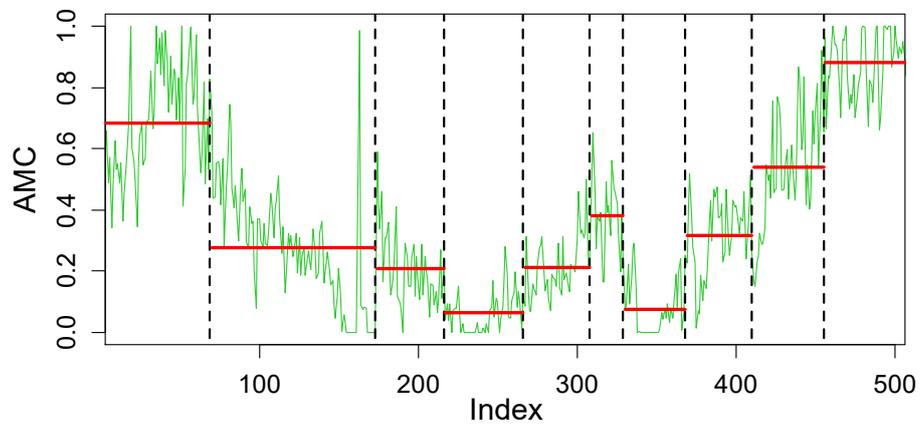
Archetype mixture coefficients (AMCs)

(b) Multiscale detection result of the distributional sequence associated with cable pair RCP5

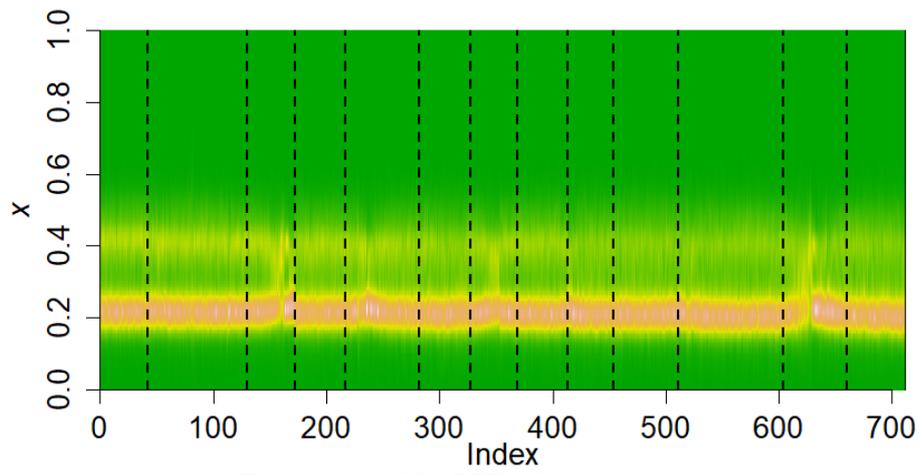

The heatmap of the PDF-valued sequence

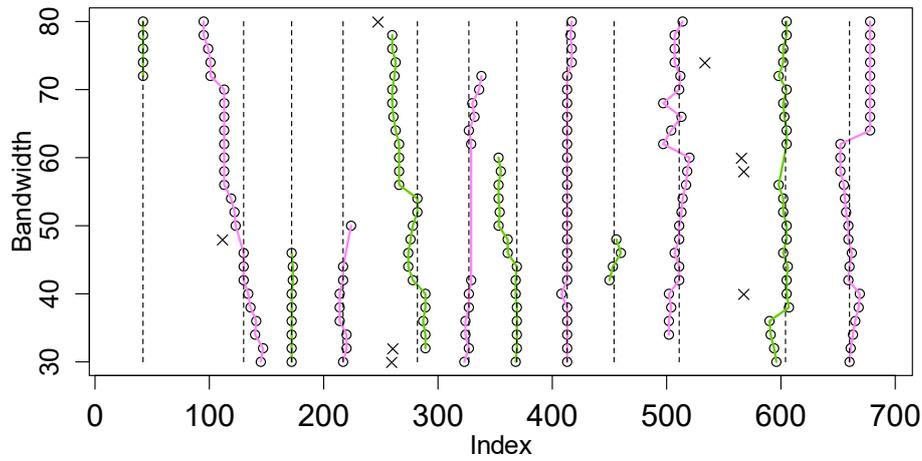

The identified change-point trajectories

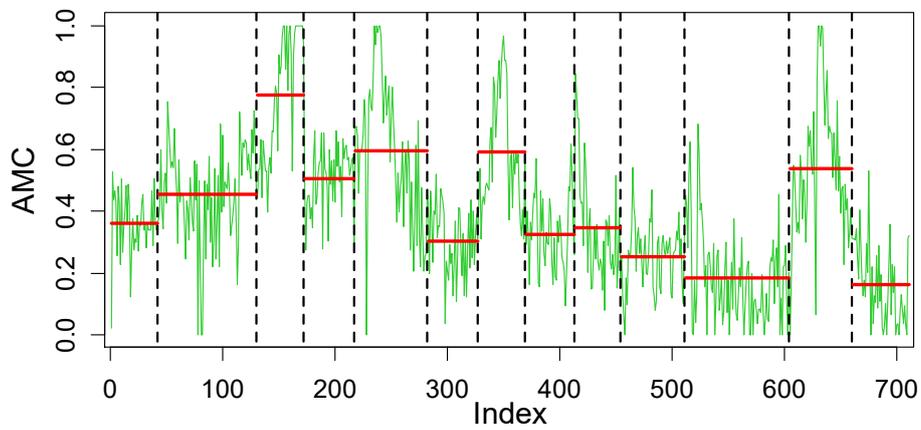

Archetype mixture coefficients (AMCs)

(c) Multiscale detection result of the distributional sequence associated with cable pair RCP10

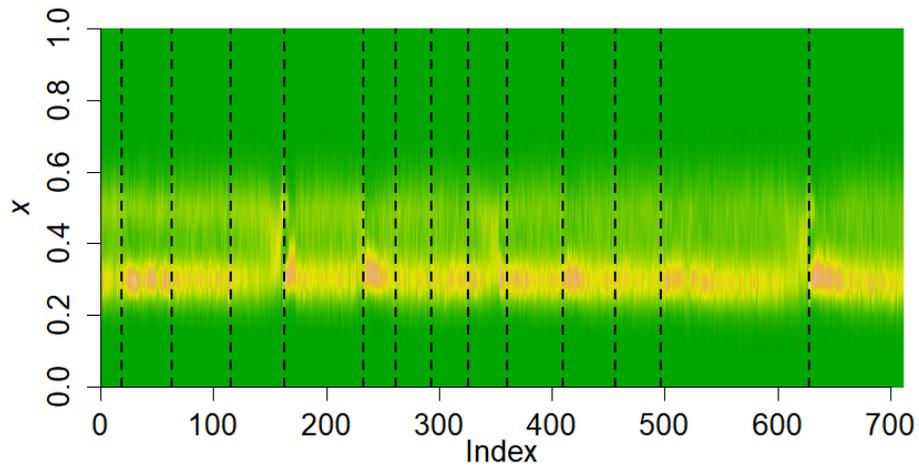
The heatmap of the PDF-valued sequence

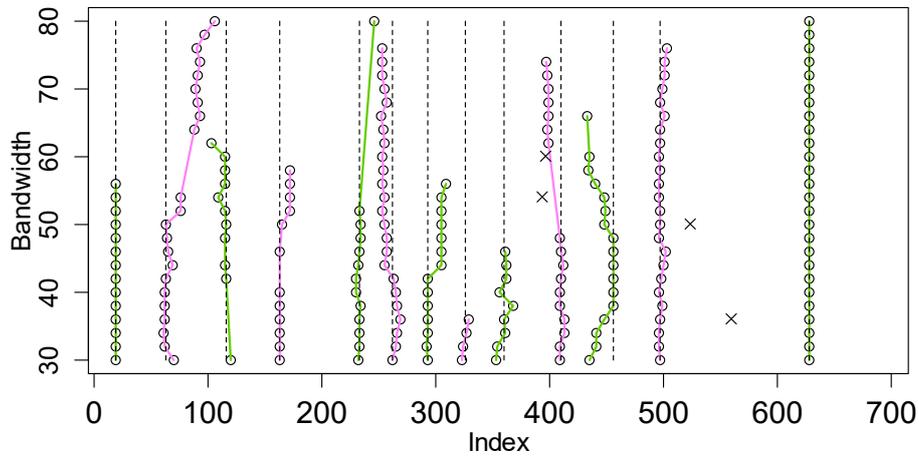
The identified change-point trajectories

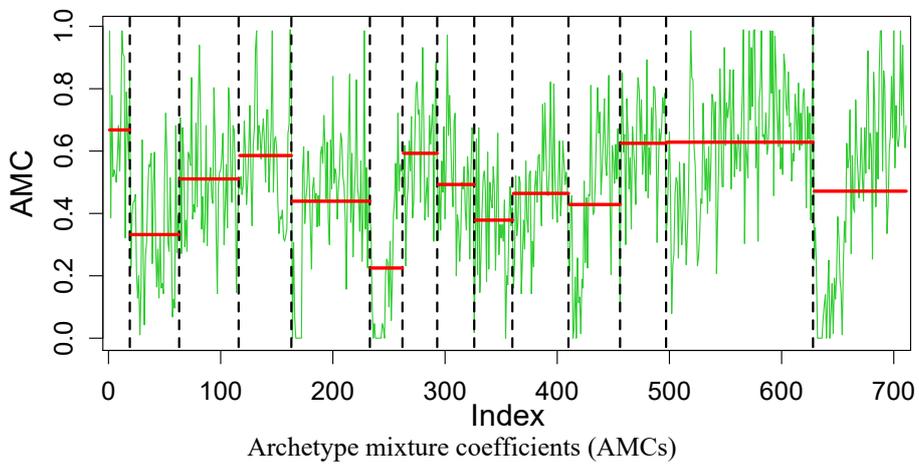
Archetype mixture coefficients (AMCs)

(d) Multiscale detection result of the distributional sequence associated with cable pair RCP16

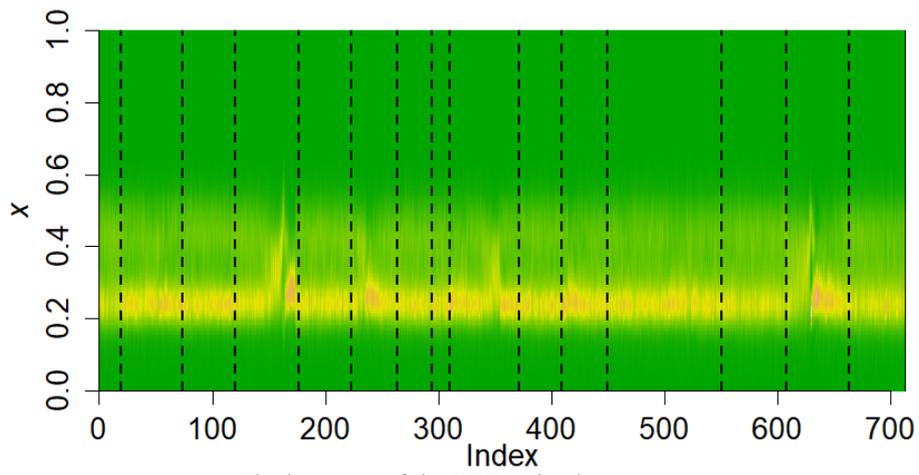

The heatmap of the PDF-valued sequence

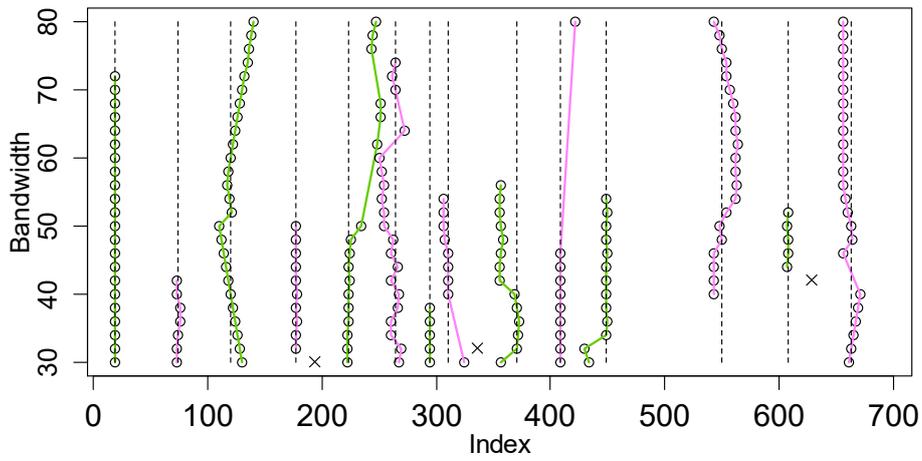

The identified change-point trajectories

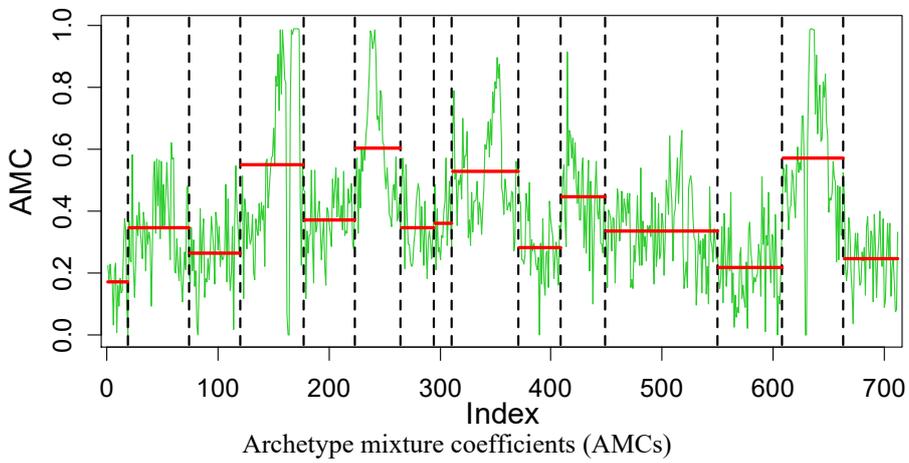

Archetype mixture coefficients (AMCs)

(e) Multiscale detection result of the distributional sequence associated with cable pair RCP19

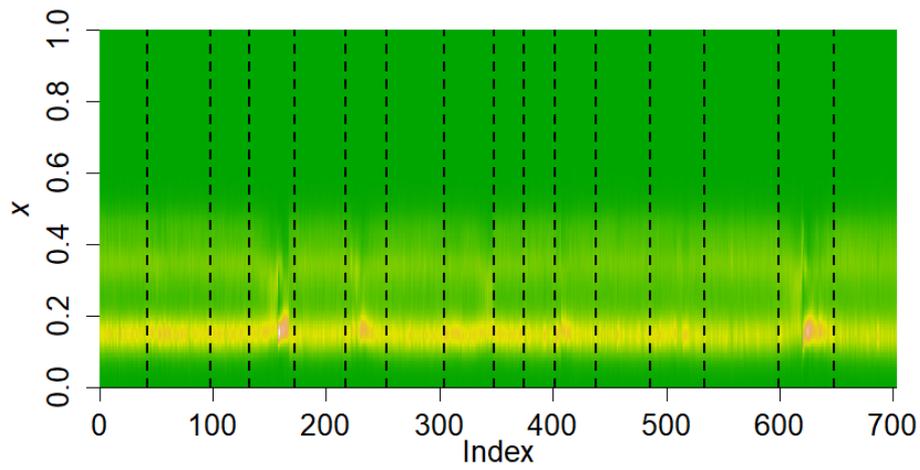
The heatmap of the PDF-valued sequence

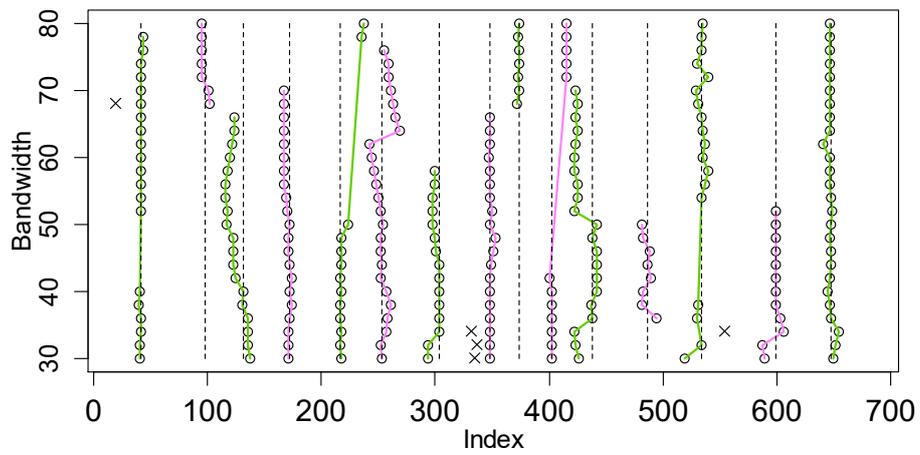
The identified change-point trajectories

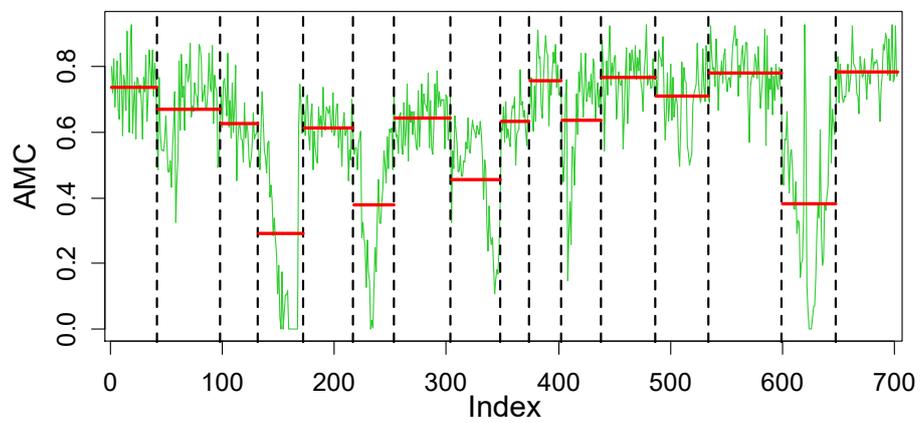
Archetype mixture coefficients (AMCs)

(f) Multiscale detection result of the distributional sequence associated with cable pair RCP32

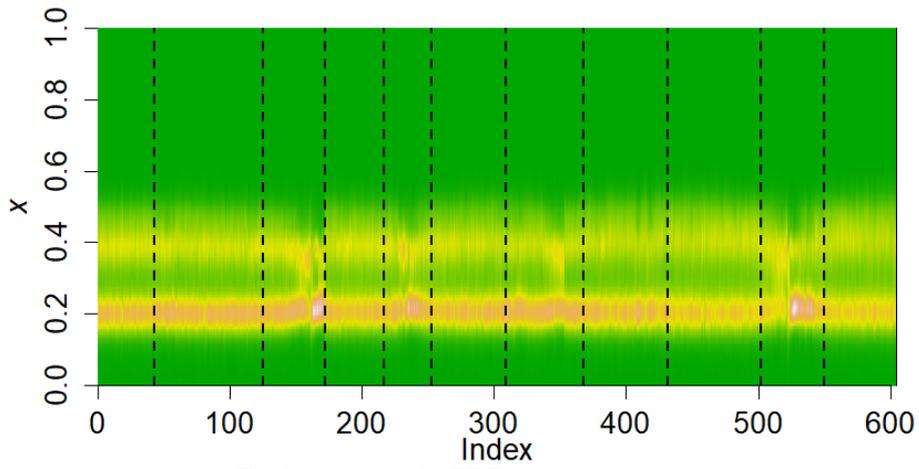

The heatmap of the PDF-valued sequence

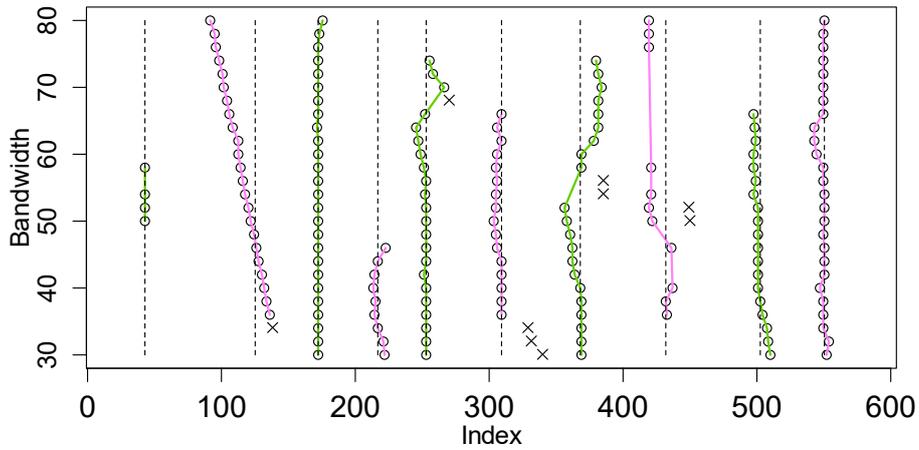

The identified change-point trajectories

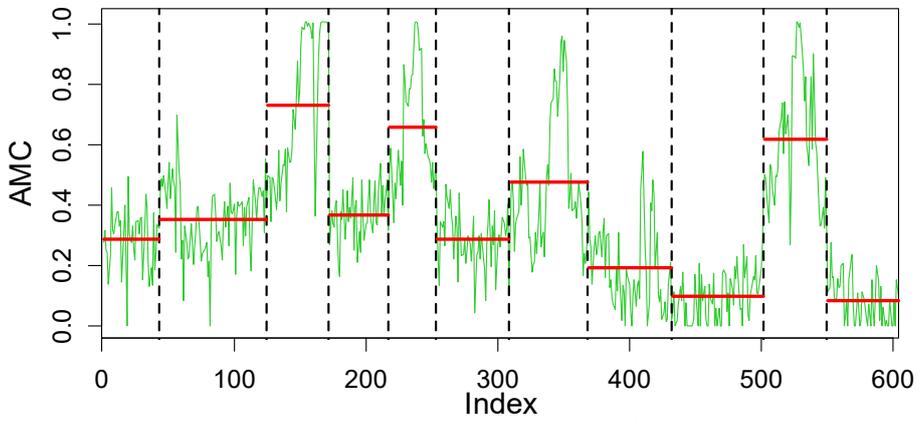

Archetype mixture coefficients (AMCs)

(g) Multiscale detection result of the distributional sequence associated with cable pair RCP42

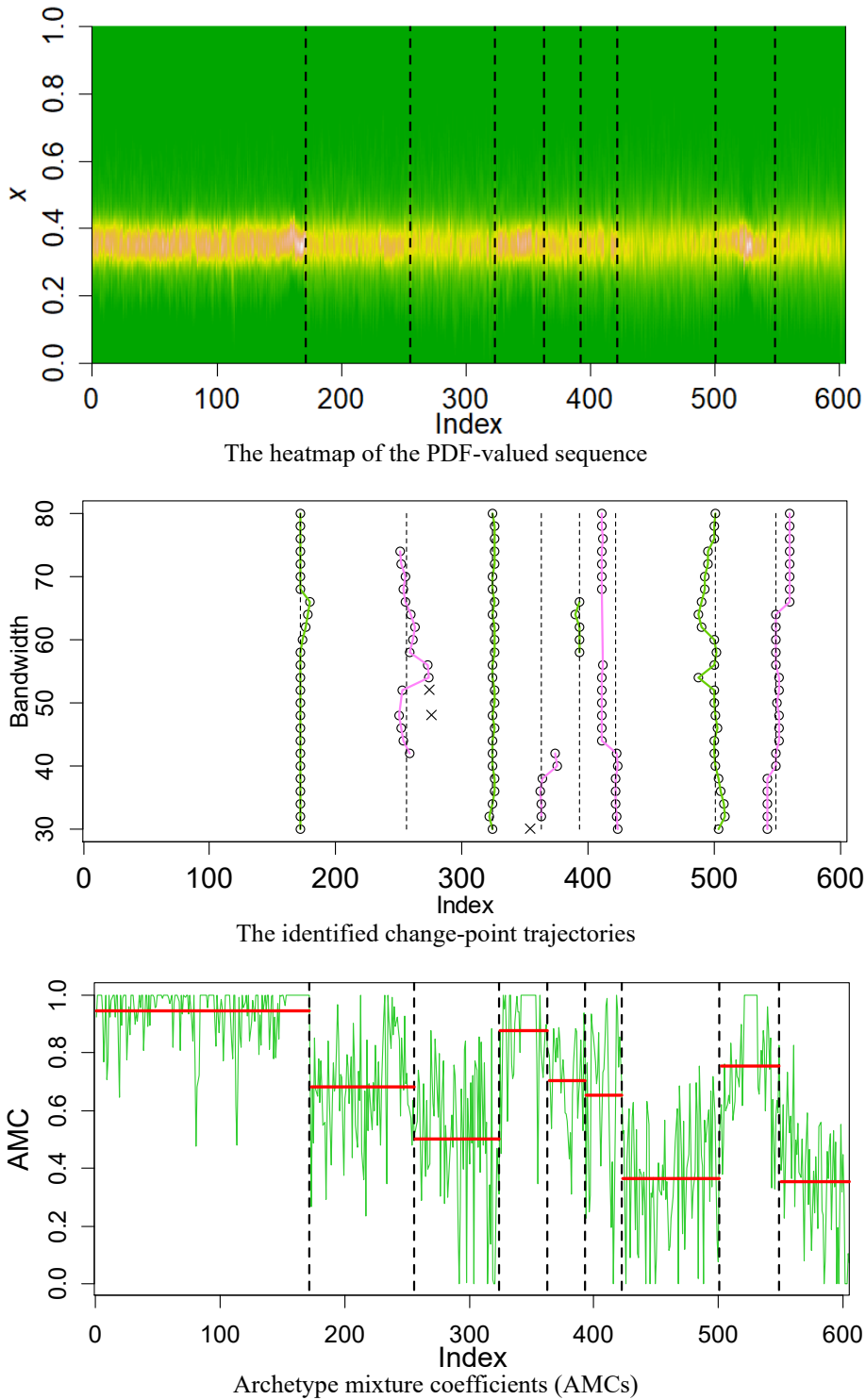

(h) Multiscale detection result of the distributional sequence associated with cable pair RCP48

**Figure S.32**. Change-point detection results obtained by the multiscale Fréchet-MOSUM method for eight representative CTR distributional sequences associated with cable pairs (a) RCP1, (b) RCP5, (c) RCP10, (d) RCP16, (e) RCP19, (f) RCP32, (g) RCP42 and (h) RCP48, respectively. The first row corresponds to the heatmap of the PDF-valued sequence, the second row corresponds to the identified change-point trajectories obtained by the multiscale detection procedure, and the third row corresponds to the archetype mixture coefficients (obtained by the archetypal analysis) used for visually confirming the detected changes (similar to the right panel of Figure 4 of the main text). The final change-point location estimates obtained by using Algorithms S.5 and S.6 are indicated by vertical dashed lines.

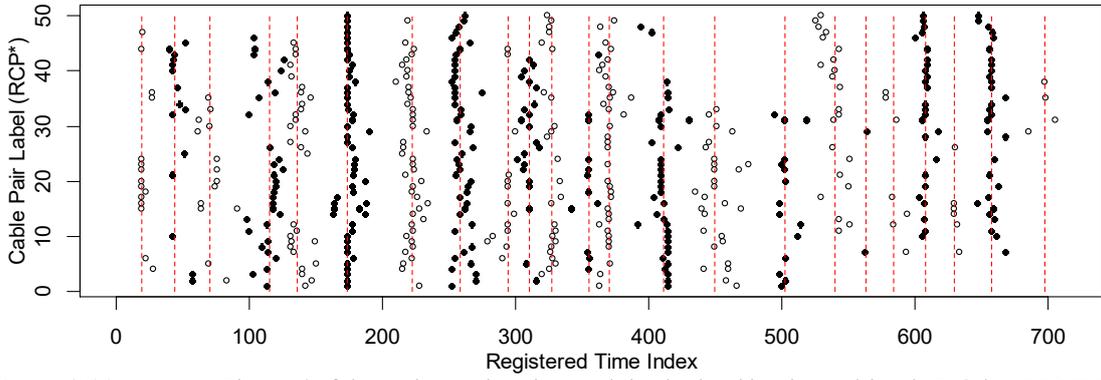

**Figure S.33**. Same as Figure 6 of the main text but the result is obtained by the multiscale Fréchet-MOSUM method.

# Appendix S.1: LQD and Inverse LQD Transformations

The log quantile density (LQD) transformation is a functional transformation first proposed by Petersen and Müller (2016) for transforming a PDF into the $L^2([0,1])$ space. This transformation can remove the inherent constraints of a PDF, and the resulting LQD-transformed function is an ordinary function residing in the $L^2([0,1])$ space.

Let $\mathcal{F}([0,1])$ denote the functional space formed by all of the univariate continuous PDFs defined on the compact interval $[0,1]$. Given a PDF $f \in \mathcal{F}([0,1])$, the LQD transformation (denoted as LQD[·]) is a map from $\mathcal{F}([0,1])$ to $L^2([0,1])$ defined as follows (Petersen and Müller 2016):

$$\psi(t) = \text{LQD}[f](t) = \log\left(\frac{dQ(t)}{dt}\right) = -\log(f(Q(t))), \forall f \in \mathcal{F}([0,1]) \quad (S.127)$$

where $Q(t) = F^{-1}(t)$ denotes the quantile function corresponding to the PDF $f(x)$. The inverse LQD transformation (denoted as $\text{LQD}^{-1}[\cdot]$) is a map from $L^2([0,1])$ to $\mathcal{F}([0,1])$ defined as follows (Petersen and Müller 2016):

$$f(x) = \text{LQD}^{-1}[\psi](x) = \theta_\psi \exp\{-\psi(F(x))\}, \ \forall \psi \in L^2([0,1]) \quad (S.128)$$

with

$$F^{-1}(t) = \theta_\psi^{-1} \int_0^t e^{\psi(s)} ds \text{ and } \theta_\psi = \int_0^1 e^{\psi(s)} ds$$

# Appendix S.2: The Thresholding Stopping Criterion for DSBE

In the backward elimination (BE) procedure summarized in Algorithm BE in Chiou et al. (2019), we replace the hypothesis testing-based stopping criterion in step (B2) with a thresholding stopping criterion.

Specifically, let $\Lambda_c = \{\hat{k}_1^*, \hat{k}_2^*, \cdots, \hat{k}_{K-1}^*\}$, such that $0 = \hat{k}_0^* < \hat{k}_1^* < \cdots < \hat{k}_{K-1}^* < \hat{k}_K^* = n$, be the collection of change points detected by Algorithm DS in Chiou et al. (2019) at the dynamic segmentation stage, and let $\hat{k}_l^* \in S_c$ be the most unlikely changepoint candidate that is selected in step (B1) of Algorithm BE. We then use the functional binary segmentation (FBS) method in Rice and Zhang (2022) to perform a change-point detection on the functional sub-sequence $\Gamma_{\hat{k}_l^*}^{\psi} = \{\psi_{\hat{k}_{l-1}^*+1}, \cdots, \psi_{\hat{k}_l^*}, \cdots, \psi_{\hat{k}_{l+1}^*}\}$ ( $\psi_j$ represents the LQD-transformed result of the PDF $f_j$). If the FBS detects no change point from the sub-sequence, we then remove $\hat{k}_l^*$ from $\Lambda_c$; otherwise, we stop the backward elimination procedure, and output the remaining change points. In this FBS-based strategy, the decision on the existence of additional change point in the sub-sequence $\Gamma_{\hat{k}_l^*}^{\psi}$ is made based on a thresholding rule detailed in Rice and Zhang (2022). Hence, this stopping criterion for the BE procedure is referred to as the thresholding stopping criterion. In our simulations, we use the default choice of the threshold recommended in Rice and Zhang (2022). In our experience, compared with the hypothesis testing-based stopping criterion used in Chiou et al. (2019), this thresholding stopping criterion has a lower risk of overestimating the number of change points.